\pdfoutput=1

\documentclass[11pt,twoside,a4paper,cmspaper,final,collab]{cms-tdr}
\def\svnVersion{270491}\def\cmsCernNo{2013-037}\def\cmsCernDate{\today}\def\cmsMessage{Published in Physical Review D as \href{http://dx.doi.org/10.1103/PhysRevD.90.112001}{\doi{10.1103/PhysRevD.90.112001.}}}

\begin{document}\cmsNoteHeader{SUS-12-005}

\hyphenation{had-ron-i-za-tion}
\hyphenation{cal-or-i-me-ter}
\hyphenation{de-vices}

\RCS$Revision: 262505 $
\RCS$HeadURL: svn+ssh://svn.cern.ch/reps/tdr2/papers/SUS-12-005/trunk/SUS-12-005.tex $
\RCS$Id: SUS-12-005.tex 262505 2014-09-30 15:51:22Z alverson $
\newlength\cmsFigWidth
\ifthenelse{\boolean{cms@external}}{\setlength\cmsFigWidth{0.85\columnwidth}}{\setlength\cmsFigWidth{0.4\textwidth}}
\newlength\cmsFigWidthBox
\ifthenelse{\boolean{cms@external}}{\setlength\cmsFigWidthBox{0.95\columnwidth}}{\setlength\cmsFigWidthBox{0.7\textwidth}}
\ifthenelse{\boolean{cms@external}}{\providecommand{\cmsLeft}{top}}{\providecommand{\cmsLeft}{left}}
\ifthenelse{\boolean{cms@external}}{\providecommand{\cmsRight}{bottom}}{\providecommand{\cmsRight}{right}}
\ifthenelse{\boolean{cms@external}}{\providecommand{\CL}{C.L.\xspace}}{\providecommand{\CL}{CL\xspace}}
\newcommand{\CLs}{\ensuremath{\mathrm{CL}_\mathrm{s}}\xspace}
\newcommand{\CLb}{\ensuremath{\mathrm{CL}_\mathrm{b}}\xspace}
\newcommand{\CLsb}{\ensuremath{\mathrm{CL}_\mathrm{s+b}}\xspace}
\newcommand{\M}{\ensuremath{M}\xspace}
\newcommand{\MR}{\ensuremath{M_R}\xspace}
\newcommand{\MTR}{\ensuremath{M_\mathrm{T}^R}\xspace}
\newcommand{\MDelta}{\ensuremath{M_\Delta}\xspace}
\newcommand{\R}{\ensuremath{R}\xspace}
\newcommand{\Rtwo}{\ensuremath{R^2}\xspace}
\newcommand{\mzero}{\ensuremath{m_0}\xspace}
\newcommand{\mhalf}{\ensuremath{m_{1/2}}\xspace}
\renewcommand{\S}{\ensuremath{S}\xspace} %shadows the section symbol
\providecommand{\logN}{\mathop{\mathrm{log}\mathcal{N}}\nolimits}
\newcolumntype{x}{D{,}{{}\pm{}}{-1}}
\cmsNoteHeader{SUS-12-005} % This is over-written in the CMS environment: useful as preprint no. for export versions
\title{Search for supersymmetry with razor variables in \texorpdfstring{$\Pp\Pp$
  collisions at $\sqrt{s} = 7$\TeV}{pp collisions at sqrt(s)=7 TeV}}

\date{\today} \abstract{The razor approach to search for R-parity
  conserving supersymmetric particles is described in detail. Two
  analyses are considered: an inclusive search for new heavy particle
  pairs decaying to final states with at least two jets and missing
  transverse energy, and a dedicated search for final states with at
  least one jet originating from a bottom quark. For both the
  inclusive study and the study requiring a bottom-quark jet, the data
  are examined in exclusive final states corresponding to
  all-hadronic, single-lepton, and dilepton events. The study is based
  on the data set of proton-proton collisions at $\sqrt{s}=7$\TeV
  collected with the CMS detector at the LHC in 2011, corresponding to
  an integrated luminosity of 4.7\fbinv.  The study consists of a shape analysis
  performed in the plane of two kinematic variables, denoted $\MR$ and
  $\Rtwo$, that correspond to the mass and transverse energy flow,
  respectively, of pair-produced, heavy, new-physics particles.  The
  data are found to be compatible with the background model, defined
  by studying event simulations and data control samples.  Exclusion
  limits for squark and gluino production are derived in the context
  of the constrained minimal supersymmetric standard model (CMSSM) and
  also for simplified-model spectra (SMS). Within the CMSSM parameter
  space considered, squark and gluino masses up to 1350\GeV are
  excluded at 95\% confidence level, depending on the model
  parameters.  For SMS scenarios, the direct production of pairs of
  top or bottom squarks is excluded for masses as high as 400\GeV.}

\hypersetup{%
pdfauthor={CMS Collaboration},%
pdftitle={Search for supersymmetry with razor variables in pp collisions at sqrt(s)=7 TeV},%
pdfsubject={CMS},%
pdfkeywords={CMS, LHC, SUSY, Razor, top+X, W+jets, Z+jets, QCD, scaling}
}

\maketitle

\section{Introduction\label{sec:prd-intro}}

Extensions of the standard model (SM) with softly broken supersymmetry
(SUSY)~\cite{Ramond,Golfand,Volkov,Wess,Fayet} predict new fundamental
particles that are superpartners of the SM particles.  Under the
assumption of $R$-parity~\cite{bib-rparity} conservation, searches for
SUSY particles at the fermilab Tevatron~\cite{:2007ww, Aaltonen:2008rv} and the
CERN LHC~\cite{Collaboration:2011xk,
  daCosta:2011qk, Aad:2011hh, Aad:2011xm, RA2, alphaT, :2011wb,
  Chatrchyan:2011bz, Chatrchyan:2012qka,
  Chatrchyan:2012mea,:2012th,:2012jx,
  ATLAS:2012ai,ATLAS:2012ag,Aad:2012rt,:2012cwa,Aad:2012cz} have
focused on event signatures with energetic hadronic jets and leptons
from the decays of pair-produced squarks $\PSq$ and gluinos
$\PSg$. Such events frequently have large missing transverse energy
(\ETm) resulting from the stable weakly interacting superpartners, one
of which is produced in each of the two decay chains.

In this paper, we present the detailed methodology of an inclusive
search for SUSY based on the razor kinematic
variables~\cite{razor2010,rogan}. A summary of the results of this
search, based on 4.7\fbinv of $\Pp\Pp$ collision data at $\sqrt{s} =
7$\TeV collected with the CMS detector at the LHC, can be found in
Ref.~\cite{PhysRevLett.111.081802}.  The search is sensitive to the
production of pairs of heavy particles, provided that the decays of
these particles produce significant {\ETm}. The jets in each event are
cast into two disjoint sets, referred to as ``megajets''.

The razor variables $\MR$ and $\Rtwo$, defined in Section~\ref{sec:prd-razor}, are calculated from the four-momenta of these megajets event-by-event, and the search is performed by determining the expected distributions of SM processes in the two-dimensional ($\MR$, $\Rtwo$) razor plane.  A critical feature of the razor variables is that they are computed in the approximate center-of-mass frame of the produced superpartner candidates.

The megajets represent the visible part of the decay chain of pair-produced superpartners, each of which decays to one or more visible SM particles and one stable, weakly interacting lightest SUSY particle (LSP), here taken to be the lightest neutralino $\chiz_1$. In this framework the reconstructed products of the decay chain of each originally produced superpartner are collected into one megajet. Every topology can then be described kinematically by the simplest example of squark-antisquark production with the direct two-body squark decay $\PSq \to \Pq \chiz_1$, denoted a ``dijet plus \ETm{}" final state, to which the razor variables strictly apply.

The strategy and execution of the search is summarized as follows:
\begin{enumerate}
\item Events with two reconstructed jets at the hardware-based first level trigger (L1) are processed by a dedicated set of algorithms in the high-level trigger (HLT).
From the jets and leptons reconstructed at the HLT level,
 the razor variables $\MR$ and $\Rtwo$ are calculated and their values are used to determine whether to retain the event for further offline processing. A looser kinematic requirement is applied for events with electrons or muons, due to the smaller rate of SM background for these processes. The correspondence between the HLT and offline reconstruction procedures allows
events of interest to be selected more efficiently than is possible with an inclusive multipurpose trigger.
\item In the offline environment, leptons and jets are reconstructed, and a tagging algorithm is applied to identify those jets likely to have originated from a bottom-quark jet ($\cPqb$ jet).
\item The reconstructed objects in each event are combined into two megajets, which are used to calculate the variables $\MR$ and $\Rtwo$.
Several baseline kinematic requirements are applied to reduce the number of misreconstructed events and to ensure that only regions of the razor plane where the trigger is efficient are selected.
\item Events are assigned to final state ``boxes'' based on the presence or absence of a reconstructed lepton.  This box partitioning scheme allows us to isolate individual SM background processes based on the final-state particle content and kinematic phase space; we are able to measure the yield and the distribution of events in the ($\MR$, $\Rtwo$) razor plane for different SM backgrounds.  Events with at least one tagged $\cPqb$ jet are considered in a parallel analysis
focusing on  a search for the superpartners of third-generation quarks. In total, we consider 12 mutually exclusive final-state boxes: dielectron events (ELE-ELE), electron-muon events (ELE-MU), dimuon events (MU-MU), single-electron events (ELE), single-muon events (MU), and events with no identified electron or muon (HAD), each inclusive or with a $\cPqb$-tagged jet.
\item For each box we use the low ($\MR$, $\Rtwo$) region of the razor plane, where negligible signal contributions are expected,
to determine the shape and normalization of the various background components. An analytic model constructed from these results is used to predict the SM background over the entire razor plane.
\item The data are compared with the prediction for the background in the sensitive regions of the razor plane and the results are used to constrain the parameter space of SUSY models.
\end{enumerate}

This paper is structured as follows. The definition of the razor variables is given in Section~\ref{sec:prd-razor}. The trigger and offline event selection are discussed in Section~\ref{sec:datana}. The features of the signal and background kinematic distributions are described in Section~\ref{sec:mc-ana}. In Section~\ref{sec:BKG2011} we describe the sources of SM background, and in Section~\ref{sec:fits} the analytic model used to characterize this background in the signal regions.  Systematic uncertainties are discussed in Section~\ref{sec:systematics}. The interpretation of the results is presented in Section~\ref{sec:INTERP2011} in terms of exclusion limits on squark and gluino production in the context both of the constrained minimal SUSY model (CMSSM)~\cite{Chamseddine:1982jx,Barbieri:1982eh,Hall:1983iz} and for some simplified model spectra (SMS)~\cite{ArkaniHamed:2007fw,Alwall-2,Alwall-1,Alwall:2008va,Alves:2011wf}.  Section~\ref{sec:summary} contains a summary. For the CMSSM, exclusion limits are provided as a function of the universal scalar and fermion mass values at the unification scale, respectively denoted $\mzero$ and $\mhalf$.  For the SMS, limits are provided in terms of the masses of the produced SUSY partner and the LSP.

\section{The razor approach\label{sec:prd-razor}}

The razor kinematic variables are designed to be sensitive to
processes involving the pair-production of two heavy particles, each
decaying to an unseen particle plus jets. Such processes include SUSY
particle production with various decay chains, the simplest example of
which is the pair production of squarks, where each squark decays to a
quark and the LSP, with the LSP assumed to be stable and weakly
interacting. In processes with two or more undetected energetic
final-state particles, it is not possible to fully reconstruct the
event kinematics. Event-by-event, one cannot make precise assignments
of the reconstructed final-state particles (leptons, jets, and
undetected neutrinos and LSPs) to each of the original superpartners
produced.  For a given event, there is not enough information to
determine the mass of the parent particles, the subprocess
center-of-mass energy $\sqrt{\hat{s}}$, the center-of-mass frame of
the colliding protons, or the rest frame of the decay of either parent
particle. As a result, it is challenging to distinguish between SUSY
signal events and SM background events with energetic neutrinos, even
though the latter involve different topologies and mass scales. It is
also challenging to identify events with instrumental sources of \ETm
that can mimic the signal topology.

The razor approach \cite{razor2010,rogan} addresses these challenges
through a novel treatment of the event kinematics. The key points of
this approach are listed below.
\begin{itemize}
\item The visible particles (leptons and jets) are used to define two
  \textit{megajets}, each representing the visible part of a parent
  particle decay. The megajet reconstruction ignores details of the
  decay chains in favor of obtaining the best correspondence between a
  signal event candidate and the presumption of a pair-produced heavy
  particle that undergoes two-body decay.
\item Lorentz-boosted reference frames are defined in terms of the
  megajets. These frames approximate, event-by-event, the
  center-of-mass frame of the signal subprocess and the rest frames of
  the decays of the parent particles.The kinematic quantities in these
  frames can be used to extract the relevant SUSY mass scales.
\item The razor variables, $\MR$, $\MTR$,
  and $\R \equiv \MTR/\MR$, are
  computed from the megajet four-momenta and the {\ETm} in the
  event. The $\MR$ variable is an estimate of an overall mass
  scale, which in the limit of massless decay products equals the mass
  of the heavy parent particle. It contains both longitudinal and
  transverse information, and its distribution peaks at the true value
  of the new-physics mass scale.  The razor variable
  $\MTR$ is defined entirely from transverse
  information: the transverse momenta ($\pt$) of the megajets and the
  \ETm. This variable has a kinematic endpoint at the same underlying
  mass scale as the $\MR$ mean value. The ratio $\R$
  quantifies the flow of energy in the plane perpendicular to the beam
  and the partitioning of momentum between visible and invisible
  particles.
\item The shapes of the distributions in the ($\MR$,
  $\Rtwo$) plane are described for the SM processes. Razor
  variable distributions exhibit peaks for most SM backgrounds, as a
  result of turn-on effects from trigger and selection thresholds as
  well as of the relevant heavy mass scales for SM processes, namely
  the top quark mass and the $\PW$ and $\cPZ$ boson masses. However,
  compared with signals involving heavier particles and new-physics
  sources of {\ETm}, the SM distributions peak at smaller values of
  the razor variables. For values of the razor variables above the
  peaks, the SM background distributions (and also the signal
  distributions) exhibit exponentially falling behavior in the
  ($\MR$, $\Rtwo$) plane.
  Hence, the asymptotic behavior of the razor variables is determined
  by a combination of the parton luminosities and the intrinsic
  sources of {\ETm}. The multijet background from processes described
  by quantum chromodynamics (QCD), which contains the smallest level
  of intrinsic {\ETm} amongst the major sources of SM background, has
  the steepest exponential fall-off. Backgrounds with energetic
  neutrinos from $\PW/\cPZ$ boson and top-quark production exhibit a
  slower fall-off and resemble each other closely in the asymptotic
  regime. Thus, razor signals are characterized by peaks in the
  ($\MR$, $\Rtwo$) plane on top of exponentially
  falling SM background distributions. Any SUSY search based on razor
  variables is then more similar to a ``bump-hunt'', e.g., a search
  for heavy resonances decaying to two jets~\cite{Harris:2011bh}, than
  to a traditional SUSY search. This justifies the use of a shape
  analysis, based on an analytic fit of the background, as described
  in Section~\ref{sec:fits}.
\end{itemize}

\subsection{Razor megajet reconstruction}

The razor megajets are defined by dividing the reconstructed jets of
each event into two partitions. Each partition contains at least one
jet. The megajet four-momenta are defined as the sum of the
four-momenta of the assigned jets. Of all the possible combinations,
the one that minimizes the sum of the squared-invariant-mass values of
the two megajets is selected. In simulated event samples, this megajet
algorithm is found to be stable against variations in the jet
definition and it provides an unbiased description of the visible part
of the two decay chains in SUSY signal events. The inclusive nature of
the megajets allows an estimate of the SM background in the razor
plane.

Reconstructed leptons in the final state can be included as visible
objects in the reconstruction of the megajets, or they can be treated
as invisible, i.e., as though they are escaping weakly interacting
particles~\cite{razor2010}.  For SM background processes such as
$\PW(\ell\cPgn)$+jets, the former choice yields more transversely
balanced megajets and lower values of $\R$.  If the leptons
are treated as invisible in these processes, the {\ETm} corresponds to
the entire $\PW$ boson $\pt$ value, similar to the case of $\cPZ (\cPgn
\cPagn)$+jets events.

\subsection{Razor variables}

To the extent that the reconstructed pair of megajets accurately
reflects the visible portion of the underlying parent particle decays,
the kinematics of the event are equivalent to that of the pair
production of heavy squarks $\PSq_1$, $\PSq_2$, with
$\PSq_i\to \cPq_i \PSGczDo$, where $\PSGczDo$ denotes the
LSP and $\cPq_i$ denotes the visible products of the decays as
represented by the megajets.

The razor analysis approximates the unknown center-of-mass and parent
particle rest frames with a razor frame defined unambiguously from
measured quantities in the laboratory frame.  Two observables
$\MR$ and $\MTR$ estimate the heavy mass scale
$\MDelta$.  Consider the two visible four-momenta written in
the rest frame of the respective parent particles:
\begin{align*}
p_{\cPq_1} &= \left(\frac{m_{\PSq}^2+m_{\cPq_1}^2-m_{\chiz_1}^2}{2m_{\PSq}},\; \frac{\MDelta}{2}\,\hat{u}_{\cPq_1}\right),\\
p_{\cPq_2} &= \left(\frac{m_{\PSq}^2+m_{\cPq_2}^2-m_{\chiz_1}^2}{2m_{\PSq}},\; \frac{\MDelta}{2}\,\hat{u}_{\cPq_2}\right).
\end{align*}
where $\hat{u}_{\cPq_i}$ (${i} = 1,2$) is a unit three-vector
and $m_\cPq{}_i$ represents the mass corresponding to the megajet,
e.g., the top-quark mass for $\PSQt \to \cPqt \PSGczDo$.
Here we have parameterized the magnitude of the three-momenta by the
mass scale $\MDelta$, where
\begin{equation}
\MDelta^2 \equiv \frac{\left[m_{\PSq}^2-(m_\cPq+m_{\chiz_1})^2\right]\left[m_{\PSq}^2-(m_\cPq-m_{\chiz_1} )^2\right]}{m_{\PSq}^2}.
\end{equation}
In the limit of massless megajets we then have $\MDelta = (m_{\PSq}^2
- m_{\chiz_1}^2)/m_{\PSq}$ and the four-momenta reduce to
\begin{align*}
p_{\cPq_1} &= \frac{\MDelta}{2}(1,\; \hat{u}_{\cPq_1}),\\
p_{\cPq_2} &= \frac{\MDelta}{2}(1,\; \hat{u}_{\cPq_2}).
\end{align*}

The razor variable $\MR$ is defined in terms of the momenta
of the two megajets by
 \begin{equation}
 \label{eq:MRstar}
 \MR \equiv
 \sqrt{(\abs{\vec{p}_{\cPq_{1}}}+\abs{\vec{p}_{\cPq_{2}}})^2 -(p^{\cPq_1}_z+p^{\cPq_2}_z)^2}.
 \end{equation}
 where $\vec{p}_{\cPq_i}$ is the momentum of megajet $\cPq_i$
 (${i} = 1,2$) and $p^{\cPq_i}_z$ is its component along the
 beam direction.

 For massless megajets, $\MR$ is invariant under a
 longitudinal boost.  It is always possible to perform a longitudinal
 boost to a razor frame where $p^{\cPq_1}_z+p^{\cPq_2}_z$
 vanishes, and $\MR$ becomes just the scalar sum of the
 megajet three-momenta added in quadrature.  For heavy particle
 production near threshold, the three-momenta in this razor frame
 do not differ significantly from the three-momenta in the actual
 parent particle rest frames. Thus, for SUSY signal events,
 $\MR$ is an estimator of $\MDelta$, and for
 simulated samples we find that the distribution of $\MR$
 indeed peaks around the true value of $\MDelta$.  This
 definition of $\MR$ is improved with respect to the one used
 in Ref.~\cite{razor2010}, to avoid configurations where the razor
 frame is unphysical.

The razor  observable  $\MTR$ is defined as
\begin{equation}
\MTR\equiv \sqrt{ \frac{\ETmiss(\pt^{\cPq_1}+\pt^{\cPq_2}) -
    \VEtmiss {\cdot}
    (\ptvec^{\,\cPq_1}+\ptvec^{\,\cPq_2})}{2}},
\end{equation}
where $\ptvec^{\,\cPq_i}$ is the transverse momentum of
megajet ${\cPq_i}$ ($i = 1,2$) and $\pt^{\cPq_i}$ is the
corresponding magnitude; similarly, $\VEtmiss$ is
the missing transverse momentum in the event and \ETm its
magnitude.

Given a global estimator $\MR$ and a transverse estimator
$\MTR$, the razor dimensionless ratio is defined as
\begin{equation}
\R \equiv \frac{\MTR}{\MR}.
\end{equation}

For signal events, $\MTR$ has a maximum value (a
kinematic endpoint) at $\MDelta$, so $\R$ has a maximum value of
approximately one.
Thus, together with the shape of $\MR$ peaking at
$\MDelta$, this behavior is in stark contrast with, for
example, QCD multijet background events, whose distributions in both
$\MR$ and $\Rtwo$ fall exponentially. These properties
allow us to identify a region of the two-dimensional (2D) razor space
where the contributions of the SM background are reduced while those of
signal events are enhanced.

\begin{figure*}[ht!]
  \centering
\includegraphics[width=0.48\textwidth]{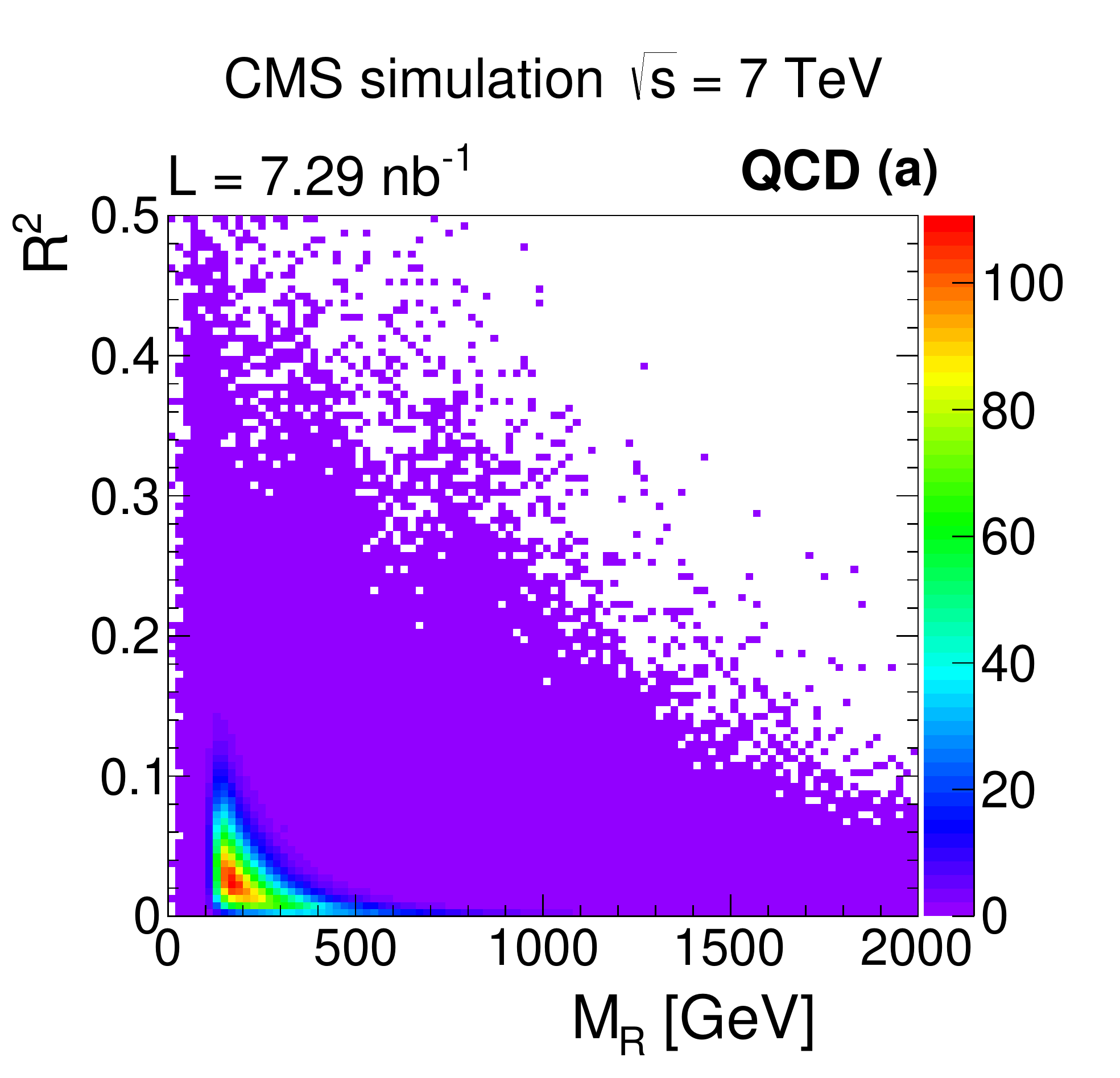}
\includegraphics[width=0.48\textwidth]{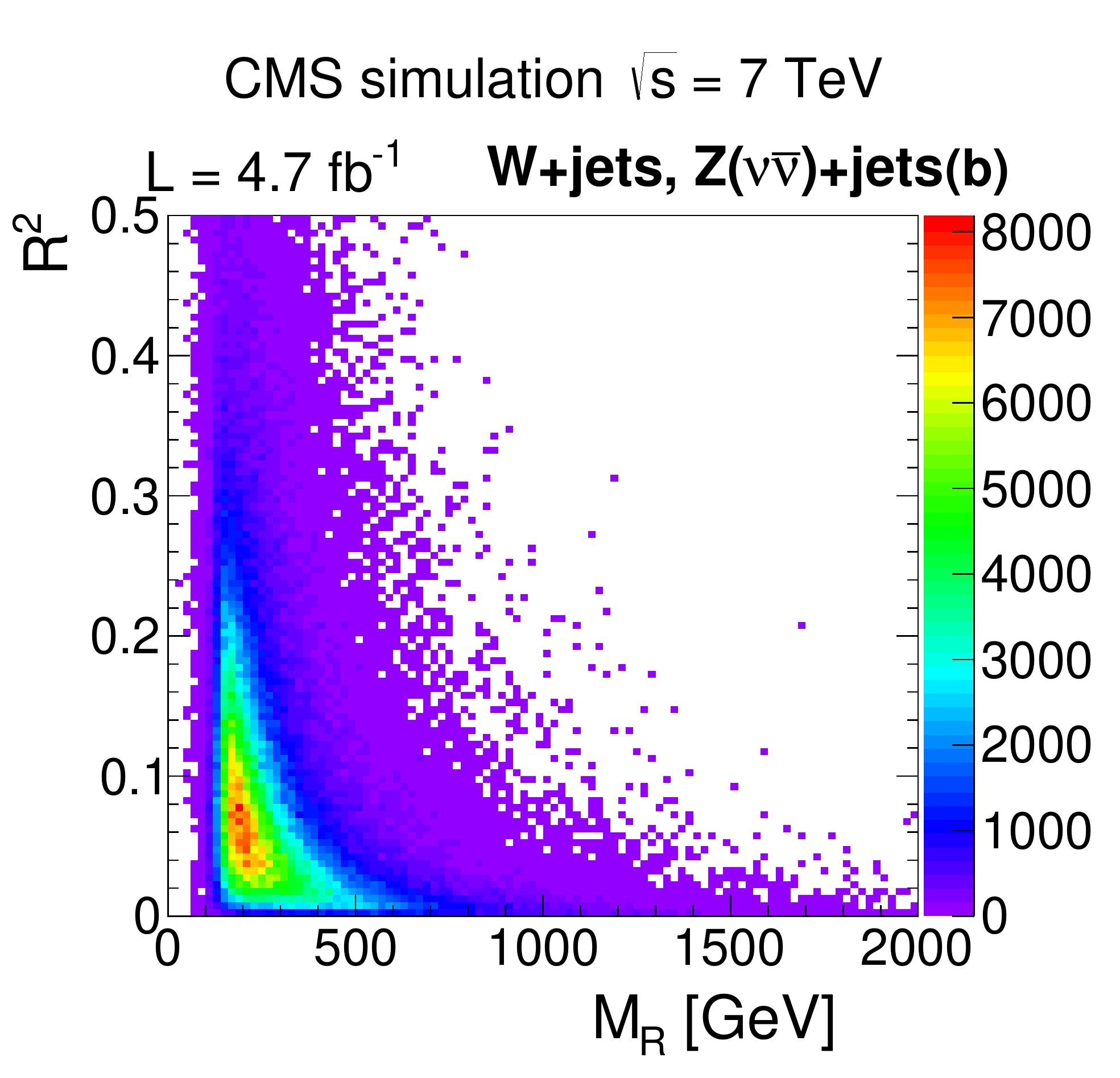}
\includegraphics[width=0.48\textwidth]{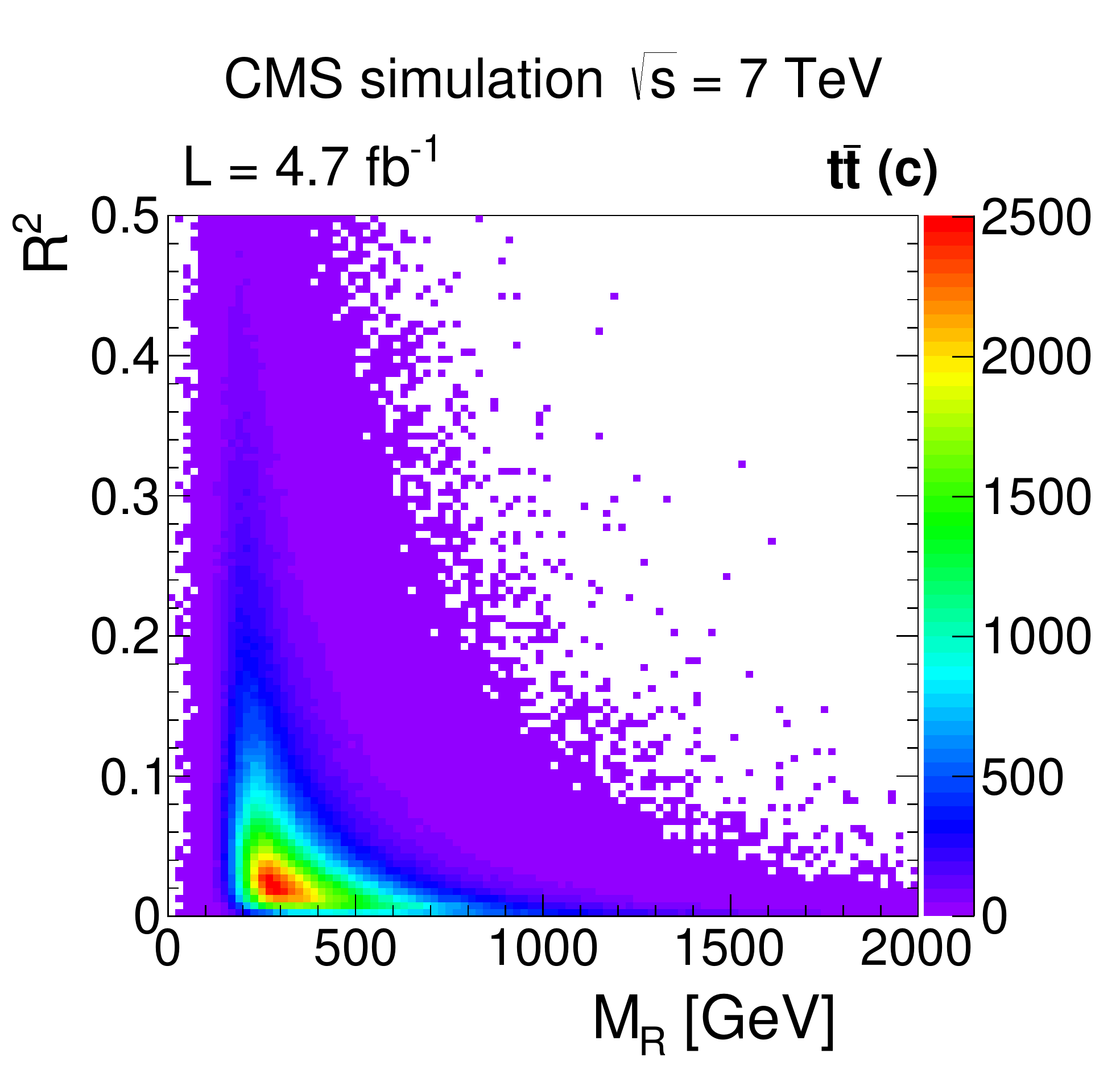}
\includegraphics[width=0.48\textwidth]{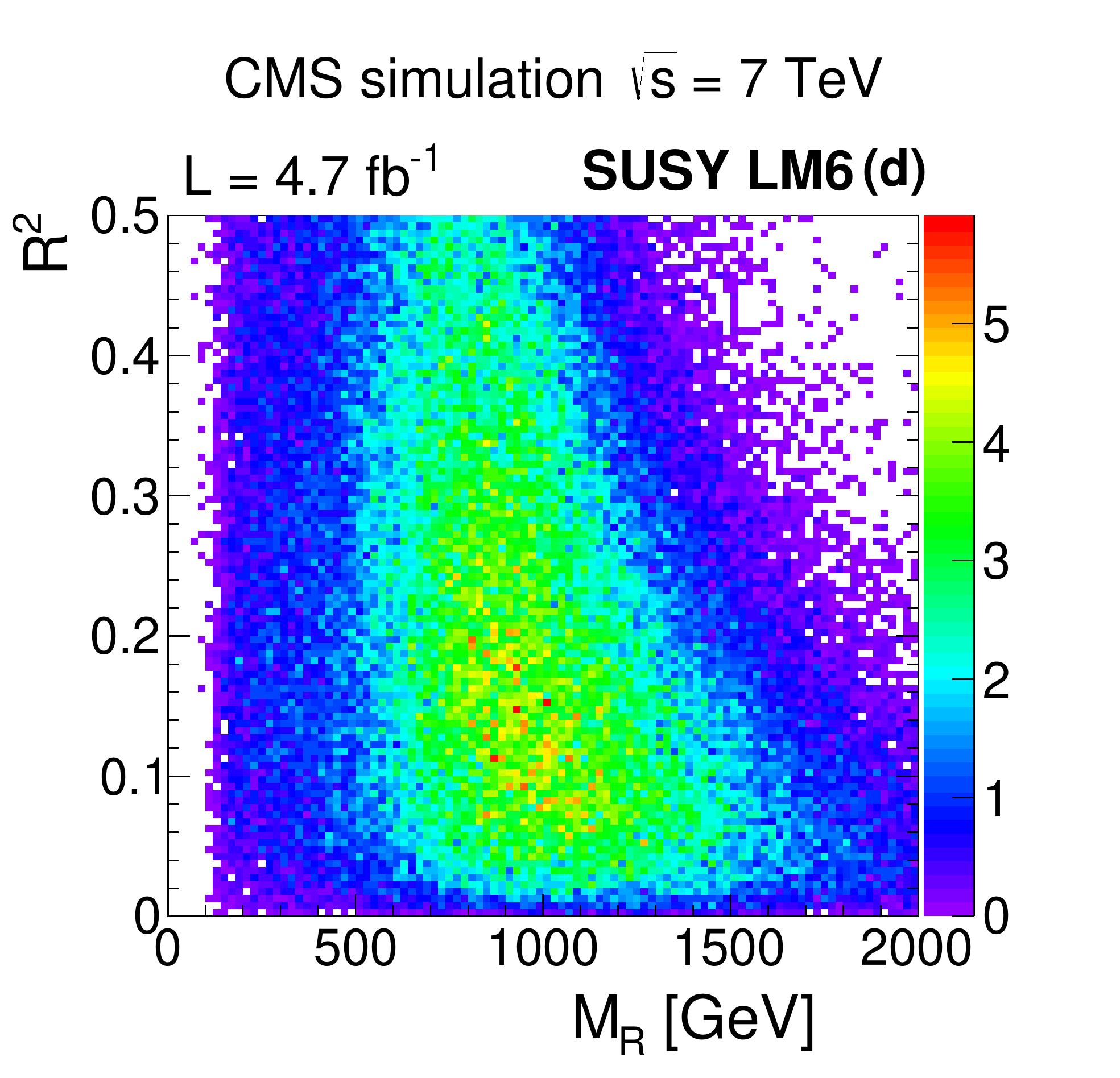}
 \caption{Razor variables $\Rtwo$ versus $\MR$ for
   simulated events: (a) QCD multijet, (b) $\PW(\ell \nu)$+jets and $\cPZ(\cPgn
   \cPagn)$+jets, (c) $\cPqt\cPaqt$, and (d) SUSY benchmark model
   LM6~\cite{PTDR2}, where the new-physics mass scale for LM6
   is $\MDelta = 831$\GeV. The yields are normalized to an
   integrated luminosity of $\sim$4.7 \fbinv except for the QCD
   multijet sample, where we use the luminosity of the generated
   sample. The bin size is 0.005 for $\Rtwo$ and 20\GeV for
   $\MR$.\label{fig:raz-plane}}

\end{figure*}

\subsection{SUSY and SM in the razor plane}

The expected distributions of the main SM backgrounds in the razor
plane, based on simulation, are shown in Fig.~\ref{fig:raz-plane},
along with the results from the CMSSM low-mass benchmark model LM6~\cite{PTDR2},
for which $\MDelta = 831$\GeV. The peaking behavior of the
signal events at $\MR\approx \MDelta$, and the
exponential fall-off of the SM distributions with increasing
$\MR$ and $\Rtwo$, are to be noted.  For both signal
and background processes, events with small values of $\MR$
are suppressed because of a requirement that there be at least two
jets above a certain threshold in \pt (Section~\ref{sec:BOX2011}).

\begin{figure*}[ht!]
  \centering
\includegraphics[width=0.48\textwidth]{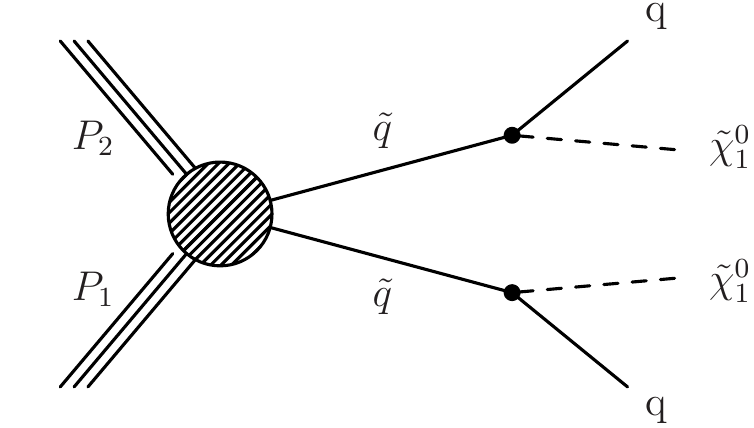}
\includegraphics[width=0.48\textwidth]{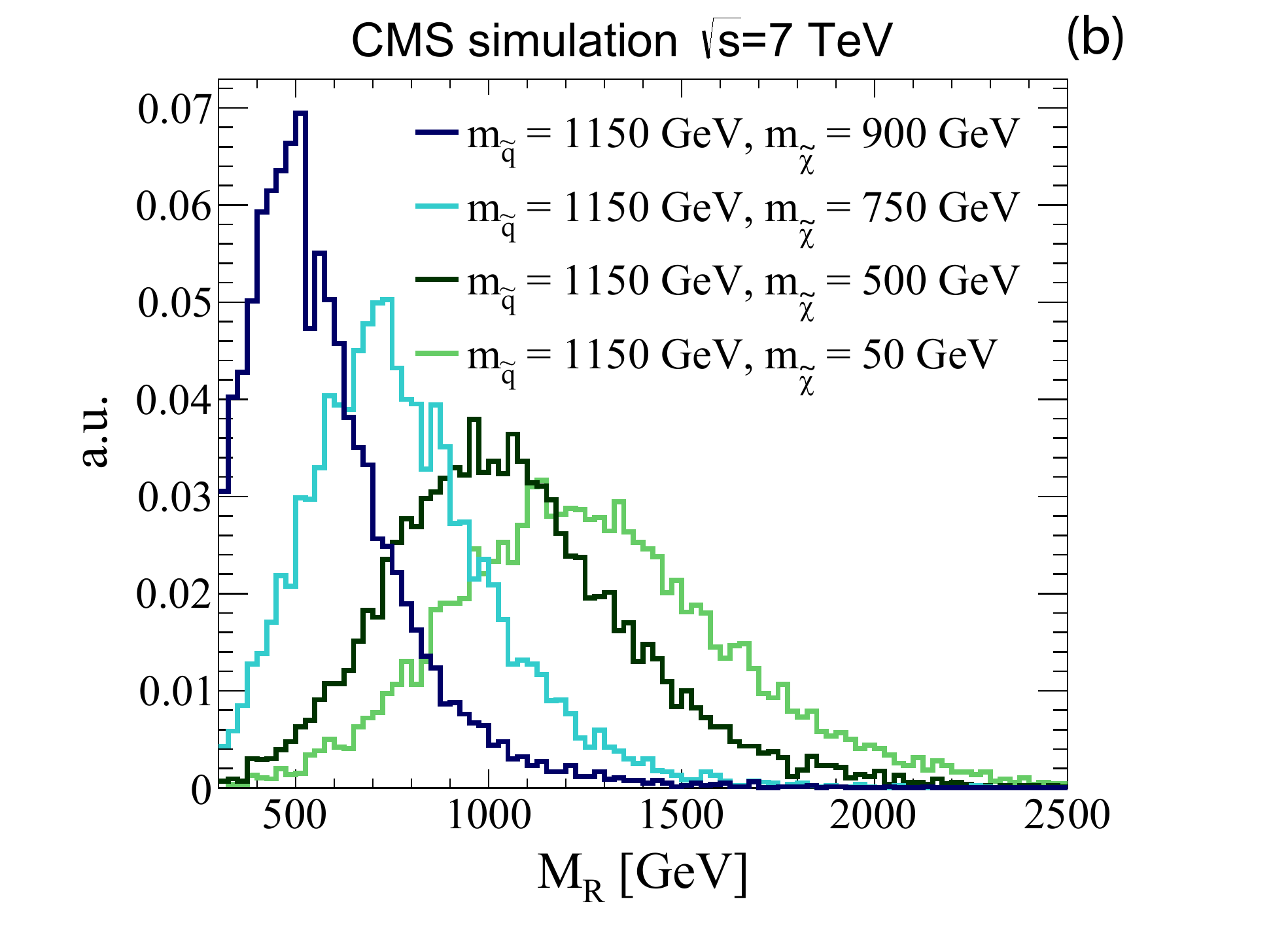}
\includegraphics[width=0.48\textwidth]{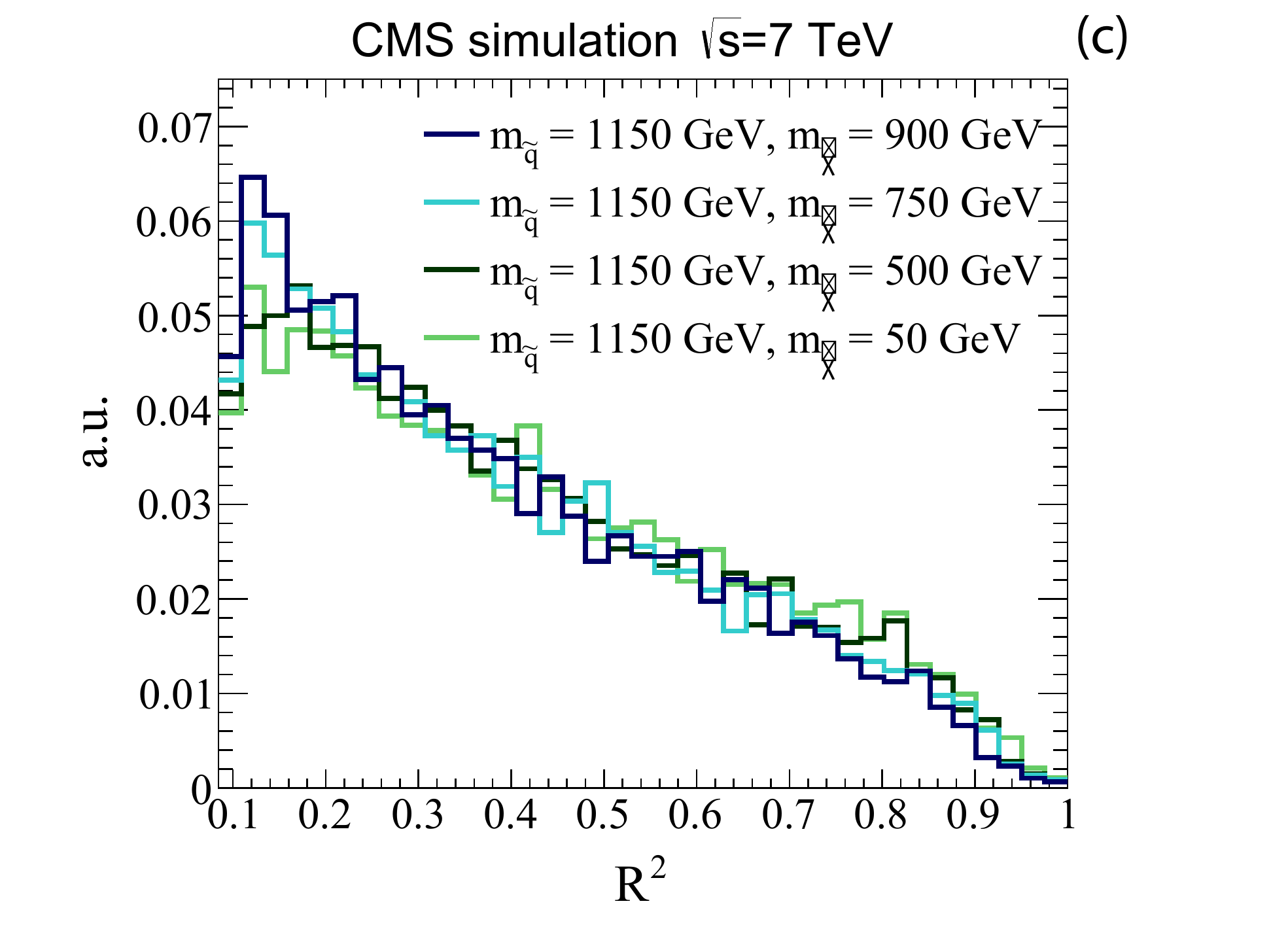}
\includegraphics[width=0.48\textwidth]{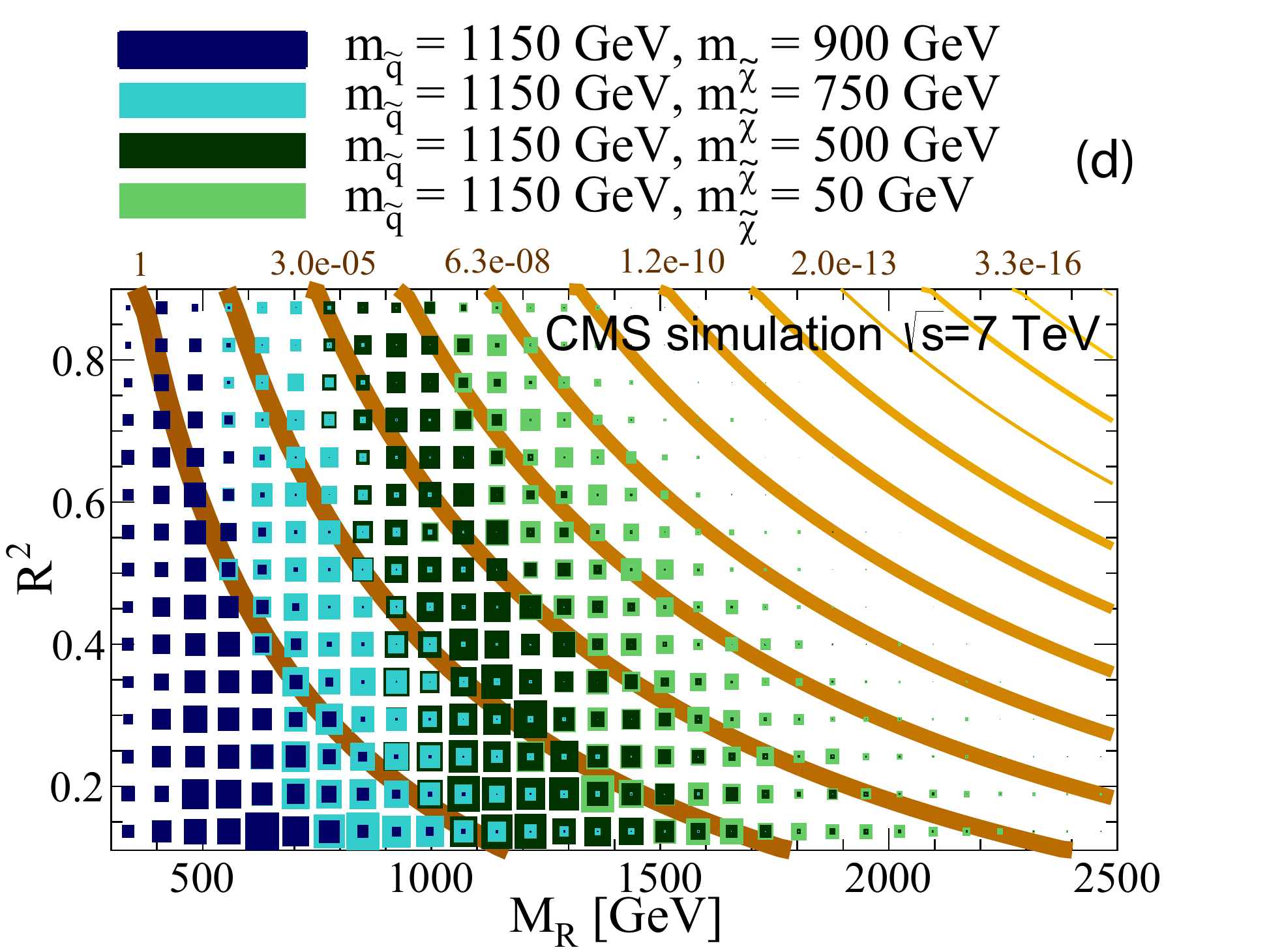}
\caption{(a) The squark-antisquark production diagram for the T2 SUSY
  SMS reference model. The distribution of (b) $\MR$ and (c) $\Rtwo$
  for different LSP masses $m_{\tilde \chi}$ in the T2 scenario. (d)
  Distribution of T2 events in the ($\MR$, $\Rtwo$) plane for the
  different LSP masses $m_{\tilde \chi}$. The orange bands represent
  contours of constant SM background.  The relative suppression
  factors corresponding to some of the bands are indicated in the
  upper part of the figure.\label{fig:T2-tutorial}}
\end{figure*}

In the context of SMS, we refer to the pair production of squark pairs
$\PSq$, $\PSq^*$, followed by $\PSq \to \Pq~\chiz_1$, as ``T2"
scenarios~\cite{SUS-11-016}, where the $\tilde\cPq^*$ state is the
charge conjugate of the $\tilde\cPq$
state. Figure~\ref{fig:T2-tutorial}~(a) shows a diagram for
heavy-squark pair production.  The distributions of $\MR$ and
$\Rtwo$ for different LSP masses are shown in
Figs.~\ref{fig:T2-tutorial} (b) and (c). Figure~\ref{fig:T2-tutorial}
(d) shows the distribution of signal events in the razor plane. The
colored bands running from top left to bottom right show the
approximate SM background constant-yield contours. The associated
numbers indicate the SM yield suppression relative to the reference
line marked ``1''. Based on these kinematic properties, a 2D
analytical description of the SM processes in the ($\MR$,
$\Rtwo$) plane is developed.

\section{Data taking and event selection\label{sec:datana}}

\subsection{The CMS apparatus\label{sec:cmsdet}}

A hallmark of the CMS detector~\cite{CMS:2008zzk} is its superconducting
solenoid magnet, of 6~m internal diameter, providing a field of
3.8~T. The silicon pixel and strip tracker, the crystal
electromagnetic calorimeter (ECAL), and the brass/scintillator hadron
calorimeter (HCAL) are contained within the solenoid. Muons are
detected in gas-ionization detectors embedded in the steel flux-return
yoke, based on three different technologies: drift tubes, resistive
plate chambers, and cathode strip chambers (CSCs).  The ECAL has an
energy resolution better than 0.5\% above 100\GeV. The combination of
the HCAL and ECAL provides jet energy measurements with a resolution
$\Delta E/E \approx 100\% / \sqrt{E/\GeV} \oplus 5\%$.

The CMS experiment uses a coordinate system with the origin located at
the nominal collision point, the $x$ axis pointing towards the center
of the LHC ring, the $y$ axis pointing up (perpendicular to the plane
containing the LHC ring), and the $z$ axis along the counterclockwise
beam direction. The azimuthal angle, $\phi$, is measured with respect
to the $x$ axis in the $(x,y)$ plane, and the polar angle, $\theta$, is
defined with respect to the $z$ axis. The pseudorapidity is $\eta =
-\ln [\tan(\theta / 2)]$.

For the data used in this analysis, the peak luminosity of the LHC
increased from \mbox{1$\times 10^{33}$ cm$^{-2}$ s$^{-1}$} to over
\mbox{4$\times 10^{33}$ cm$^{-2}$ s$^{-1}$}.  For the data collected
between \mbox{(1--2)$\times 10^{33}$ cm$^{-2}$ s$^{-1}$}, the increase
was achieved by increasing the number of bunches colliding in the
machine, keeping the average number of interactions per crossing at
about 7.  For the rest of the data, the increase in the instantaneous
luminosity was achieved by increasing the number and density of the
protons in each bunch, leading to an increase in the average number of
interactions per crossing from around 7 to around 17. The presence of
multiple interactions per crossing was taken into account in the CMS
Monte Carlo (MC) simulation by adding a random number of minimum bias
events to the hard interactions, with the multiplicity distribution
matching that in data.

\subsection{Trigger selection\label{sec:TRIGGER2011}}
The CMS experiment uses a two-stage trigger system, with events
flowing from the L1 trigger at a rate up to 100\unit{kHz}. These events are
then processed by the HLT computer farm. The HLT software selects
events for storage and offline analysis at a rate of a few hundred
Hz. The HLT algorithms consist of sequences of offline-style
reconstruction and filtering modules.

The 2010 CMS razor-based inclusive search for SUSY~\cite{razor2010}
used triggers based on the scalar sum of jet $\pt$, $\HT$,
for hadronic final states and single-lepton triggers for leptonic
final states. Because of the higher peak luminosity of the LHC in
2011, the corresponding triggers for 2011 had higher thresholds. To
preserve the high sensitivity of the razor analysis, CMS designed a
suite of dedicated razor triggers, implemented in the spring of
2011. The total integrated luminosity collected with these triggers
was 4.7\fbinv at $\sqrt{s} = 7$\TeV.

The razor triggers apply thresholds to the values of $\MR$
and $\R$ driven by the allocated bandwidth. The algorithms
used for the calculation of $\MR$ and $\R$ are based
on calorimetric objects. The reconstruction of these objects is fast
enough to satisfy the stringent timing constraints imposed by the HLT.

Three trigger categories are used: hadronic triggers, defined by
applying moderate requirements on $\MR$ and $\R$ for
events with at least two jets with $\pt>$ 56\GeV; electron triggers,
similar to the hadronic triggers, but with looser requirements for
$\MR$ and $\R$ and requiring at least one electron
with $\pt>$10\GeV satisfying loose isolation criteria; and muon
triggers, with similar $\MR$ and $\R$ requirements
and at least one muon with $\abs{\eta}<2.1$ and $\pt>$ 10\GeV. All these
triggers have an efficiency of $(98 \pm 2)\%$ in the kinematic regions
used for the offline selection.

In addition, control samples are defined using several non-razor
triggers.  These include prescaled inclusive hadronic triggers,
hadronic multijet triggers, hadronic triggers based on $\HT$,
and inclusive electron and muon triggers.

\subsection{Physics object reconstruction\label{sec:RECO2011}}

Events are required to have at least one reconstructed interaction
vertex \cite{TRK-10-005}. When multiple vertices are found, the one
with the highest scalar sum of charged track $\pt^2$ is taken to be
the event interaction vertex. Jets are reconstructed offline from
calorimeter energy deposits using the infrared-safe
anti-\kt~\cite{antikt} algorithm with a distance parameter
$\R=0.5$. Jets are corrected for the non-uniformity of the
calorimeter response in energy and $\eta$ using corrections derived
from data and simulations and are required to have $\pt> 40$\GeV and
$\abs{\eta} < 3.0$~\cite{JES}.  To match the trigger requirements, the
$\pt$ of the two leading jets is required to be greater than 60\GeV.
The jet energy scale uncertainty for these corrected jets is 5\%
\cite{JES}.  The $\ETmiss$ is defined as the
negative of the vector sum of the transverse energies (\ET) of all the
particles found by the particle-flow algorithm \cite{PFMET}.

Electrons are reconstructed using a combination of shower shape
information and matching between tracks and electromagnetic
clusters~\cite{CMS_e}.  Muons are reconstructed using information from
the muon detectors and the silicon tracker and are required to be
consistent with the reconstructed primary vertex \cite{CMS_mu}.

The selection criteria for electrons and muons are considered to be
{\it tight} if the electron or muon candidate is isolated, satisfies
the selection requirements of Ref.~\cite{CMS_WZ}, and lies within
$\abs{\eta}<2.5$ and $\abs{\eta}<2.1$, respectively. {\it Loose} electron and
muon candidates satisfy relaxed isolation requirements.

\subsection{Selection of good quality data}

The 4.7\fbinv integrated luminosity used in this analysis is certified
as having a fully functional detector. Events with various sources of
noise in the ECAL or HCAL detectors are rejected using either
topological information, such as unphysical charge sharing between
neighboring channels, or timing and pulse shape information. The last
requirement exploits the difference between the shapes of the pulses
that develop from particle energy deposits in the calorimeters and
from noise events~\cite{METJINST}. Muons produced from proton
collisions upstream of the detector (beam halo) can mimic
proton-proton collisions with large $\ETm$ and are identified using
information obtained from the CSCs. The geometry of the CSCs allows
efficient identification of beam halo muons, since halo muons that
traverse the calorimetry will mostly also traverse one or both CSC
endcaps. Events are rejected if a significant amount of energy is lost
in the masked crystals that constitute approximately 1\% of the ECAL,
using information either from the separate readout of the L1 hardware
trigger or by measuring the energy deposited around the masked
crystals. We select events with a well-reconstructed primary vertex and
with the scalar $\sum \pt$ of tracks associated to it greater than
10\% of the scalar $\sum \pt$ of all jet transverse momenta. These
requirements reject 0.003\%~of an otherwise good inclusive sample of
proton-proton interactions (minimum bias events).

\subsection{Event selection and classification\label{sec:BOX2011}}

Electrons enter the megajet definition as ordinary jets.
Reconstructed muons are not included in the megajet grouping because,
unlike electrons, they are distinguished from jets in the HLT.  This
choice also allows the use of $\PW(\Pgm\cPgn)$+jets events to
constrain and study the shape of $\cPZ(\cPgn\cPagn)$+jets events in
fully hadronic final states.

The megajets are constructed as the sum of the four-momenta of their
constituent objects.  After considering all possible partitions into
two megajets, the combination is selected that has
the smallest sum of megajet squared-invariant-mass values.

The variables $\MR$ and $\Rtwo$ are calculated from
the megajet four-momenta. The events are assigned to one of the six
final state {\it boxes} according to whether the event has zero, one,
or two isolated leptons, and according to the lepton flavor (electrons
and muons), as shown in Table~\ref{tab:boxes}.  The lepton $\pt$,
$\MR$, and $\Rtwo$ thresholds for each of the boxes
are chosen so that the trigger efficiencies are independent of
$\MR$ and $\Rtwo$.

\begin{table}[ht!]
  \topcaption{Definition of the {\it full analysis regions} for the mutually exclusive boxes, based on the
    $\MR $ and $\Rtwo$ values, and, for the categories  with leptons, on their \pt
    value, listed according to the hierarchy followed in the
    analysis, the ELE-MU (HAD) being the first (last).\label{tab:boxes}}
\centering
\begin{scotch}{l c}
\multicolumn{2}{l}{Lepton boxes $\MR > 300$\GeV, $0.11< \Rtwo < 0.5$ }\\
\hline
 ELE-MU (loose-tight)  & $\pt > 20$\GeV, $\pt > 15$\GeV\\
MU-MU (loose-loose) & $\pt > 15$\GeV, $\pt > 10$\GeV\\
ELE-ELE (loose-tight) & $\pt > 20$\GeV, $\pt > 10$\GeV\\
MU (tight) & $\pt > 12$\GeV\\
ELE (loose) & $\pt > 20$\GeV \\
\hline
  \multicolumn{2}{l}{HAD box $\MR > 400$\GeV, $0.18< \Rtwo < 0.5$ }\\
\end{scotch}
\end{table}

The requirements given in Table~\ref{tab:boxes} determine the {\it
  full analysis regions} of the ($\MR$, $\Rtwo$) plane
for each box.  These regions are large enough to allow an accurate
characterization of the background, while maintaining efficient
triggers.  To prevent ambiguities when an event satisfies the
selection requirements for more than one box, the boxes are arranged
in a predefined hierarchy. Each event is uniquely assigned to the
first box whose criteria the event satisfies. Table~\ref{tab:boxes}
shows the box-filling order followed in the analysis.

Six additional boxes are formed with the requirement that at least one
of the jets with $\pt> 40$\GeV and $\abs{\eta} < 3.0$ be tagged as a b
jet, using an algorithm that orders the tracks in a jet by their
impact parameter significance and discriminates using the track with
the second-highest significance~\cite{BTAG}. This algorithm has a
tagging efficiency of about 60\%, evaluated using $\cPqb$ jets
containing muons from semileptonic decays of $\cPqb$ hadrons in data,
and a misidentification rate of about 1\% for jets originating from u,
d, and s quarks or from gluons, and of about 10\% for jets coming from
c quarks~\cite{BTAG}. The combination of these six boxes defines an
inclusive event sample with an enhanced heavy-flavor content.

\section{Signal and standard model background
  modeling\label{sec:mc-ana}}

The razor analysis is guided by studies of MC event samples generated
with the \PYTHIA v6.426 \cite{pythia} (with Z2 tune) and {\MADGRAPH v4.22} \cite{Maltoni:2002qb} programs, using the CTEQ6
parton distribution functions (PDF)~\cite{Pumplin:2002vw}. Events
generated with {\MADGRAPH} are processed with
\PYTHIA~\cite{pythia} to provide parton showering,
hadronization, and the underlying event description.  The matrix
element/parton shower matching is performed using the approach
described in Ref.~\cite{mlm}. Generated events are processed with the
\GEANTfour \cite{G4} based simulation of the CMS detector, and then
reconstructed with the same software used for data.

The simulation of the $\ttbar$, $\PW$+jets, $\cPZ$+jets, single-top
($s$, $t$, and $\cPqt$--$\PW$ channels), and diboson samples is
performed using \MADGRAPH. The events containing top-quark
pairs are generated accompanied by up to three extra partons in the
matrix-element calculation~\cite{Alwall:2007st}. Multijet samples from
QCD processes are produced using \PYTHIA.

To generate SUSY signal MC events in the context of the CMSSM, the
mass spectrum is first calculated with the \textsc{softsusy} program
\cite{softsusy} and the decays with the \textsc{sus-hit} \cite{Susyhit}
package. The \PYTHIA generator is used with the SUSY Les
Houches Accord (SLHA) interface~\cite{SLHA} to generate the
events. The generator-level cross sections and the K-factors for the
next-to-leading-order (NLO) cross sections are computed
using \textsc{prospino}~\cite{prospino}.

We also use SMS MC simulations in the interpretation of the
results. In an SMS simulation, a limited set of hypothetical particles
is introduced to produce a given topological signature.  The
amplitude describing the production and decay of these particles is
parameterized in terms of the particle masses.
Compared with the constrained SUSY models, SMS provide benchmarks that
focus on one final-state topology at a time, with a broader variation
in the masses determining the final-state kinematics. The SMS are thus
useful for comparing search strategies as well as for identifying
challenging areas of parameter space where search methods may
lack sensitivity. Furthermore, by providing a tabulation of both the
signal acceptance and the 95\% confidence level (\CL) exclusion limit
on the signal cross section as a function of the SMS mass parameters,
SMS results can be used to place limits on a wide variety of
theoretical models beyond SUSY.

The considered SMS scenarios produce multijet final states with or
without leptons and $\cPqb$-tagged jets~\cite{SUS-11-016}. While the
SUSY terminology is employed, interpretations of SMS scenarios are not
restricted to SUSY scenarios.

In the SMS scenarios considered here, each produced particle decays directly
to the LSP and SM particles through a two-body or three-body decay.
Simplified models that are relevant to inclusive hadronic jets+\MET
analyses are gluino pair production with the direct three-body decay
$\sGlu \to \cPq \cPaq \chiz_1$ (T1), and squark-antisquark
production with the direct two-body decay $\sQua \to \cPq
\chiz_1$ (T2).  For $\cPqb$-quark enriched final states, we have
considered two additional gluino SMS scenarios, where each gluino is
forced into the three-body decay $\sGlu \to \cPqb \cPaqb
\chiz_1$ with 100\%\ branching fraction (T1bbbb), or where each gluino
decays through $\sGlu \to \cPqt \cPaqt \chiz_1$ (T1tttt).  For
$\cPqb$-quark enriched final states we also consider SMS that describe
the direct pair production of bottom or top squarks, with the two-body
decays $\sBot \to \cPqb \chiz_1$ (T2bb) and $\sTop \to
\cPqt \chiz_1$ (T2tt).

Note that first-generation $\sQua\sQua$ production (unlike
$\sQua\sQua^*$ production) is not part of the simplified models used
for the interpretation of the razor results, even though it is often
the dominant process in the CMSSM for low values of the scalar-mass
parameter $m_0$. This is because of the additional complication that
the production rate depends on the gluino mass.  However, the
acceptance for $\sQua\sQua$ production is expected to be somewhat
higher than for $\sQua\sQua^*$, so the limits from T2 can be
conservatively applied to $\sQua\sQua$ production with analogous
decays.

For each SMS, simulated samples are generated for a range of masses of
the particles involved, providing a wider spectrum of mass spectra
than allowed by the CMSSM.  A minimum requirement of ${\cal
  O}(100~\GeV)$ on the mass difference between the mother particle and
the LSP is applied, to remove phase space where the jets from
superpartner decays become soft and the signal is detected only when
it is given a boost by associated jet production. By restricting
attention to SMS scenarios with large mass differences, we avoid the
region of phase space where accurate modeling of initial- and
final-state radiation from quarks and gluons is required, and where
the description of the signal shape has large uncertainties.

The production of the primary particles in each SMS is modeled with
SUSY processes in the appropriate decoupling limit of the other
superpartners.  In particular, for $\sQua\,\sQua^*$ production, the
gluino mass is set to a very large value so that it has a minimal
effect on the kinematics of the squarks. The mass spectrum and decay
modes of the particles in a specific SMS point are fixed using the
SLHA input files, which are processed with \PYTHIA v6.426 with
Tune Z2~\cite{Field:2010bc,UEPAS} to produce signal events as an input
to a parameterized fast simulation of the CMS
detector~\cite{CMS-DP-2010-039}, resulting in simulated samples of
reconstructed events for each choice of masses for each SMS.  These
samples are used for the direct calculation of the signal efficiency,
and together with the background model are used to determine the 95\% \CL upper
bound on the allowed production cross section.

\section{Standard model backgrounds in the \texorpdfstring{($\boldmath{\MR}$,
  $\boldmath{\Rtwo}$)}{(MR MR2)} razor plane\label{sec:BKG2011}}

The distributions of SM background events in both the MC simulations
and the data are found to be described by the sum of exponential
functions of $\MR$ and $\Rtwo$ over a large part of the ($\MR$,
$\Rtwo$) plane.  Spurious instrumental effects and QCD multijet
production are challenging backgrounds due to difficulties in modeling
the high $\pt$ and \ETm tails. Nevertheless, these event classes
populate predictable regions of the ($\MR$, $\Rtwo$) plane, which
allows us to study them and reduce their contribution to negligible
levels. The remaining backgrounds in the lepton, dilepton, and
hadronic boxes are processes with genuine \ETm due to energetic
neutrinos and charged leptons from vector boson decay, including $\PW$
bosons from top-quark and diboson production.  The analysis uses
simulated events to characterize the shapes of the SM background
distributions, determine the number of independent parameters needed
to describe them, and to extract initial estimates of the values of
these parameters.  Furthermore, for each of the main SM backgrounds a
control data sample is defined using ${\approx}250$\pbinv of data
collected at the beginning of the run. These events cannot be used in
the search, as the dedicated razor triggers were not
available. Instead, events in this run period were collected using
inclusive non-razor hadronic and leptonic triggers, thus defining
kinematically unbiased data control samples.  We use these control
samples to derive a data-driven description of the shapes of the
background components and to build a background representation using
statistically independent data samples; this is used as an input to a
global fit of data selected using the razor triggers in a signal-free
region of the ($\MR$, $\Rtwo$) razor plane.

The two-dimensional probability density function
$P_{j}(\MR,\Rtwo)$ describing the $\Rtwo$
versus $\MR$ distribution of each considered SM process $j$
is found to be well approximated by the same family of functions
$F_{j}(\MR,\Rtwo)$:
\ifthenelse{\boolean{cms@external}}{
\begin{multline}\label{eqn:2dpdf}
F_{j}(\MR, \Rtwo) =
\left[ k_{j}(\MR - M^0_{\R,j})(\Rtwo - \R_{0,j}^2)-1\right]\\ \times
\re^{-k_{j}(\MR - M^0_{\R,j})(\Rtwo - \R_{0,j}^2)}.
\end{multline}
}{
\begin{equation}\label{eqn:2dpdf}
F_{j}(\MR, \Rtwo) =
\left[ k_{j}(\MR - M^0_{\R,j})(\Rtwo - \R_{0,j}^2)-1\right] \times
\re^{-k_{j}(\MR - M^0_{\R,j})(\Rtwo - \R_{0,j}^2)}.
\end{equation}
}
where $k_{j}$, $\M^0_{\R,j}$, and $\R_{0,j}^2$ are free
parameters of the background model. After applying a baseline
selection in the razor kinematic plane,
$\MR>{\MR^\text{min}}$ and $\Rtwo >
{\Rtwo_\text{min}}$, this function exhibits an exponential
behavior in $\Rtwo$ ($\MR$), when integrated over
$\MR$ ($\Rtwo$):
\begin{align}
\int_{{\Rtwo_\text{min}}}^{+\infty} F_{j}(\MR, \Rtwo) \rd\Rtwo &\sim
\re^{-(a+b\times {\Rtwo_\text{min}})\MR}, \\
\int_{{\MR^\text{min}}}^{+\infty} F_{j}(\MR, \Rtwo)
\rd{}M_{\R} &\sim \re^{-(c+d\times {\MR^\text{min}})\Rtwo},
\end{align}
where $a = -k_j \times \R_{0,j}^2$, $c = -k_j \times
M^0_{\R,j}$, and $b = d = k_j$.  The fact that the function in
Eq.~(\ref{eqn:2dpdf}) depends on $\Rtwo$ and not simply on
$\R$ motivates the choice of $\Rtwo$ as the kinematic
variable quantifying the transverse imbalance. The values of
$\M^0_{\R,j}$, $\R_{0,j}$, $k_{j}$, and the
normalization constant are floated when fitting the function to the
data or simulation samples.

The function of Eq.~(\ref{eqn:2dpdf}) describes the QCD multijet, the
lepton+jets (dominated by $\PW$+jets and \ttbar events), and the
dilepton+jets (dominated by \ttbar and Z+jets events) backgrounds in
the simulation and data control samples. The initial filtering of the
SM backgrounds is performed at the trigger level and the analysis
proceeds with the analytical description of the SM backgrounds.

\subsection{QCD multijet background\label{sec:qcd}}

The QCD multijet control sample for the hadronic box is obtained using
events recorded with prescaled jet triggers.  The trigger used in this
study requires at least two jets with average uncorrected $\pt$
thresholds of 60\GeV.  The QCD multijet background samples provide
$\ga 95\%$ of the events with low $\MR$, allowing the study
of the $\MR$ shapes with different thresholds on
$\Rtwo$, which we denote ${\Rtwo_\text{min}}$.  The study was
repeated using datasets collected with many jet trigger thresholds and
prescale factors during the course of the 2011 LHC data taking, with
consistent results.

\begin{figure}[htpb]
\centering
\includegraphics[width=0.495\textwidth]{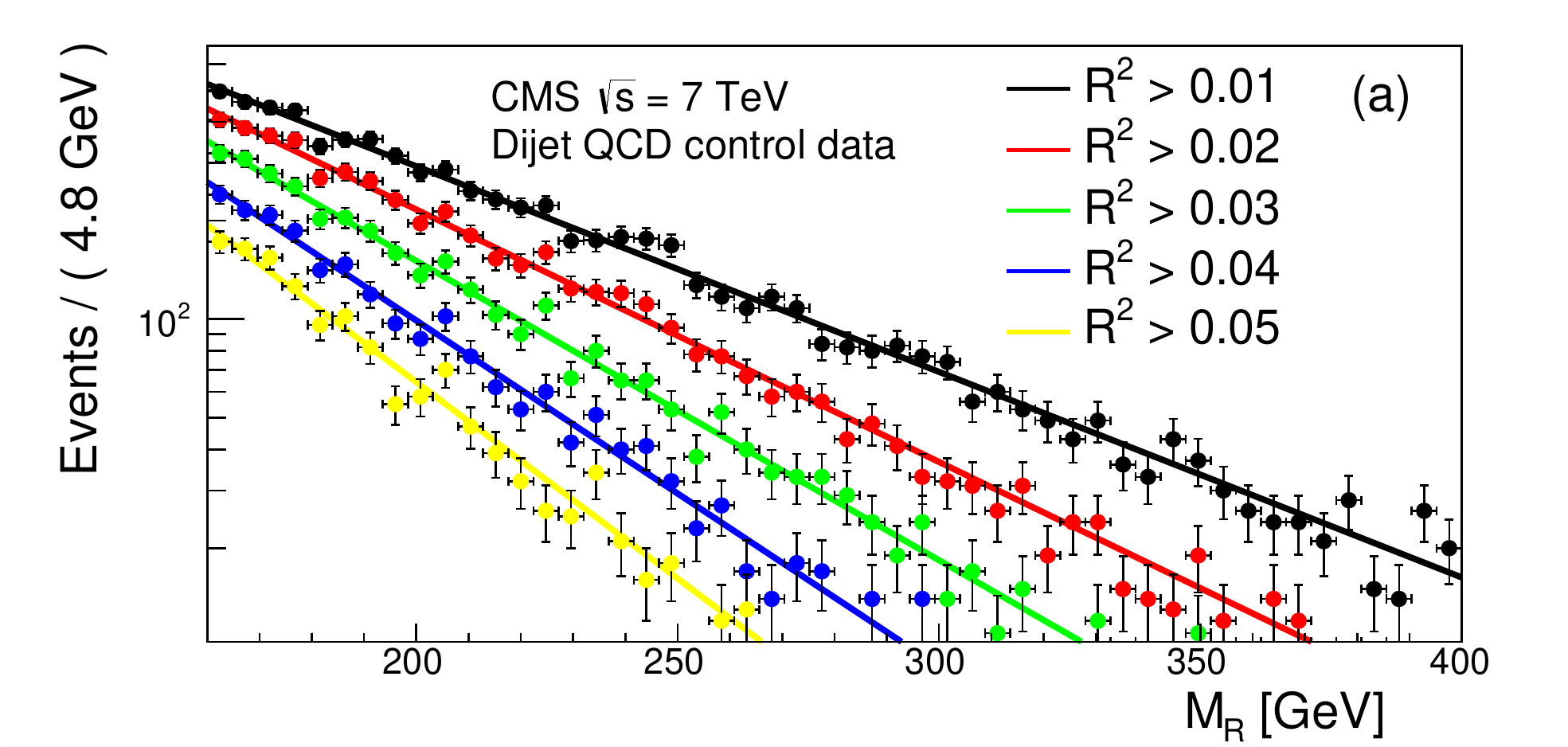}
\includegraphics[width=0.495\textwidth]{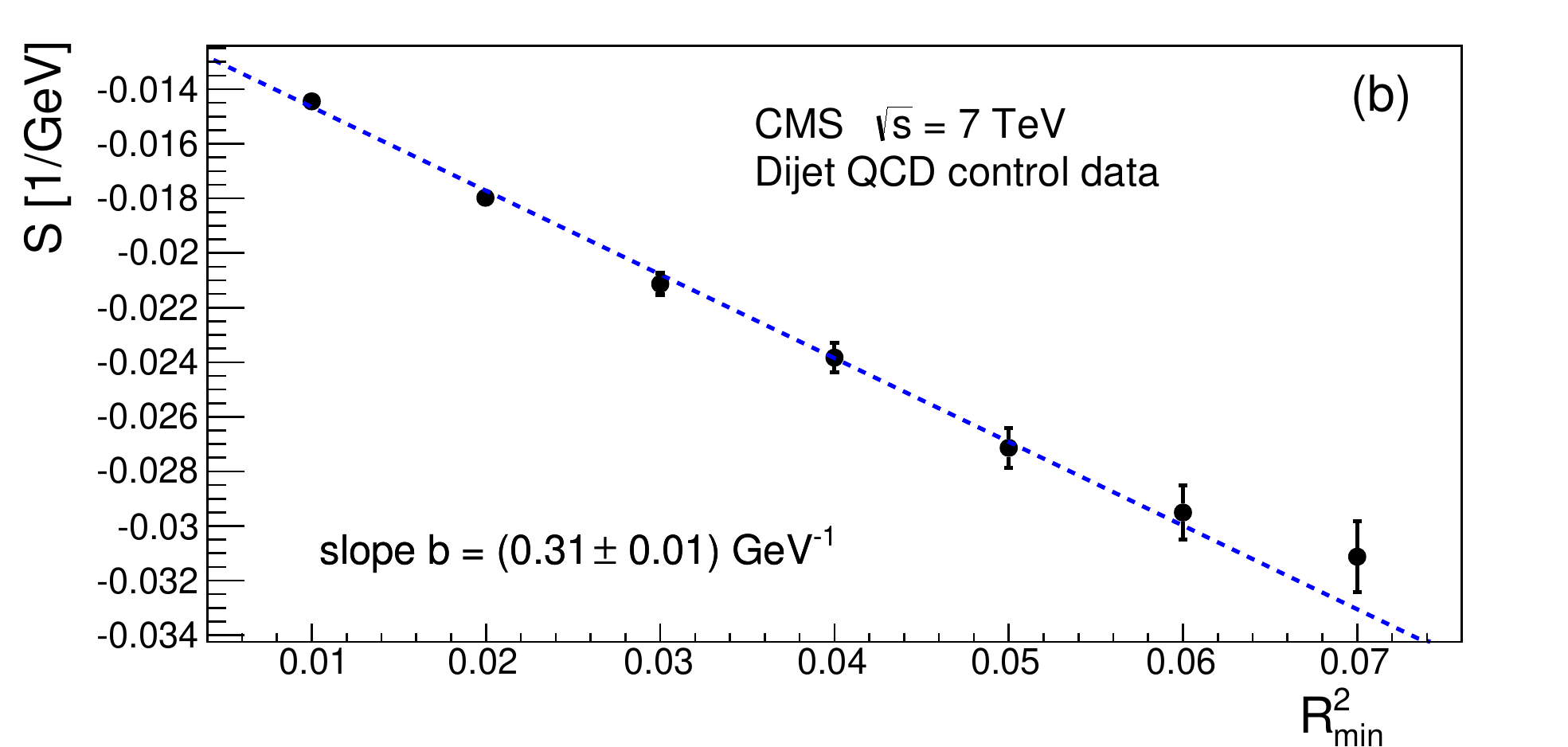}
\caption{(a) The $\MR$ distribution for different values of
  $\Rtwo_{\text{min}}$ for events in the HAD box of a multijet
  control sample, fit to an exponential function. (b) The coefficient
  in the exponent $\S$ from fits to the $\MR$
  distributions, as a function of $\Rtwo_\text{min}$. \label{fig:QCD_calo1}}
\includegraphics[width=0.495\textwidth]{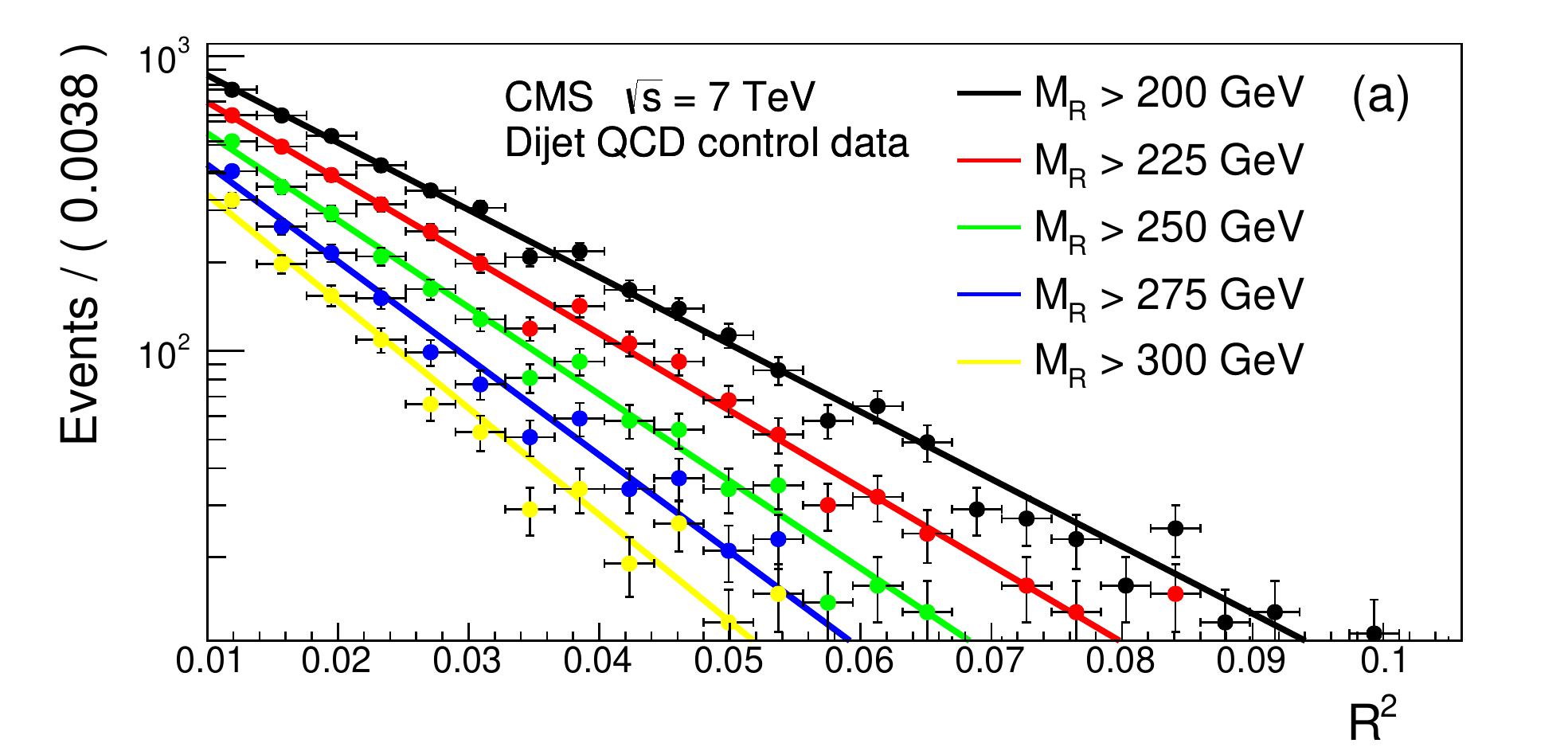}
\includegraphics[width=0.495\textwidth]{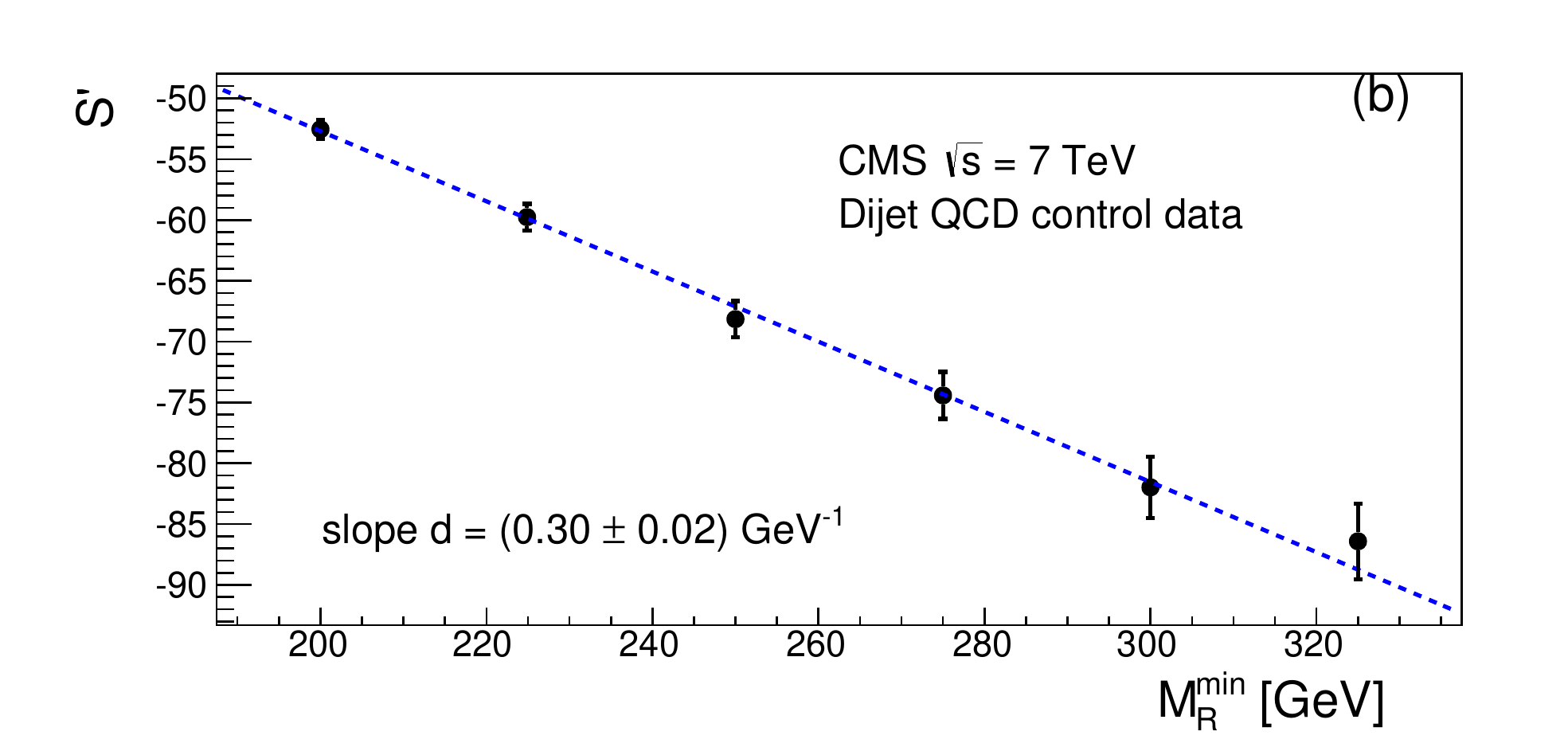}
\caption{(a) The $\Rtwo$ distributions for different values of
  ${\MR^\text{min}}$ for events in data selected in the HAD box
  of a multijet control sample, fit to an exponential function. (b) The coefficient in the
  exponent $\S'$ from fits to the $\Rtwo$ distributions, as a function of
  ${\MR^\text{min}}$. \label{fig:QCD_calo2}}
\end{figure}

The $\MR$ distributions for events satisfying the HAD box
selection in this multijet control data sample are shown for different
values of the $\R_\text{min}$ threshold in
Fig.~\ref{fig:QCD_calo1} (a). The $\MR$ distribution is
exponentially falling, except for a turn-on at low $\MR$
resulting from the $\pt$ threshold requirement on the jets entering
the megajet calculation. The exponential region of these distributions
is fitted for each value of ${\Rtwo_\text{min}}$ to extract the
absolute value of the coefficient in the exponent, denoted $\S$.  The
value of $\S$ that maximizes the likelihood in the exponential fit is
found to be a linear function of ${\Rtwo_\text{min}}$, as shown
in Fig~\ref{fig:QCD_calo1} (b). Fitting $\S$ to the form $\S = -a -
b{\Rtwo_\text{min}}$ determines the values of $a$ and $b$.

The $\R_\text{min}^2$ distributions are shown for different
values of the $\MR$ threshold in
Fig.~\ref{fig:QCD_calo2} (a). The $\Rtwo$ distribution is
exponentially falling, except for a turn-on at low $\Rtwo$. The
exponential region of these distributions is fitted for each value of
${\MR^\text{min}}$ to extract the absolute value of the
coefficient in the exponent, denoted by $\S'$.  The value of $\S'$ that
maximizes the likelihood in the exponential fit is found to be a
linear function of ${\MR^\text{min}}$  as shown in
Fig.~\ref{fig:QCD_calo2} (b).  Fitting $\S'$ to the form $\S' = - c -
dM^\text{min}_\R$ determines the values of $c$ and $d$.  The
slope $d$ is found to be equal to the slope $b$ to within a few per cent,
as seen from the values of these parameters listed in
Figs.~\ref{fig:QCD_calo1} (b) and \ref{fig:QCD_calo2} (b), respectively.
The equality $d=b$ is essential for building the 2D probability
density function that analytically describes the $\Rtwo$
versus $\MR$ distribution, as it reduces the number of possible
2D functions to the function given in Eq.~(\ref{eqn:2dpdf}). Note that
in Eq.~(\ref{eqn:2dpdf}) the $k_j$ parameters are the $b_j$, $d_j$
parameters used in the description of the SM backgrounds.

\subsection{Lepton+jets backgrounds}
\label{V+jetsBKG}

\begin{figure}[htpb]
\centering
\includegraphics[width=0.495\textwidth]{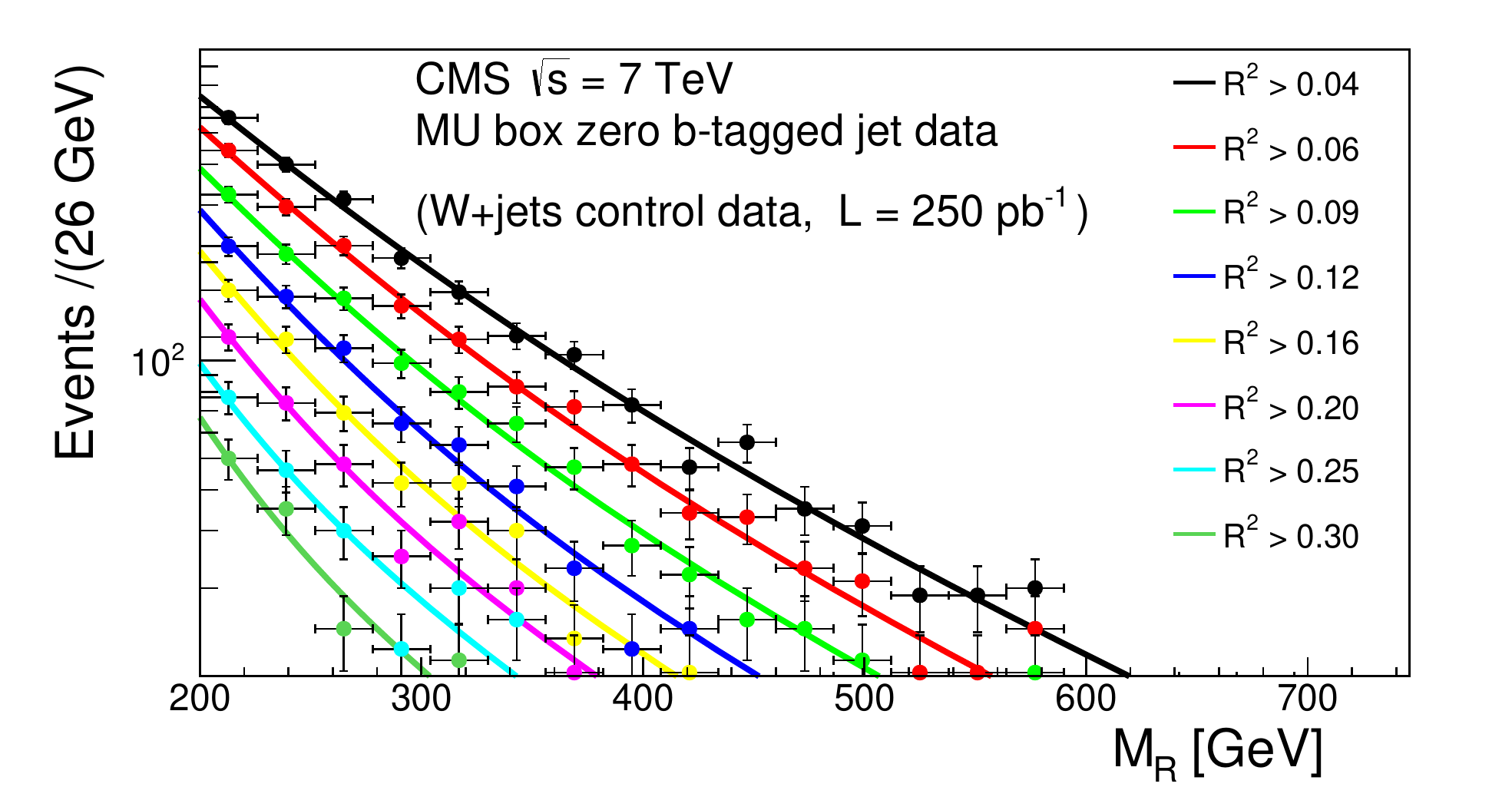}
\caption{The $\MR$ distribution for different values of
  ${\Rtwo_\text{min}}$ for events in the MU box, with the requirement
 of zero $\cPqb$-tagged jets. The curves show the results of fits
 of a sum of two exponential distributions.\label{fig:data_W_MR}}
\includegraphics[width=0.495\textwidth]{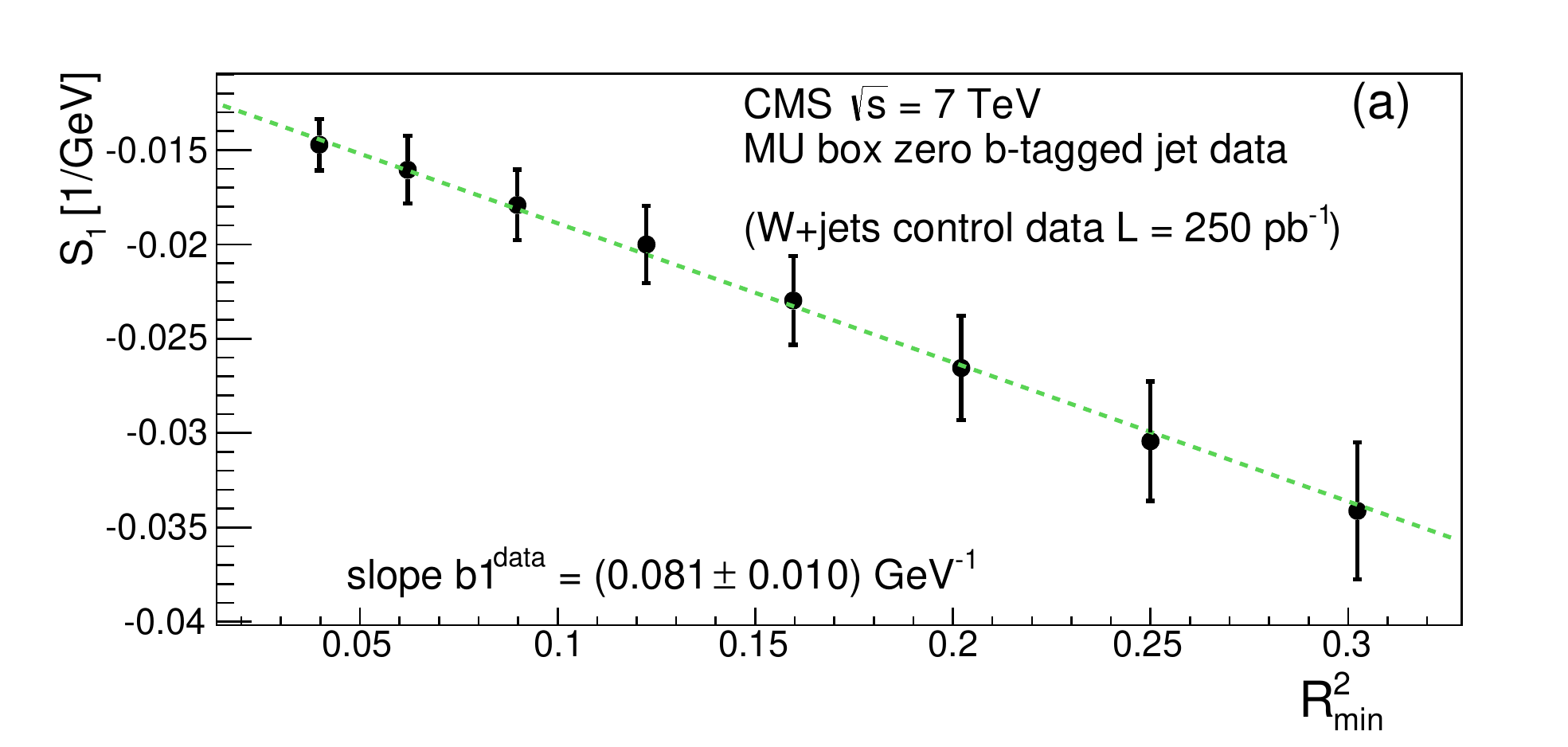}
\includegraphics[width=0.495\textwidth]{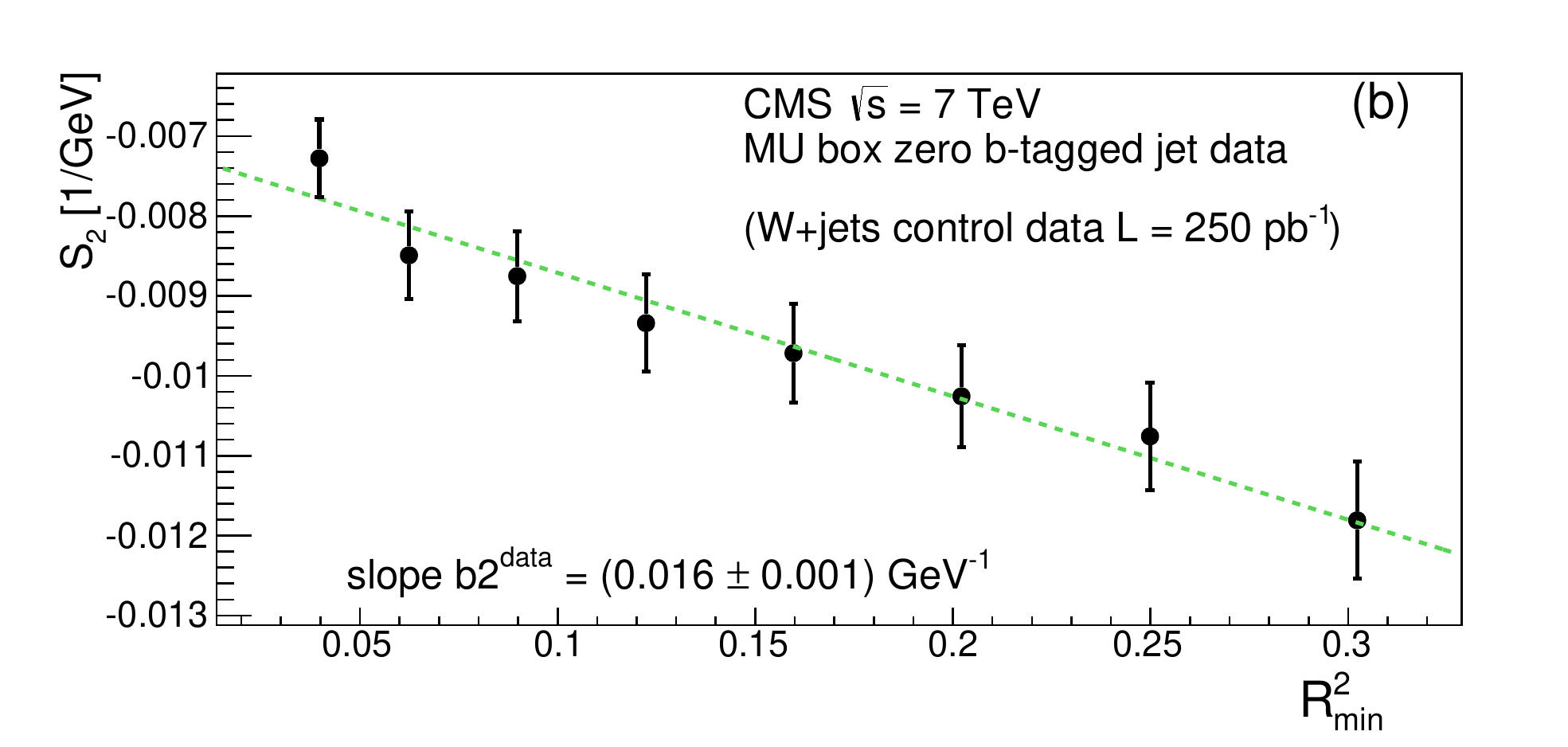}
\caption{Value of (a) the coefficient in the first exponent,
  $\S_1$, and (b) the coefficient in the second exponent,
  $\S_2$, from fits to the $\MR$ distribution, as a
  function of ${\Rtwo_\text{min}}$, for events in the MU box, with the
  requirement of zero $\cPqb$-tagged jets.\label{fig:data_W_MRs}}
\end{figure}

\begin{figure}[htpb]
\centering
\includegraphics[width=0.495\textwidth]{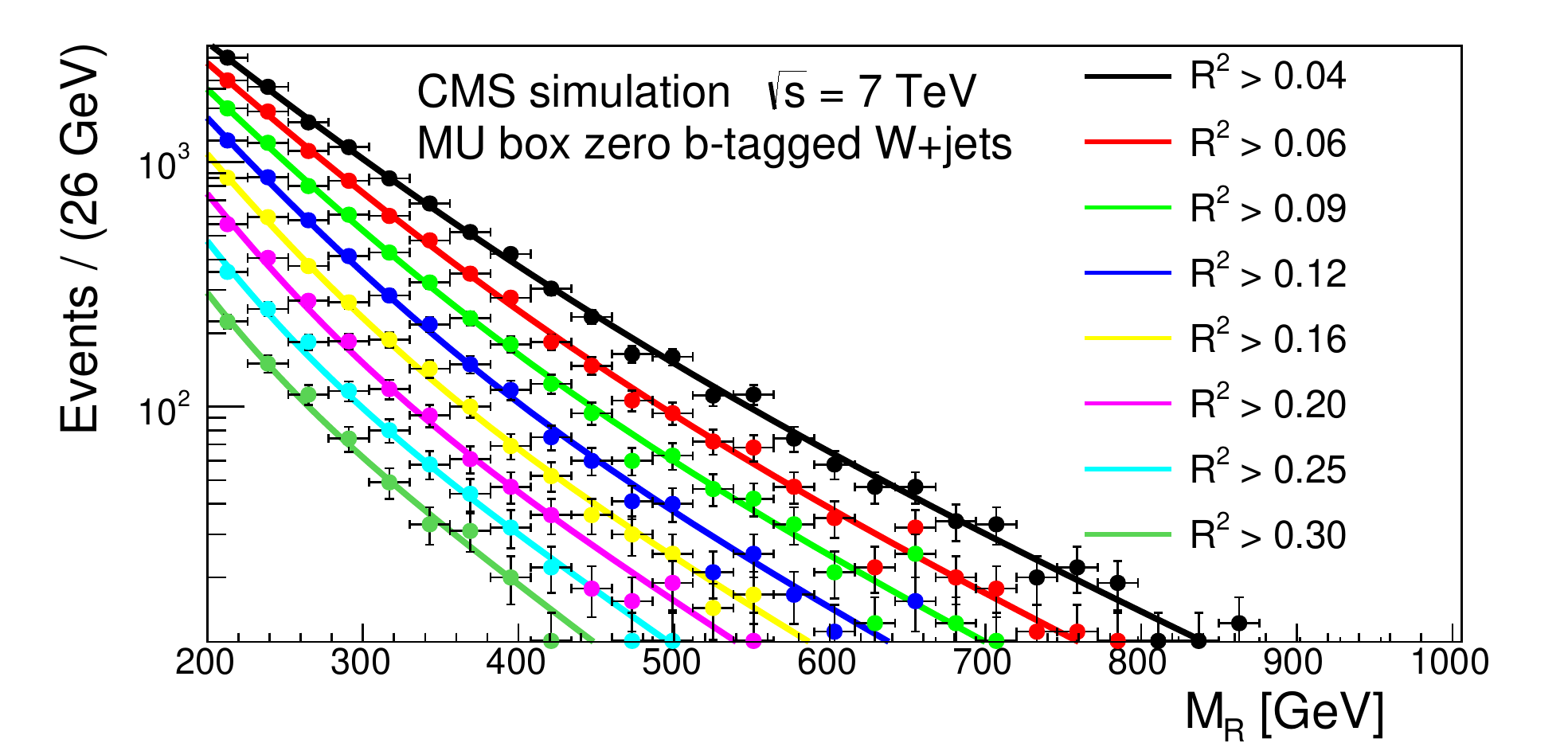}
\caption{The $\MR$ distributions for different values of
  ${\Rtwo_\text{min}}$ for $\PW$+jets simulated events in the MU box
  with the requirement of zero $\cPqb$-tagged jets. The curves show
  the results of fits of a sum of two exponential
  distributions.\label{fig:MC_W_MR}}
\includegraphics[width=0.495\textwidth]{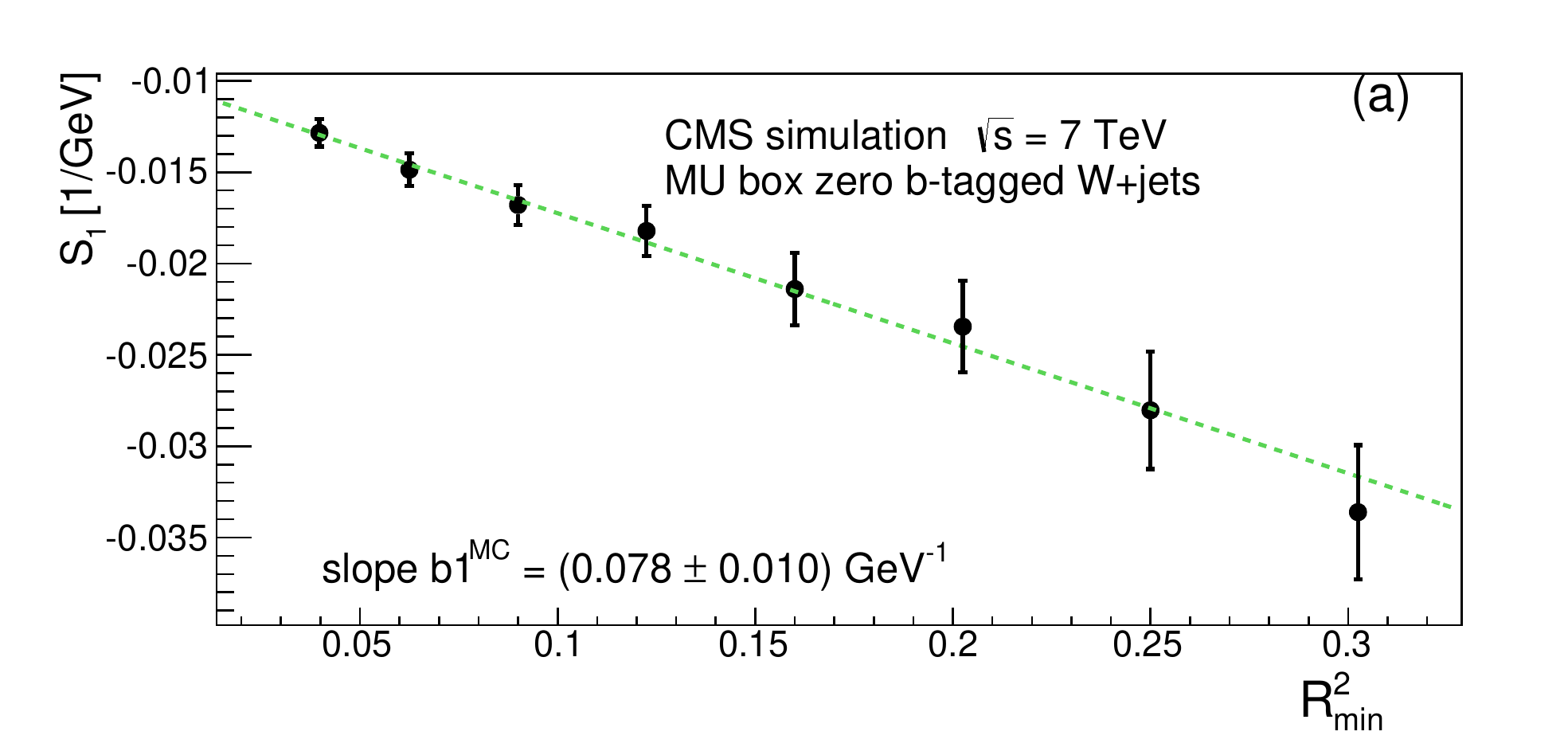}
\includegraphics[width=0.495\textwidth]{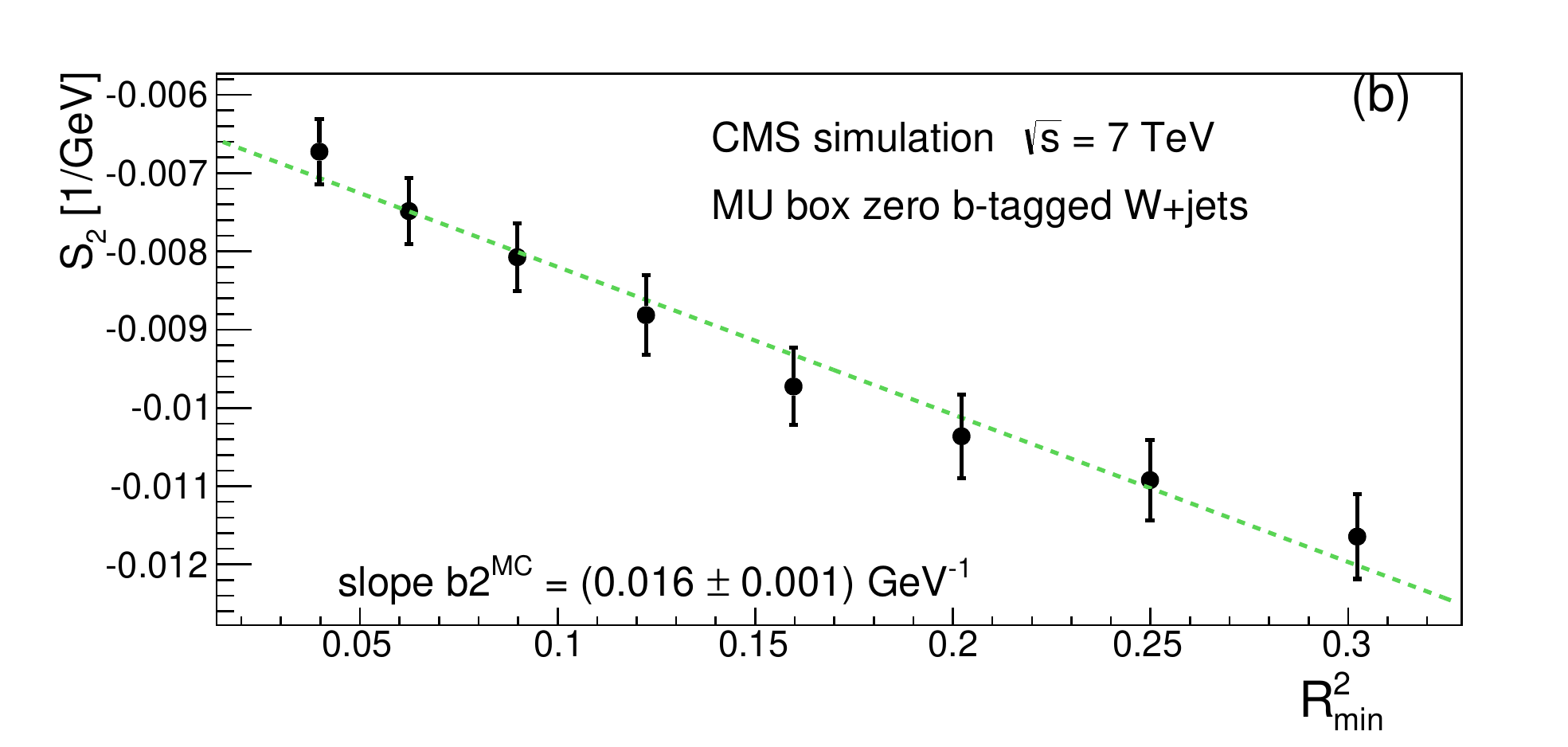}
\caption{Value of (a) the coefficient in the first exponent,
  $\S_1$, and (b) the coefficient in the second exponent,
  $\S_2$, from fits to the $\MR$ distribution, as a
  function of ${\Rtwo_\text{min}}$, for simulated $\PW$+jets events in
  the MU box with the requirement of zero $\cPqb$-tagged
  jets.\label{fig:MC_W_MRs}}
\end{figure}

The major SM backgrounds with leptons and jets in the final state are
($\PW/\cPZ$)+jets, $\ttbar$, and single-top-quark production. These
events can also contain genuine \ETm.  In both the simulated and the
data events in the MU and ELE razor boxes, the $\MR$ distribution is
well described by the sum of two exponential components. One
component, which we denote the ``first" component, has a steeper slope
than the other, ``second" component, i.e., $\abs{\S_1} > \abs{\S_2}$,
and thus the second component is dominant in the high-$\MR$
region. The relative normalization of the two components is considered
as an additional degree of freedom. Both the $\S_1$ and $\S_2$ values,
along with their relative and absolute normalizations, are determined
in the fit. The $\MR$ distributions are shown as a function of
$\R_\text{min}^2$ in Fig.~\ref{fig:data_W_MR} for the zero $\cPqb$-jet
MU data, which is dominated by W+jets events. The dependence of $\S_1$
and $\S_2$ on $\R_\text{min}^2$ is shown in Fig.~\ref{fig:data_W_MRs}.

The corresponding results from simulation are shown in
Figs.~\ref{fig:MC_W_MR} and~\ref{fig:MC_W_MRs}.  It is seen that the
values of the slope parameters $b_1$ and $b_2$ from simulation, given
in Fig.~\ref{fig:MC_W_MRs}, agree within the uncertainties with the
results from data, given in Fig. ~\ref{fig:data_W_MRs}.

\begin{figure}[htpb]
\centering
\includegraphics[width=0.495\textwidth]{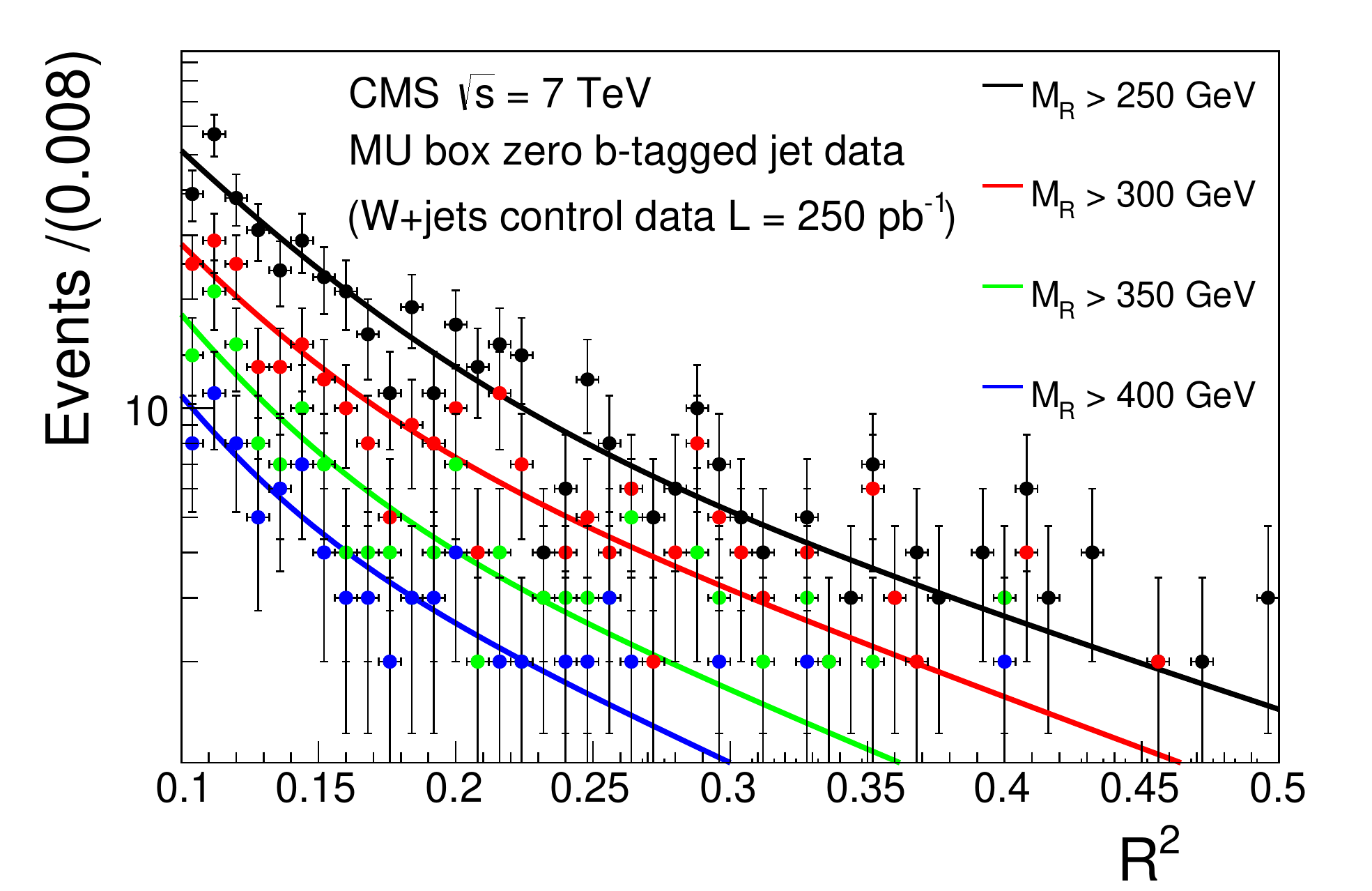}
\caption{The $\Rtwo$ distributions for different values of
  ${\MR^\text{min}}$ for events in the MU box, with the
  requirement of zero $\cPqb$-tagged jets. The curves show the results
  of fits of a sum of two exponential
  distributions.\label{fig:data_W_R}}
\includegraphics[width=0.495\textwidth]{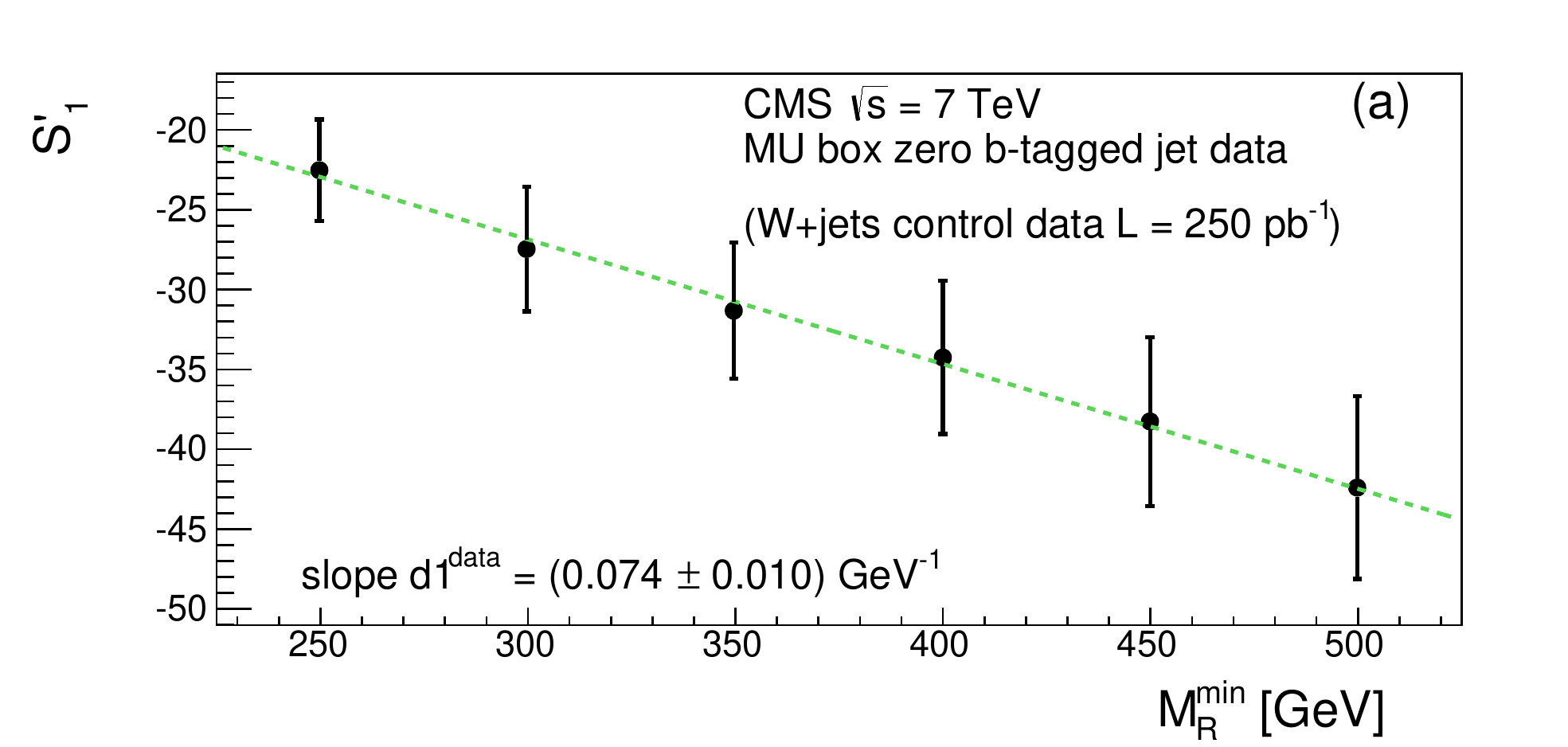}
\includegraphics[width=0.495\textwidth]{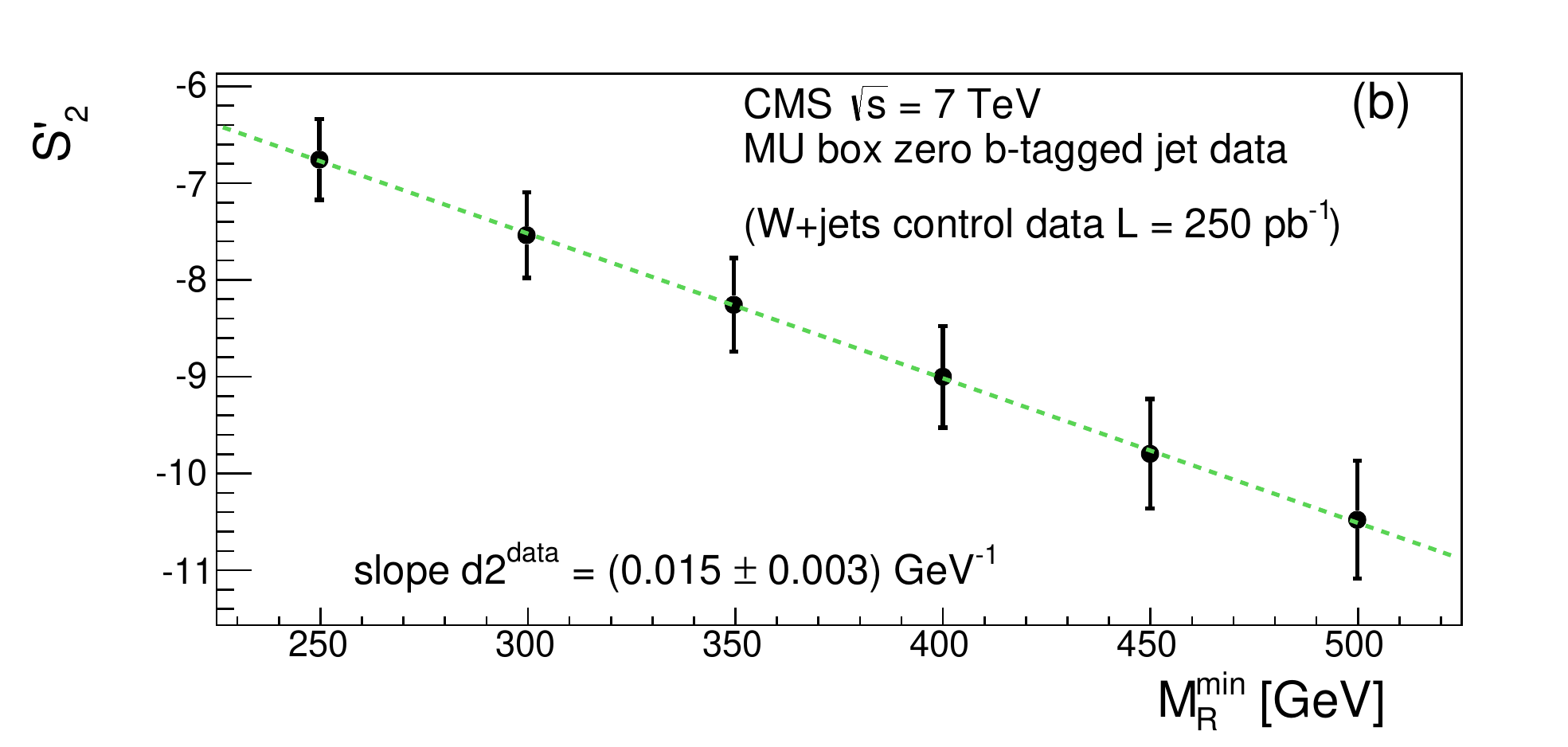}
\caption{Value of (a) the coefficient in the first exponent,
  $\S'_1$, and (b) the coefficient in the second exponent,
  $\S'_2$, from fits to the $\Rtwo$ distribution, as a
  function of ${\MR^\text{min}}$, for events in the MU box, with the
  requirement of zero $\cPqb$-tagged jets.\label{fig:data_W_Rs}}
\end{figure}

\begin{figure}[htpb]
\centering
\includegraphics[width=0.495\textwidth]{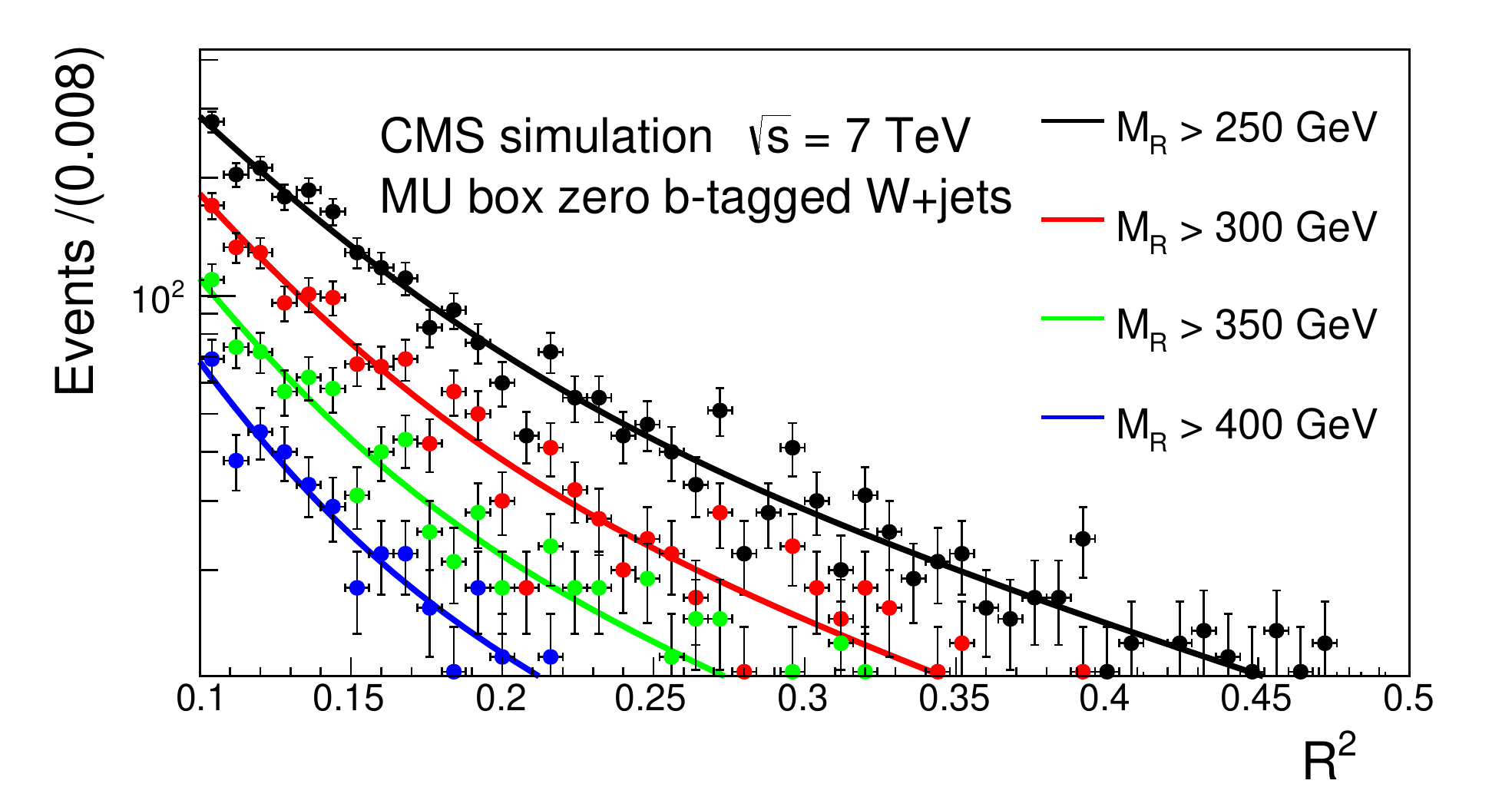}
\caption{The $\Rtwo$ distributions for different values of
  ${\MR^\text{min}}$ for $\PW$+jets simulated events in the MU box
  with the requirement of zero $\cPqb$-tagged jets. The curves show
  the results of fits of a sum of two exponential
  distributions.\label{fig:MC_W_R}}
\includegraphics[width=0.495\textwidth]{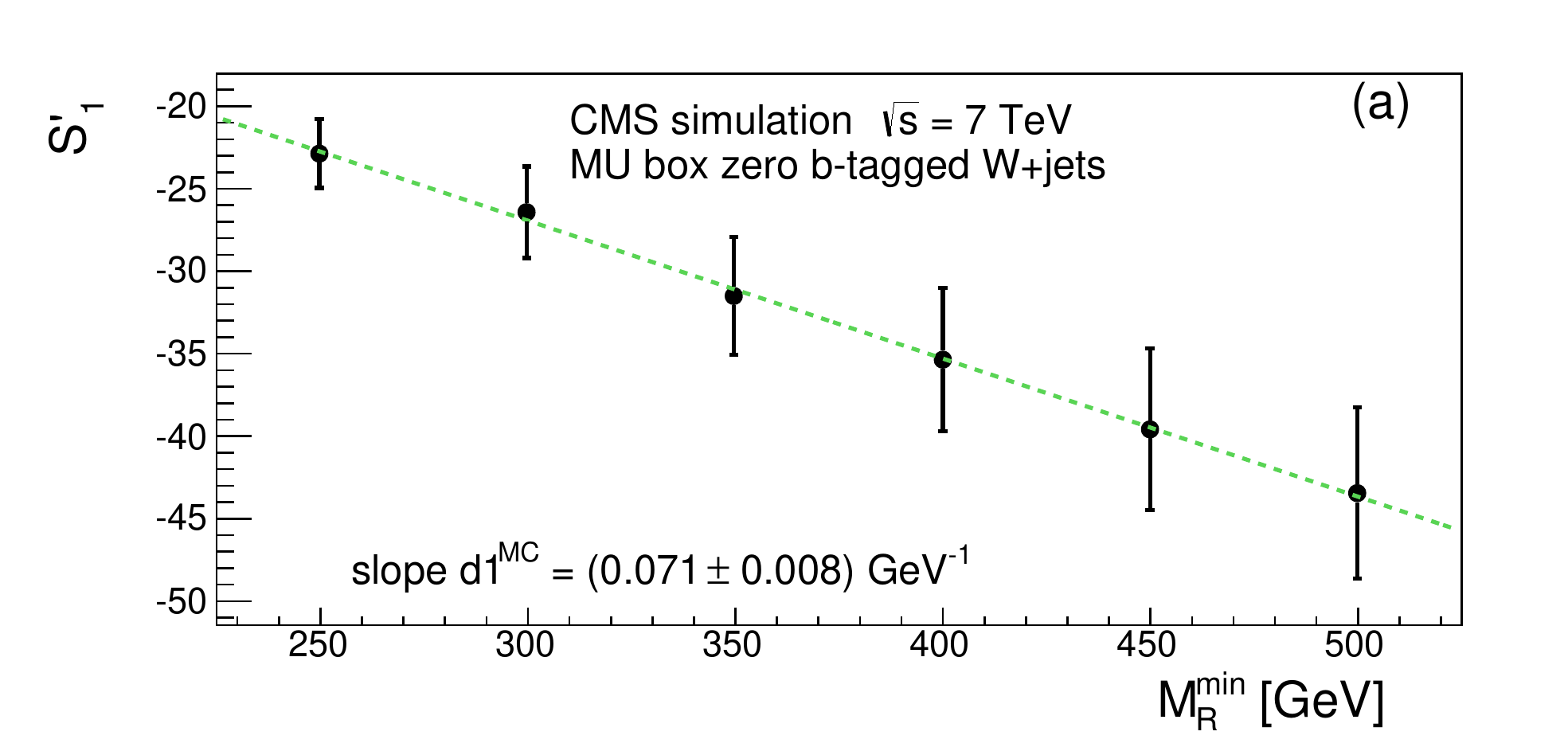}
\includegraphics[width=0.495\textwidth]{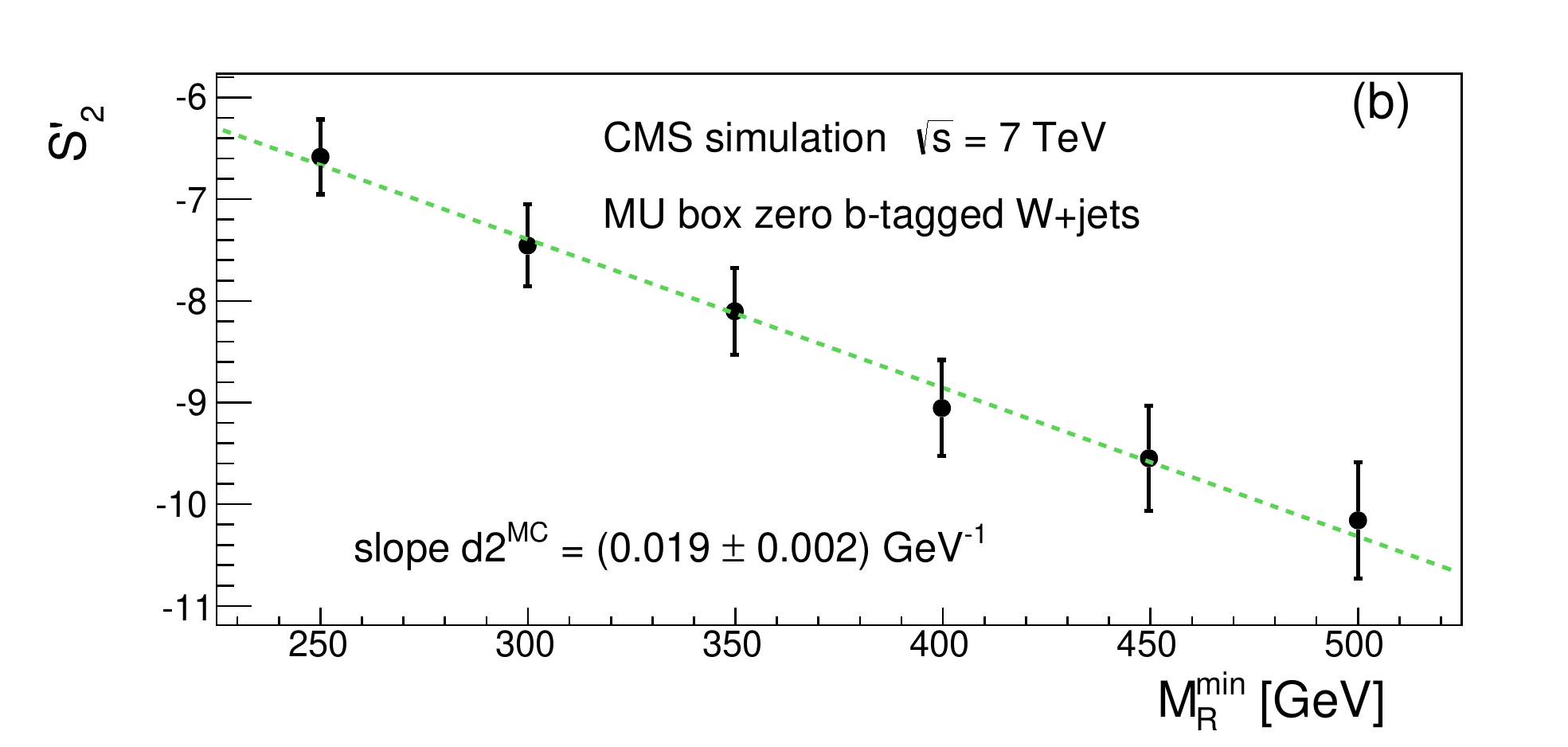}
\caption{Value of (a) the coefficient in the first exponent,
  $\S'_1$, and (b) the coefficient in the second exponent,
  $\S'_2$, from fits to the $\Rtwo$ distribution, as a
  function of ${\MR^\text{min}}$, for $\PW$+jets simulated events in
  the MU box with the requirement of zero $\cPqb$-tagged
  jets.\label{fig:MC_W_Rs}}
\end{figure}

The $\Rtwo$ distributions as a function of $\MR^\text{min}$ for the data are shown in Fig.~\ref{fig:data_W_R} for the MU
box with the requirement of zero b-tagged jets.  The $\S'_1$
and $\S'_2$ parameters characterizing the exponential behavior
of the first and second $\PW(\mu\nu)$+jets components are shown in
Fig.~\ref{fig:data_W_Rs}.  The corresponding results from simulation
are shown in Figs.~\ref{fig:MC_W_R} and~\ref{fig:MC_W_Rs}.  The
results for the slopes $d_1$ and $d_2$ from simulation, listed in
Fig.~\ref{fig:MC_W_Rs}, are seen to be in agreement with the measured
results, listed in Fig.~\ref{fig:data_W_Rs}. Furthermore, the
extracted values of $d_1$ and $d_2$ are in agreement with the
extracted values of $b_1$ and $b_2$, respectively. This is the
essential ingredient to build a 2D template for the
($\MR$,$\Rtwo$) distributions, starting with the
function of Eq.~(\ref{eqn:2dpdf}).

The corresponding distributions for the $\ttbar$ MC simulation with
$\ge$1 $\cPqb$-tagged jet are presented in
Appendix~\ref{sec:ttbarAppendix}, for events selected in the HAD box.


\subsection{Dilepton backgrounds}

The SM contributions to the ELE-ELE and MU-MU boxes are expected to be
dominated by $\cPZ$+jets events, and the SM contribution to the ELE-MU
box by $\ttbar$ events, all at the level of $\ga$95\%.  We find that
the $\MR$ distributions as a function of
${\Rtwo_\text{min}}$, and the $\Rtwo$ distribution as a
function of ${\MR^\text{min}}$, are independent of the lepton-flavor
combination for both the ELE-ELE and MU-MU boxes, as determined using
simulated $\ttbar(2\ell 2\nu$+jets$)$ events. In addition, the
asymptotic second component is found to be process-independent.

\section{Background model and fits}\label{sec:fits}

As described earlier, the full 2D SM background representation is
built using statistically independent data control samples. The
parameters of this model provide the input to the final fit performed
in the {\it fit region} (FR) of the data samples, defining an
extended, unbinned maximum likelihood (ML) fit with the \textsc{RooFit}
fitting package \cite{Verkerke:2003ir}. The fit region is defined for
each of the razor boxes as the region of low $\MR$ and small
$\Rtwo$, where signal contamination is expected to have
negligible impact on the shape fit.  The 2D model is extrapolated to
the rest of the ($\MR$, $\Rtwo$) plane, which is
sensitive to new-physics signals and where the search is performed.

For each box, the fit is conducted in the signal-free FR of the
($\MR$, $\Rtwo$) plane; their definition can be found
in
Figs.~\ref{fig:had-blue-plot},~\ref{fig:ele-blue-plot},~\ref{fig:mu-blue-plot},
~\ref{fig:eleele-blue-plot},~\ref{fig:mumu-blue-plot},
and~\ref{fig:muele-blue-plot}. These regions are used to provide a
full description of the SM background in the entire ($\MR$,
$\Rtwo$) plane in each box.  The likelihood function for a
given box is written as \cite{Barlow:1990vc}:
\begin{equation}
\label{eq:Lb}
\mathcal{L}_{b} =  \frac{\re^{-(\sum_{j \in SM} N_{j})}}{N !}
\prod_{i=1}^{N} \left[\sum_{j \in SM} N_{j} P_{j}(M_{R,i},\Rtwo_i)\right],
\end{equation}
where $N$ is the total number of events in the FR region of the box,
the sum runs over all the SM processes relevant for that box, and the
$N_{j}$ are normalization parameters for each SM process involved in
the considered box.

We find that each SM process in a given final state box is well
described in the ($\MR$, $\Rtwo$) plane by the function $P_{j}$
defined as
\begin{equation}
  P_{j}(M_R, \Rtwo) =  (1-f^{j}_2)\times F^{1st}_{j}(M_R, \Rtwo) +
  f^{j}_2\times F^{2nd}_{j}(M_R, \Rtwo),
\label{eq:twocomponents}
\end{equation}
where the {\it first} ($F^{1st}_{j}$) and {\it second} ($F^{2nd}_{j}$)
components are defined as in Eq.~(\ref{eqn:2dpdf}), and $f^{j}_2$ is
the normalization fraction of the second component with respect to the
total.  When fitting this function to the data, the shape parameters
of each $F_{j}(M_R, \Rtwo)$ function, the absolute
normalization, and the relative fraction $f^{j}_2$ are floated in the
fit. Studies of simulated events and fits to data control samples with
either a $\cPqb$-jet requirement or a $\cPqb$-jet veto indicate that
the parameters corresponding to the first components of these
backgrounds (with steeper slopes at low $\MR$ and
$\Rtwo$) are box-dependent. The parameters describing the
second components are box-independent, and at the current precision of
the background model, they are identical between the dominant
backgrounds considered in these final states.

\begin{figure}[htpb]
\centering
\includegraphics[width=0.495\textwidth]{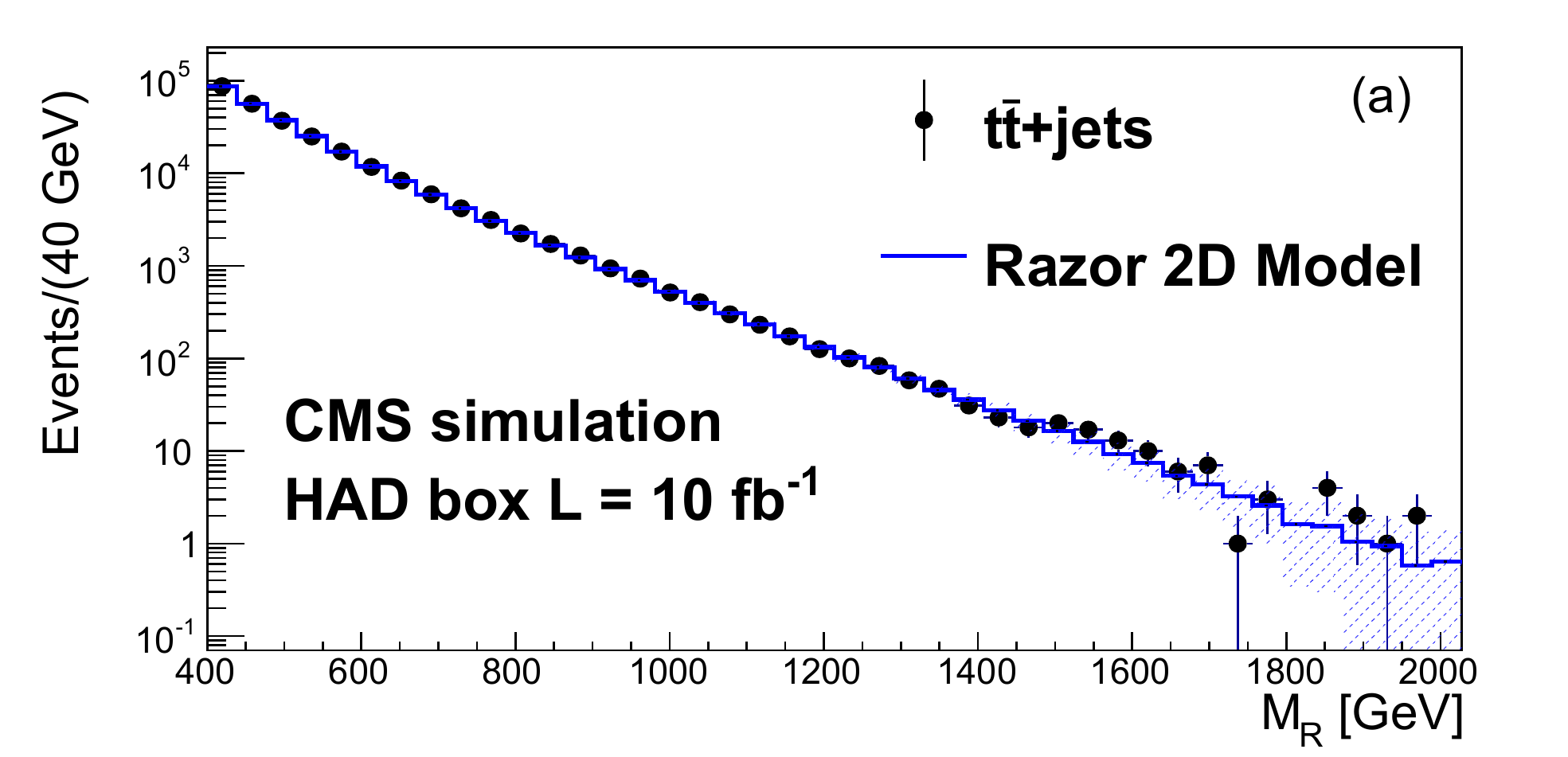}
\includegraphics[width=0.495\textwidth]{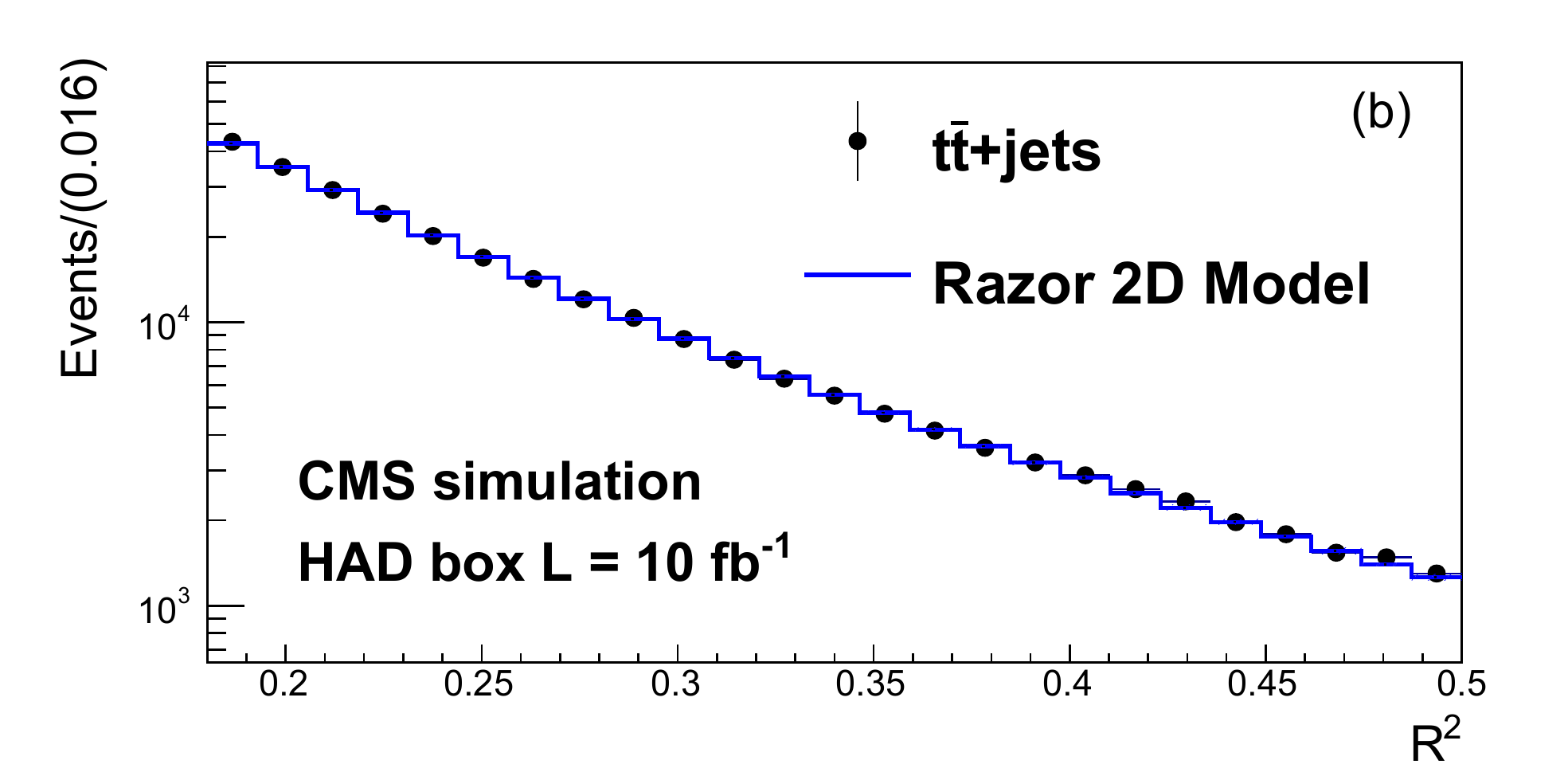}
\caption{Projection of the 2D fit result on (a) $\MR$ and (b)
  $\Rtwo$ for the HAD box in $\ttbar$ MC simulation. The
  continuous histogram is the 2D model prediction obtained from a
  single pseudo-experiment based on the 2D fit. The fit is performed
  in the ($\MR$,$\Rtwo$) fit region and projected into
  the full analysis region.  Only the statistical uncertainty band in
  the background prediction is drawn in these projections. The points
  show the distribution for the MC simulated
  events.\label{fig:had-tthstat}}
\end{figure}

We validate the choice of the background shape by use of a sample of
$\ttbar$ MC simulated events corresponding to an integrated luminosity
of 10\fbinv. Besides being the dominant background in the $\ge$1
$\cPqb$-tag search, $\ttbar$ events are the dominant background for
the inclusive search for large values of $\MR$ and
$\Rtwo$.  The result for the HAD box in the inclusive razor
path is shown in Fig.~\ref{fig:had-tthstat} expressed as the
projection of the 2D fit on $\MR$ and $\Rtwo$.  As the
same level of agreement is found in all boxes both in the inclusive
and in the $\ge$1 $\cPqb$-tagged razor path, we proceed to fit all
the SM processes with this shape.

\subsection{Fit results and validation}\label{sec:datafits}

The shape parameters in Eq.~(\ref{eqn:2dpdf}) are determined for each
box via the 2D fit. The likelihood of Eq.~(\ref{eq:Lb}) is multiplied
by Gaussian {\it penalty terms}~\cite{penalty} to account for the
uncertainties of the shape parameters $k_{j}$, $\M^0_{\R,j}$, and
$\R_{0,j}^2$. The central values of the Gaussians are derived from
analogous 2D fits in the low-statistics data control sample.  The
penalty terms pull the fit to the local minimum closer to the shape
derived from the data control samples. Using pseudo-experiments, we
verified that this procedure does not bias the determination of the
background shape. As an example, the $k_{j}$ parameter uncertainties
are typically $\sim$30\%.  Additional background shape uncertainties
due to the choice of the functional form were considered and found to
be negligible, as discussed in Appendix~\ref{altBKG}.

\begin{figure}[ht!]
  \centering
\includegraphics[width=0.495\textwidth]{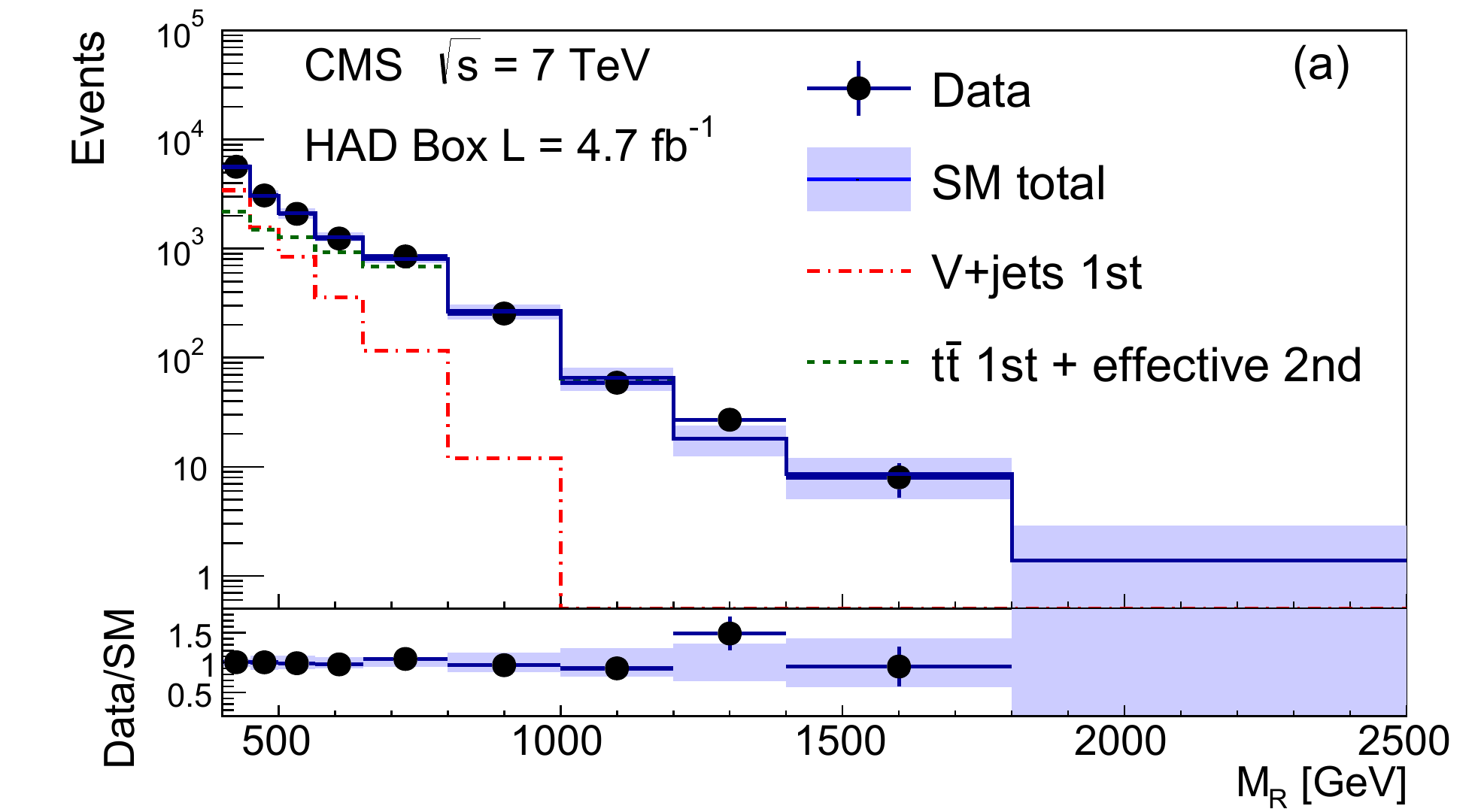}
\includegraphics[width=0.495\textwidth]{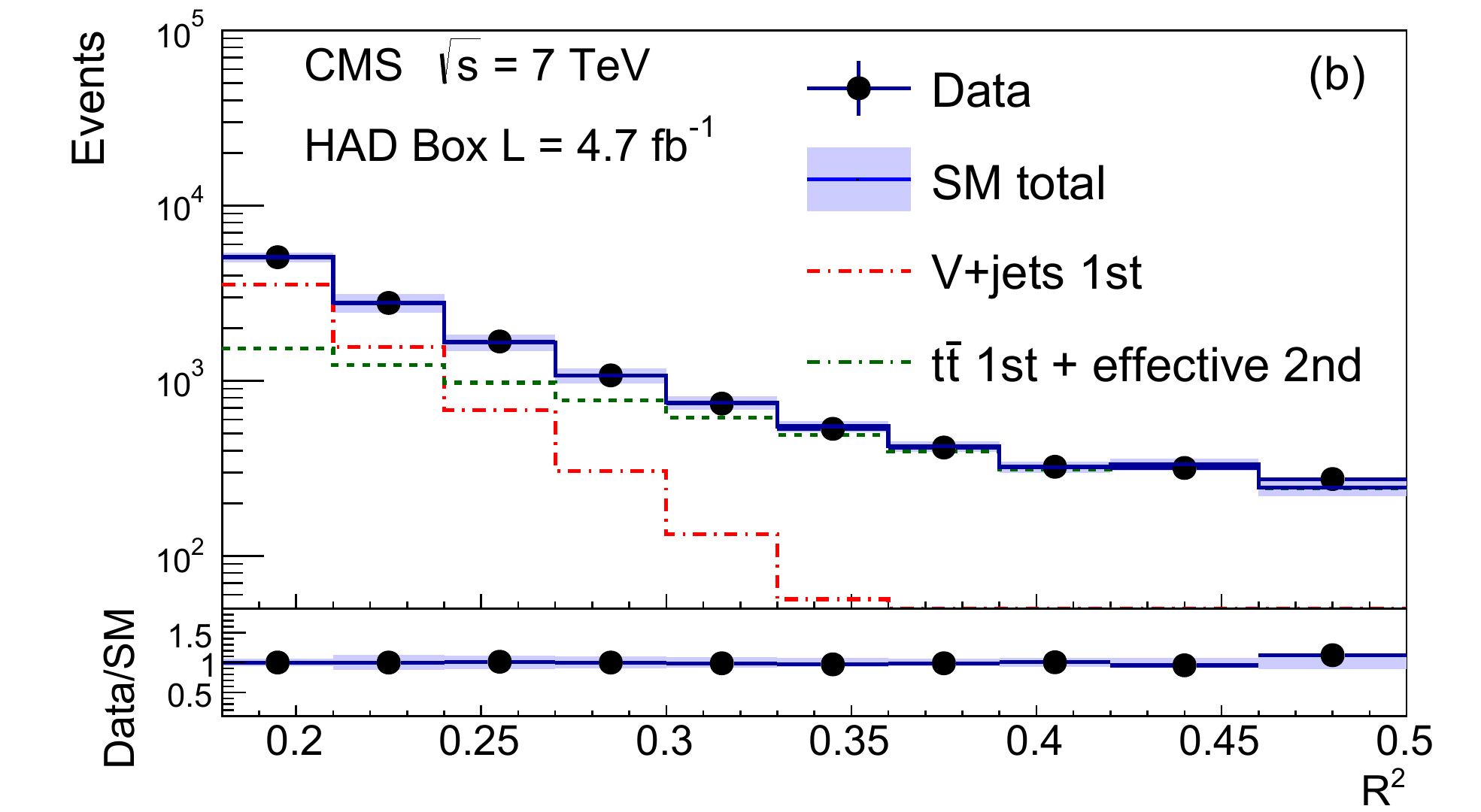}
\caption{Projection of the 2D fit result on (a) $\MR$ and (b)
  $\Rtwo$ for the inclusive HAD box. The continuous histogram
  is the total SM prediction. The dash-dotted and dashed histograms
  are described in the text. The fit is performed in the ($\MR$,
  $\Rtwo$) fit region (shown in Fig.~\ref{fig:had-blue-plot})
  and projected into the full analysis region. The full uncertainty in
  the total background prediction is drawn in these projections,
  including the one due to the variation of the background shape
  parameters and normalization.\label{fig:had-box}}
\end{figure}

\begin{figure}[ht!]
\centering
\includegraphics[width=\cmsFigWidthBox]{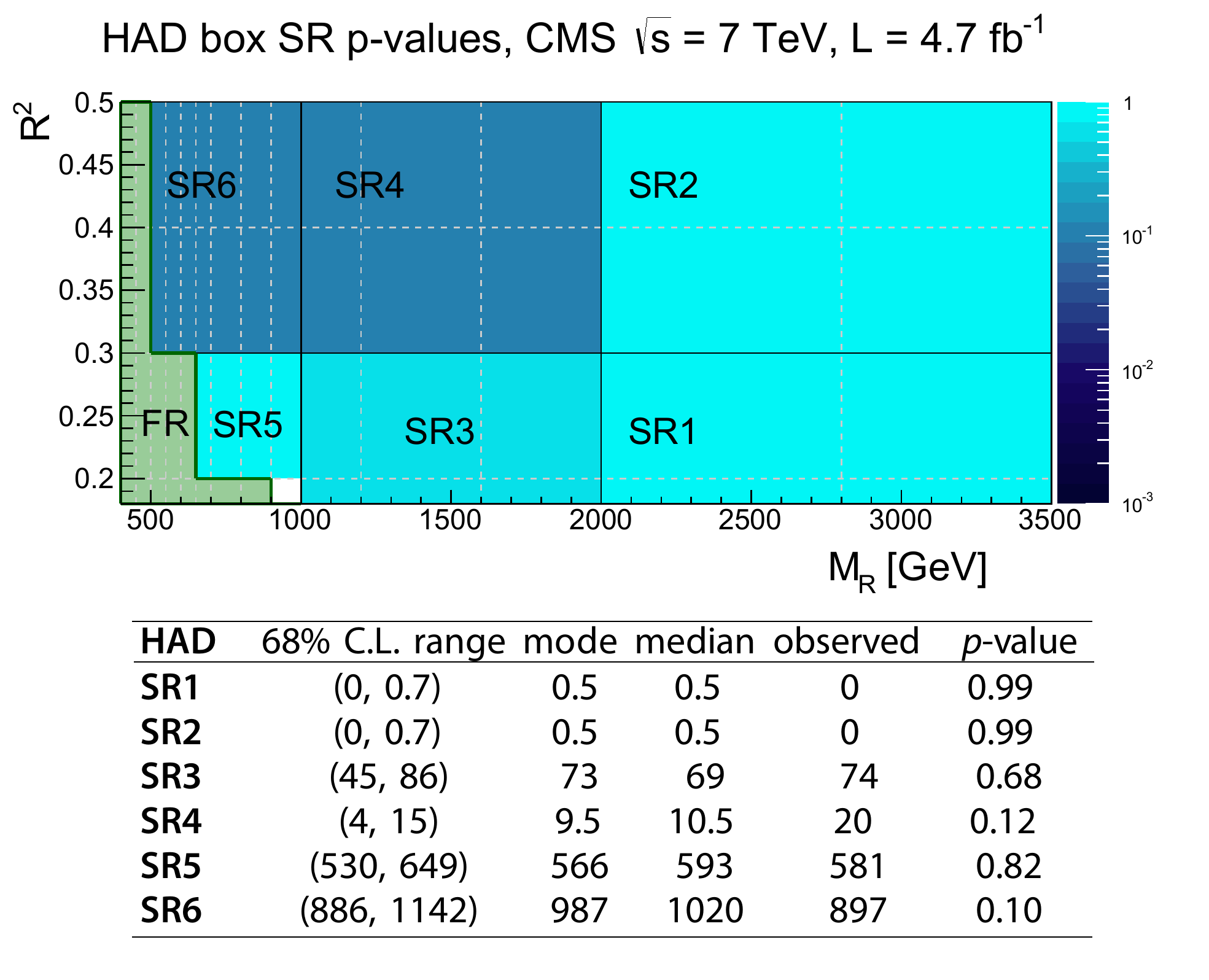}
\caption{The fit region, FR, and signal regions, SR$i$, are defined in
  the ($\MR$, $\Rtwo$) plane for the HAD box. The
  color scale gives the $p$-values corresponding to the observed
  number of events in each SR$i$, computed from background
  parameterization derived in the FR.  The $p$-values are also given in
  the table, together with the observed number of events, the median
  and the mode of the yield distribution, and a 68\% \CL
  interval.\label{fig:had-blue-plot}}
\end{figure}

\begin{figure}[ht!]
\centering
\includegraphics[width=0.495\textwidth]{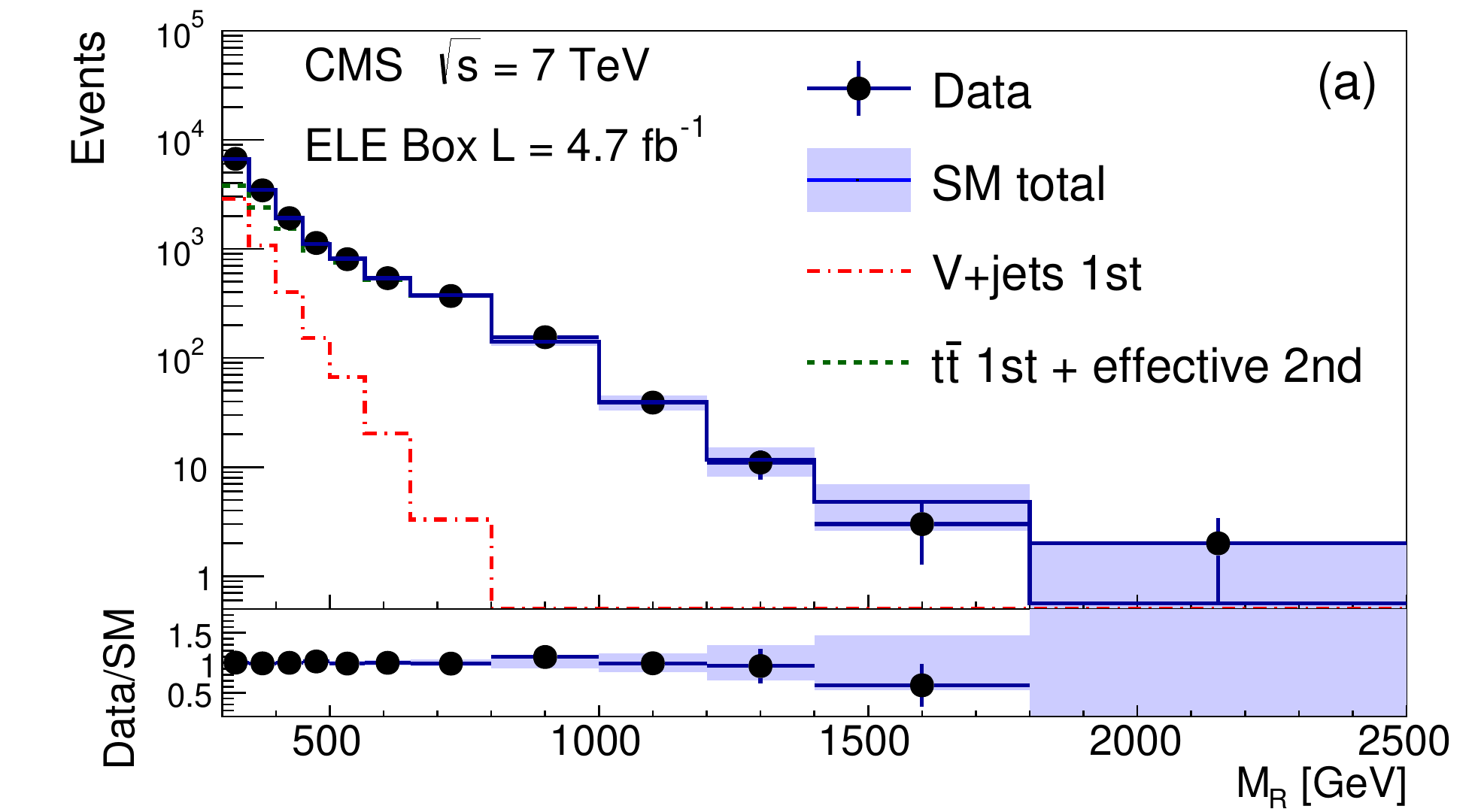}
\includegraphics[width=0.495\textwidth]{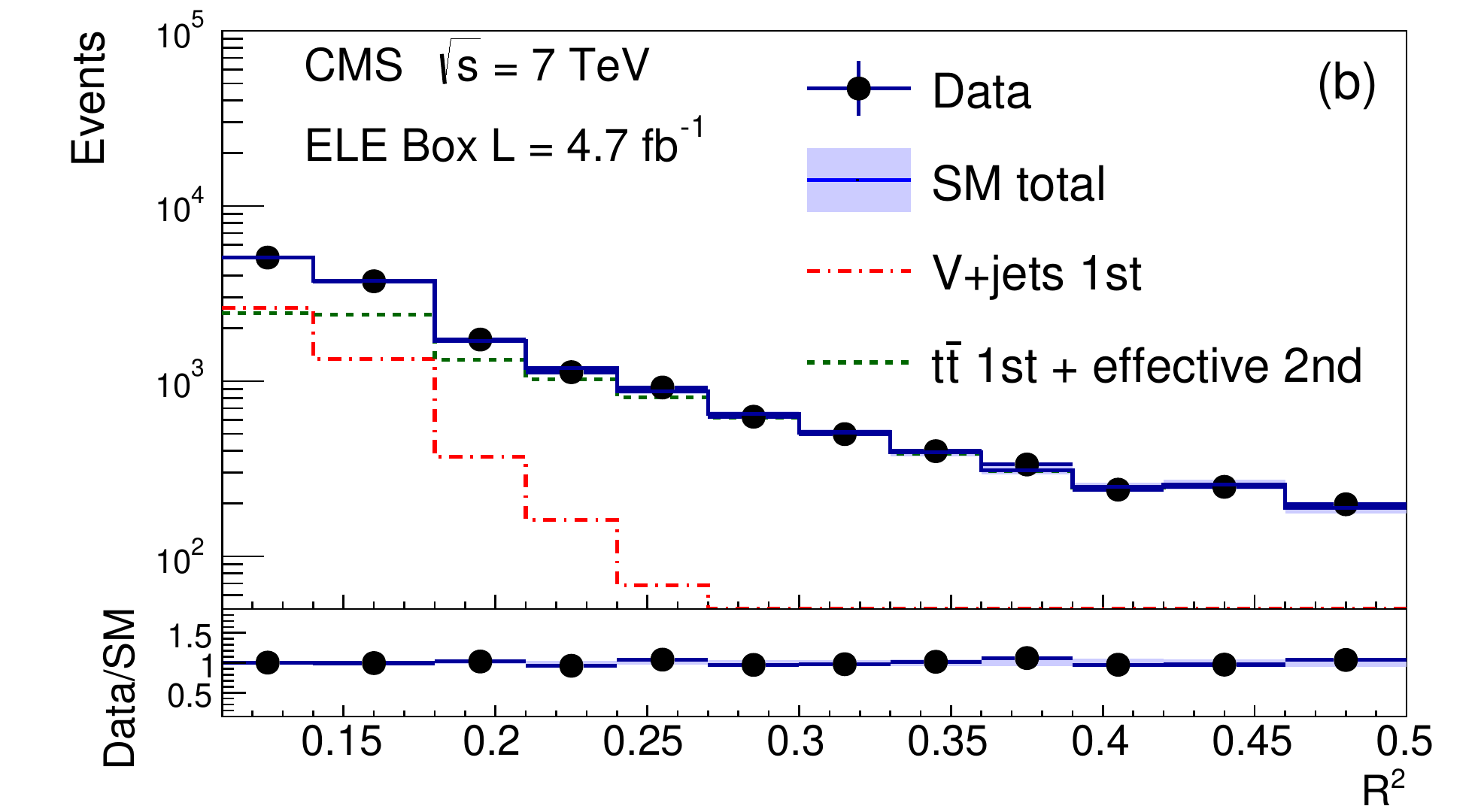}
\caption{Projection of the 2D fit result on (a) $\MR$ and (b)
  $\Rtwo$ for the inclusive ELE box. The fit is performed in the
  ($\MR$, $\Rtwo$) fit region (shown in
  Fig.~\ref{fig:ele-blue-plot}) and projected into the full analysis
  region. The histograms are described in the text.\label{fig:ele-box}}
\end{figure}
\begin{figure}[ht!]
\centering
\includegraphics[width=\cmsFigWidthBox]{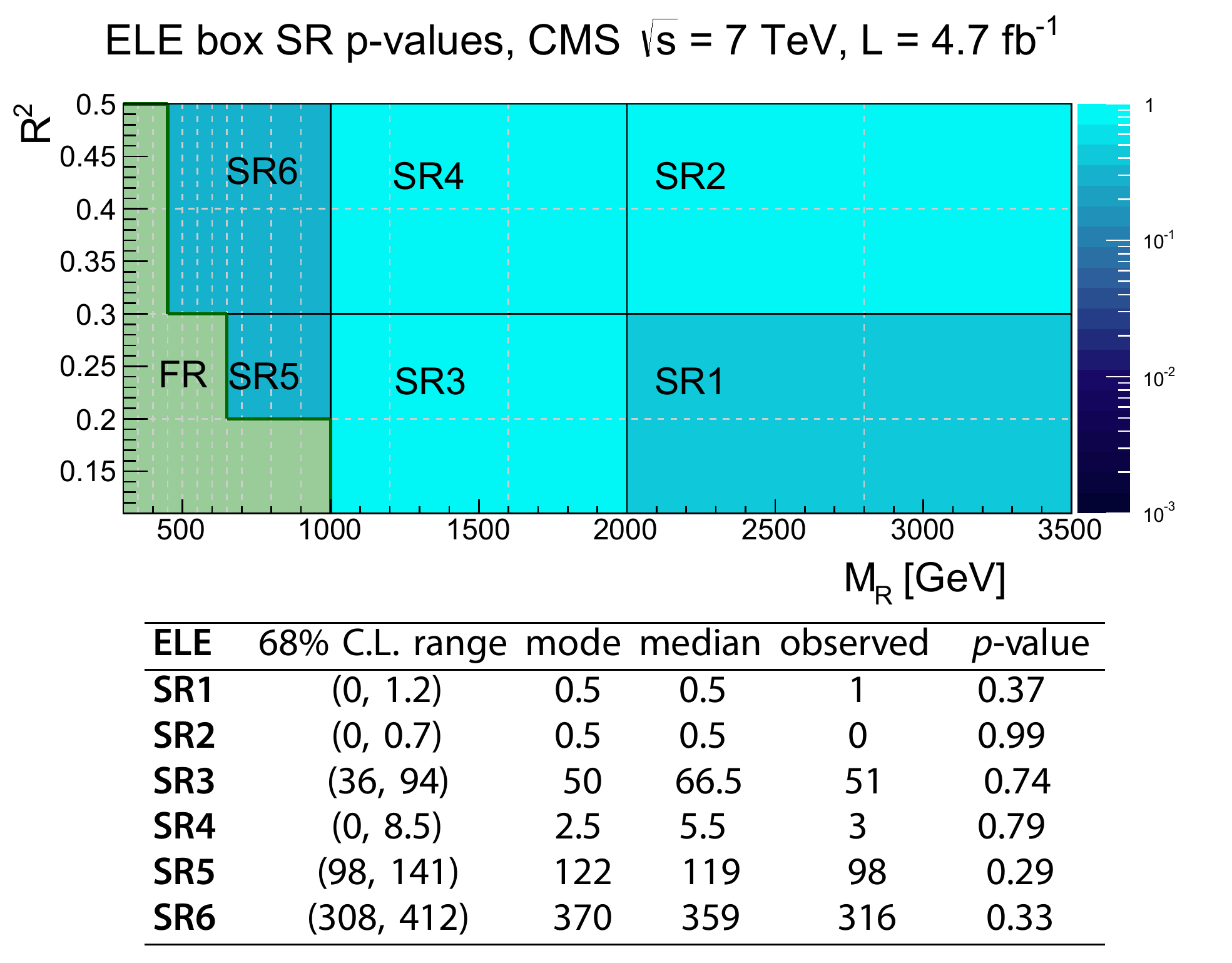}
\caption{The fit region, FR, and signal regions, SR$i$, are defined in
  the ($\MR$, $\Rtwo$) plane for the ELE box. The
  color scale gives the $p$-values corresponding to the observed
  number of events in each SR$i$. Further explanation is given in the
  Fig.~\ref{fig:had-blue-plot} caption.\label{fig:ele-blue-plot}}
\end{figure}

\begin{figure}[ht!]
\centering
\includegraphics[width=0.495\textwidth]{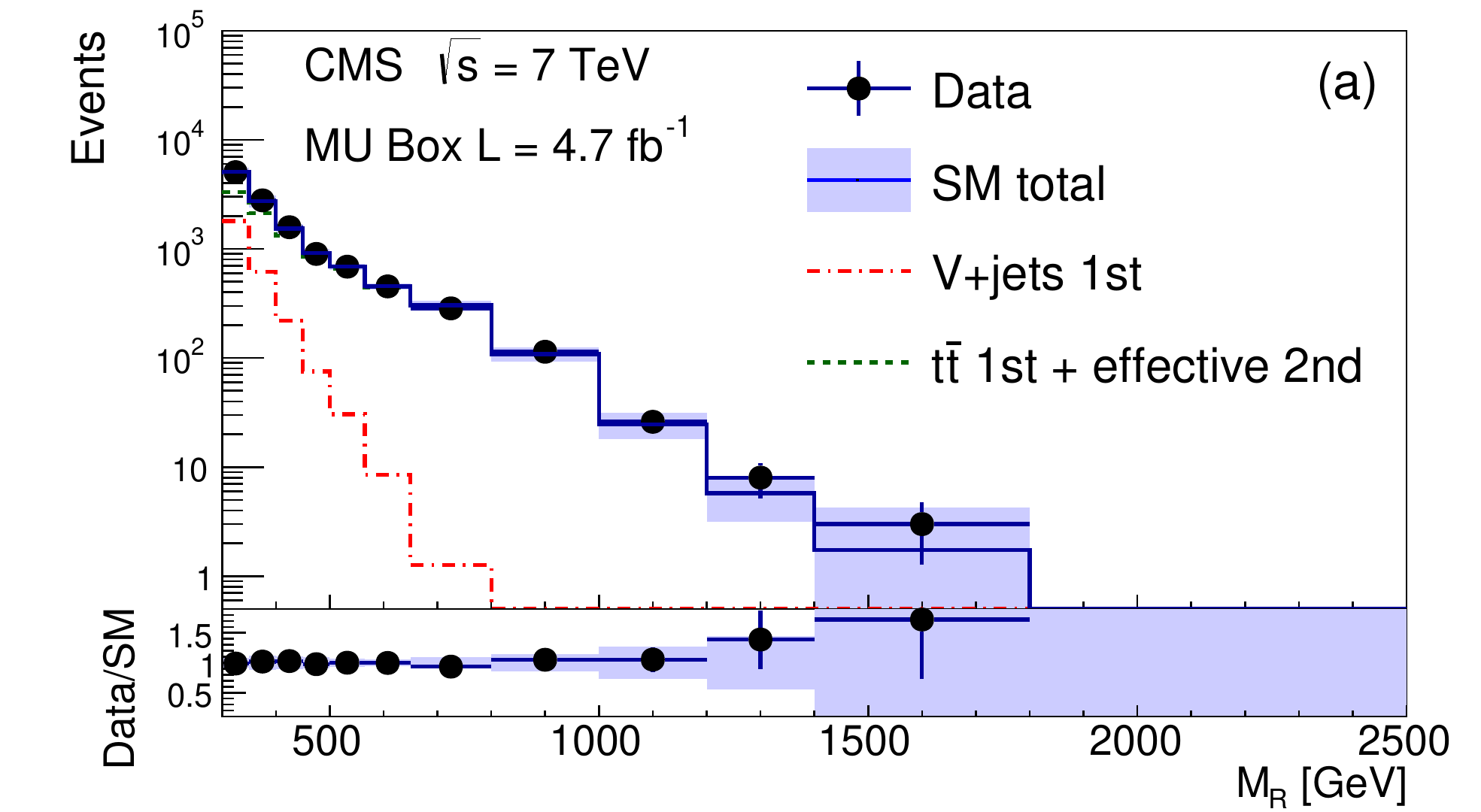}
\includegraphics[width=0.495\textwidth]{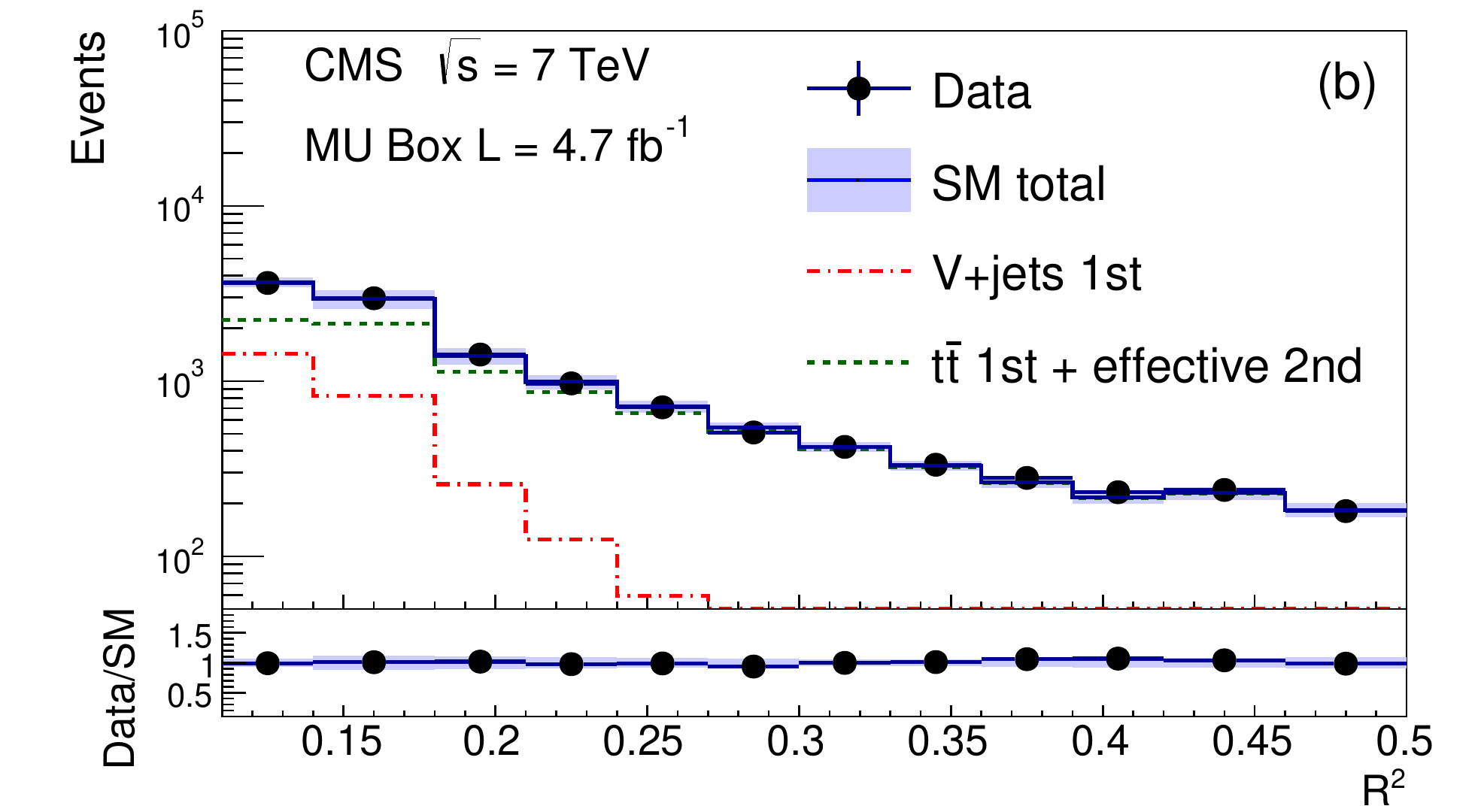}
\caption{Projection of the 2D fit result on (a) $\MR$ and (b)
  $\Rtwo$ for the inclusive MU box. The fit is performed in the
  ($\MR$, $\Rtwo$) fit region (shown in
  Fig.~\ref{fig:mu-blue-plot}) and projected into the full analysis
  region. The histograms are described in the text.\label{fig:mu-box}}
\end{figure}
\begin{figure}[ht!]
\centering
\includegraphics[width=\cmsFigWidthBox]{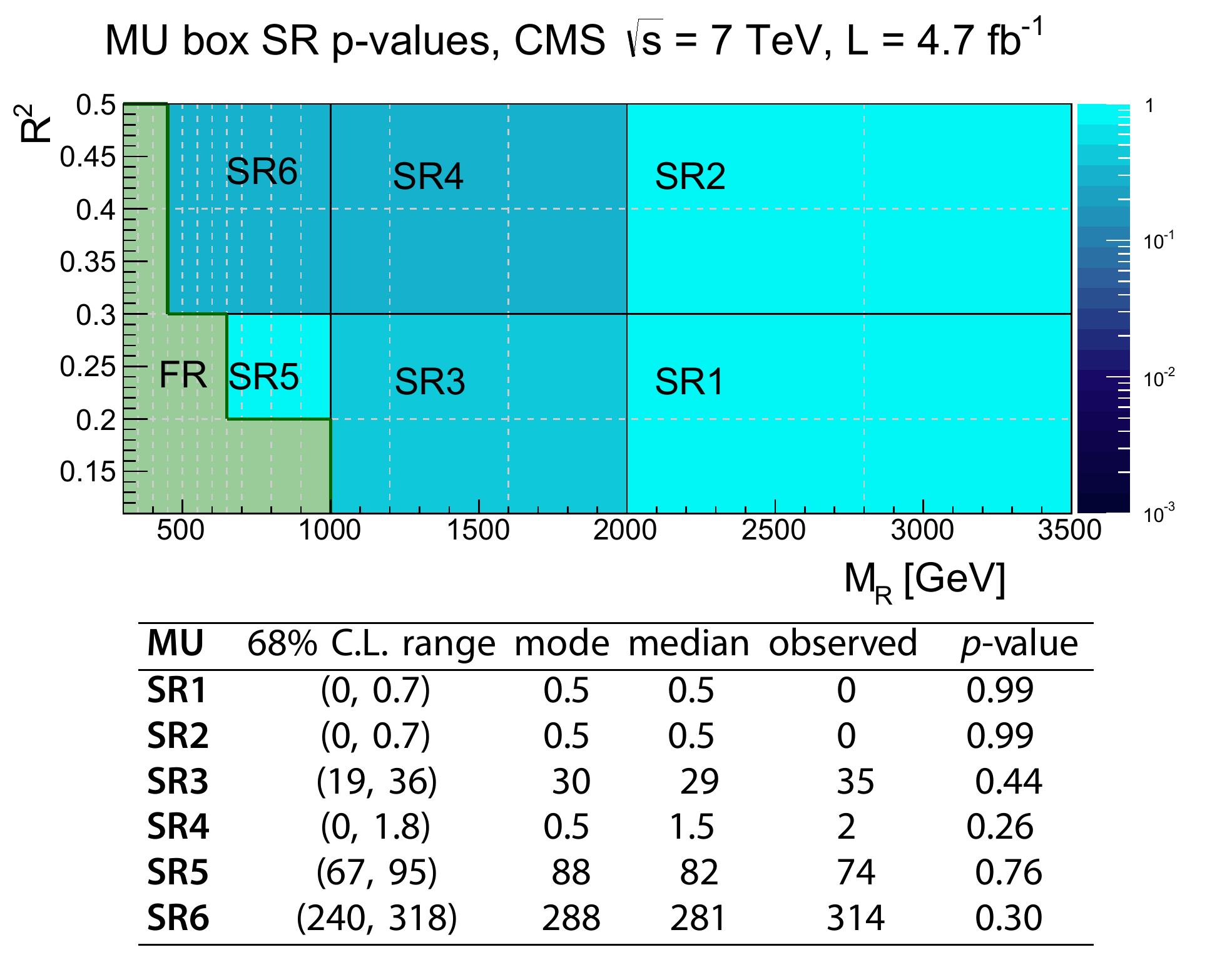}
\caption{The fit region, FR, and signal regions, SR$i$, are defined in
  the ($\MR$, $\Rtwo$) plane for the MU box. The
  color scale gives the $p$-values corresponding to the observed
  number of events in each SR$i$. Further explanation is given in the
  Fig.~\ref{fig:had-blue-plot} caption.\label{fig:mu-blue-plot}}
\end{figure}

\begin{figure}[ht!]
\centering
\includegraphics[width=0.495\textwidth]{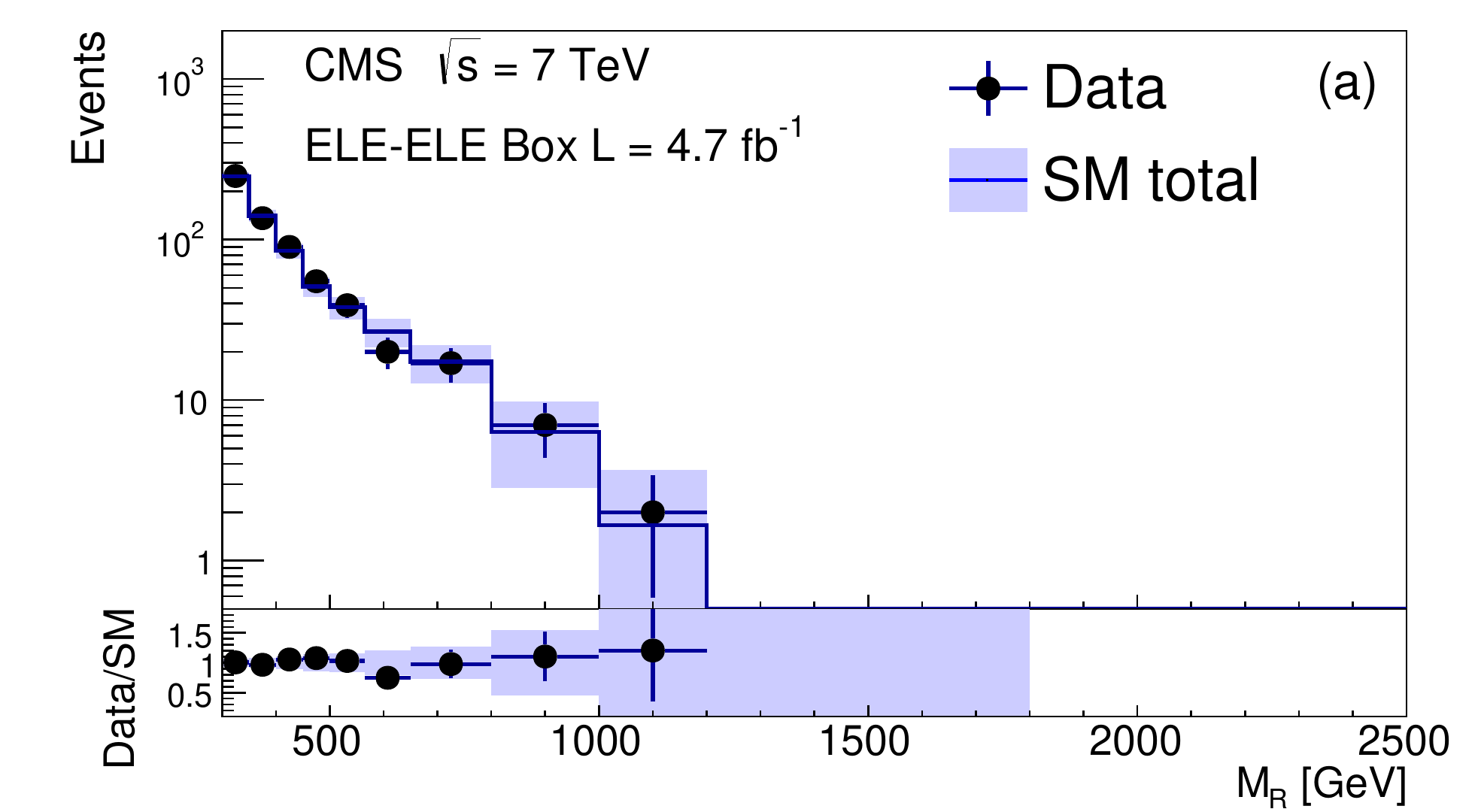}
\includegraphics[width=0.495\textwidth]{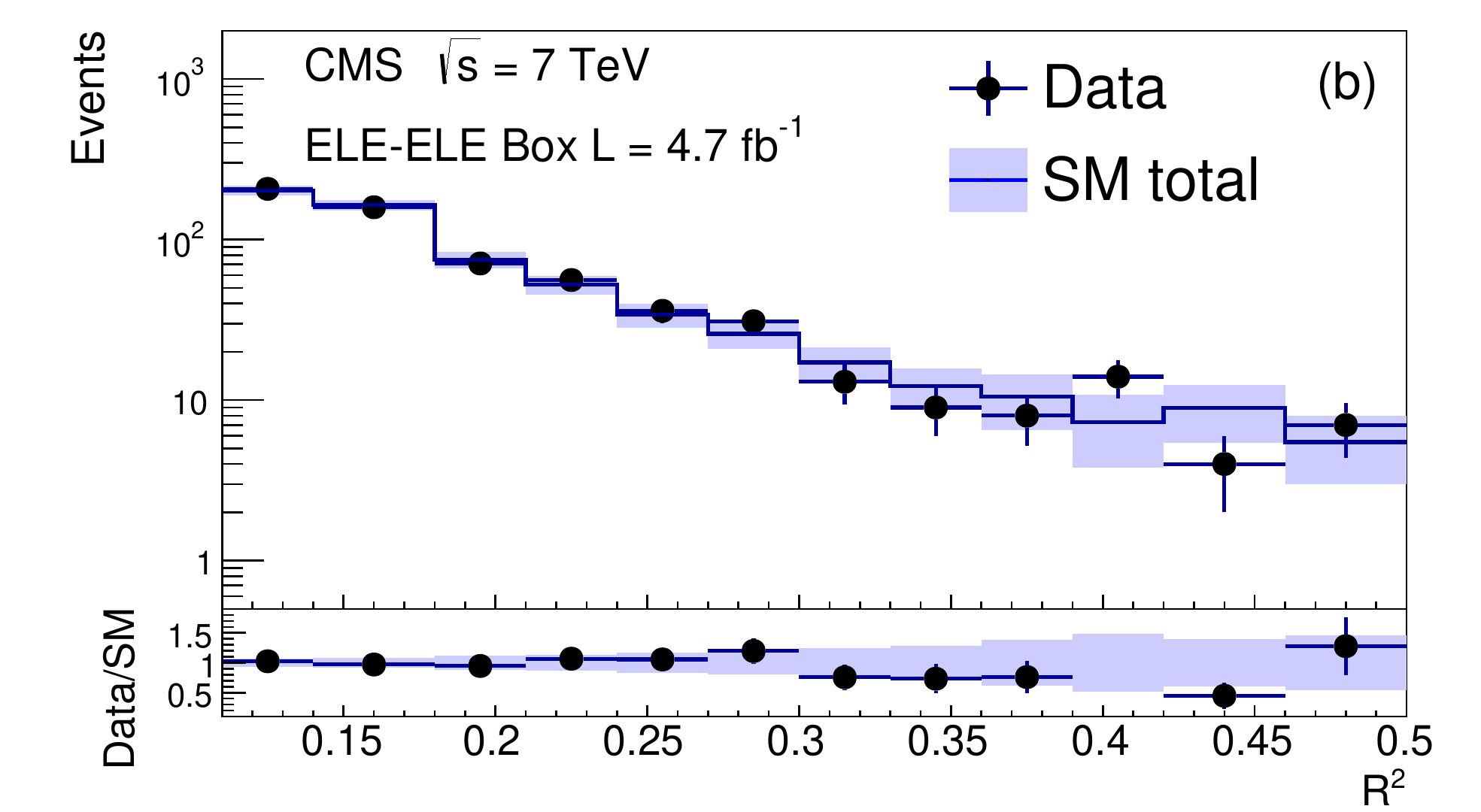}
\caption{Projection of the 2D fit result on (a) $\MR$ and (b)
  $\Rtwo$ for the ELE-ELE box. The continuous histogram is the
  total standard model prediction. The histogram is described in the
  text.\label{fig:eleele-box}}
\end{figure}
\begin{figure}[ht!]
\centering
\includegraphics[width=\cmsFigWidthBox]{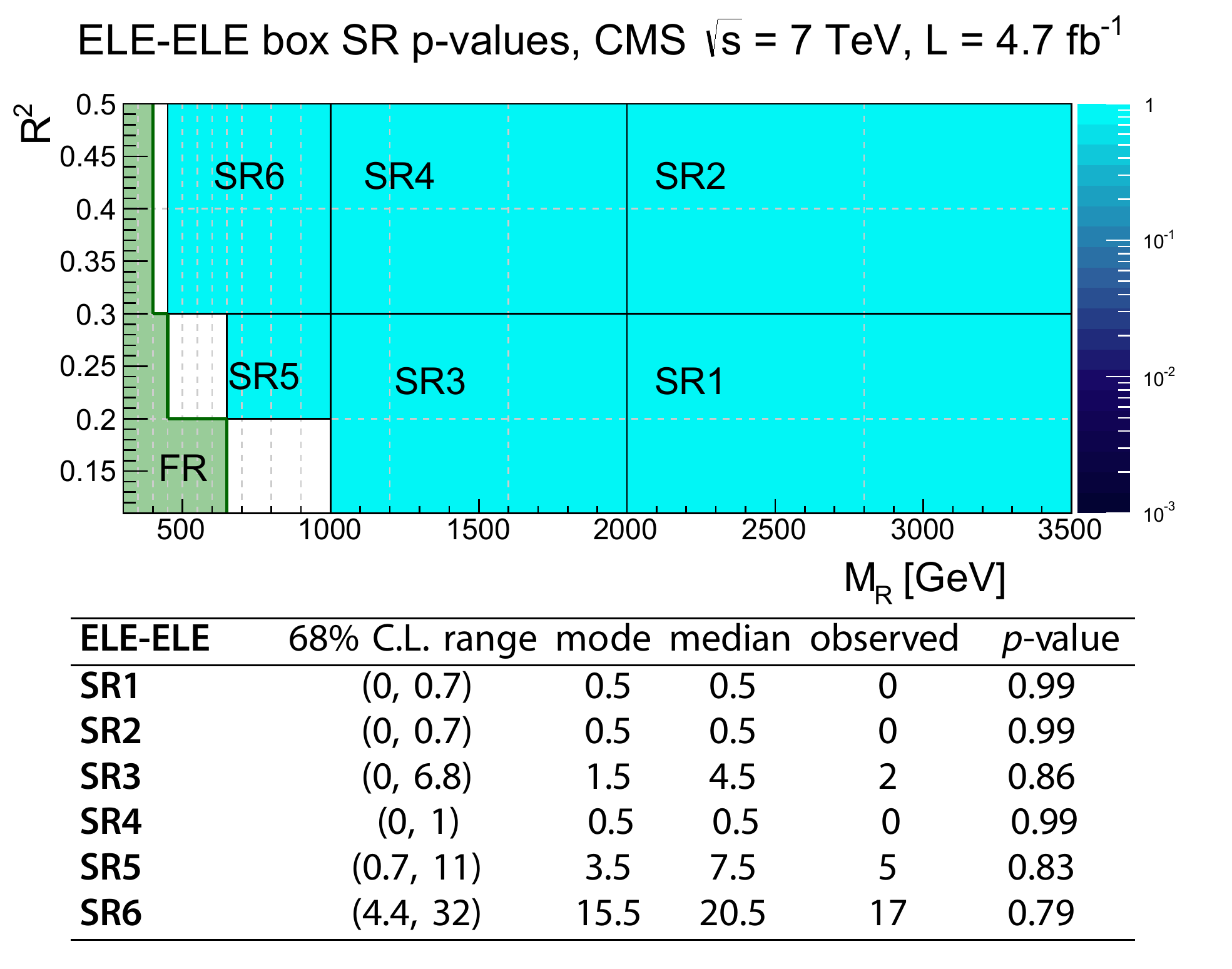}
\caption{The fit region, FR, and signal regions, SR$i$, are defined in
  the ($\MR$, $\Rtwo$) plane for the ELE-ELE box. The
  color scale gives the $p$-values corresponding to the observed
  number of events in each SR$i$. Further explanation is given in the
  Fig.~\ref{fig:had-blue-plot} caption.\label{fig:eleele-blue-plot}}

\end{figure}

\begin{figure}[ht!]
\centering
\includegraphics[width=0.495\textwidth]{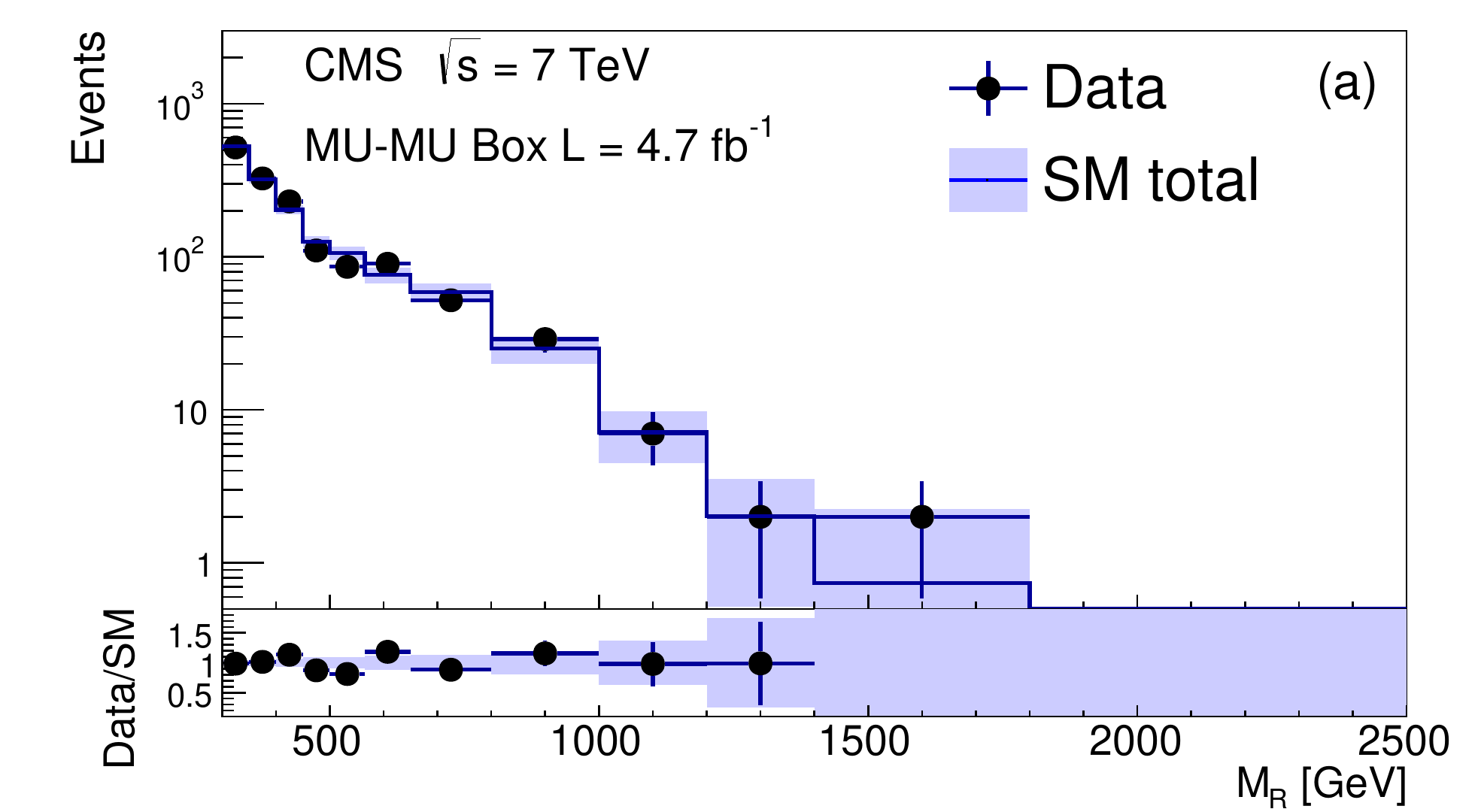}
\includegraphics[width=0.495\textwidth]{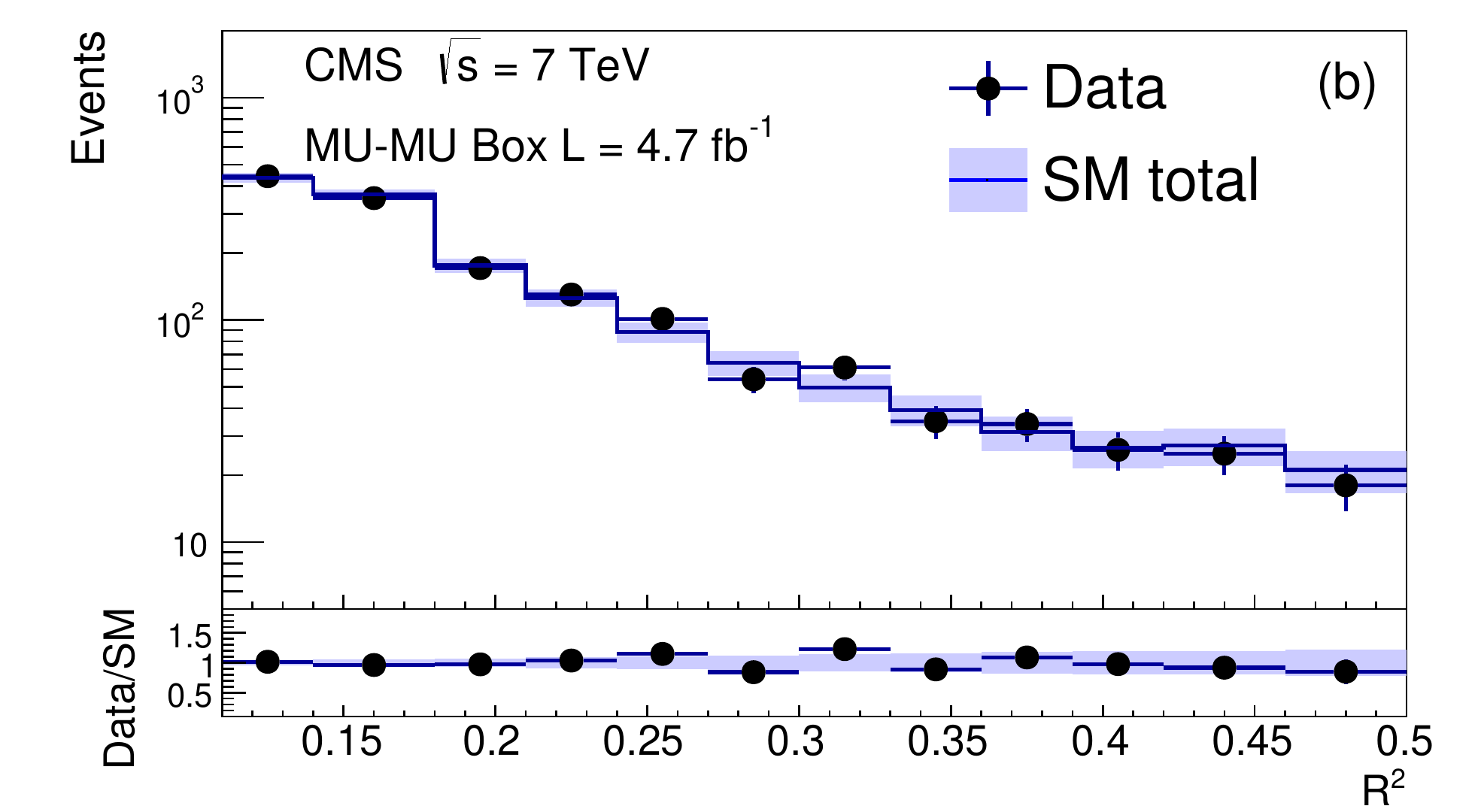}
\caption{Projection of the 2D fit result on (a) $\MR$ and (b)
  $\Rtwo$ for the inclusive MU-MU box. The fit is performed in the
  ($\MR$, $\Rtwo$) fit region (shown in
  Fig.~\ref{fig:mumu-blue-plot}) and projected into the full analysis
  region. The histogram is described in the
  text.\label{fig:mumu-box}}

\end{figure}
\begin{figure}[ht!]
\centering
\includegraphics[width=\cmsFigWidthBox]{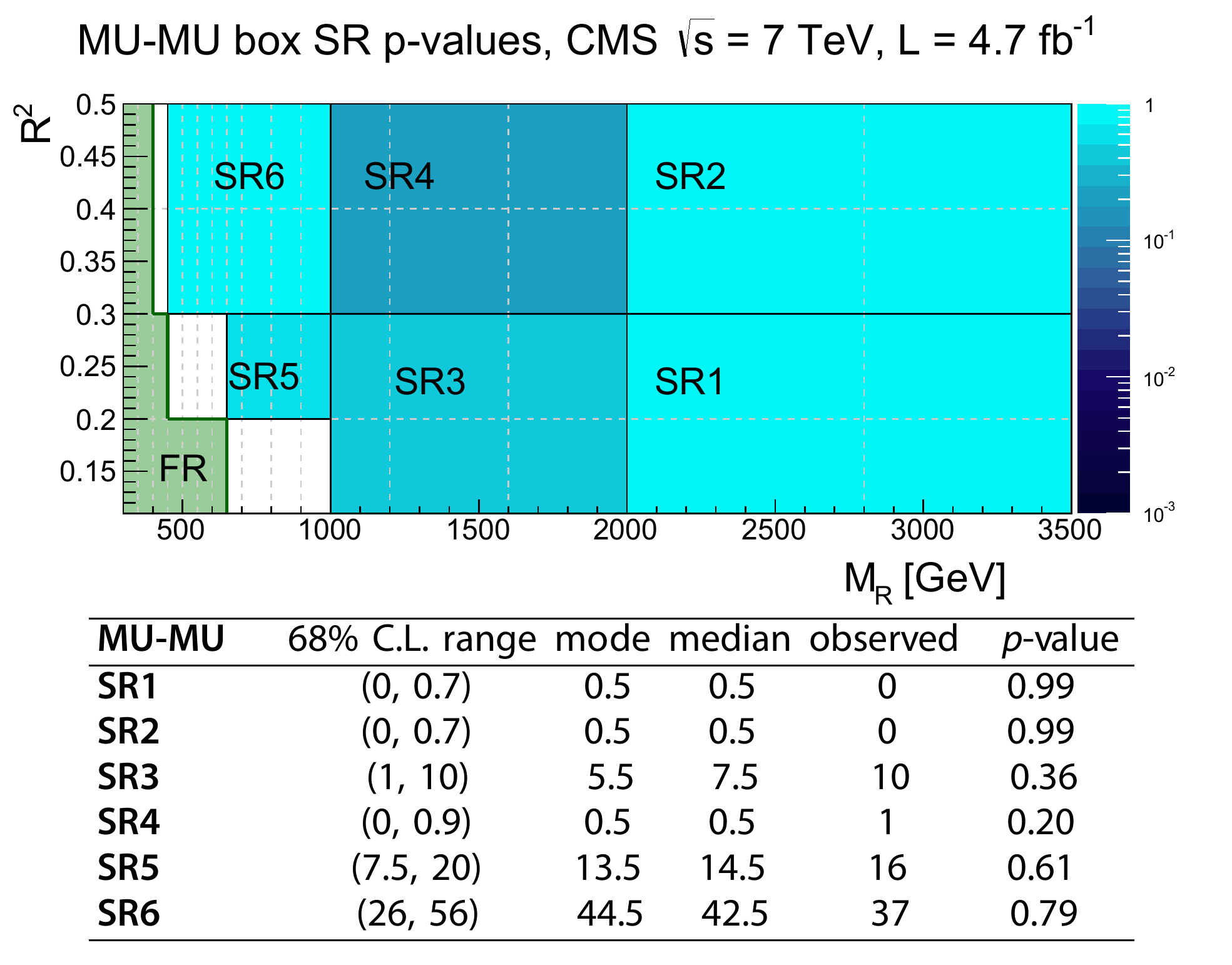}
\caption{The fit region, FR, and signal regions, SR$i$, are defined in
  the ($\MR$, $\Rtwo$) plane for the MU-MU box. The
  color scale gives the $p$-values corresponding to the observed
  number of events in each SR$i$. Further explanation is given in the
  Fig.~\ref{fig:had-blue-plot} caption.\label{fig:mumu-blue-plot}}
\end{figure}

\begin{figure}[ht!]
\centering
\includegraphics[width=0.495\textwidth]{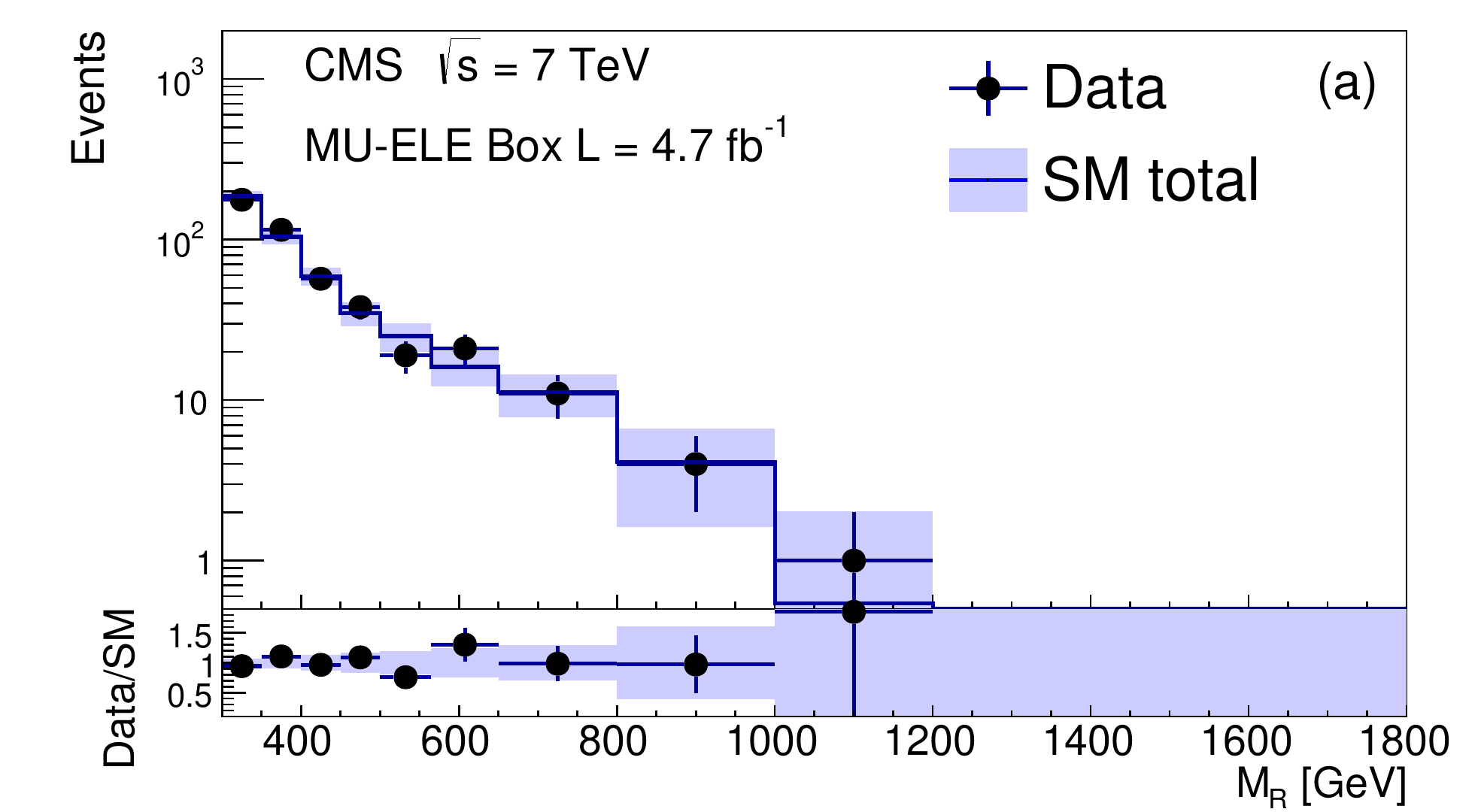}
\includegraphics[width=0.495\textwidth]{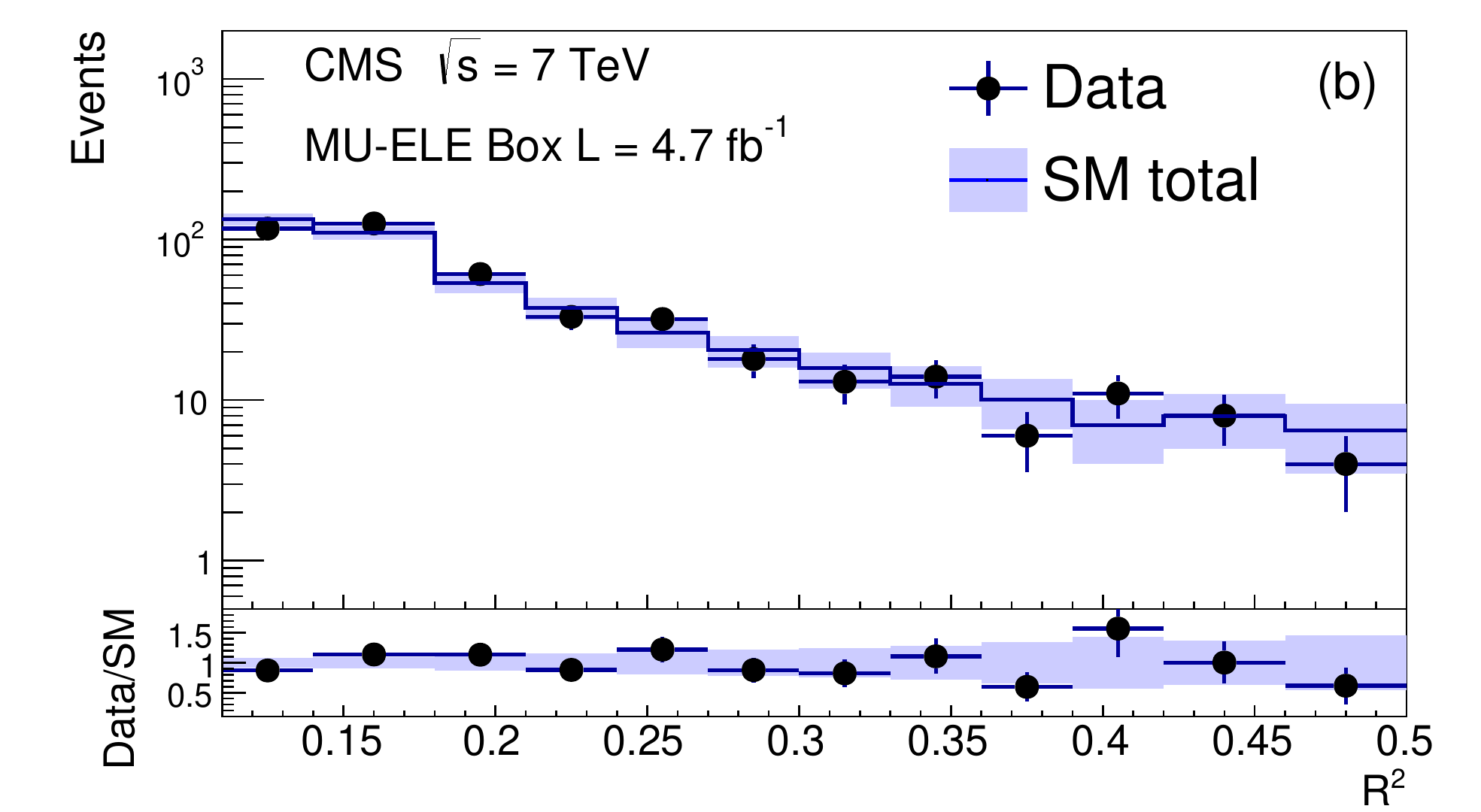}
\caption{Projection of the 2D fit result on (a) $\MR$ and (b)
  $\Rtwo$ for the inclusive MU-ELE box. The fit is performed in the
  ($\MR$, $\Rtwo$) fit region (shown in
  Fig.~\ref{fig:muele-blue-plot}) and projected into the full analysis
  region. The histogram is described in the
  text.\label{fig:muele-box}}
\end{figure}
\begin{figure}[ht!]
\centering
\includegraphics[width=\cmsFigWidthBox]{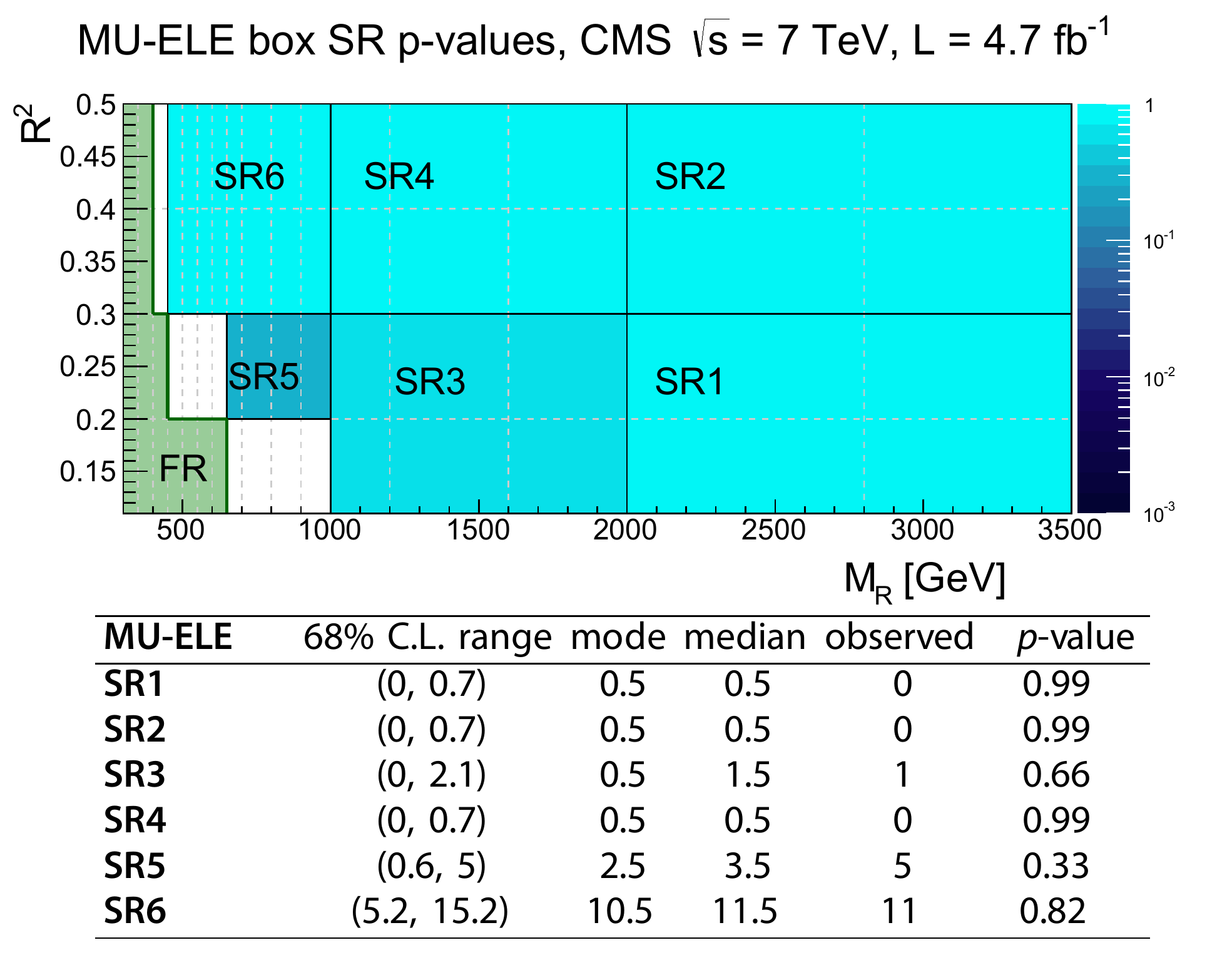}
\caption{The fit region, FR, and signal regions, SR$i$, are defined in
  the ($\MR$, $\Rtwo$) plane for the MU-ELE box. The
  color scale gives the $p$-values corresponding to the observed
  number of events in each SR$i$. Further explanation is given in the
  Fig.~\ref{fig:had-blue-plot} caption. \label{fig:muele-blue-plot}}

\end{figure}

The result of the ML fit projected on $\MR$ and $\Rtwo$ is shown in
Fig.~\ref{fig:had-box} for the inclusive HAD box.  No significant
discrepancy is observed between the data and the fit model for any of
the six boxes.  In order to establish the compatibility of the
background model with the observed dataset, we define a set of signal
regions (SR${i}$) in the tail of the SM background distribution. Using
the 2D background model determined using the ML fit, we derive the
distribution of the expected yield in each SR${i}$ using
pseudo-experiments, accounting for correlations and uncertainties in
the parameters describing the background model. In order to correctly
account for the uncertainties in the parameters describing the
background model and their correlations, the shape parameters used to
generate each pseudo-experiment dataset are sampled from the
covariance matrix returned by the ML fit performed on the actual
dataset. The actual number of events in each dataset is drawn from a
Poisson distribution centered on the yield returned by the covariance
matrix sampling. For each pseudo-experiment dataset, the number of
events in the SR$i$ is found. For each of the SR${i}$, the
distribution of the number of events derived by the pseudo-experiments
is used to calculate a two-sided $p$-value (as shown for the HAD box
in Fig.~\ref{fig:had-blue-plot}), corresponding to the probability of
observing an equal or less probable outcome for a counting experiment
in each signal region. The result of the ML fit and the corresponding
$p$-values are shown in Figs.~\ref{fig:ele-box} and
\ref{fig:ele-blue-plot} for the ELE box, Figs.~\ref{fig:mu-box} and
\ref{fig:mu-blue-plot} for the MU box, Figs.~\ref{fig:eleele-box} and
\ref{fig:eleele-blue-plot} for the ELE-ELE box,
Figs.\ref{fig:mumu-box} and \ref{fig:mumu-blue-plot} for the MU-MU
box, and Figs.~\ref{fig:muele-box} and \ref{fig:muele-blue-plot} for
the MU-ELE box.  We note that the background shapes in the
single-lepton and hadronic boxes are well described by the sum of two
functions: a single-component function with a steeper-slope component,
denoted as the $\mathrm{V}$+jets first component, obtained by fixing
$f_2=0$ in Eq.~(\ref{eq:twocomponents}); and a two-component function
as in Eq.~(\ref{eq:twocomponents}), with the first component
describing the steeper-slope core of the \ttbar and single-top
background distributions (generically referred to as \ttbar), and the
{\it effective} second component modeling the sum of the
indistinguishable tails of different SM background processes. In the
dilepton boxes we show the total SM background, which is composed of
$\mathrm{V}$+jets and \ttbar events in the ELE-ELE and MU-MU boxes and
of \ttbar events in the MU-ELE boxes. The corresponding results for
the $\geq$1 $\cPqb$-tagged samples are presented in
Appendix~\ref{sec:datafitsBTAGAppendix}.

\section{Signal systematic uncertainties}\label{sec:systematics}
We evaluate the impact of systematic uncertainties on the shape of the
signal distributions, for each point of each SUSY model, using the
simulated signal event samples. The following systematic uncertainties
are considered, with the approximate size of the uncertainty given in
parentheses: (i) PDFs (up to 30\%, evaluated
point-by-point))~\cite{Bourilkov:2006cj}; (ii) jet-energy scale (up to
1\%, evaluated point-by-point)~\cite{JES}; (iii) lepton
identification, using the ``tag-and-probe'' technique based on $\cPZ
\to \ell \ell$ events~\cite{CMS:2011aa} ($\ell=\Pe,\mu$, 1\% per
lepton). In addition, the following uncertainties, which affect the
signal yield, are considered: (i) luminosity
uncertainty~\cite{lumi-moriond} (2.2\%); (ii) theoretical cross
section~\cite{Kramer:2012bx} (up to 15\%, evaluated point-by-point);
(iii) razor trigger efficiency (2\%); (iv) lepton trigger efficiency
(3\%). An additional systematic uncertainty is considered for the
$\cPqb$-tagging efficiency~\cite{BTAG} (between 6\% and 20\% in $\pt$
bins).  We consider variations of the function modeling, the signal
uncertainty (log-normal versus Gaussian), and the binning, and find
negligible deviations in the results.  The systematic uncertainties
are included using the best-fit shape to compute the likelihood values
for each pseudo-experiment, while sampling the same pseudo-experiment
from a different function, derived from the covariance matrix
of the fit to the data. This procedure is repeated for both the
background and signal probability density functions.

\section{Interpretation of the results\label{sec:INTERP2011}}

In order to evaluate exclusion limits for a given SUSY model, its
parameters are varied and an excluded cross section at the 95\% \CL is
associated with each configuration of the model parameters, using the
hybrid version of the \CLs method~\cite{CLs1,CLs2,CLs3}, described below.

For each box, we consider the test statistic given by the logarithm of
the likelihood ratio $\ln Q = \ln[\mathcal{L}(s+b|H_i)/\mathcal{L}(b|H_i)]$, where $H_i~(i=0,2)$ is the
hypothesis under test: $H_1$ (signal-plus-background) or $H_0$
(background-only).  The likelihood function for the background-only
hypothesis is given by Eq.~(\ref{eq:Lb}). The likelihood corresponding
to the signal-plus-background hypothesis is written as
\ifthenelse{\boolean{cms@external}}{
\begin{multline}
\label{eq:Lsb}
\mathcal{L}_{s+b} =  \frac{\re^{-(\sum_{j \in SM} N_{j})}}{N !}
\prod_{i=1}^{N} \Bigg[\sum_{j \in SM} N_{j} P_{j}(M_{R,i},\Rtwo_i) \\+ \sigma
\times L \times \epsilon \ P_S (M_{R,i},\Rtwo_i) \Bigg],
\end{multline}
}{
\begin{equation}
\label{eq:Lsb}
\mathcal{L}_{s+b} =  \frac{\re^{-(\sum_{j \in SM} N_{j})}}{N !}
\prod_{i=1}^{N} \left[\sum_{j \in SM} N_{j} P_{j}(M_{R,i},\Rtwo_i) + \sigma
\times L \times \epsilon \ P_S (M_{R,i},\Rtwo_i) \right],
\end{equation}
}
where $\sigma$ is the signal cross section, \ie, the parameter of
interest; $L$ is the integrated luminosity; $\epsilon$ is the signal
acceptance times efficiency; and $P_S (M_{R,i},\Rtwo_i)$ is the
two-dimensional probability density function for the signal, computed
numerically from the distribution of simulated signal events.  The
signal and background shape parameters, and the normalization factors
$L$ and $\epsilon$, are the nuisance parameters.

For each analysis (inclusive razor or inclusive $\cPqb$-jet razor) we
sum the test statistics of the six corresponding boxes to compute the
combined test statistic.

The distribution of $\ln Q$ is derived numerically with a MC
technique. The values of the nuisance parameters in the likelihood are
randomized for each iteration of the MC generation, to reflect the
corresponding uncertainty. Once the likelihood is defined, a sample of
events is generated according to the signal and background probability
density functions. The value of $\ln Q$ for each generated
sample is then evaluated, fixing each signal and background parameter
to its expected value. This procedure corresponds to a numerical
marginalization of the nuisance parameters.

Given the distribution of $\ln Q$ for the background-only and
the signal-plus-background pseudo-experiments, and the value of
$\ln Q$ observed in the data, we calculate \CLsb
and $1-\CLb$ ~\cite{CLs1}. From these values,
$\CLs=\CLsb/\CLb$ is computed for that model point.
The procedure is independently applied to each of the two analyses
(inclusive razor and inclusive $\cPqb$-jet razor).

\begin{figure*}[htpb]
  \centering
    \includegraphics[width=0.495\textwidth]{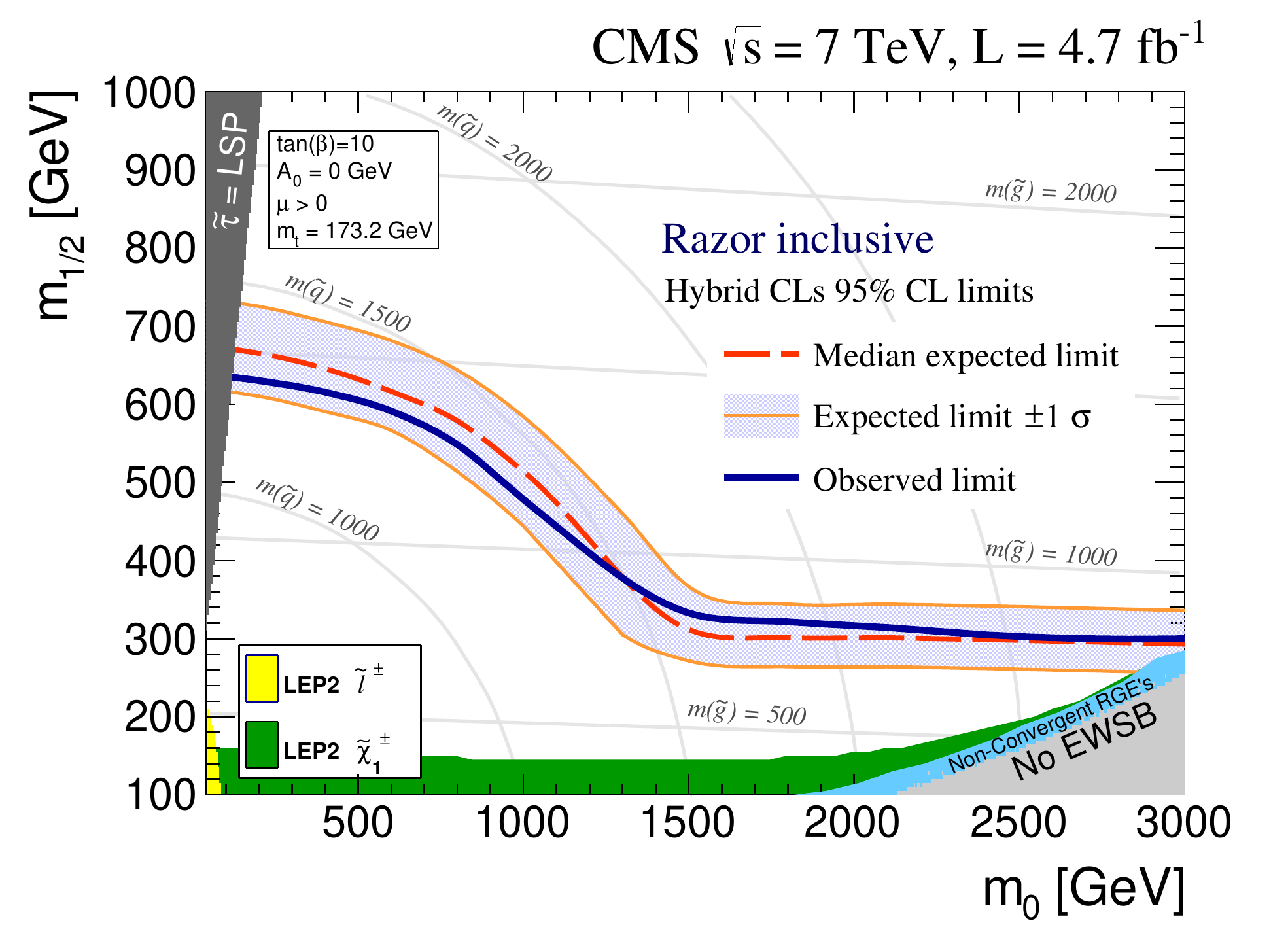}
    \includegraphics[width=0.495\textwidth]{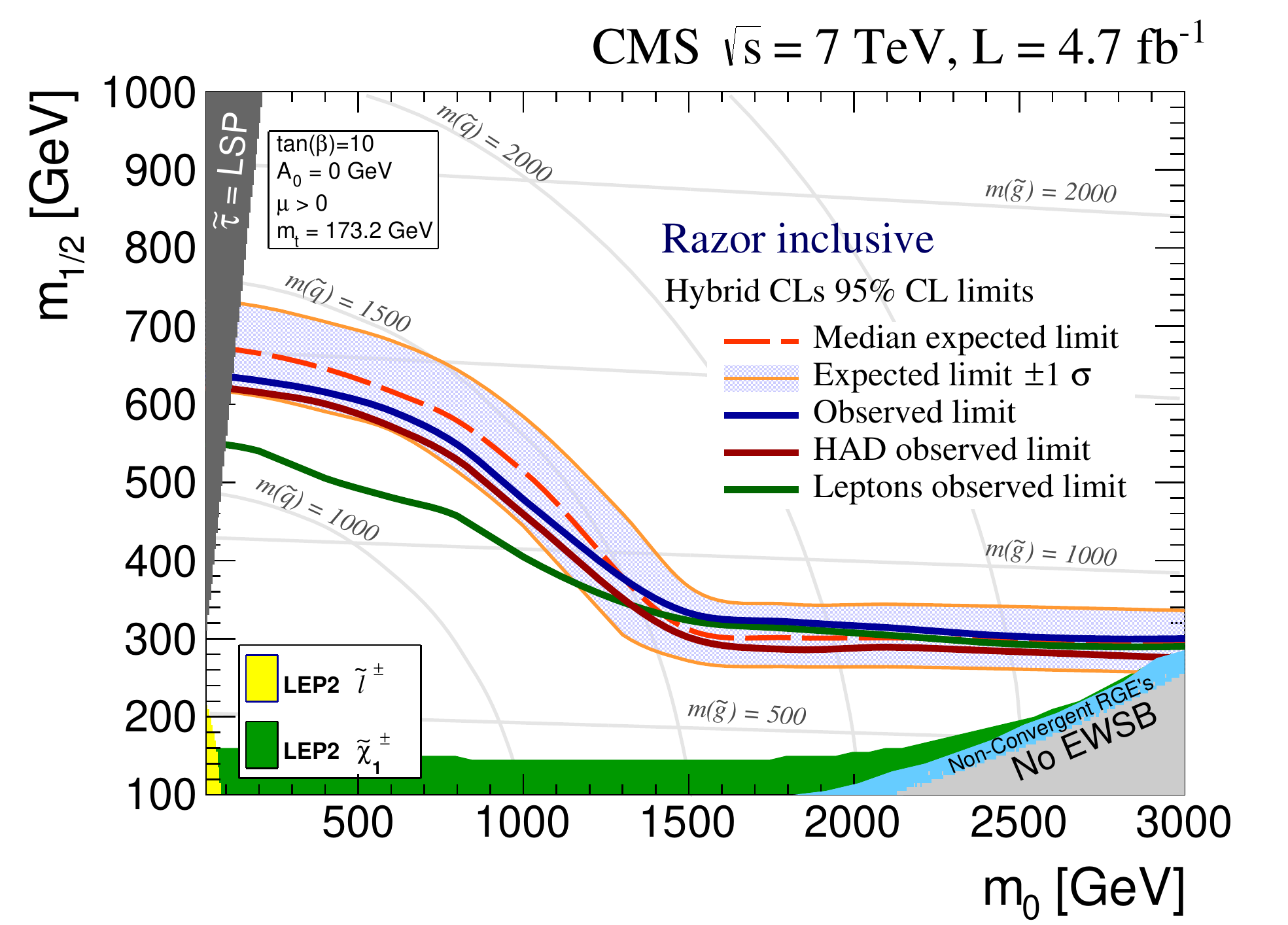}
    \includegraphics[width=0.495\textwidth]{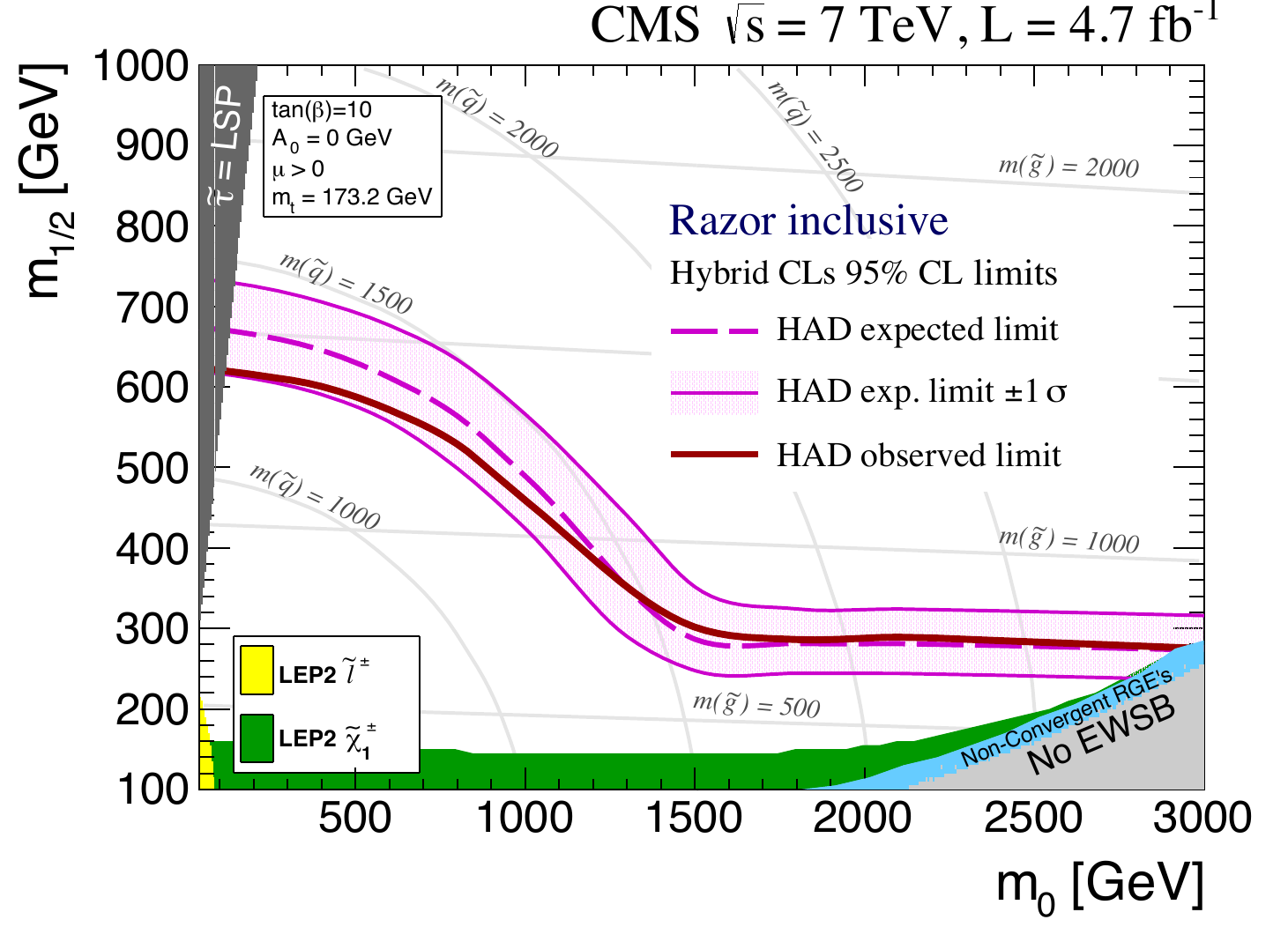}
    \includegraphics[width=0.495\textwidth]{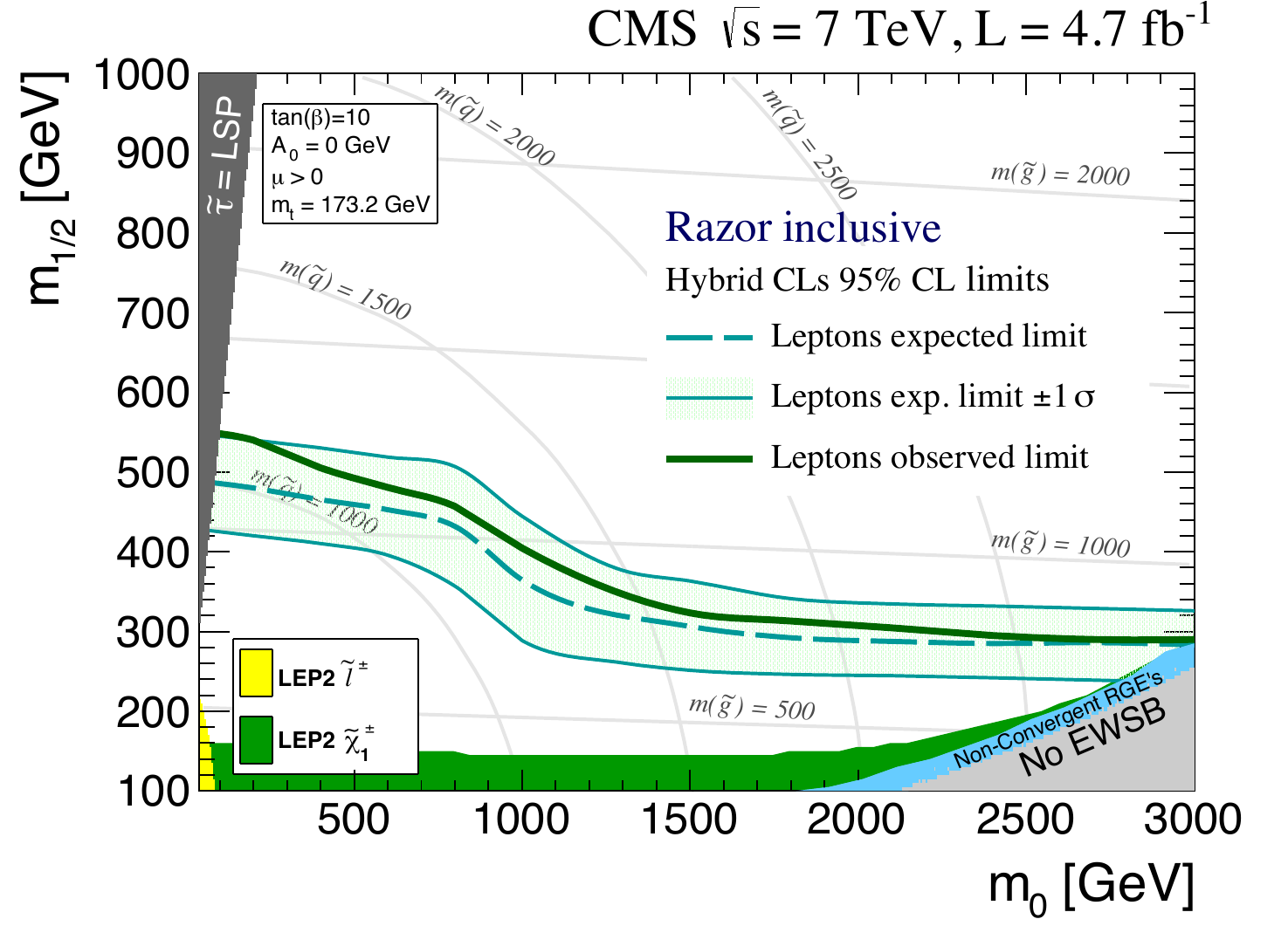}
 \caption{(upper left) Observed (solid curve) and median expected (dashed
   orange curve) 95\% \CL limits in the ($\mzero$, $\mhalf$) CMSSM
   plane (drawn according to Ref.~\cite{Rony}) with $\tan\beta=10$, $A_{0}
   = 0$, and $\sgn(\mu) = {+}1$. The ${\pm}$1
   standard-deviation equivalent variations due to the uncertainties are shown as a
   band around the median expected limit. (upper right) The observed HAD-only
   (solid red) and leptonic-only (solid green) 95\% \CL limits are
   shown, compared to the combined limit (solid blue curve).  The
   expected (dashed curve) and observed (solid curve) limits for the
   (lower left) HAD-only and (lower right) leptonic boxes only are also shown.
   \label{fig:all-result}}
\end{figure*}

The CMSSM model is studied in the (${m_{0}}$,
$\mhalf$) plane, fixing $\tan\beta=10$, $A_{0} = 0$, and
$\sgn(\mu) = +1$. A point in the plane is excluded at
the 95\% \CL if $\CLs < 0.05$. The result obtained for the
inclusive razor analysis is shown in Fig.~\ref{fig:all-result} (a).
The shape of the observed exclusion curves reflects the changing
relevant SUSY strong-production processes across the parameter space,
with squark-antisquark and gluino-gluino production dominating at low
and high $\mzero$, respectively. The observed limit is less
constraining than the median expected limit at lower $\mzero$
due to a local excess of events at large $\Rtwo$ in the
hadronic box.

For large values of $\mzero$, boxes with leptons in the final
state have a sensitivity comparable to that of the hadronic boxes, as
cascade decays of gluinos yield leptons production.
Figure~\ref{fig:all-result} parts (b)-(d) show the CMSSM exclusion
limits based on the HAD box only and on the leptonic boxes only.

\begin{figure*}[htpb]
\centering
\includegraphics[width=0.45\textwidth]{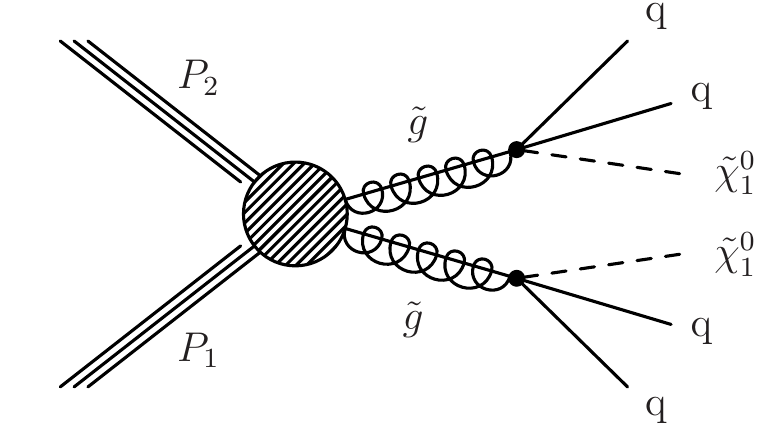}
\includegraphics[width=0.45\textwidth]{T2.pdf}
\includegraphics[width=0.45\textwidth]{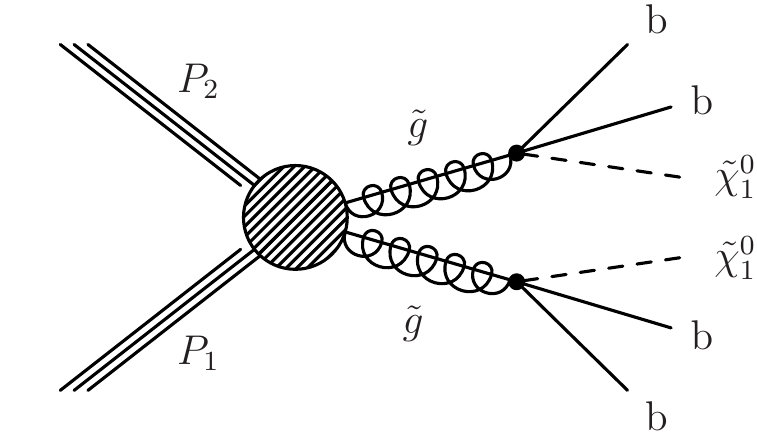}
\includegraphics[width=0.45\textwidth]{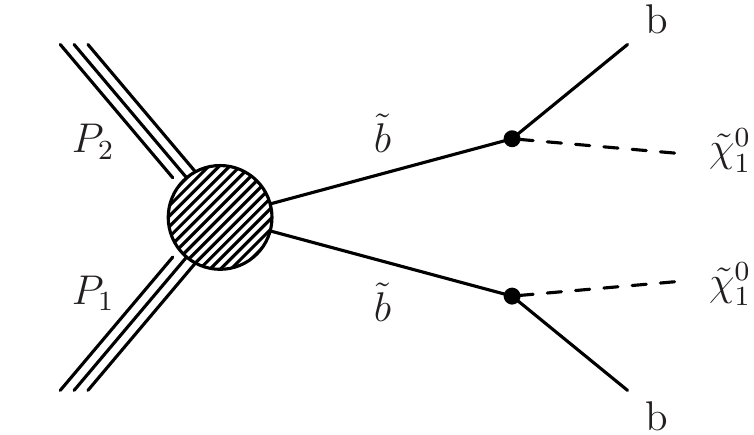}
\includegraphics[width=0.45\textwidth]{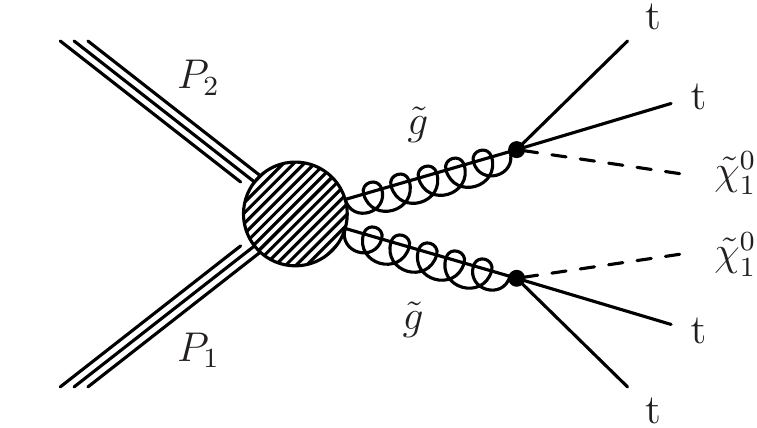}
\includegraphics[width=0.45\textwidth]{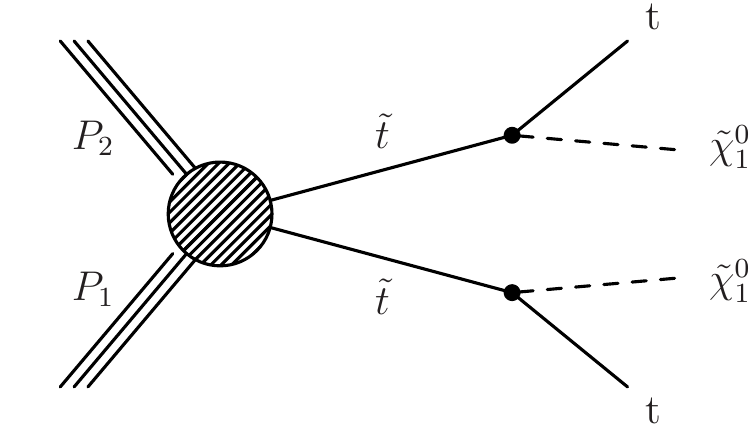}
\caption{The diagrams corresponding to the SMS models considered in this analysis.\label{fig:SMSdecay}}

\end{figure*}

\begin{figure*}[htpb]
\centering
\includegraphics[width=0.495\textwidth]{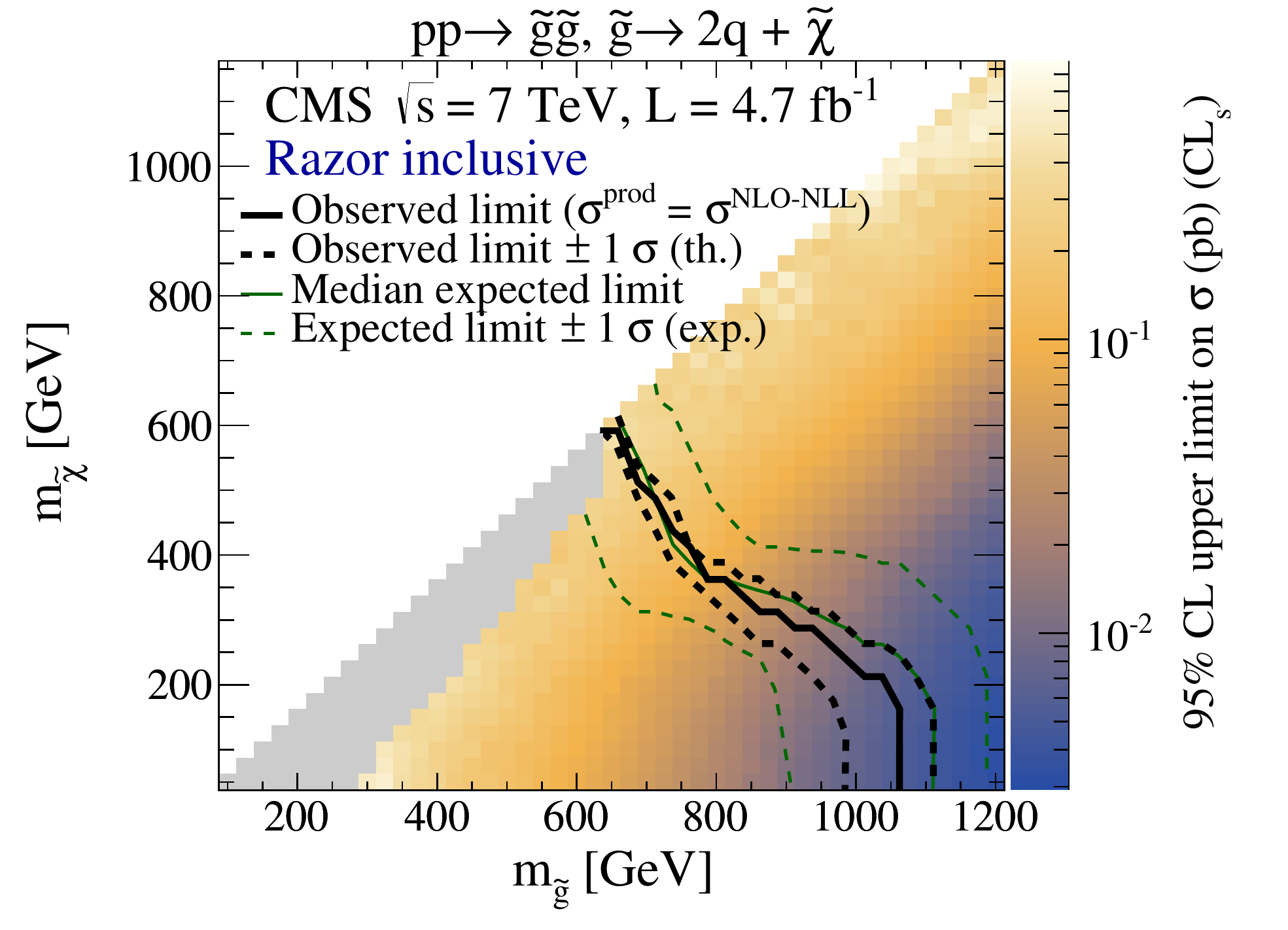}
\includegraphics[width=0.495\textwidth]{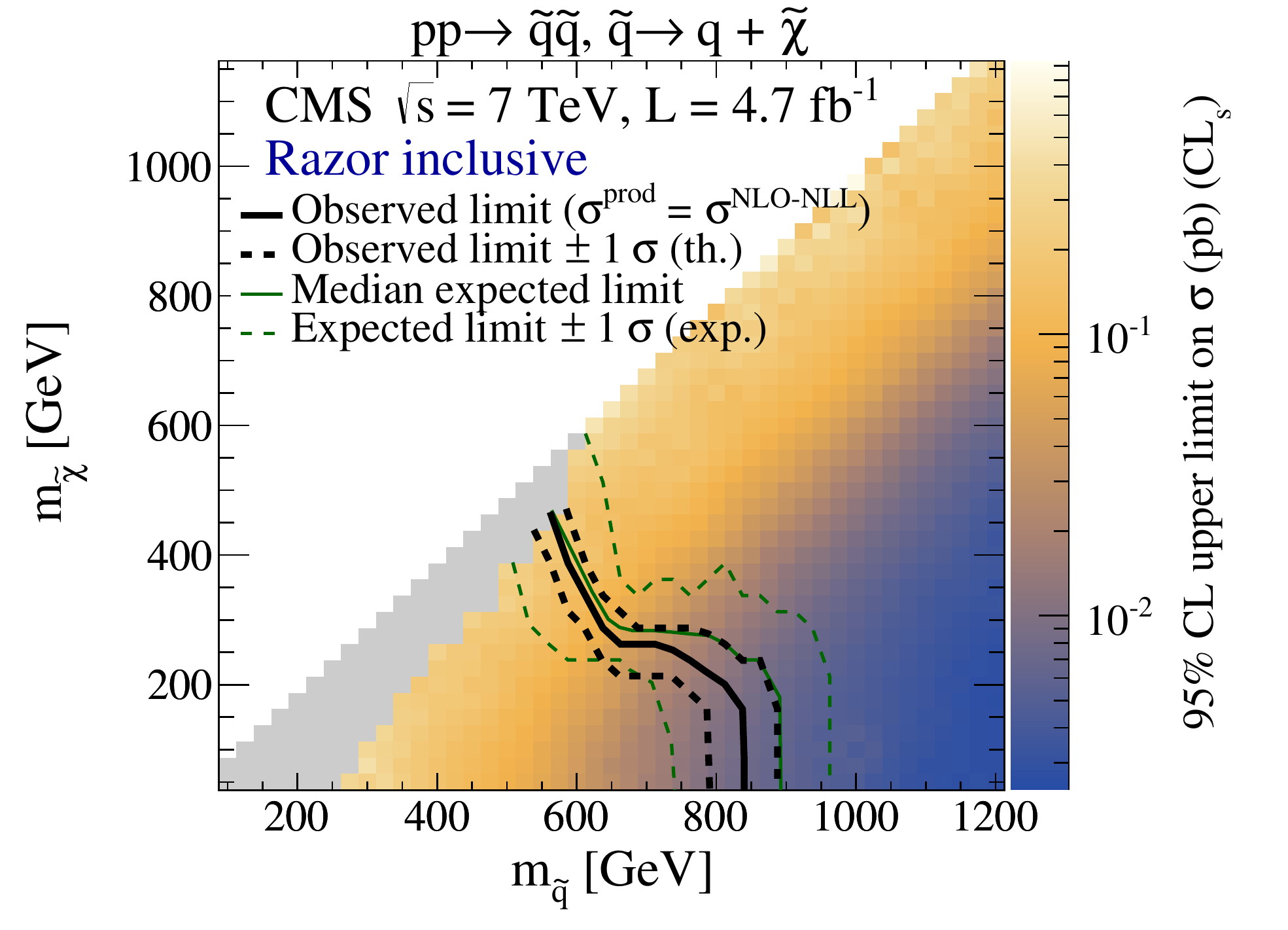}
\includegraphics[width=0.495\textwidth]{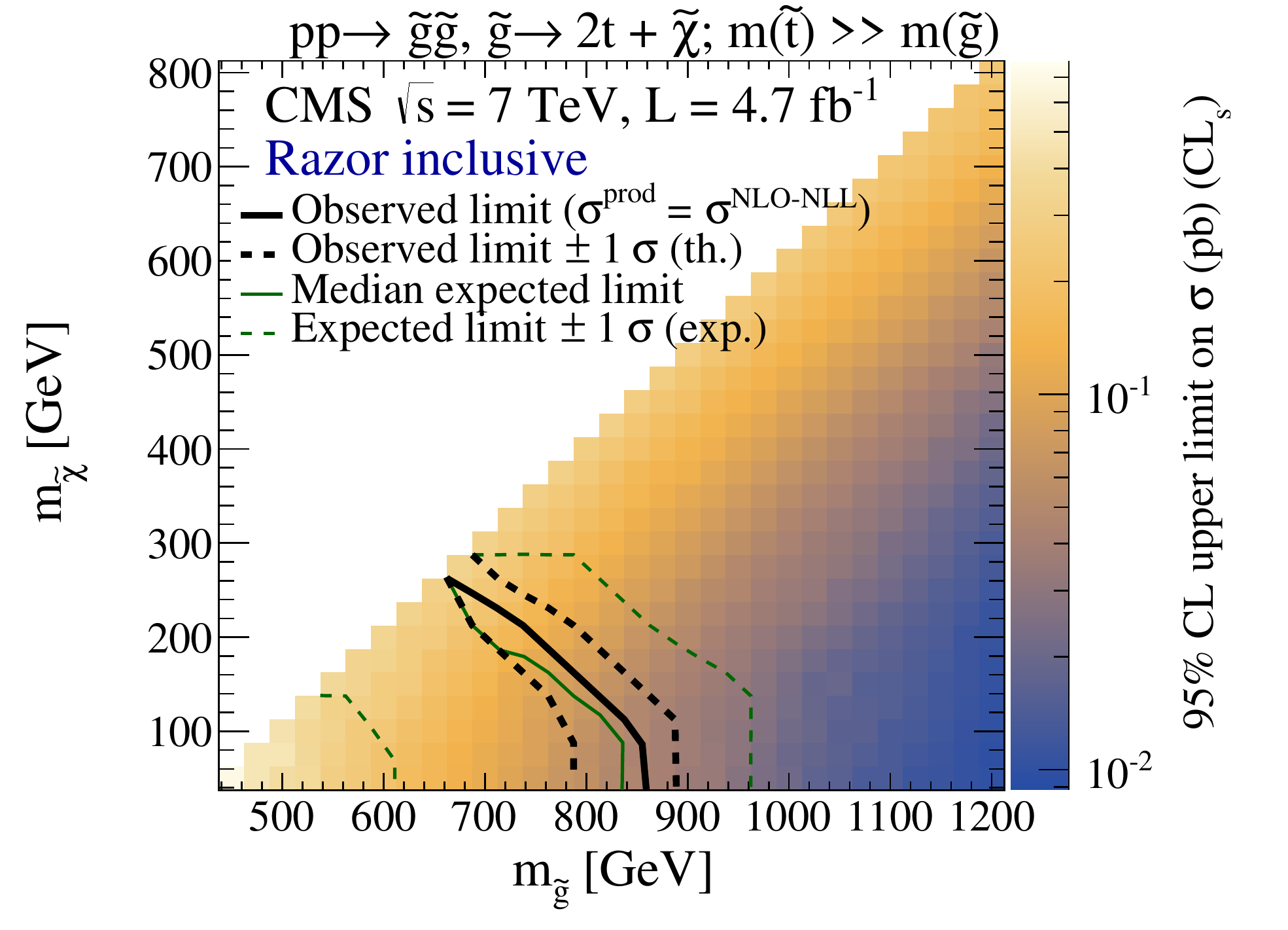}
\includegraphics[width=0.495\textwidth]{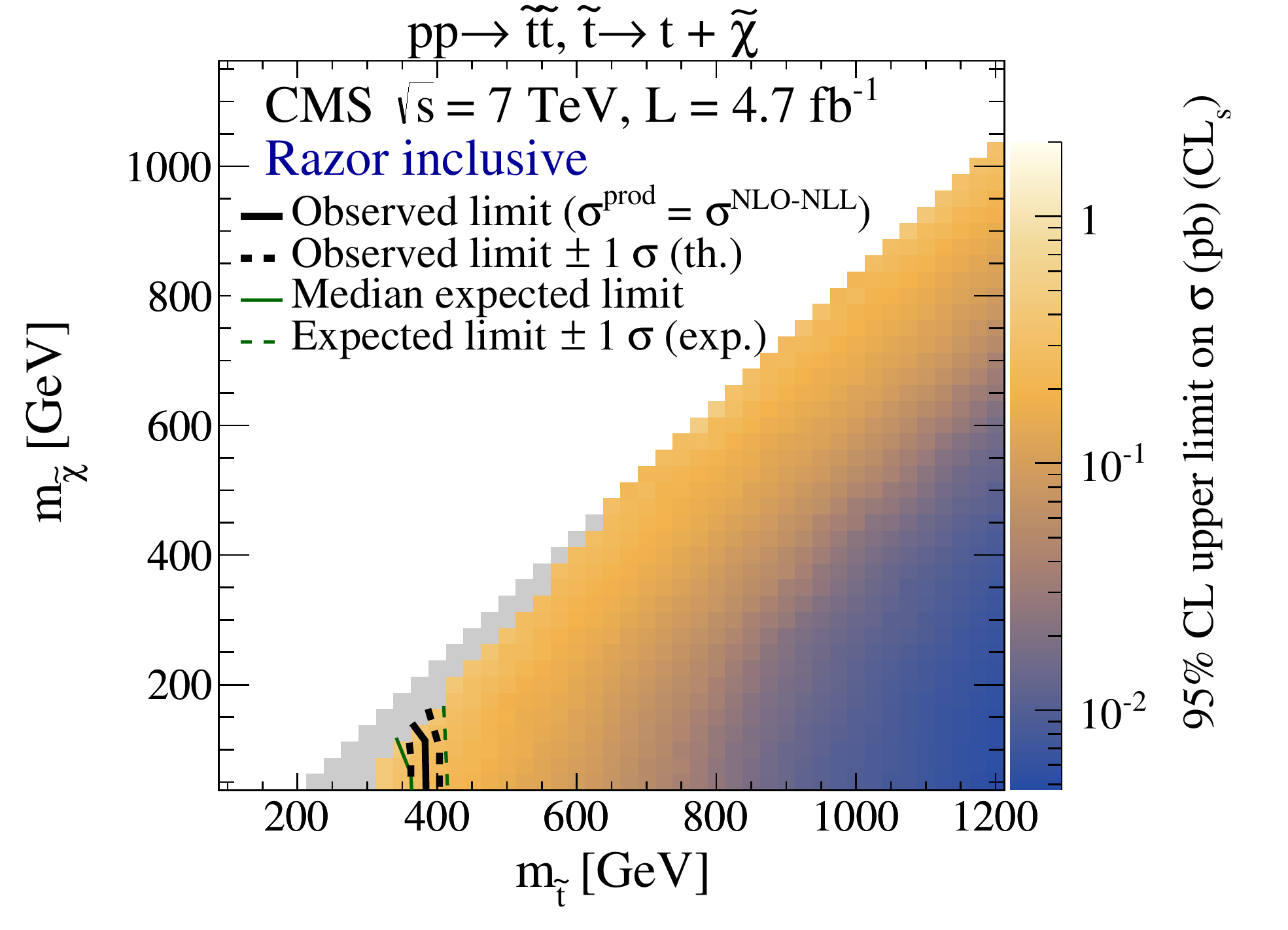}
\caption{Cross section upper limits, in pb, at 95\% \CL (color scale),
   in the mass plane of the produced superparticles for (a) T1, (b)
   T2, (c) T1tttt, and (d) T2tt, for the inclusive razor analysis.
   The solid black line indicates the observed exclusion region,
   assuming the nominal NLO+NLL SUSY production cross section. The
   dotted black lines show the observed exclusion taking ${\pm}1$
   standard deviation theoretical uncertainties around the nominal
   cross section. The solid green line indicates the median expected
   exclusion region, with dotted green lines indicating the expected
   exclusion with ${\pm}1$ standard deviation experimental
   uncertainties. The solid gray region indicates model points where
   the selection efficiency is found to have dependence on ISR
   modeling in the simulation of signal events above a predefined
   tolerance; no interpretation is presented for these model
   points.\label{fig:sms-incl0}}
\end{figure*}

\begin{figure*}[ht!]
\centering
\includegraphics[width=0.495\textwidth]{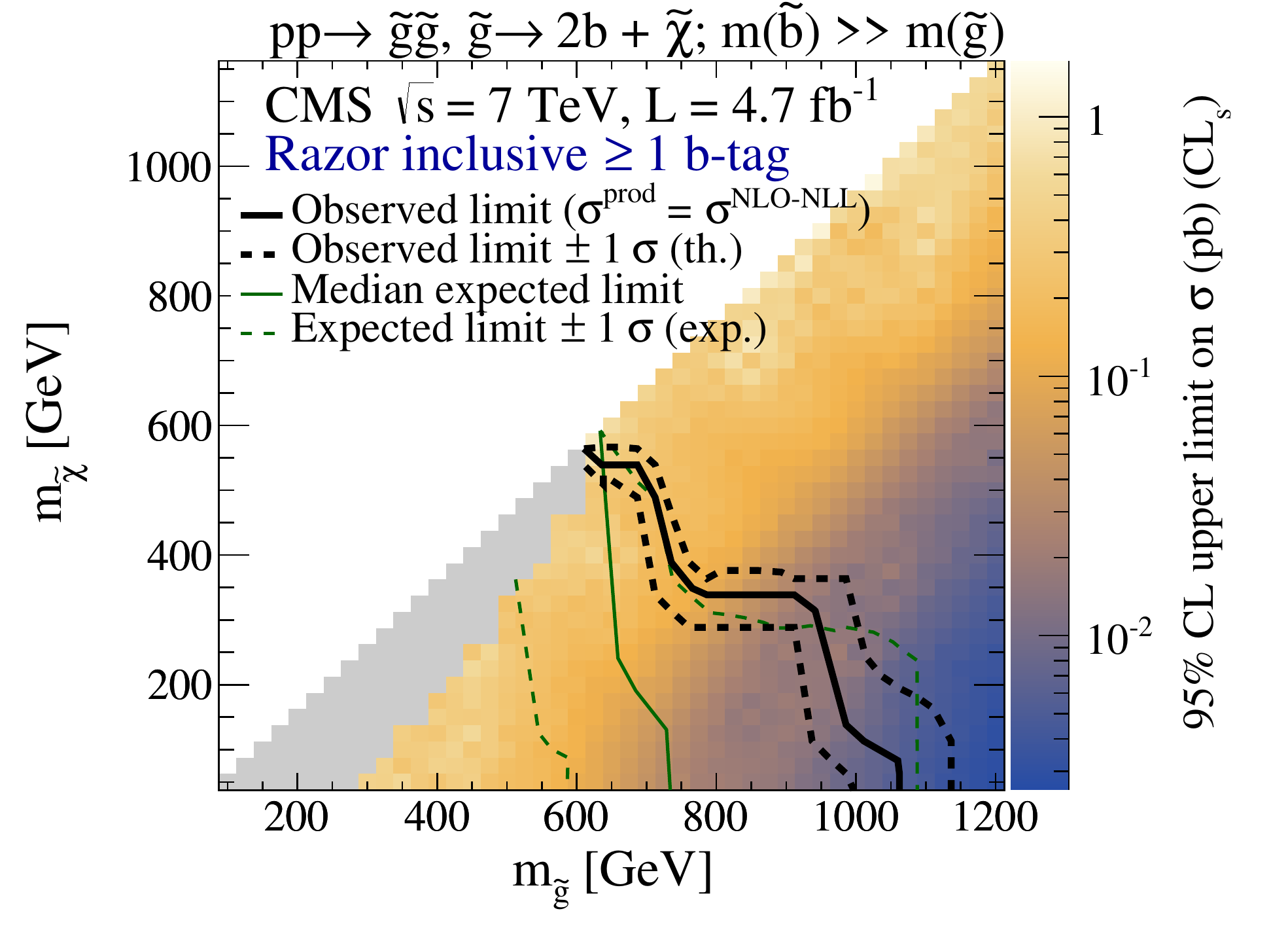}
\includegraphics[width=0.495\textwidth]{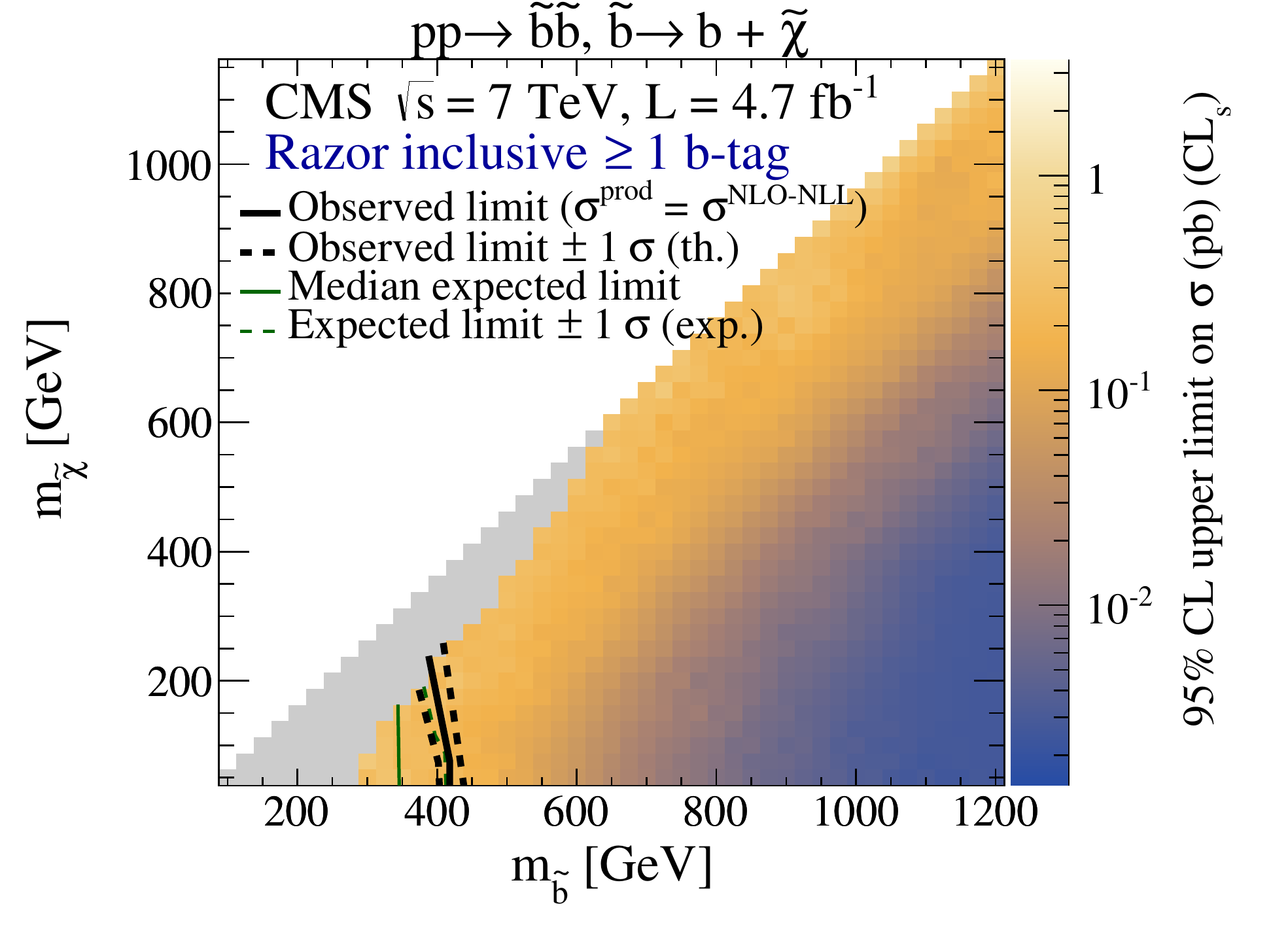}
\includegraphics[width=0.495\textwidth]{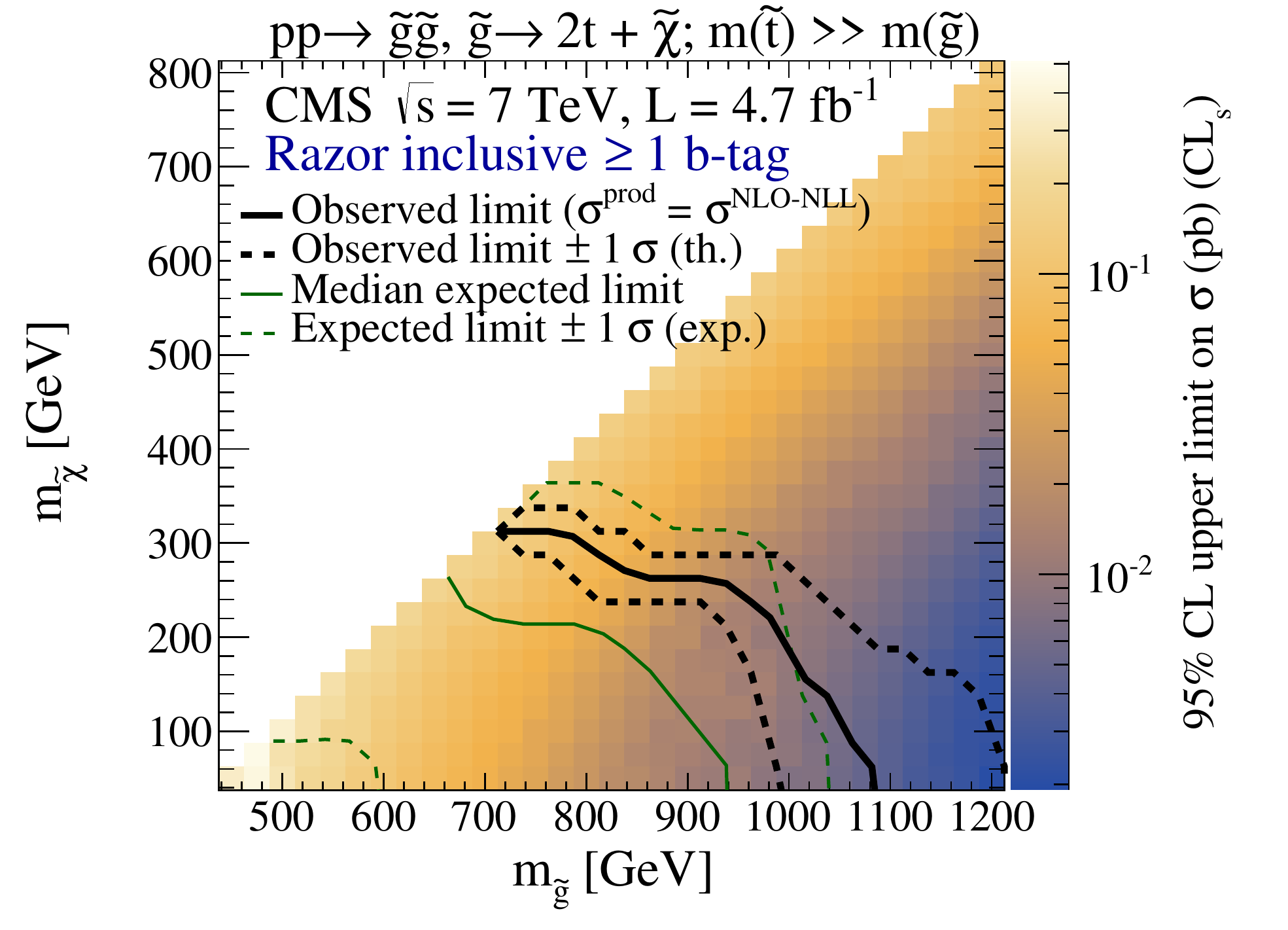}
\includegraphics[width=0.495\textwidth]{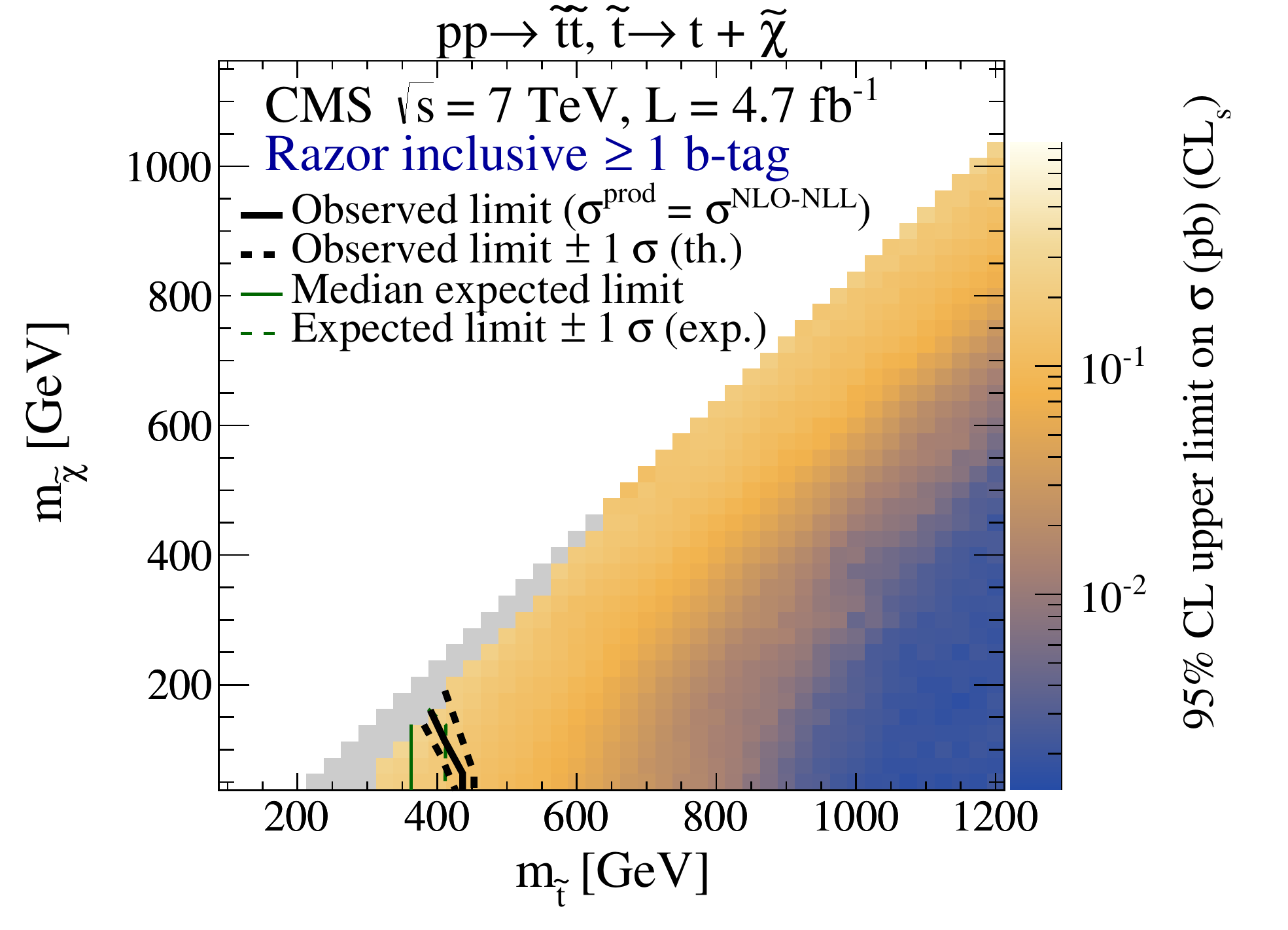}
\caption{Cross section upper limits, in pb, at 95\% \CL (color scale),
   in the mass plane of the produced superparticles for (a) T1bbbb,
   (b) T2bb, (c) T1tttt, and (d) T2tt, for the $\ge$1 $\cPqb$-tag
   razor analysis.  The solid black line indicates the observed
   exclusion region, assuming the nominal NLO+NLL SUSY production
   cross section. The dotted black lines show the observed exclusion
   taking ${\pm}1$ standard deviation theoretical uncertainties around
   the nominal cross section. The solid green line indicates the
   median expected exclusion region, with the dotted green lines
   indicating the expected exclusion with ${\pm}1$ standard deviation
   experimental uncertainties. The solid gray region indicates model
   points where the selection efficiency is found to have dependence
   on ISR modeling in the simulation of signal events above a
   predefined tolerance; no interpretation is presented for these
   model points.\label{fig:sms-bincl0}}
\end{figure*}

The results are also interpreted as cross section limits on a number
of simplified
models~\cite{ArkaniHamed:2007fw,Alwall-2,Alwall-1,Alwall:2008va,Alves:2011wf}
where a limited set of hypothetical particles and decay chains are
introduced to produce a given topological signature. For each model
studied, we derive the maximum allowed cross section at the 95\% \CL as a
function of the mass of the produced particles (gluinos or squarks,
depending on the model) and the LSP mass, as well as the exclusion
limit corresponding to the SUSY cross section.
We study several SMS benchmark scenarios~\cite{SUS-11-016}:
\begin{itemize}
\item gluino-gluino production with four light-flavor jets+\ETm in
  the superpartner decays, T1 in Fig.~\ref{fig:SMSdecay}.
\item squark-antisquark production with two light-flavor jets+\ETm
  in the superpartner decays, T2 in Fig.~\ref{fig:SMSdecay}.
\item gluino-gluino production with four $\cPqb$ jets+\ETm in the
  superpartner decays, T1bbbb in Fig.~\ref{fig:SMSdecay}.
\item squark-antisquark production with two $\cPqb$ jets+\ETm in the
  superpartner decays, T2bb in Fig.~\ref{fig:SMSdecay}.
\item gluino-gluino production with four top quarks+\ETm in the
  superpartner decays, T1tttt in Fig.~\ref{fig:SMSdecay}.
\item squark-antisquark production with two top quarks+\ETm in the
  superpartner decays, T2tt in Fig.~\ref{fig:SMSdecay}.
\end{itemize}
In all cases, additional jets in the final state can arise from
initial- and final-state radiation (ISR and FSR), simulated by
  \PYTHIA. We show in Figs.~\ref{fig:sms-incl0} and
~\ref{fig:sms-bincl0} the excluded cross section at 95\% \CL as a
function of the mass of the produced particle (gluinos or squarks,
depending on the model) and the LSP mass, as well as the exclusion
curve corresponding to the NLO+NLL SUSY cross
section~\cite{NLONLL1,NLONLL2,NLONLL3,NLONLL4,NLONLL5}, where NLL
indicates the next-to-leading-logarithmic. A result is not quoted for
the region of the SMS plane in which the signal efficiency strongly
depends on the ISR and FSR modeling (gray area), as a consequence of
the small mass difference between the produced superpartner and the
LSP and the consequent small \pt for the jets produced in the cascade.

\begin{figure}[ht!]
\centering
\includegraphics[width=0.90\columnwidth]{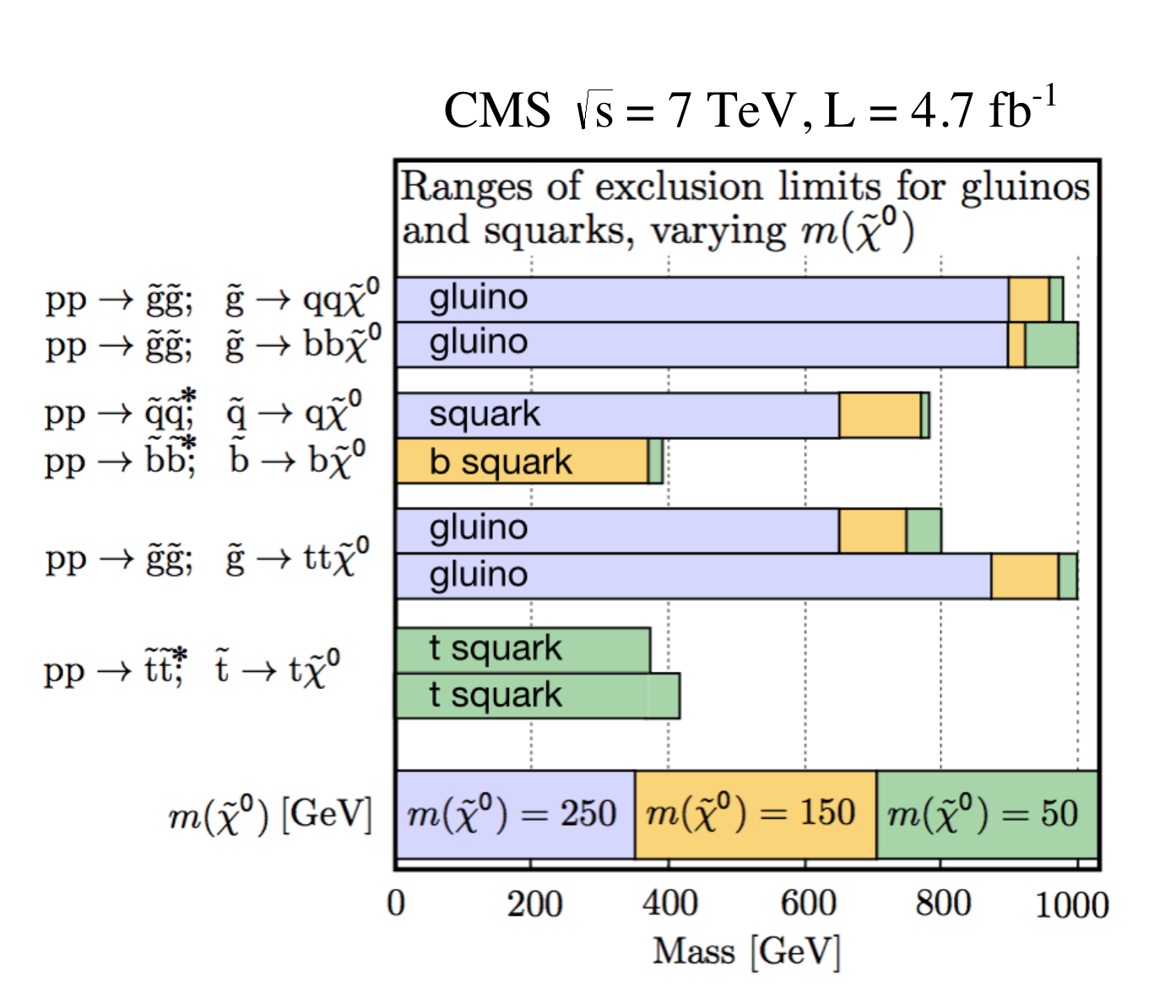}
\caption{Summary of the 95\% \CL excluded largest parent mass for each
  of the simplified models studied, for various LSP masses. The
  results from the $\cPqb$-jet razor analysis are shown immediately
  below those from the inclusive razor analysis for each of the four
  categories of events indicated.\label{fig:sum-sms}}
\end{figure}

In Fig.~\ref{fig:sum-sms}, we present a summary of the 95\% \CL
excluded largest parent mass for various LSP masses in each of the
simplified models studied, showing separately the results from the
inclusive razor analysis and the inclusive $\cPqb$-jet razor analysis.
A comparison of the razor results with those obtained from other
approaches is given in Ref.~\cite{Mahbubani:2012qq}.

\section{Summary}\label{sec:summary}
Using a data sample of $\sqrt{s}$ = 7\TeV proton-proton collisions
collected by the CMS experiment at the LHC in 2011, corresponding to
an integrated luminosity of 4.7\fbinv, we have performed a search for
pair-produced supersymmetric particles such as squarks and gluinos in
the razor-variable plane. A 2D shape description of the relevant
standard model processes determined from data control samples and
validated with simulated events has been used, and no significant
excess over the background expectations has been observed.  The
results are presented as a 95\% \CL limit in the ($\mzero$, $\mhalf$)
CMSSM parameter space.  We exclude squark and gluino masses up to
1350\GeV for $m(\PSq)\sim m(\PSg)$, while for $m(\PSq)>m(\PSg)$ we
exclude gluino masses up to 800\GeV. For simplified models, we exclude
gluino masses up to 1000\GeV, and first- and second- generation squark
masses up to 800\GeV. The direct production of top or bottom squarks
is excluded for squark masses up to 400\GeV.

\begin{acknowledgments}
We congratulate our colleagues in the CERN accelerator departments for
the excellent performance of the LHC and thank the technical and
administrative staffs at CERN and at other CMS institutes for their
contributions to the success of the CMS effort. In addition, we
gratefully acknowledge the computing centers and personnel of the
Worldwide LHC Computing Grid for delivering so effectively the
computing infrastructure essential to our analyses. Finally, we
acknowledge the enduring support for the construction and operation of
the LHC and the CMS detector provided by the following funding
agencies: BMWF and FWF (Austria); FNRS and FWO (Belgium); CNPq, CAPES,
FAPERJ, and FAPESP (Brazil); MES (Bulgaria); CERN; CAS, MoST, and NSFC
(China); COLCIENCIAS (Colombia); MSES and CSF (Croatia); RPF (Cyprus);
MoER, SF0690030s09 and ERDF (Estonia); Academy of Finland, MEC, and
HIP (Finland); CEA and CNRS/IN2P3 (France); BMBF, DFG, and HGF
(Germany); GSRT (Greece); OTKA and NIH (Hungary); DAE and DST (India);
IPM (Iran); SFI (Ireland); INFN (Italy); NRF and WCU (Republic of
Korea); LAS (Lithuania); MOE and UM (Malaysia); CINVESTAV, CONACYT,
SEP, and UASLP-FAI (Mexico); MBIE (New Zealand); PAEC (Pakistan); MSHE
and NSC (Poland); FCT (Portugal); JINR (Dubna); MON, RosAtom, RAS and
RFBR (Russia); MESTD (Serbia); SEIDI and CPAN (Spain); Swiss Funding
Agencies (Switzerland); NSC (Taipei); ThEPCenter, IPST, STAR and NSTDA
(Thailand); TUBITAK and TAEK (Turkey); NASU (Ukraine); STFC (United
Kingdom); DOE and NSF (USA).

Individuals have received support from the Marie-Curie program and
the European Research Council and EPLANET (European Union); the
Leventis Foundation; the A. P. Sloan Foundation; the Alexander von
Humboldt Foundation; the Belgian Federal Science Policy Office; the
Fonds pour la Formation \`a la Recherche dans l'Industrie et dans
l'Agriculture (FRIA-Belgium); the Agentschap voor Innovatie door
Wetenschap en Technologie (IWT-Belgium); the Ministry of Education,
Youth and Sports (MEYS) of Czech Republic; the Council of Science and
Industrial Research, India; the Compagnia di San Paolo (Torino); the
HOMING PLUS programme of Foundation for Polish Science, cofinanced by
EU, Regional Development Fund; and the Thalis and Aristeia programmes
cofinanced by EU-ESF and the Greek NSRF.
\end{acknowledgments}
\bibliography{auto_generated}

\clearpage
\section*{Appendices}
\appendix
\section{Additional standard model backgrounds in the \texorpdfstring{($\boldmath{\MR}$,
  $\boldmath{\Rtwo}$)}{(MR MR2)} razor plane\label{sec:ttbarAppendix}}

\begin{figure}[htpb]
\centering
\includegraphics[width=0.495\textwidth]{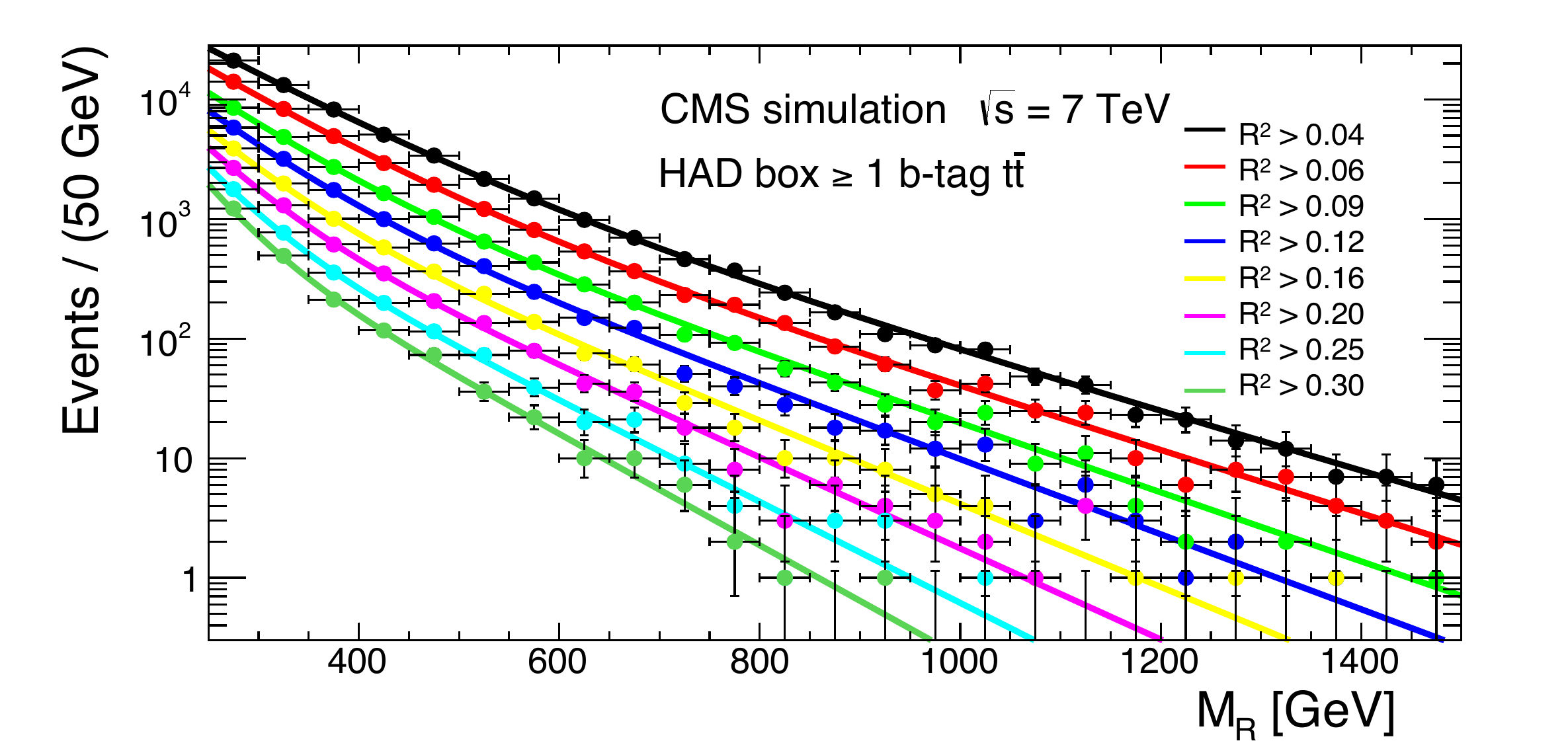}
\caption{The $\MR$ distributions for different values of
  ${\Rtwo_\text{min}}$ for $\ttbar$ simulated events in the HAD box
  with the requirement of $\ge$1 $\cPqb$-tagged jets. The curves show
  the results of fits of a sum of two exponential
  distributions.\label{fig:MC_ttj_MR}}
\includegraphics[width=0.495\textwidth]{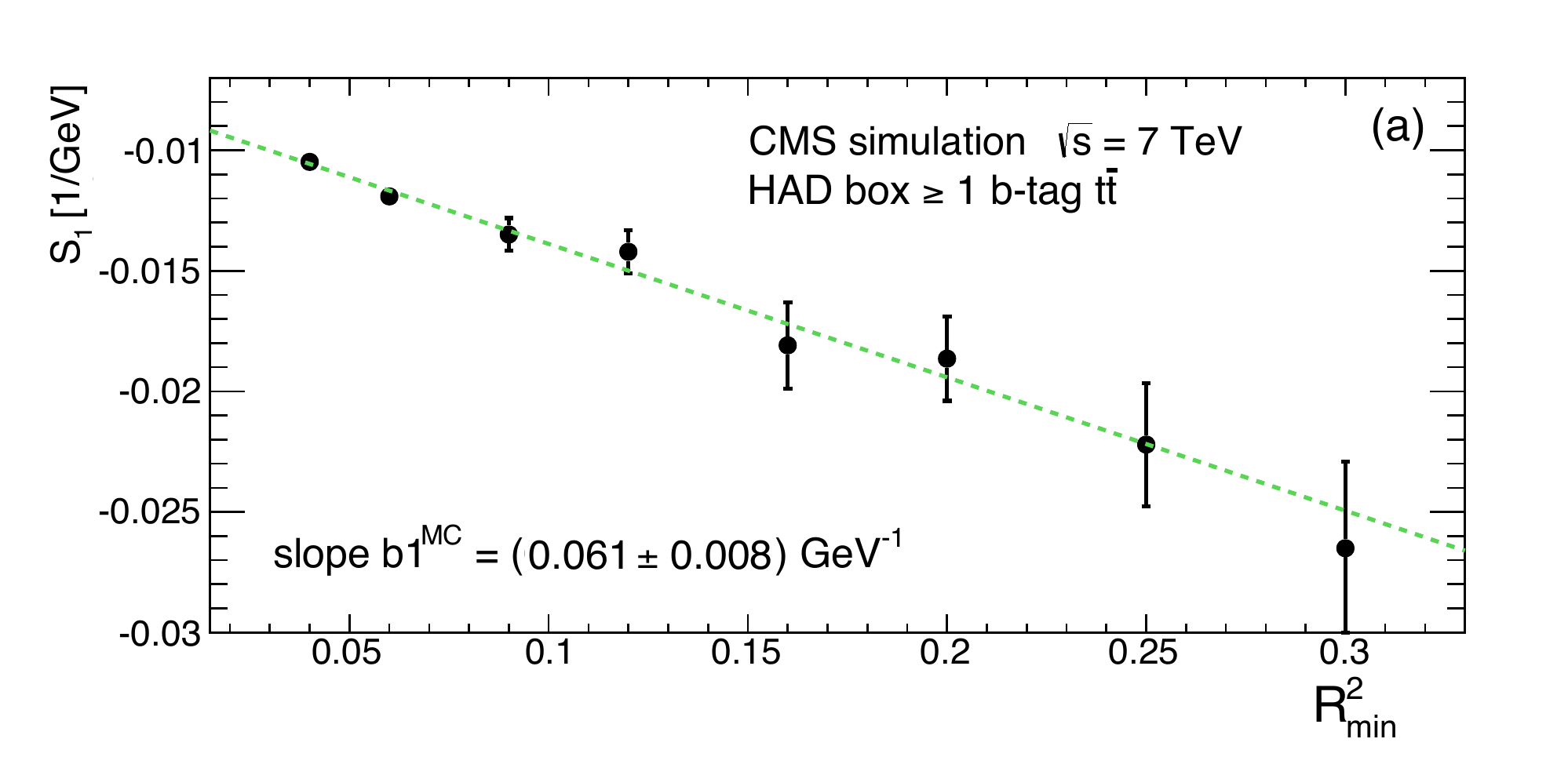}
\includegraphics[width=0.495\textwidth]{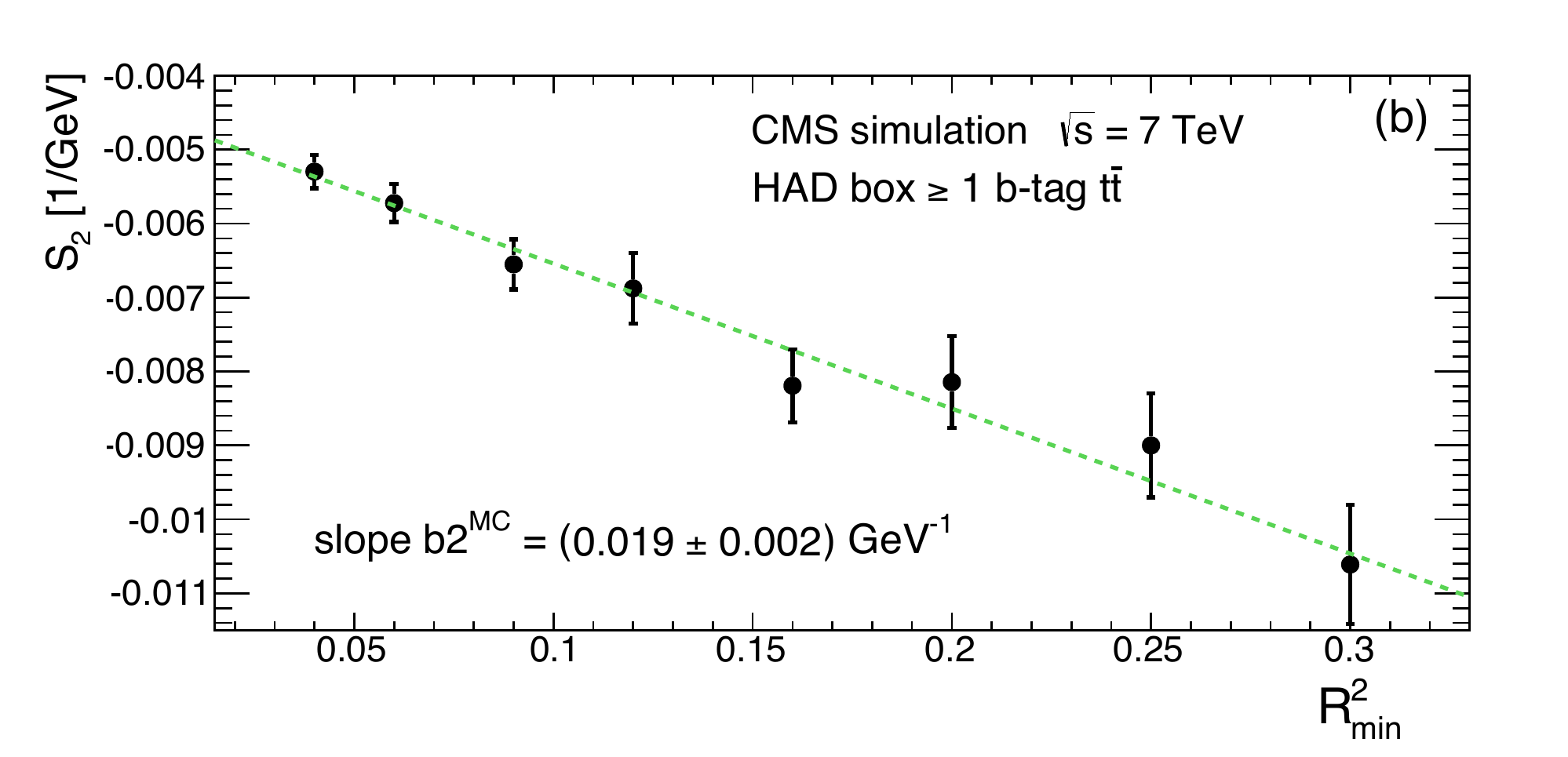}
\caption{Value of (a) the coefficient in the first exponent,
  $\S_1$, and (b) the coefficient in the second exponent,
  $\S_2$, from fits to the $\MR$ distribution, as a
  function of ${\Rtwo_\text{min}}$, for $\ttbar$ simulated events in
  the HAD box with the requirement of $\ge$1
  $\cPqb$-tagged jets.\label{fig:MC_ttj_MRs}}

\end{figure}

\begin{figure}[htpb]
\centering
\includegraphics[width=0.495\textwidth]{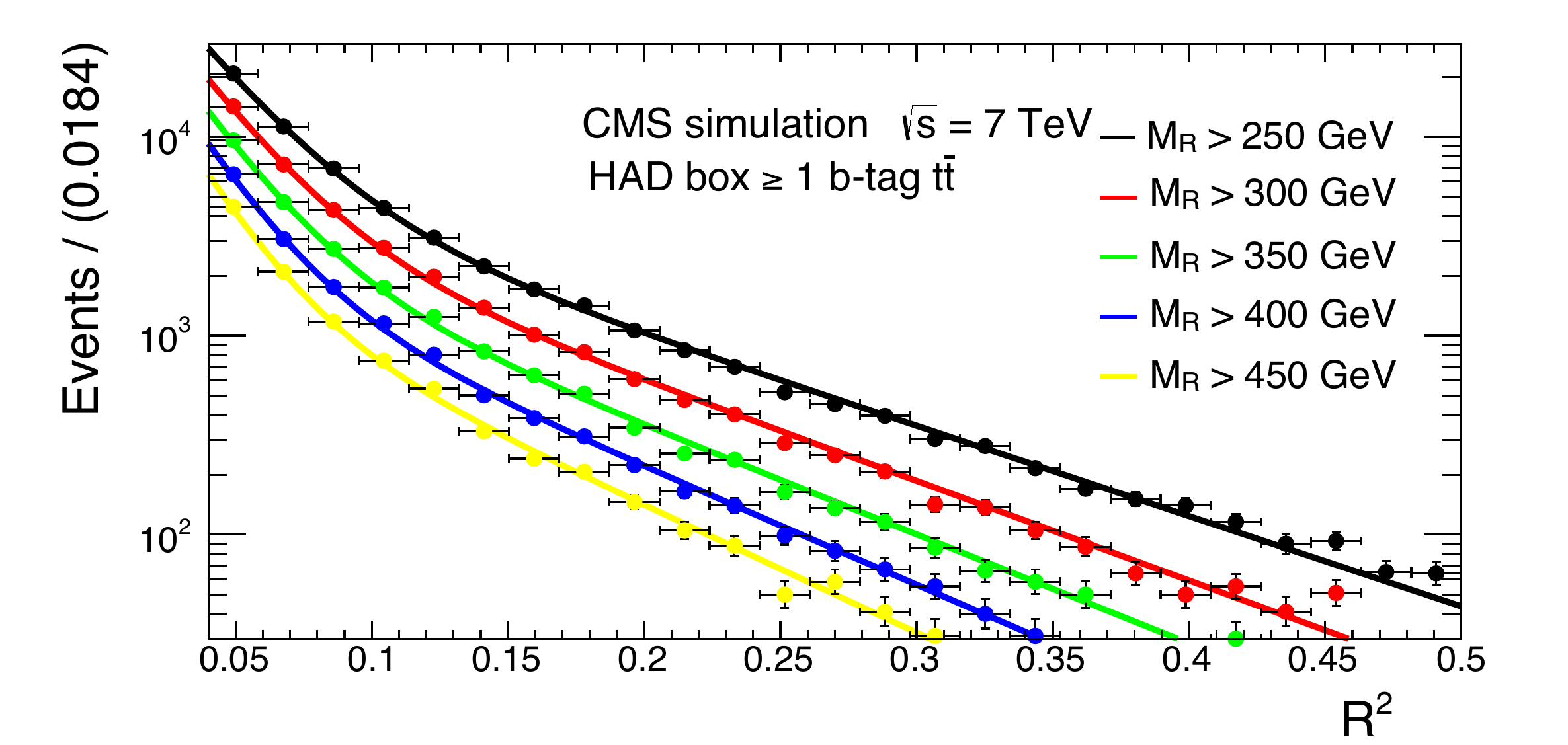}
\caption{The $\Rtwo$ distributions for different values of
  ${\MR^\text{min}}$ for $\ttbar$ simulated events in the HAD box
  with the requirement of $\ge$1 $\cPqb$-tagged jets. The curves show
  the results of fits of a sum of two exponential
  distributions.\label{fig:MC_ttj_R}}
\includegraphics[width=0.495\textwidth]{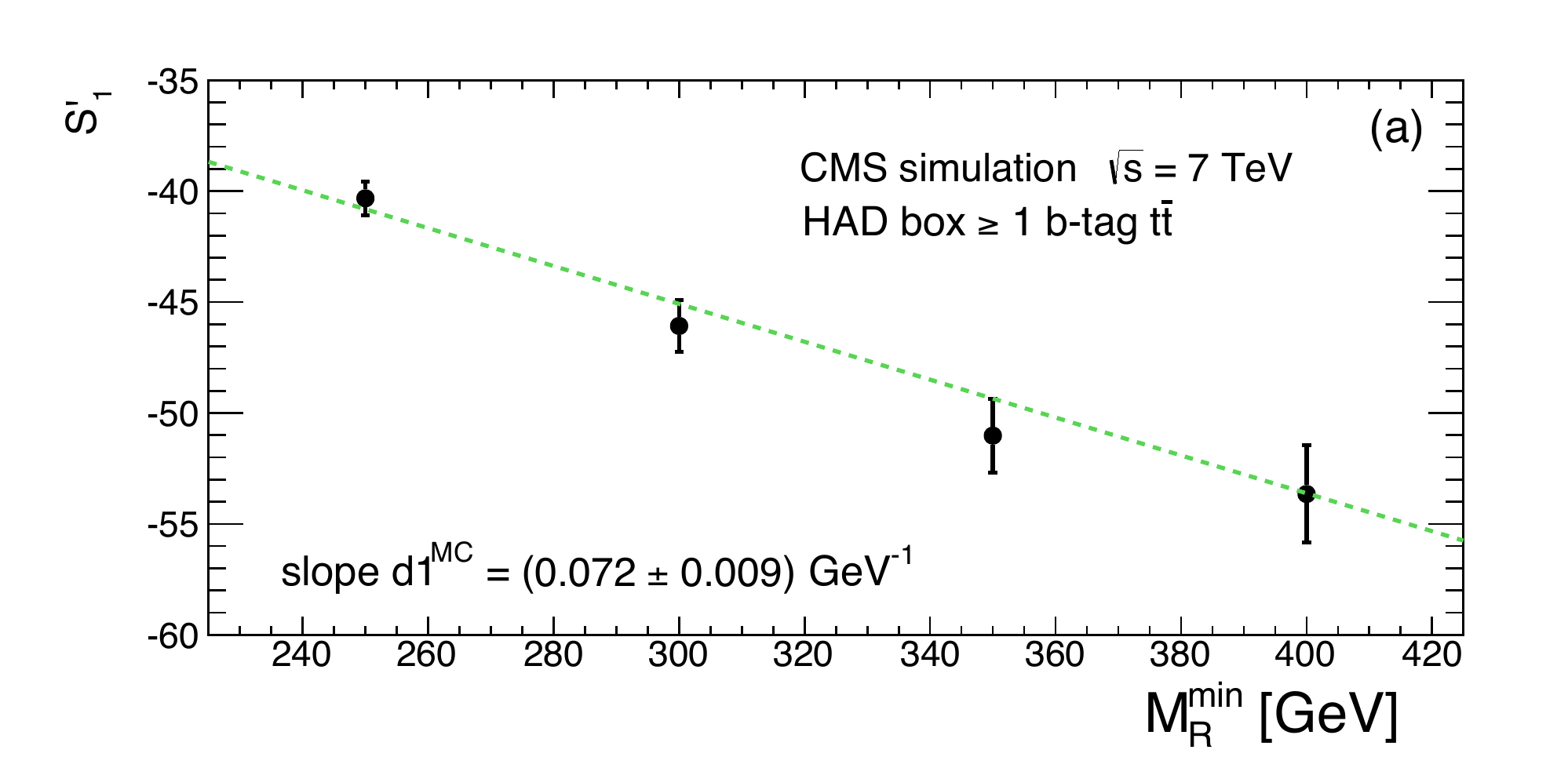}
\includegraphics[width=0.495\textwidth]{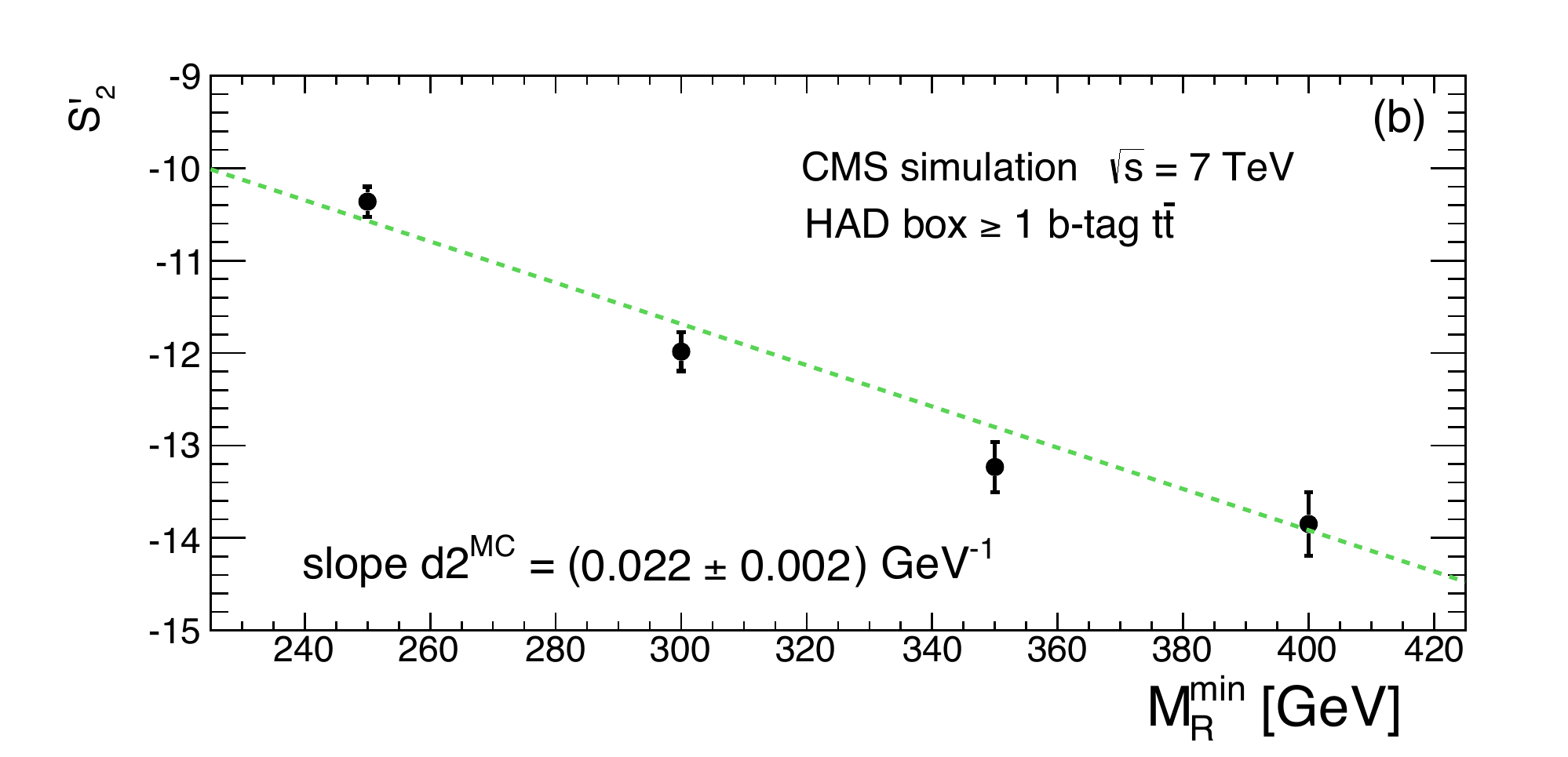}
\caption{Value of (a) the coefficient in the first exponent,
  $\S'_1$, and (b) the coefficient in the second exponent,
  $\S'_2$, from fits to the $\Rtwo$ distribution, as a
  function of ${\MR^\text{min}}$, for $\ttbar$ simulated events in
  the HAD box with the requirement of $\ge$1 $\cPqb$-tagged
  jets.\label{fig:MC_ttj_Rs}}

\end{figure}

Figure~\ref{fig:MC_ttj_MR} shows the $\MR$ distribution as a
function of ${\Rtwo_\text{min}}$ for $\ttbar$ MC events with $\ge$1
$\cPqb$-tagged jets in the HAD box. The $\S_1$ and
$\S_2$ parameters characterizing the exponential behavior of
the first and second $\PW(\mu\nu)$+jets components are shown in
Fig.~\ref{fig:MC_ttj_MRs}.  The corresponding distributions for
$\Rtwo$, and for the $\S'_1$ and $\S'_2$
parameters, are shown in
Figs.~\ref{fig:MC_ttj_R}~and~\ref{fig:MC_ttj_Rs}, respectively. The
conclusions derived from the data and MC studies of
Section~\ref{sec:BKG2011} hold also for $\ttbar$ MC events .

\section{Alternative background shape analysis}\label{altBKG}

In order to quantify a systematic uncertainty associated with the
choice of the fit function, we first generalize our 2D function to
allow for deviations from the exponential behavior, once projected onto
$\MR$ or $\Rtwo$.  To do this, we i) identify a set of
functions that describe the data, ii) use one as a default description,
iii) use the rest to quantify the systematic variation, iv)
randomly choose one of the three functions when generating the
pseudo-experiments used to set limits, and v) use the nominal function
when evaluating the likelihood.

For a 1D fit of the $\MR$ distribution, an obvious choice is
\begin{equation}
  (\MR) = A \re^{-b \MR^\beta},
\end{equation}
where $\beta \neq 1$ accounts for deviations from the
exponential function.  In this analysis, we need a 2D function of
$\MR$ and $\Rtwo$ that allows us to measure the
deviation from the nominal shape on the projections. For this purpose,
we introduce a generalization of the razor 2D function:
\ifthenelse{\boolean{cms@external}}{
\begin{multline}
F_\mathrm{SYS}(\MR, \Rtwo) = \left[ b(\MR - M^0_\R)^{1/n}(\Rtwo - \R_0^2)
  ^{1/n}-n\right] \\ \times \re^{-b(\MR - M^0_\R) ^{1/n} (\Rtwo - \R_0^2) ^{1/n}},
\end{multline}
}{
\begin{equation}
F_\mathrm{SYS}(\MR, \Rtwo) = \left[ b(\MR - M^0_\R)^{1/n}(\Rtwo - \R_0^2)
  ^{1/n}-n\right]  \times \re^{-b(\MR - M^0_\R) ^{1/n} (\Rtwo - \R_0^2) ^{1/n}},
\end{equation}
}
which has the two following properties:
\begin{align}
\int_{{\Rtwo_\text{min}}}^{+\infty} F_\mathrm{SYS}(\MR, \Rtwo)\, \rd\Rtwo &\sim \re^{-k_{\MR}(\MR - \MR^0) ^{1/n} }, \\
\int_{{\MR^\text{min}}}^{+\infty} F_\mathrm{SYS}(\MR, \Rtwo)\, \rd\MR &\sim \re^{-k_{\Rtwo}(\Rtwo - \Rtwo_0) ^{1/n}},
\end{align}
where
\begin{align}
k_{\MR} &= (k^0_{\MR} + b {\Rtwo_\text{min}})^{1/n}, \\
k_{\Rtwo} &= (k^0_{\Rtwo} + b {\MR^\text{min}})^{1/n} ,
\end{align}
with ${\MR^\text{min}}$ and ${\Rtwo_\text{min}}$ respectively the
thresholds applied on $\MR$ and $\Rtwo$ before
projecting onto $\Rtwo$ and $\MR$.  Using this
function to evaluate systematic uncertainties corresponds to the 2D
generalization needed here. We proceed as follows:
\begin{itemize}
\item we repeat the fit in the fit region of each box, using $F_\mathrm{SYS}(\MR, \Rtwo)$ rather than $F(\MR,
  \Rtwo)$ for the second component of the background model (the
  one that extrapolates to the signal region), with $n$ floated in the
  fit. We determine $n_\text{fit} \pm \sigma_n$ in this fit.
\item we assign an allowed range to the difference $n-1$ taking the
  larger of $n_\text{fit}-1$ and $\sigma_n$, which we refer to as $[n_\text{min},n_\text{max}]$.
\item we repeat the fit in the fit region fixing $n$ to first to
  $n_\text{min}$ and then to $n_\text{max}$ and we take these fits as
  the alternative background descriptions.
\end{itemize}

In particular, we find that the fit returns values of $n_\text{fit}$ that are
very close to $n$. Following the prescription outlined above, we take
the fit uncertainty as the shift in $n$.

The main conclusion of the study is that the systematic uncertainty in
the choice of the function is already covered by the large uncertainty
in the fit parameters and that the effect corresponds to an increase
of about 15\% in the 68\% \CL range, once this contribution is summed
in quadrature with the already quoted uncertainty.

\begin{figure}[htpb]
  \centering
    \includegraphics[width=0.48\textwidth]{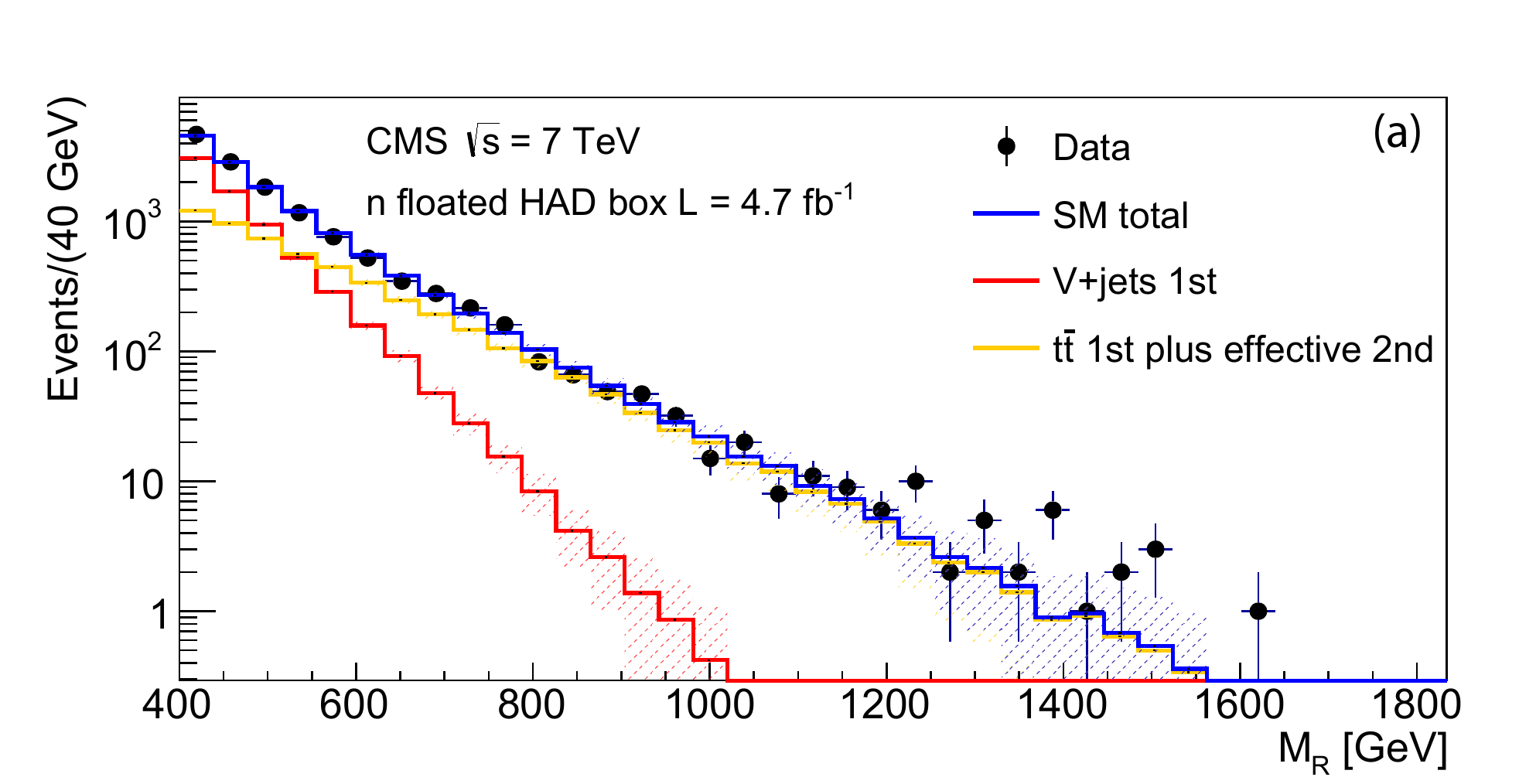}
    \includegraphics[width=0.48\textwidth]{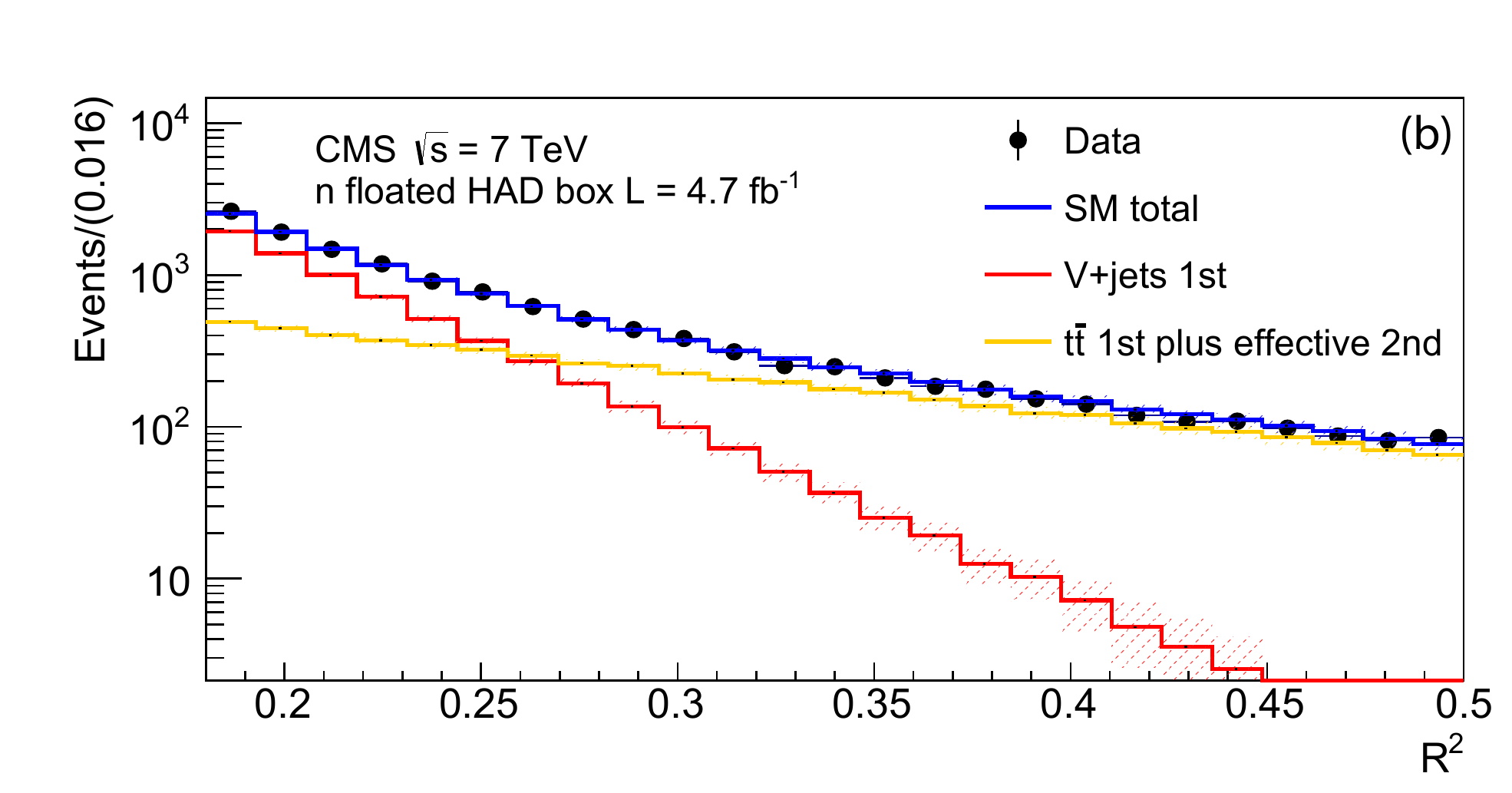}
 \caption{\label{fig:fitHADsys} Projection of the fit result on the
   (a) $\MR$ and (b) $\Rtwo$ axis for the HAD box,
   obtained as explained in the text.}
\end{figure}

As an example, we present the results of the above procedure for the
bins in the HAD box. Fig.~\ref{fig:fitHADsys} shows the fit result
with $n$ floated in the full region of the HAD box, projected onto
$\MR$ and $\Rtwo$. The quality of the fit is similar
to that of the nominal procedure. We find $n= 0.96 \pm 0.04$. We then
take $n_\text{min} = 0.96$ and $n_\text{max} = 1.04$.  We show in
Table~\ref{tab:predSYSHad} the bin-by-bin background prediction for
the nominal fit and the two alternative fits. We use a finer binning
than the one used to compute the $p$-values in the nominal analysis. For
comparison, we also show the values obtained with $n$ floated in the
fit. For all cases, we quote the predicted background as the center of
the 68\% probability range and the associated uncertainty
corresponds to half the range. The range is defined by integrating the
background distribution (derived from the pseudo-experiments) using
the probability value as the ordering algorithm. Similar results are
obtained for all boxes.

\begin{table}[!ht]
\centering
  \topcaption{The bin-by-bin background prediction for the nominal fit,
    the two alternative fits, and with $n$ floated, for the HAD
    box. \label{tab:predSYSHad}}
\begin{scotch}{lxxxx}
Bin & \multicolumn{1}{l}{$n=1$} & \multicolumn{1}{l}{$n=n_\text{min}$} & \multicolumn{1}{l}{$n=n_\text{max}$} & \multicolumn{1}{l}{$n$ floated} \\
\hline
HAD\_1\_1 &  1558,69 & 1527,109 & 1509,111 & 1511,126 \\
HAD\_1\_2 &  2898,80 &  2888,89 & 2868,98 & 2866,99 \\
HAD\_1\_3 &  711,35 & 729,45 & 714,43 & 726,49 \\
HAD\_1\_4 &  329,37 & 338,31 & 328,32 & 337,34 \\
HAD\_2\_1 &  1785,64 & 1787,75 & 1759,69 & 1774,67 \\
HAD\_2\_2 &  3301,82 & 3336,104 & 3313,112 & 3349,118 \\
HAD\_2\_3 &  945,46 & 957,47 & 957,47 & 964,48 \\
HAD\_2\_4 &  432,36 & 423,35  & 454,37 & 424,38 \\
HAD\_3\_1 &  251,26 & 263,28 & 259,31 & 260,29 \\
HAD\_3\_2 &  537,47 & 544,45 & 561,50 & 550,49 \\
HAD\_3\_3 &  173,36 & 157,29 & 182,33 & 162,34 \\
HAD\_3\_4 &  58,18 &  52,17 & 66,19 & 51,18 \\
HAD\_4\_1 &  39,9 &  37,11 & 43,9 & 38,9 \\
HAD\_4\_2 &  86,23 &  74,17 & 90,24 & 76,21 \\
HAD\_4\_3 &  20,7 & 14,6 & 22,9 & 14,7 \\
HAD\_4\_4 &  4.2,2.9 & 2.7,2.3  & 4.9,3.1 & 2.4,2.4 \\
HAD\_5\_1 &  4.7,2.8 & 3.9,2.5 & 5.3,3.1 & 4.1,2.9 \\
HAD\_5\_2 &  8.3,4.7 & 6.0,3.7 & 9.5,4.7 & 5.9,4.0 \\
HAD\_5\_3 &  1.2,1.2 & 0.8,0.8 & 1.5,1.5 & 0.8,0.8 \\
HAD\_5\_4 &  0.4,0.4 & 0.4,0.4 & 0.5,0.5  & 0.4,0.4 \\
HAD\_6\_1 &  0.8,0.8 & 0.6,0.6 & 0.9,0.9 & 0.6,0.6 \\
HAD\_6\_2 &  1.0,1.0 & 0.7,0.7 & 1.2,1.2 & 0.8,0.8 \\
HAD\_6\_3 &  0.4,0.4 & 0.3,0.3 & 0.4,0.4 & 0.4,0.4 \\
\end{scotch}
\end{table}

\section{Fit results and validations for \texorpdfstring{$\ge$1 $\cPqb$-tagged}{>1 b-tagged}
  events}\label{sec:datafitsBTAGAppendix}

\begin{figure}[ht!]
\centering
\includegraphics[width=0.495\textwidth]{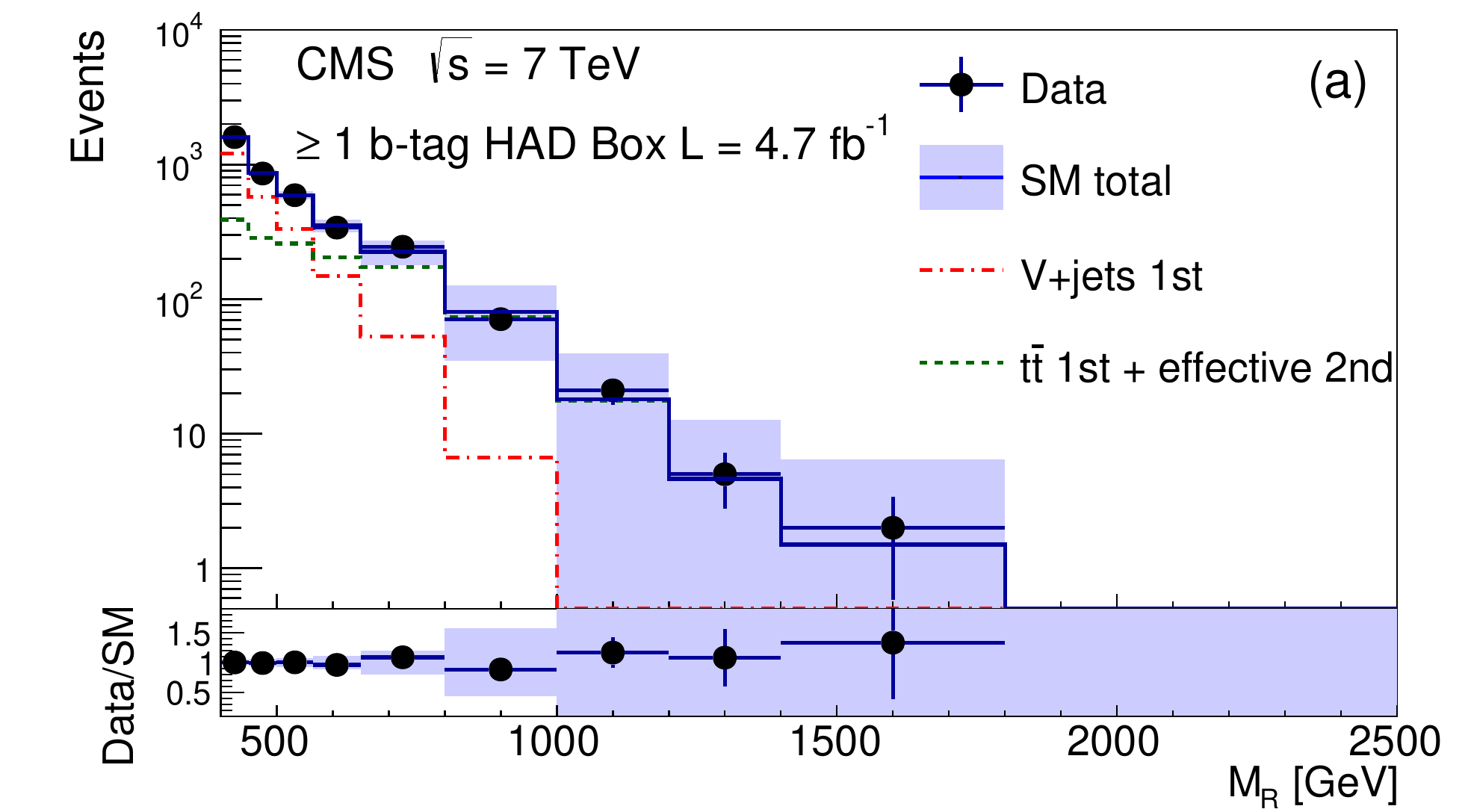}
\includegraphics[width=0.495\textwidth]{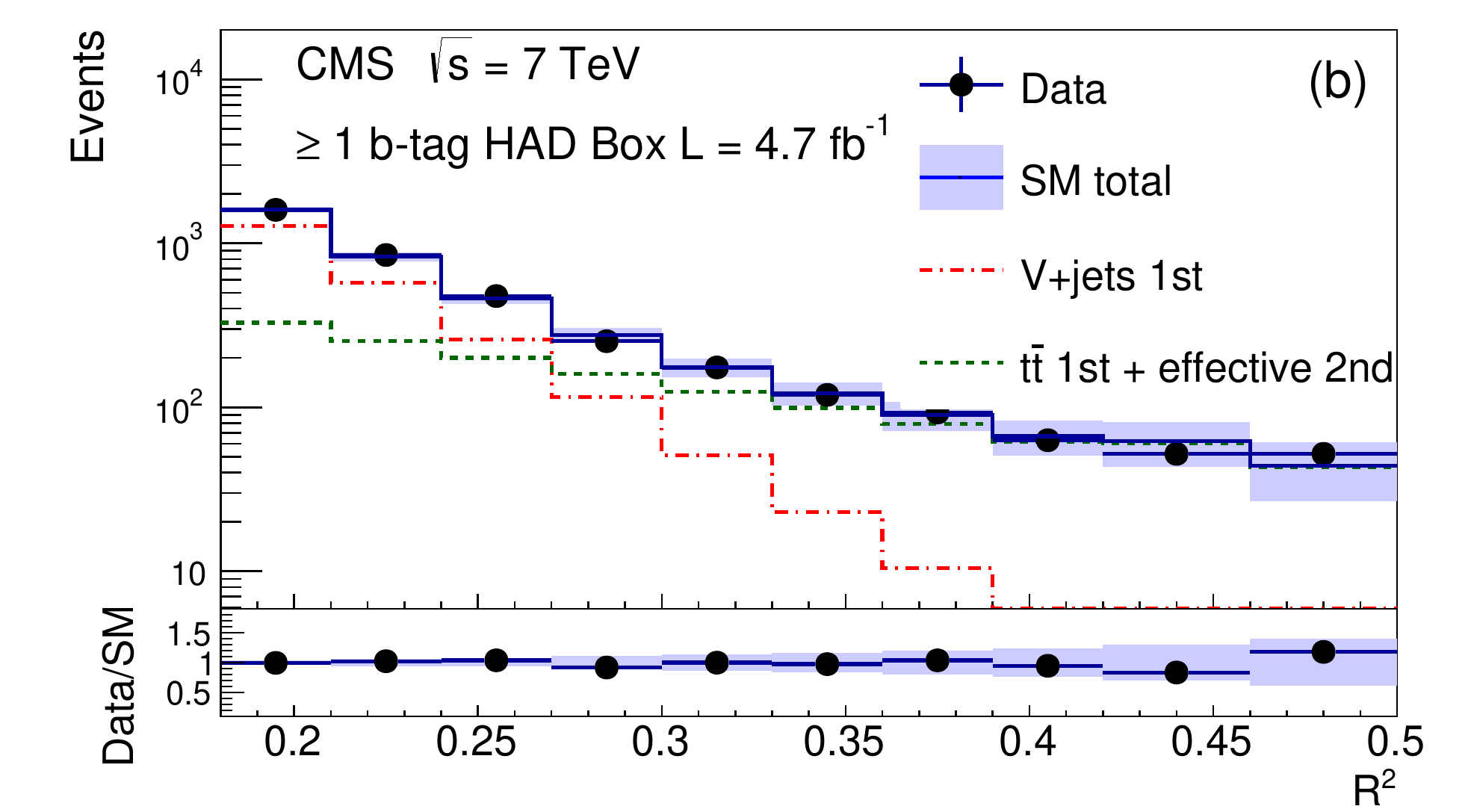}
\caption{Projection of the 2D fit result on (a) $\MR$ and
  (b) $\Rtwo$  for the HAD box in the $\ge$1 $\cPqb$-tag analysis
  path. \label{fig:bhad-box}}

\end{figure}

\begin{figure}[ht!]
\centering
\includegraphics[width=\cmsFigWidthBox]{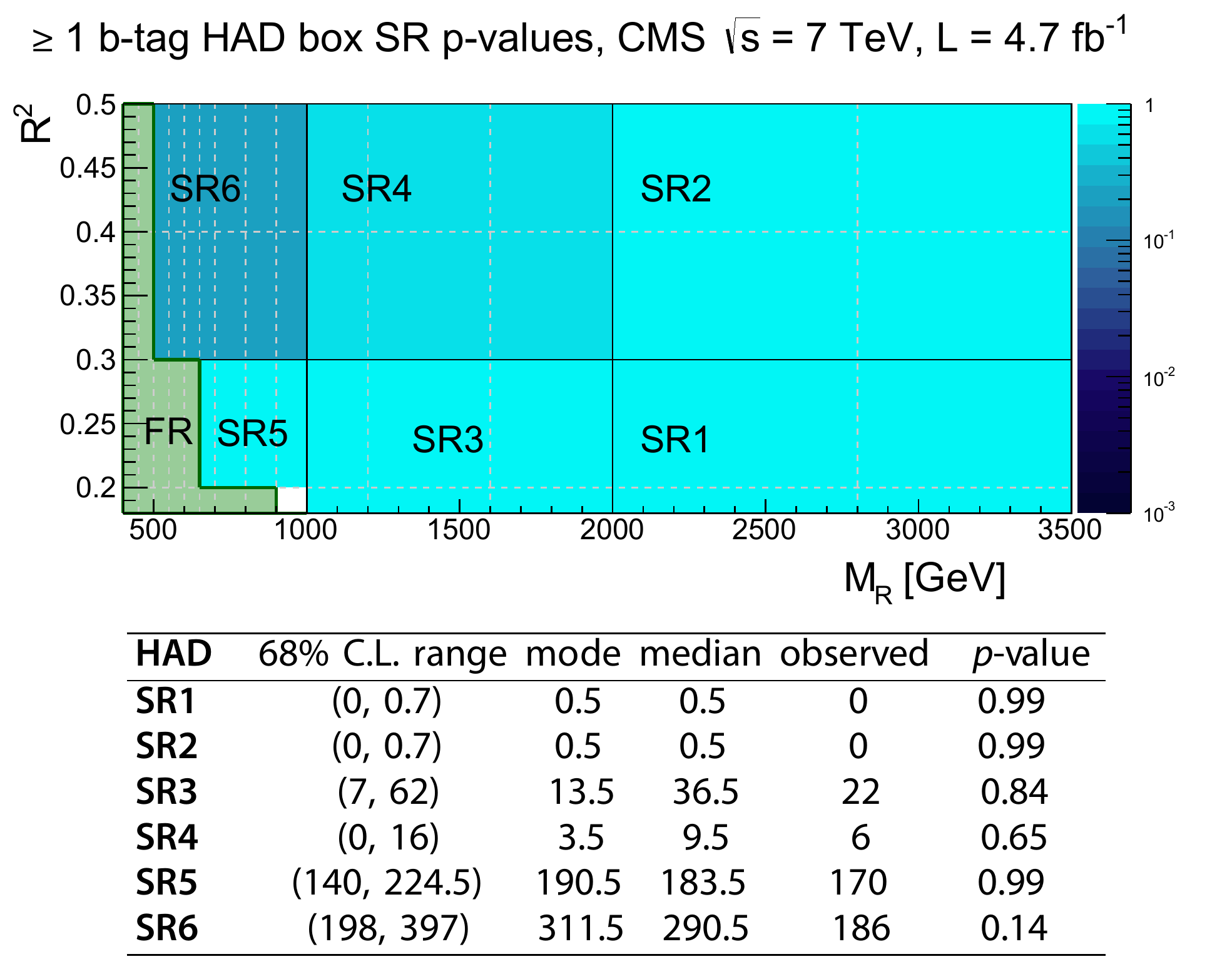}
\caption{The $p$-values corresponding to the observed number of events
  in the $\ge$1 $\cPqb$-tag HAD box signal regions (SR$i$).
\label{fig:bhad-blue-plot}}

\end{figure}

\begin{figure}[ht!]
\centering
\includegraphics[width=0.495\textwidth]{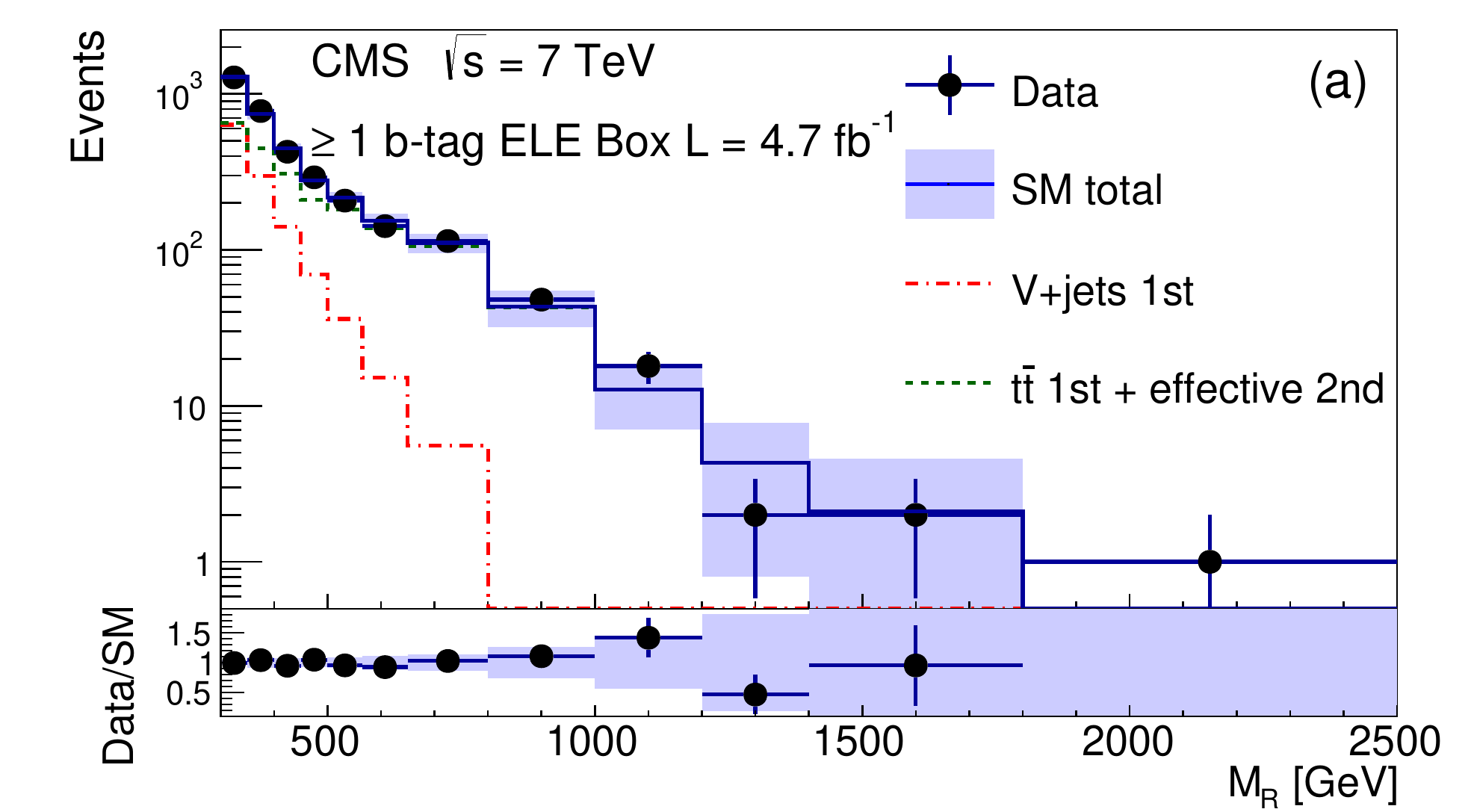}
\includegraphics[width=0.495\textwidth]{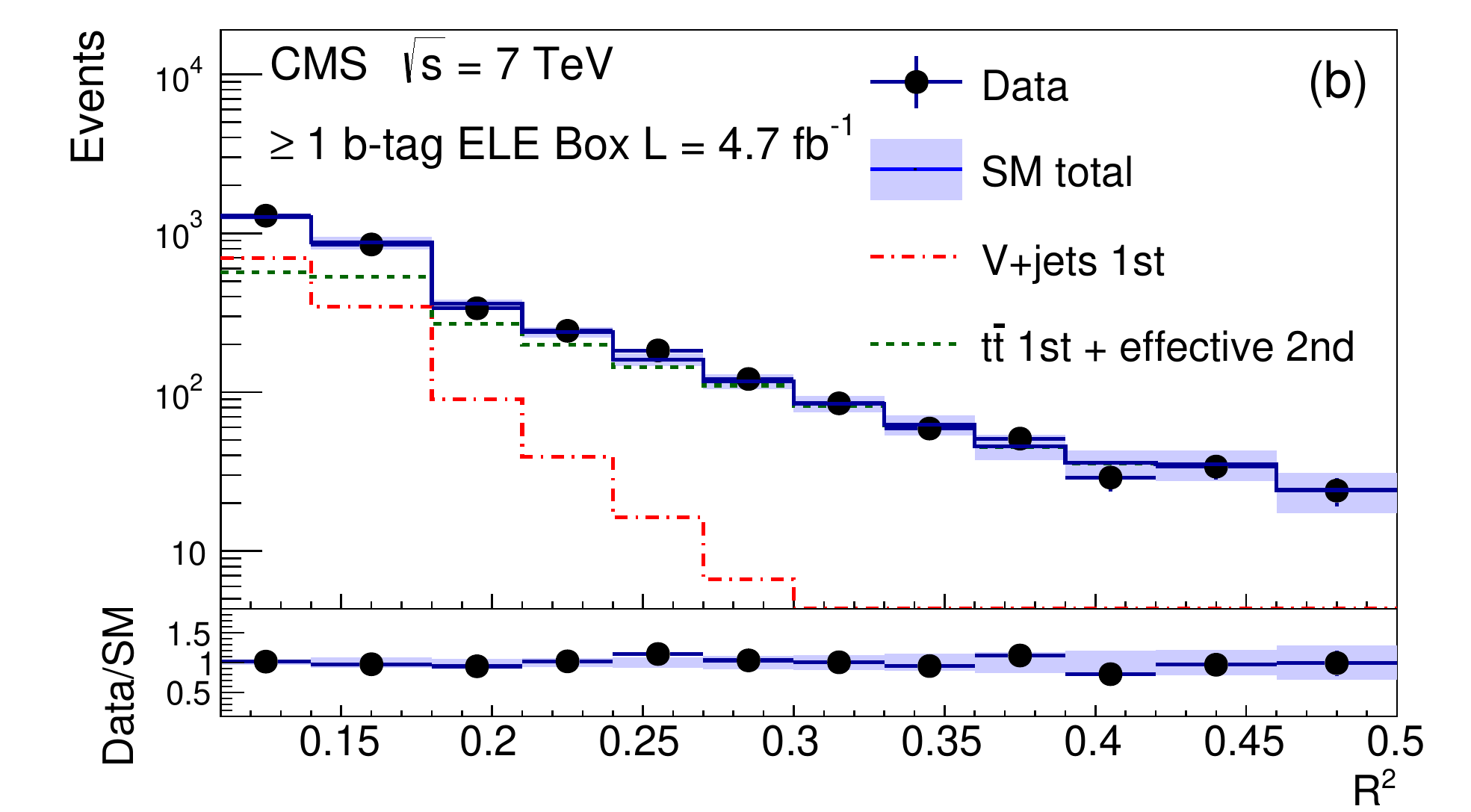}
\caption{Projection of the 2D fit result on (a) $\MR$ and (b) $\Rtwo$  for the ELE box in the $\ge$1 $\cPqb$-tag analysis path. \label{fig:bele-box}}

\end{figure}

\begin{figure}[ht!]
\centering
\includegraphics[width=\cmsFigWidthBox]{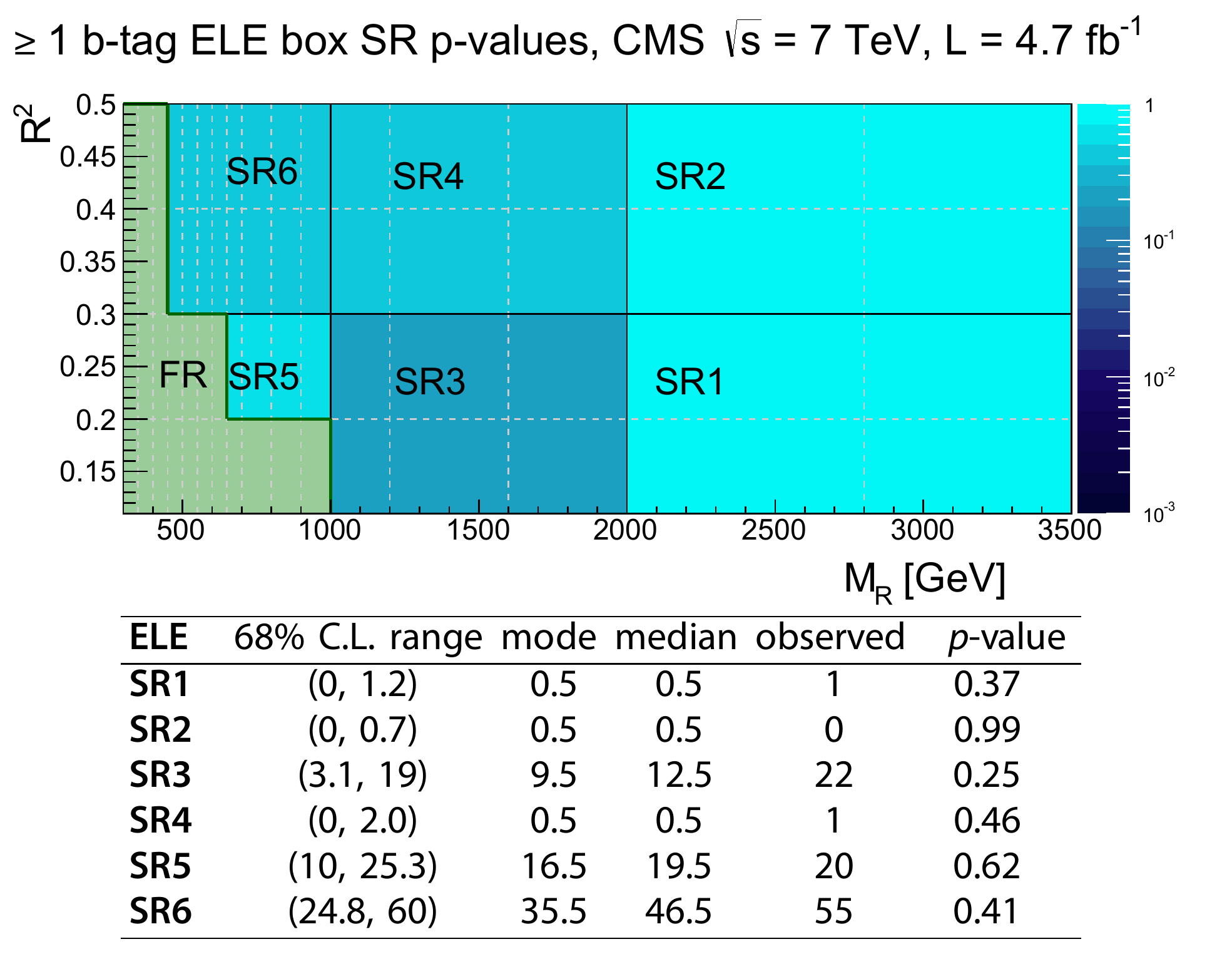}
\caption{The $p$-values corresponding to the observed number of events
  in the $\ge$1 $\cPqb$-tag  ELE box signal regions (SR$i$).
\label{fig:bele-blue-plot}}

\end{figure}

\begin{figure}[ht!]
\centering
\includegraphics[width=0.495\textwidth]{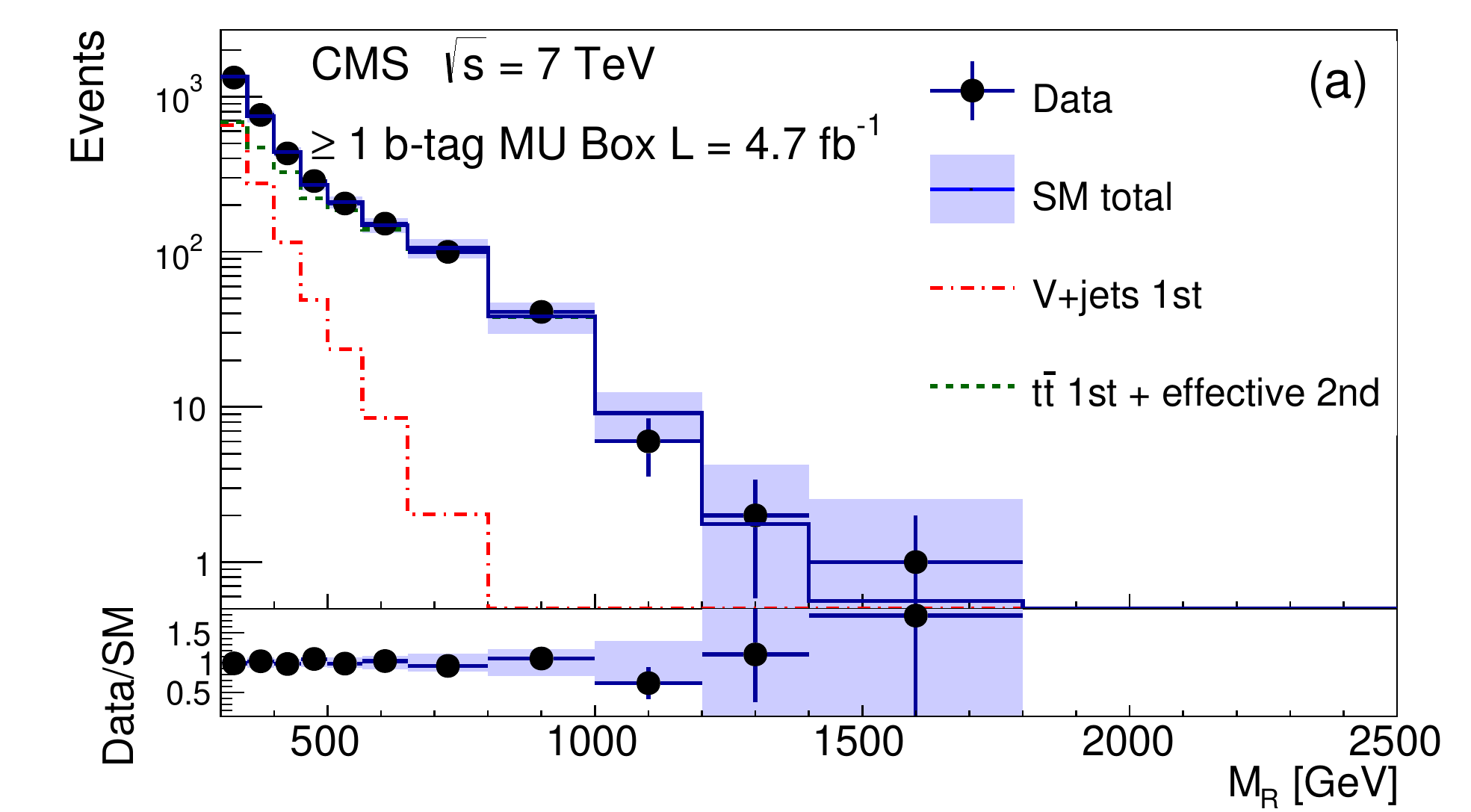}
\includegraphics[width=0.495\textwidth]{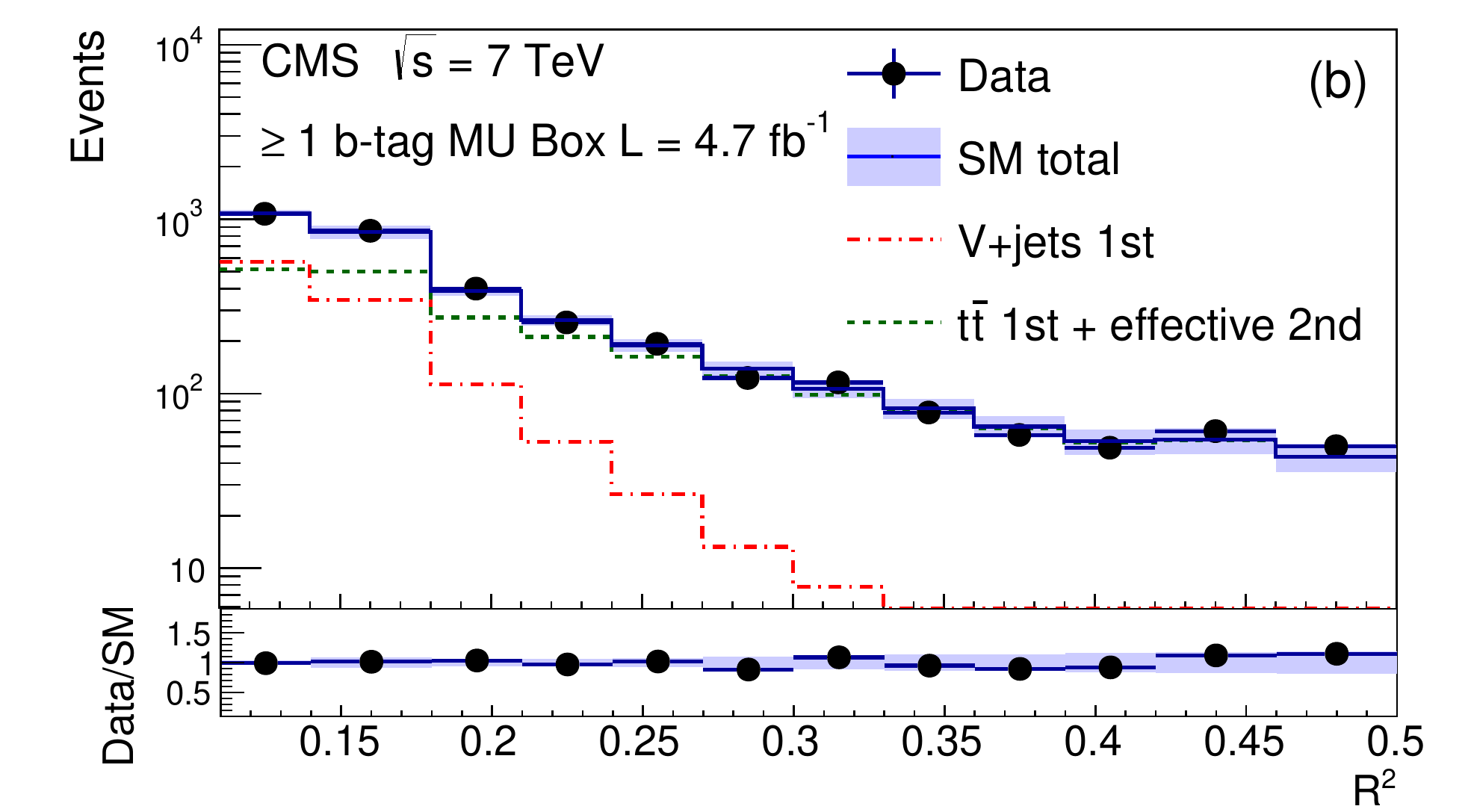}
\caption{Projection of the 2D fit result on (a) $\MR$ and (b) $\Rtwo$  for the MU box  in the $\ge$1 $\cPqb$-tag analysis path.\label{fig:bmu-box}}

\end{figure}

\begin{figure}[ht!]
\centering
\includegraphics[width=\cmsFigWidthBox]{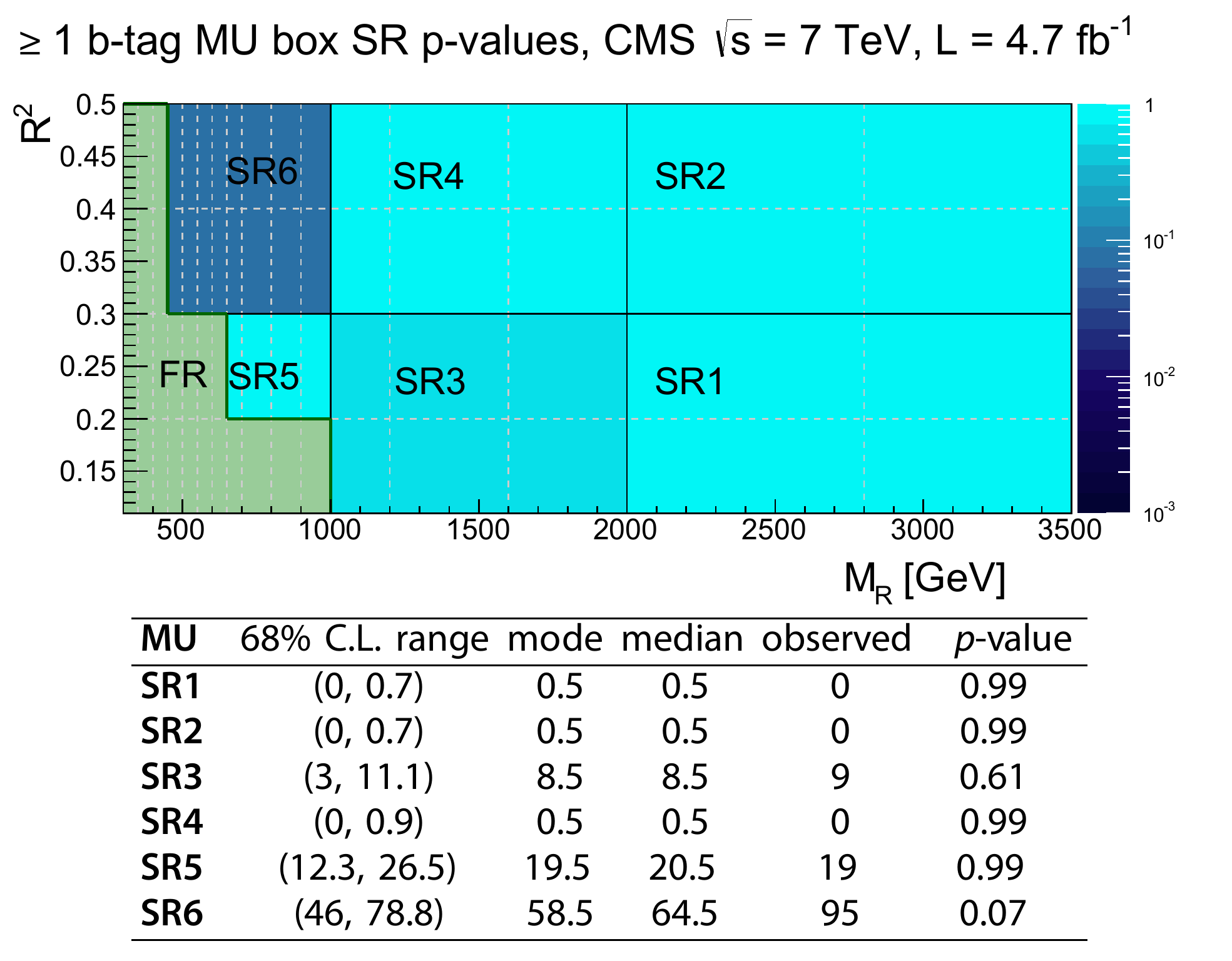}
\caption{The $p$-values corresponding to the observed number of events
  in the $\ge$1 $\cPqb$-tag  MU box signal regions (SR$i$).
\label{fig:bmu-blue-plot}}

\end{figure}

\begin{figure}[ht!]
\centering
\includegraphics[width=0.495\textwidth]{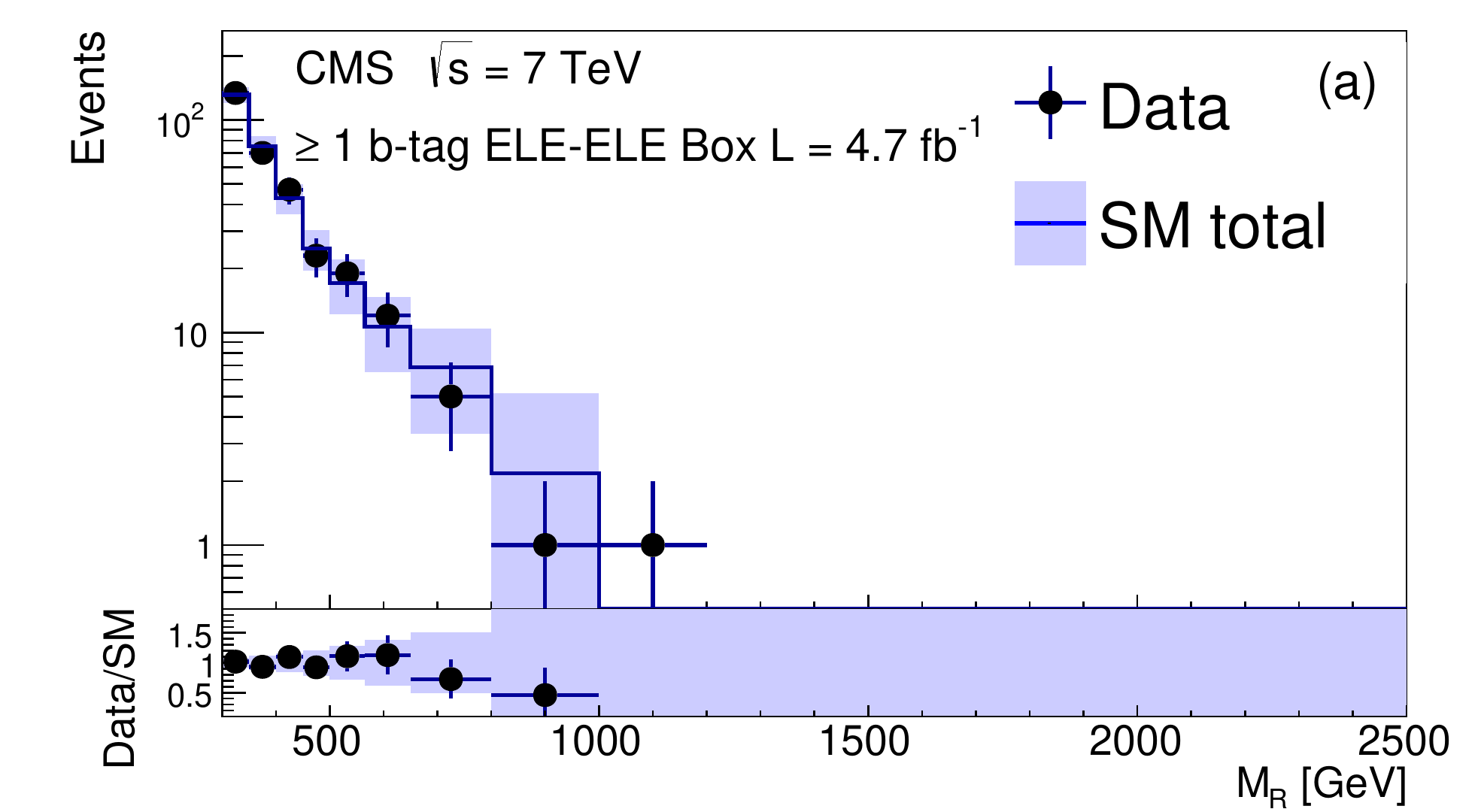}
\includegraphics[width=0.495\textwidth]{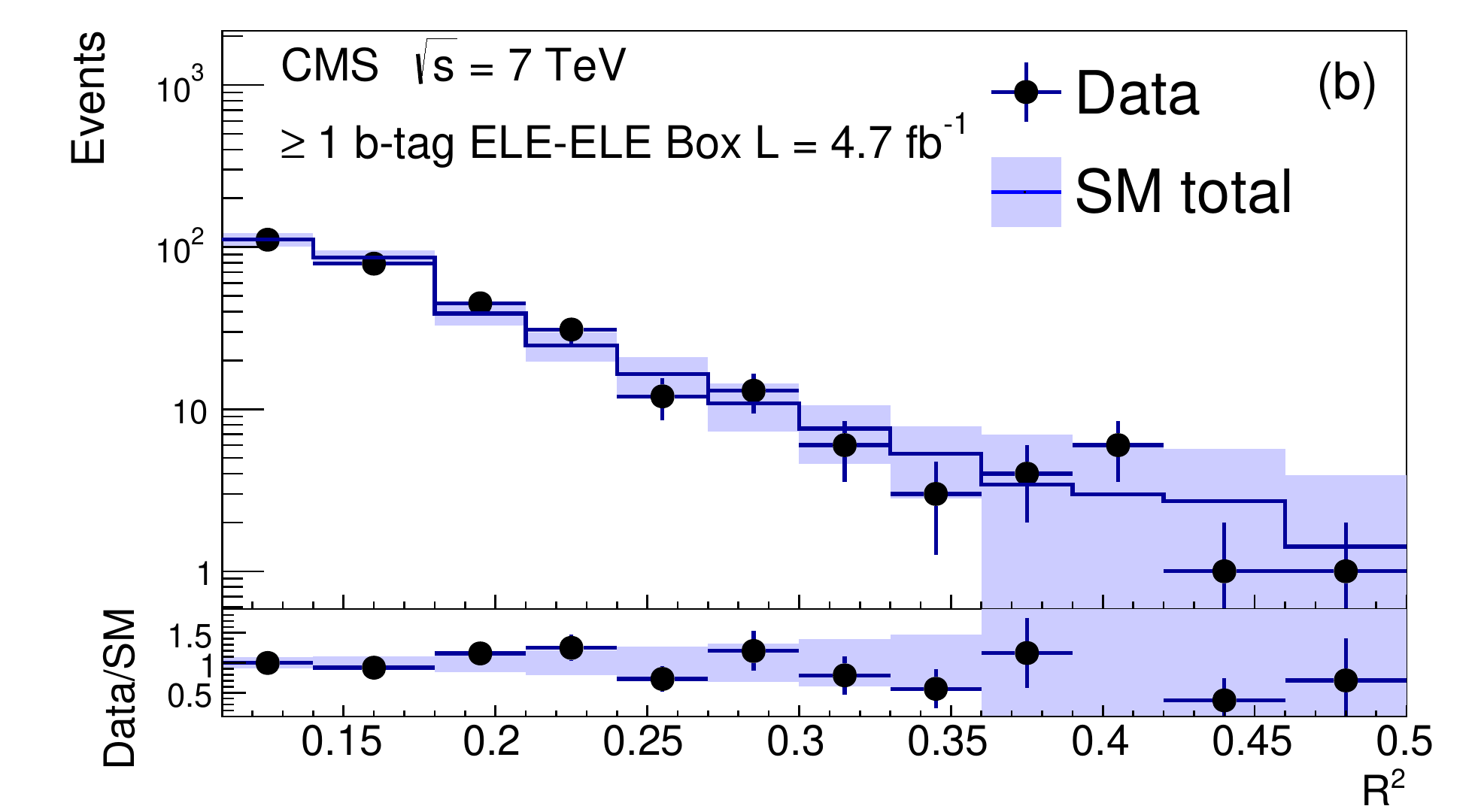}
\caption{Projection of the 2D fit result on (a) $\MR$ and (b) $\Rtwo$  for the ELE-ELE box  in the $\ge$1 $\cPqb$-tag analysis path. \label{fig:beleele-box}}

\end{figure}

\begin{figure}[htpb]
\centering
\includegraphics[width=\cmsFigWidthBox]{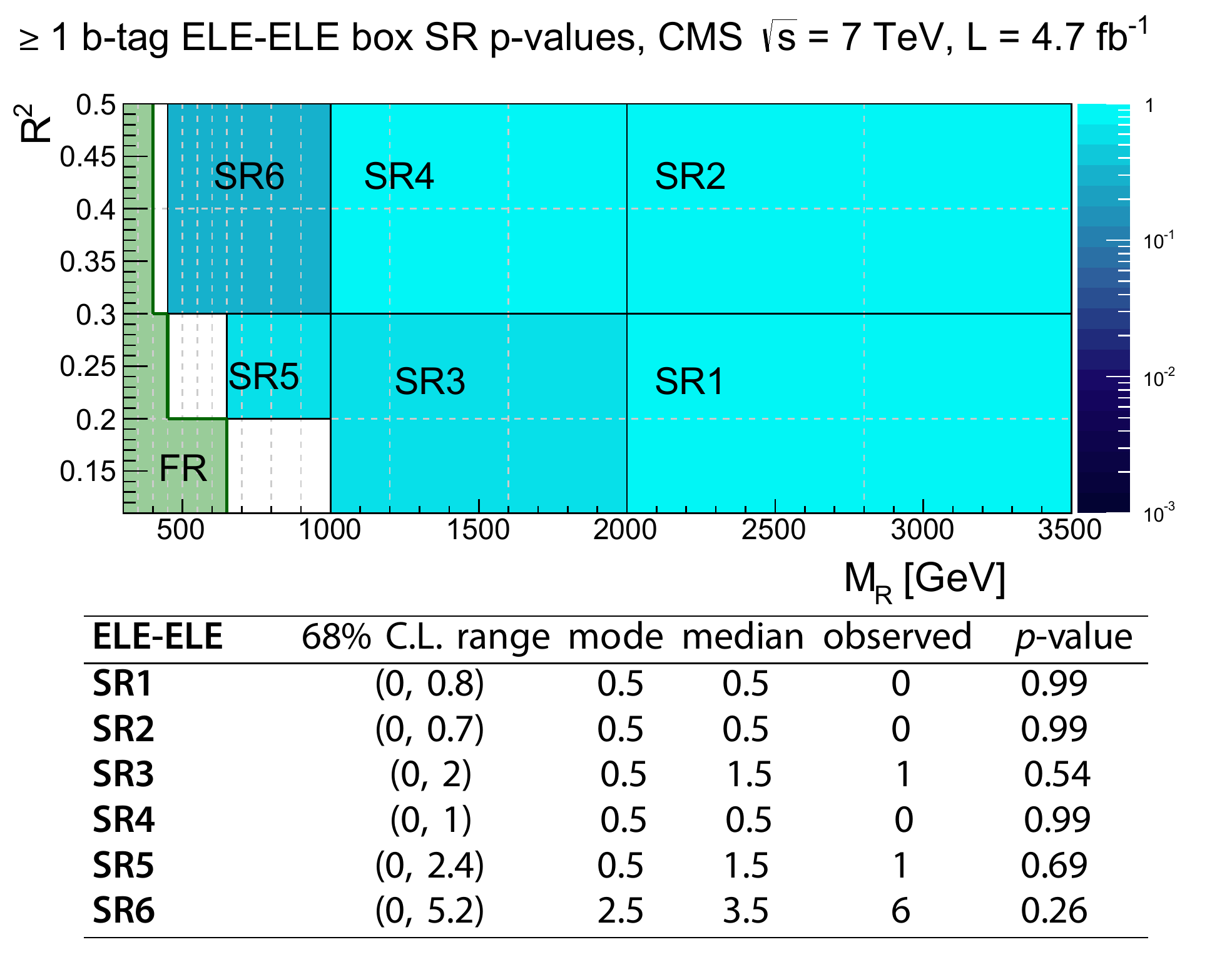}
\caption{The $p$-values corresponding to the observed number of events
  in the $\ge$1 $\cPqb$-tag ELE-ELE box signal regions (SR$i$).
  \label{fig:beleele-blue-plot}}

\end{figure}

\begin{figure}[ht!]
\centering
\includegraphics[width=0.495\textwidth]{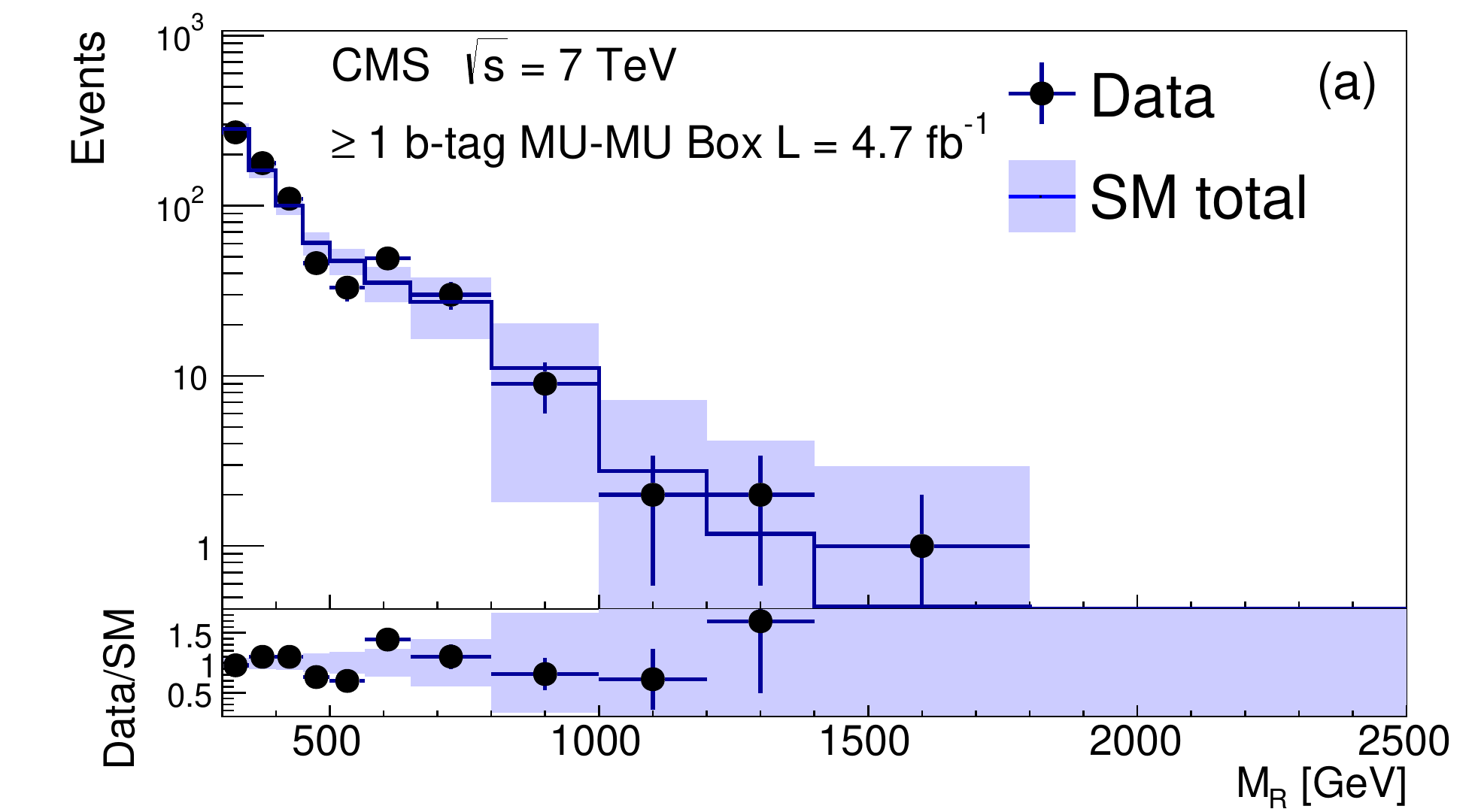}
\includegraphics[width=0.495\textwidth]{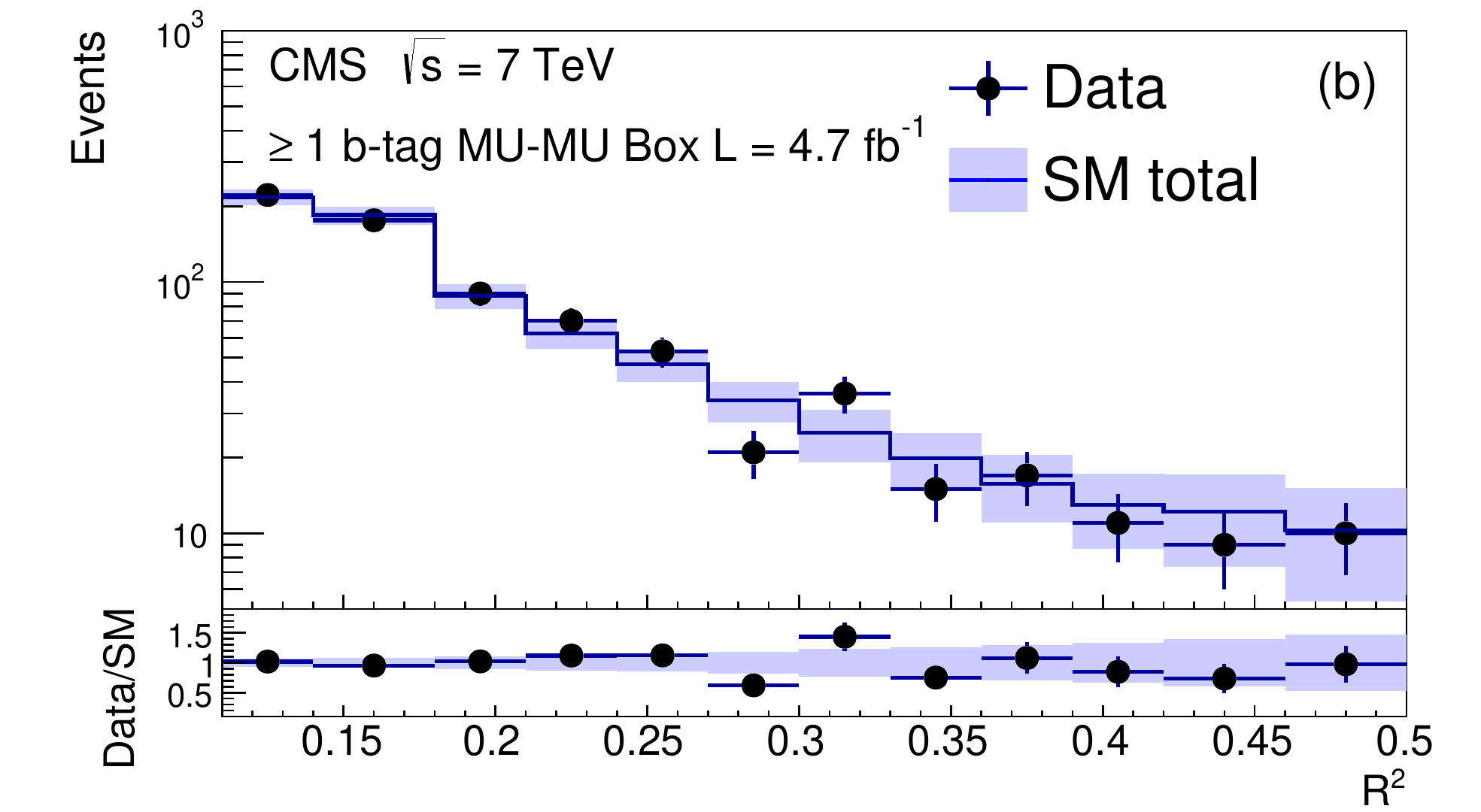}
\caption{Projection of the 2D fit result on (a) $\MR$ and (b)
  $\Rtwo$ for the MU-MU box in the $\ge$1 $\cPqb$-tag analysis
  path. \label{fig:bmumu-box}}
\end{figure}

\begin{figure}[htpb]
\centering
\includegraphics[width=\cmsFigWidthBox]{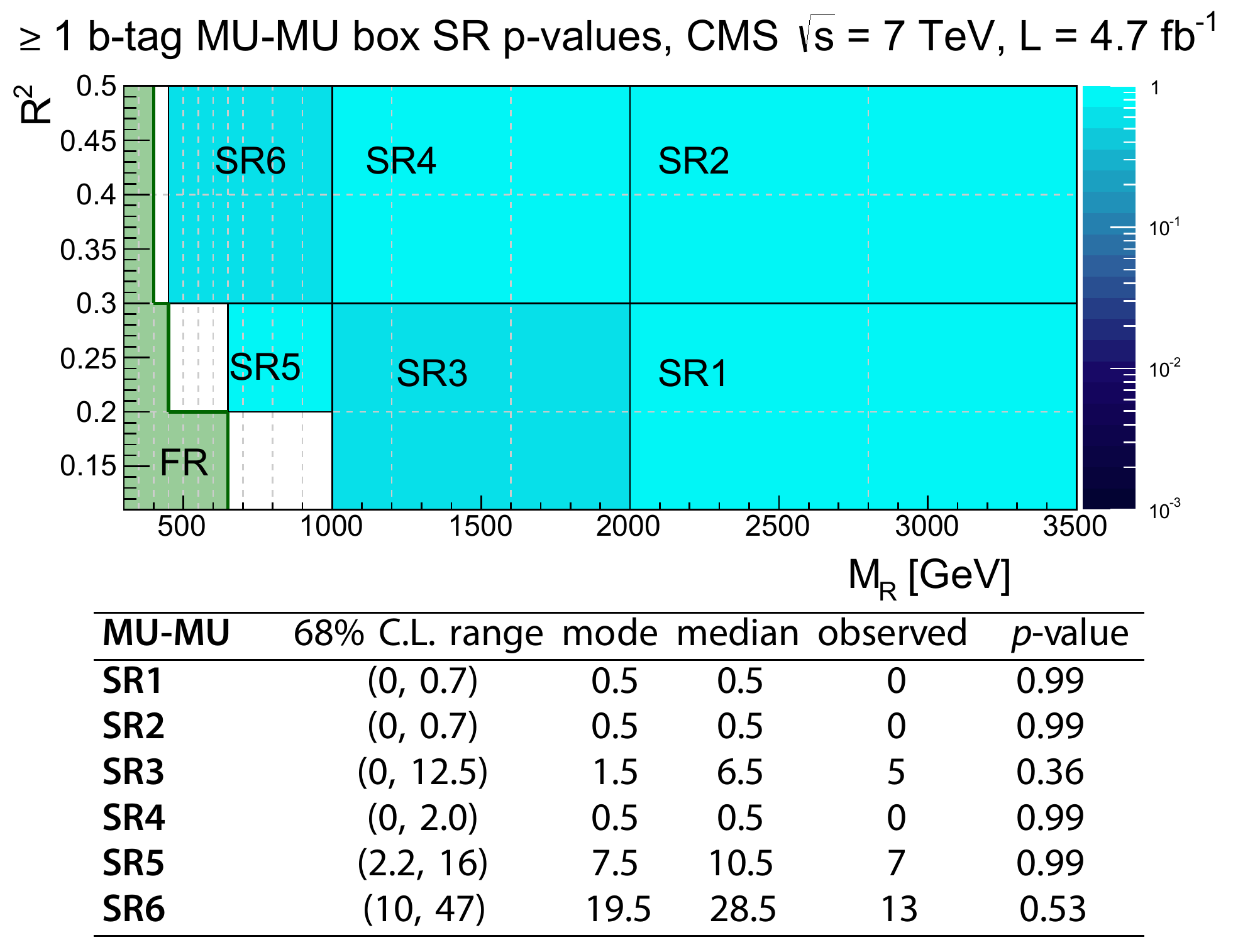}
\caption{The $p$-values corresponding to the observed number of events
  in the $\ge$1 $\cPqb$-tag  MU-MU box signal regions (SR$i$).
\label{fig:bmumu-blue-plot}}

\end{figure}

\begin{figure}[ht!]
\centering
\includegraphics[width=0.495\textwidth]{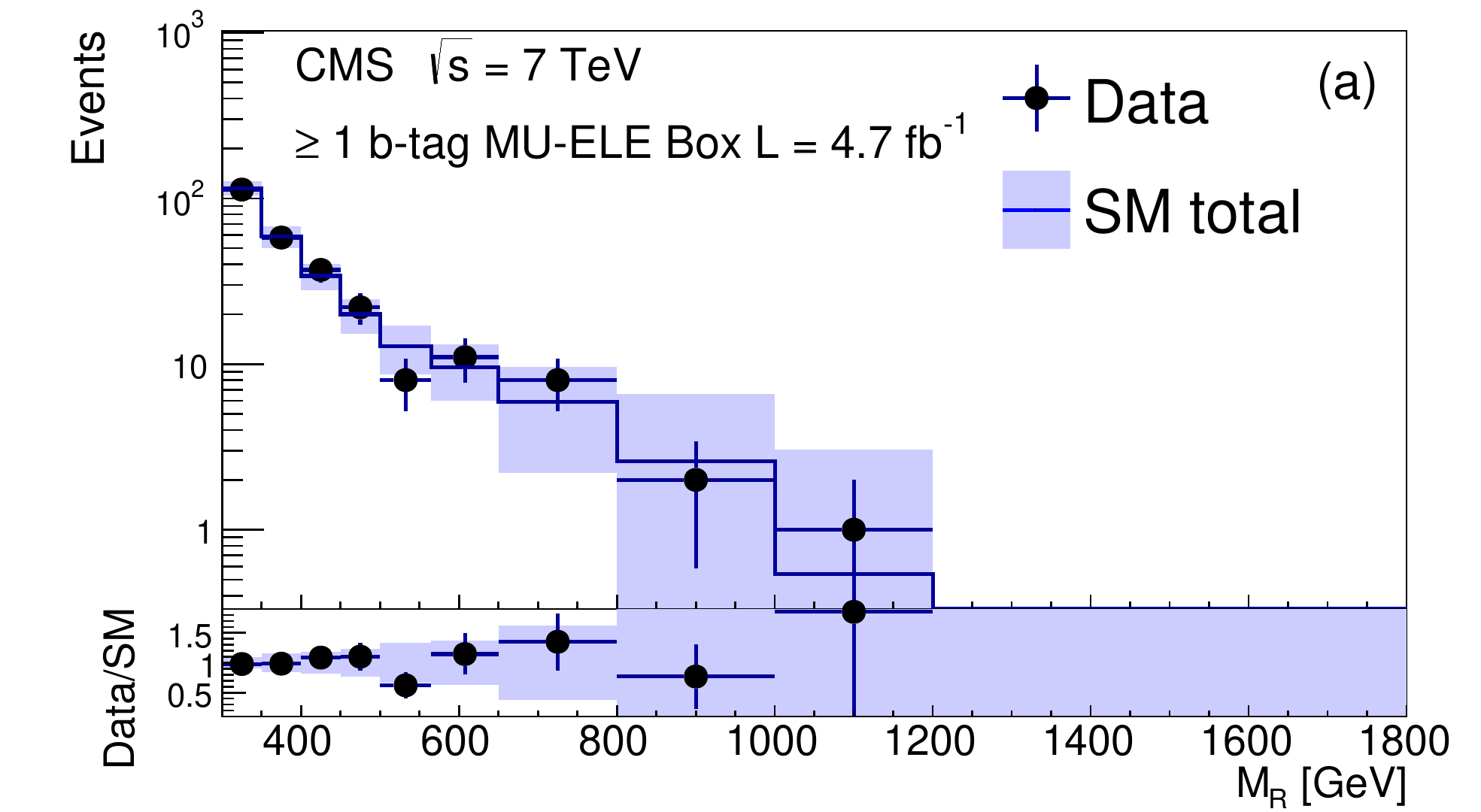}
\includegraphics[width=0.495\textwidth]{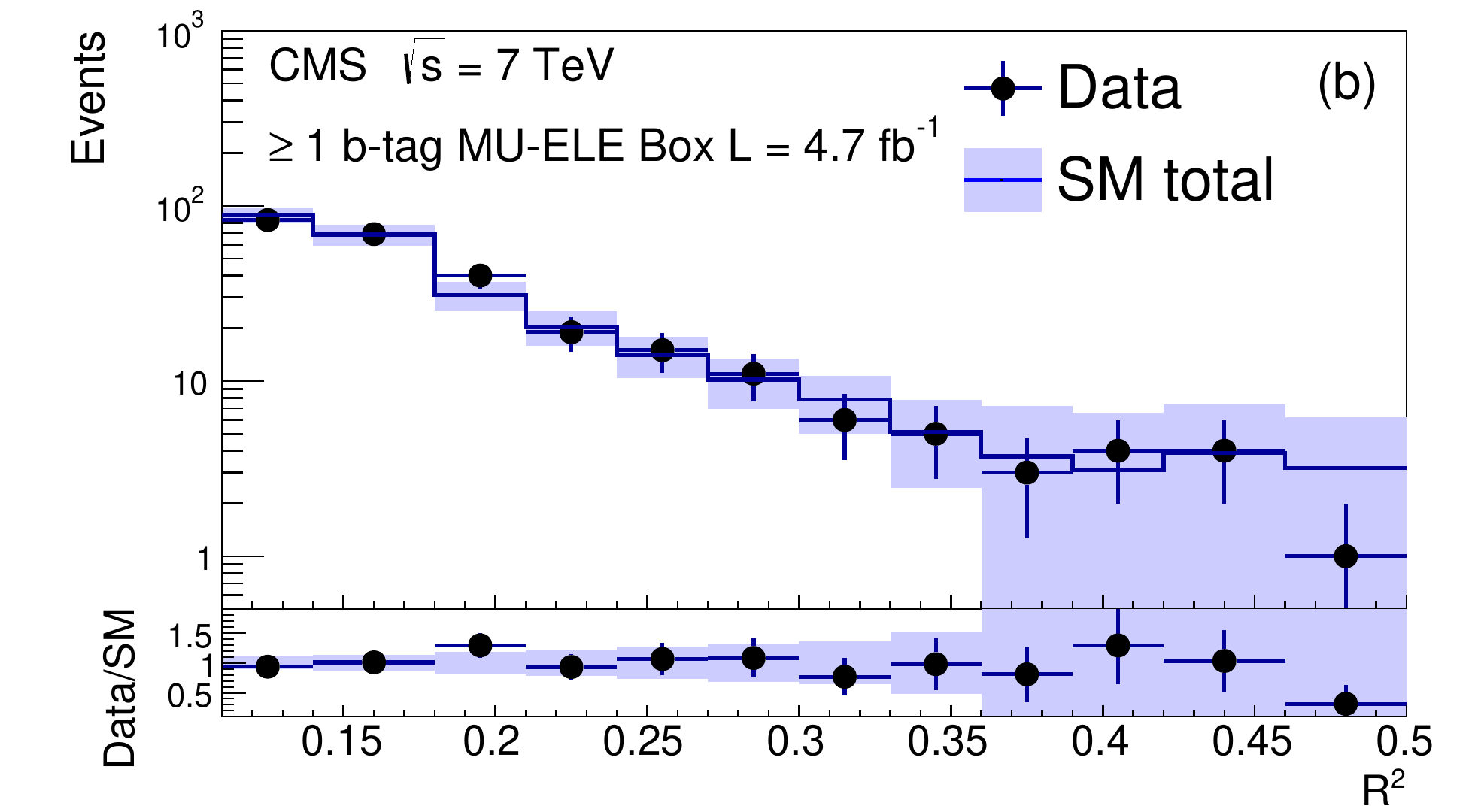}
\caption{Projection of the 2D fit result on (a) $\MR$ and (b)
  $\Rtwo$ for the MU-ELE box in the $\ge$1 $\cPqb$-tag
  analysis path. \label{fig:bmuele-box}}
\end{figure}

\begin{figure}[ht!]
\centering
\includegraphics[width=\cmsFigWidthBox]{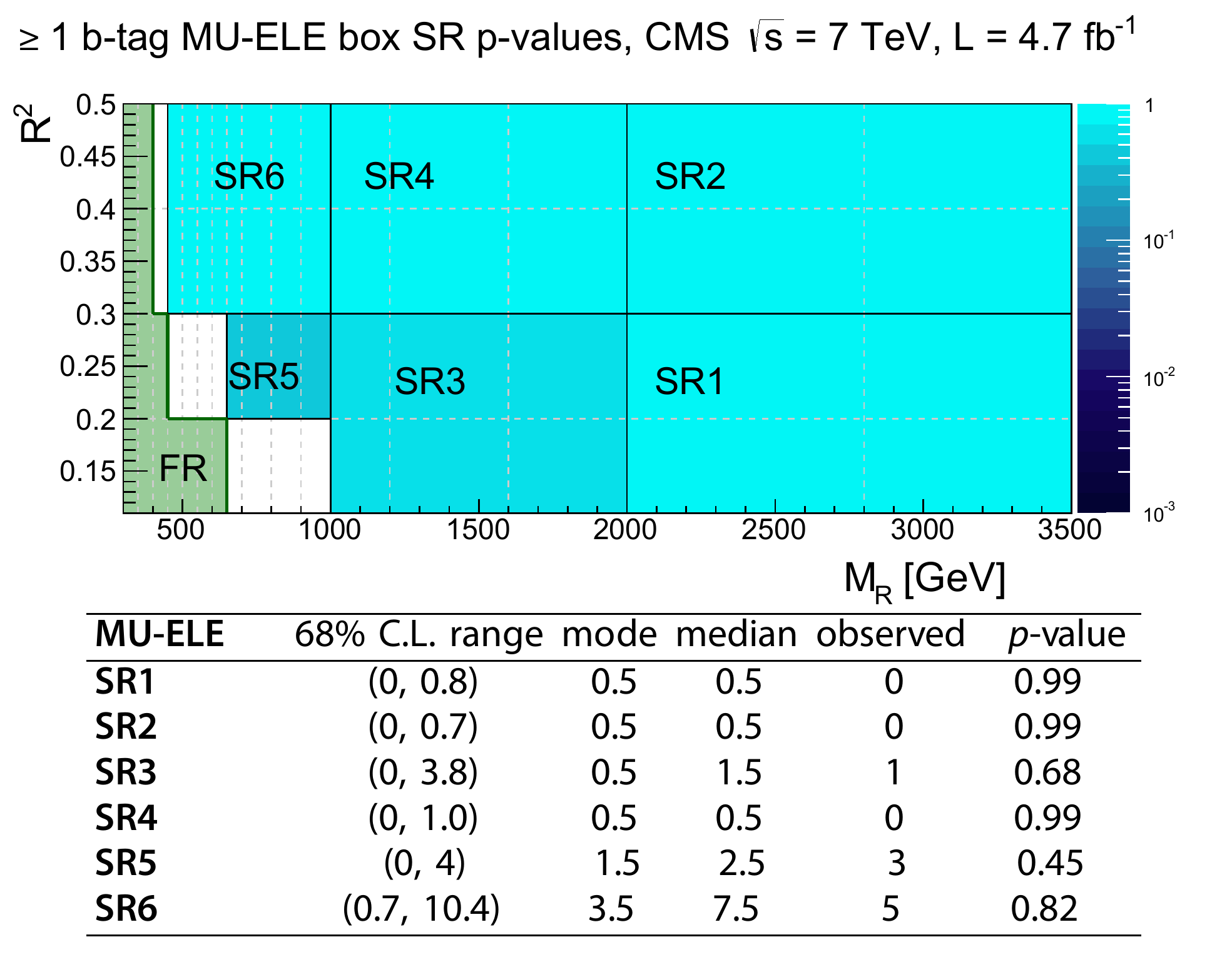}
\caption{The $p$-values corresponding to the observed number of events
  in the $\ge$1 $\cPqb$-tag  MU-ELE box signal regions (SR$i$).
\label{fig:bmuele-blue-plot}}

\end{figure}

Figures~\ref{fig:bhad-box}-\ref{fig:bmuele-blue-plot} show the results
for the $\ge$1 $\cPqb$-tagged jet analysis corresponding to the
results presented in Section~\ref{sec:datafits} for the inclusive
analysis.






\clearpage

\section{Guide on emulating the razor analysis for additional studies}

In this appendix, we provide a guide to facilitate use of the razor
analysis results for the interpretation of signal scenarios not
considered here. We assume the existence of an event generator that
can simulate LHC collisions for a given theoretical model. We also
assume that this event generator is interfaced to a parton shower
simulation, such that a list of produced particles at the generator
level is available. The procedure described in this appendix
represents a simplification of the analysis, giving conservative
limits within the $\pm 1$ standard deviation band of the nominal
result.

The following classes of stable particles are relevant to this
analysis: i) invisible particles (neutrinos and any weakly interacting
stable new particles, for example the LSP in SUSY models); ii)
electrons; iii) muons; iv) all other stable electrically charged
SM particles; and v) all other stable electrically neutral SM particles.
It is possible to emulate the razor analysis as follows:
\begin{itemize}
\item all the visible stable particles are clustered into
  generator-level jets using the anti-\kt algorithm with a
  distance parameter of 0.5.
\item the generator-level \ETm is computed as $E^\text{miss}_\text{T, Gen} =
  -\sum_p \pt^{p}$, where the sum runs over all the
  visible stable particles $p$.
\item the detector resolution is applied to electrons and muons
  according to a simplified Gaussian resolution function. The RMS of
  the Gaussian smearing depends on the $\eta$ and $\pt$ values of the
  lepton, as well as its flavor. Similarly, the \ETm and jet
  momenta are smeared according to a Gaussian response model.
\item the detector efficiency is applied to electrons and muons
  generating unweighted events from the reconstruction efficiency,
  interpreted as a probability (see Section~\ref{sec:appemu}). The
  efficiency depends on the $\eta$ and $\pt$ values of the lepton, its
  flavor, and its generator-level isolation, as computed from the
  stable particles in the event.
\item the analysis selection and box classification is applied.
\end{itemize}
This procedure allows us to estimate the $\Rtwo$ versus
$\MR$ distribution for a signal model and the efficiency in
each box. This is the information that is needed to associate a 95\%
\CL upper limit to a given input model. The procedure matches the full
simulation of CMS to within 20\% and in general provides a result
that is yet closer to the CMS full simulation. The result is in
general conservative, since the computation of the upper limit starts
from a simplified binned likelihood, which reduces the sensitivity to
a signal. This procedure is not expected to correctly simulate the
special case of slowly moving electrically charged particles (e.g.,
staus). The remainder of this appendix describes each step of the
razor emulation in more detail, including the calculation of the
exclusion limit.

\subsection{Emulation of reconstructed electrons and muons}\label{sec:appemu}

The emulation of reconstructed electrons and muons consists of two independent
steps: the accounting for the detector resolution and for the
reconstruction efficiency.

\begin{figure}[tpb]
  \centering
    \includegraphics[width=0.49\textwidth]{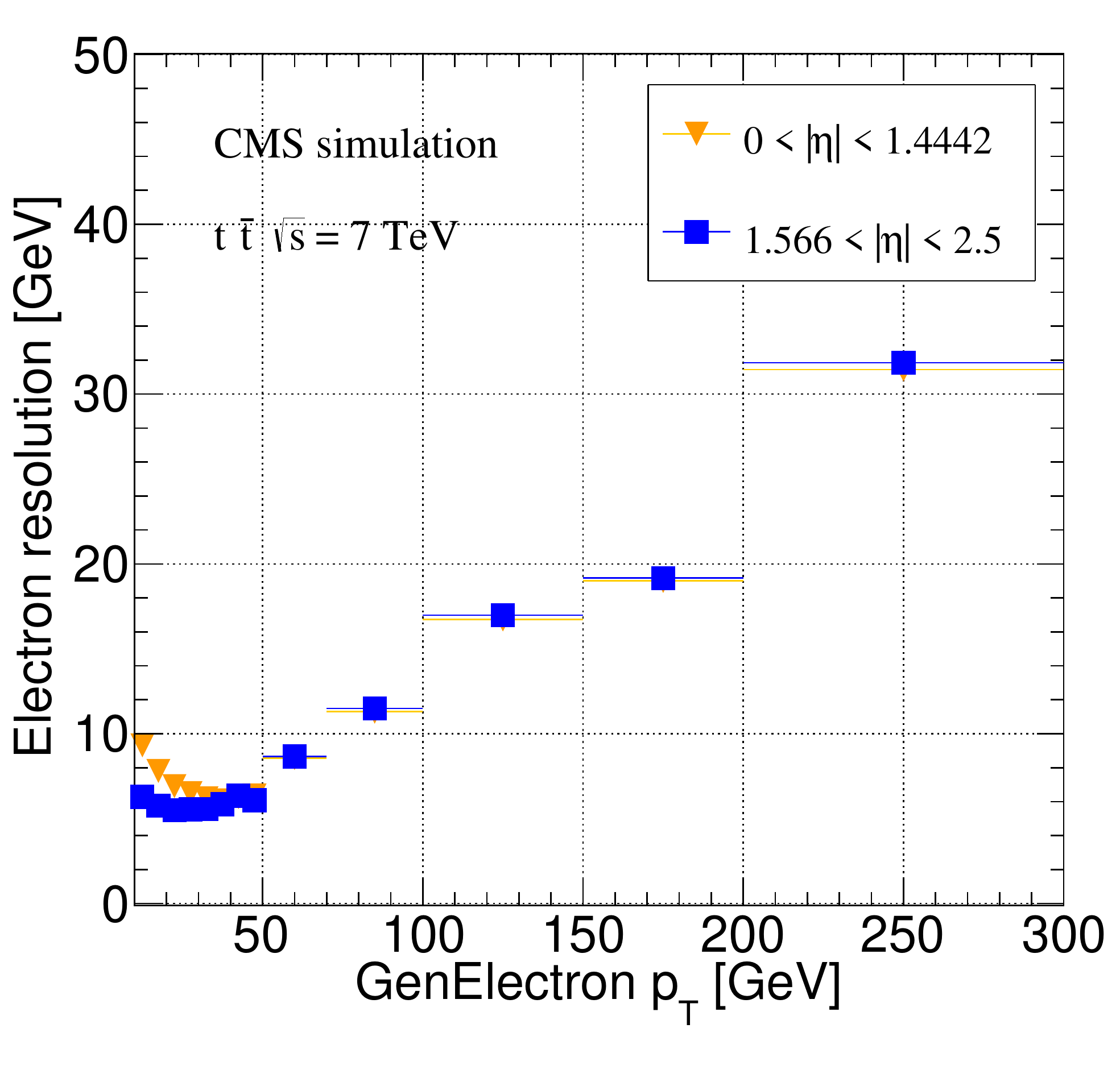}
    \includegraphics[width=0.49\textwidth]{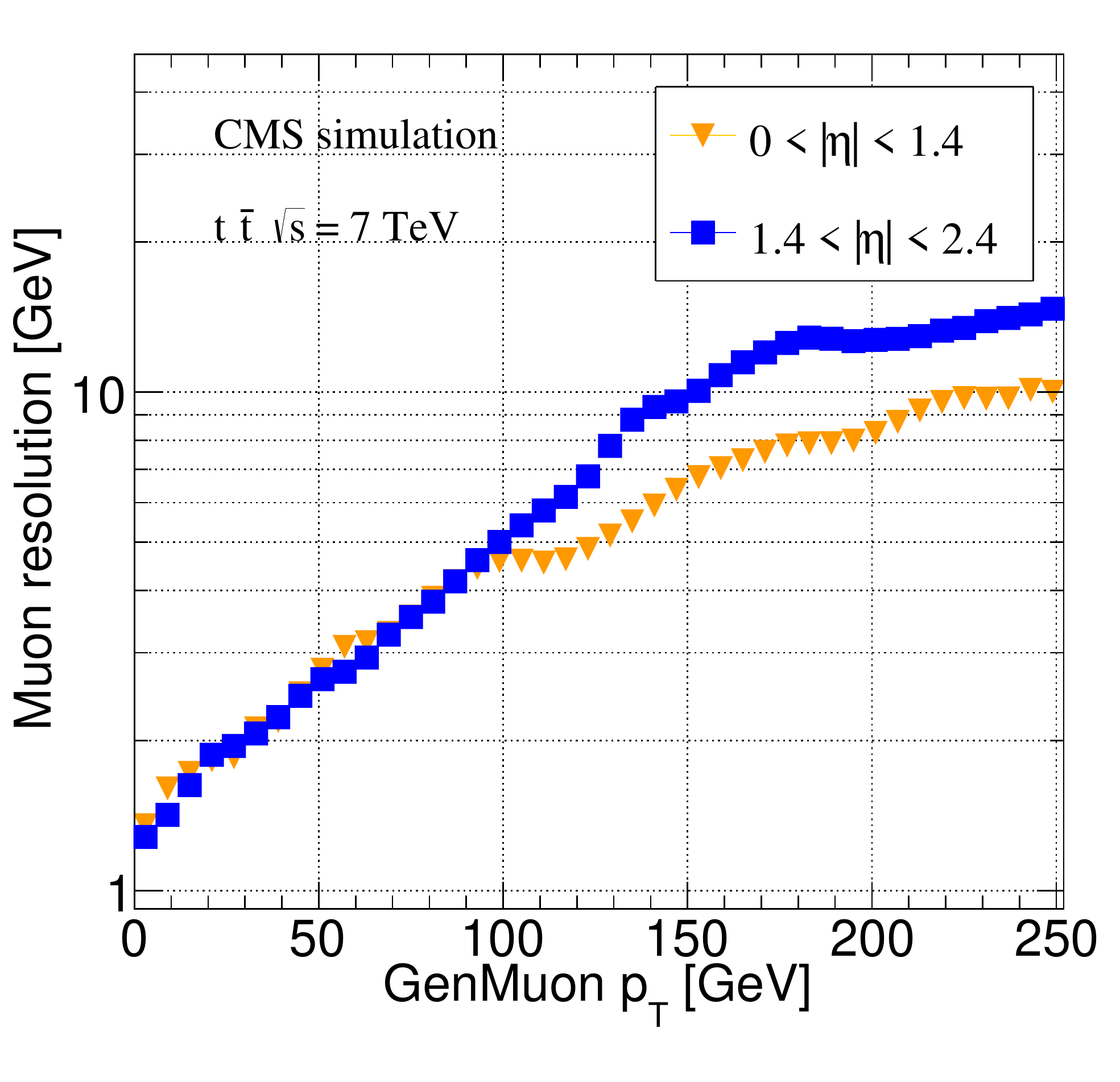}
 \caption{\label{fig:leptonResolution} Momentum resolution for
  (\cmsLeft) electrons and (\cmsRight) muons within the barrel region of the CMS
  detector (squares) and in the endcaps (triangles).}
\end{figure}

The effects of detector resolution can be incorporated through a
Gaussian smearing of the genuine $\pt$ of a given lepton, while the
lepton $\eta$ and $\phi$ can be considered to be unaffected by the
detector resolution. The generated lepton is then replaced by the
reconstructed one, having the same flight direction with a $\pt$
value randomly extracted according to a Gaussian distribution centered at
$\pt^\text{Gen}$ and with $\sigma(\pt^\text{Gen})$ taken from
Fig.~\ref{fig:leptonResolution}. Any lepton outside the two
$\eta$ ranges considered in Fig.~\ref{fig:leptonResolution} should be
discarded from the analysis.

To account for the reconstruction efficiency of a given lepton, the
generator-level isolation is computed as follows:
\begin{equation}
\text{GenIso}(\ell) = \frac{\sum_{p \neq \ell} \pt^{p}}{\pt^\ell},
\end{equation}
where the sum runs over all the stable charged and neutral visible
particles $p$ within a distance $\Delta R
=\sqrt{(\Delta\eta)^2+(\Delta\phi)^2}<0.5$ from the lepton.

\begin{figure*}[htpb]
\includegraphics[width=0.49\textwidth]{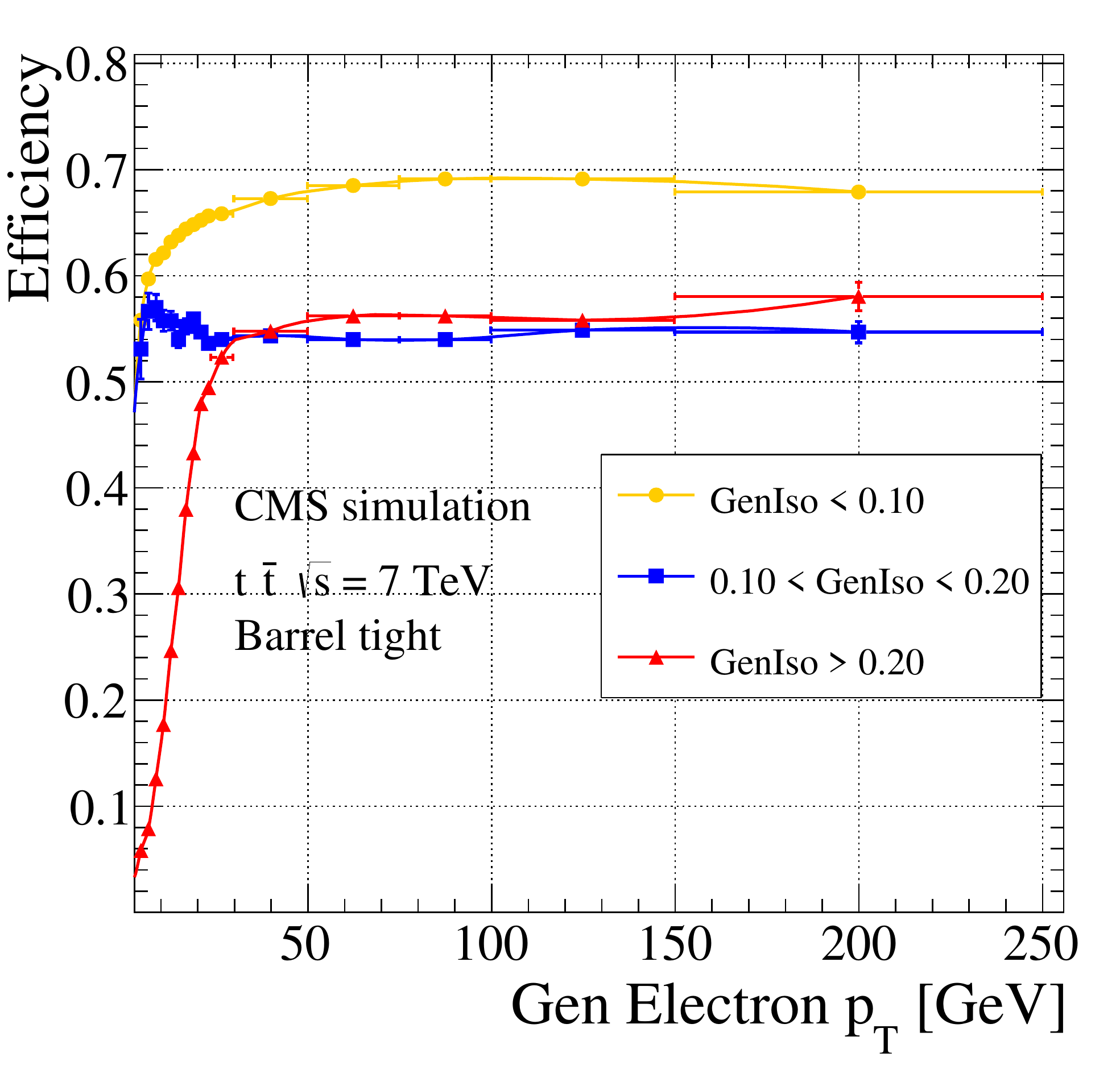}
\includegraphics[width=0.49\textwidth]{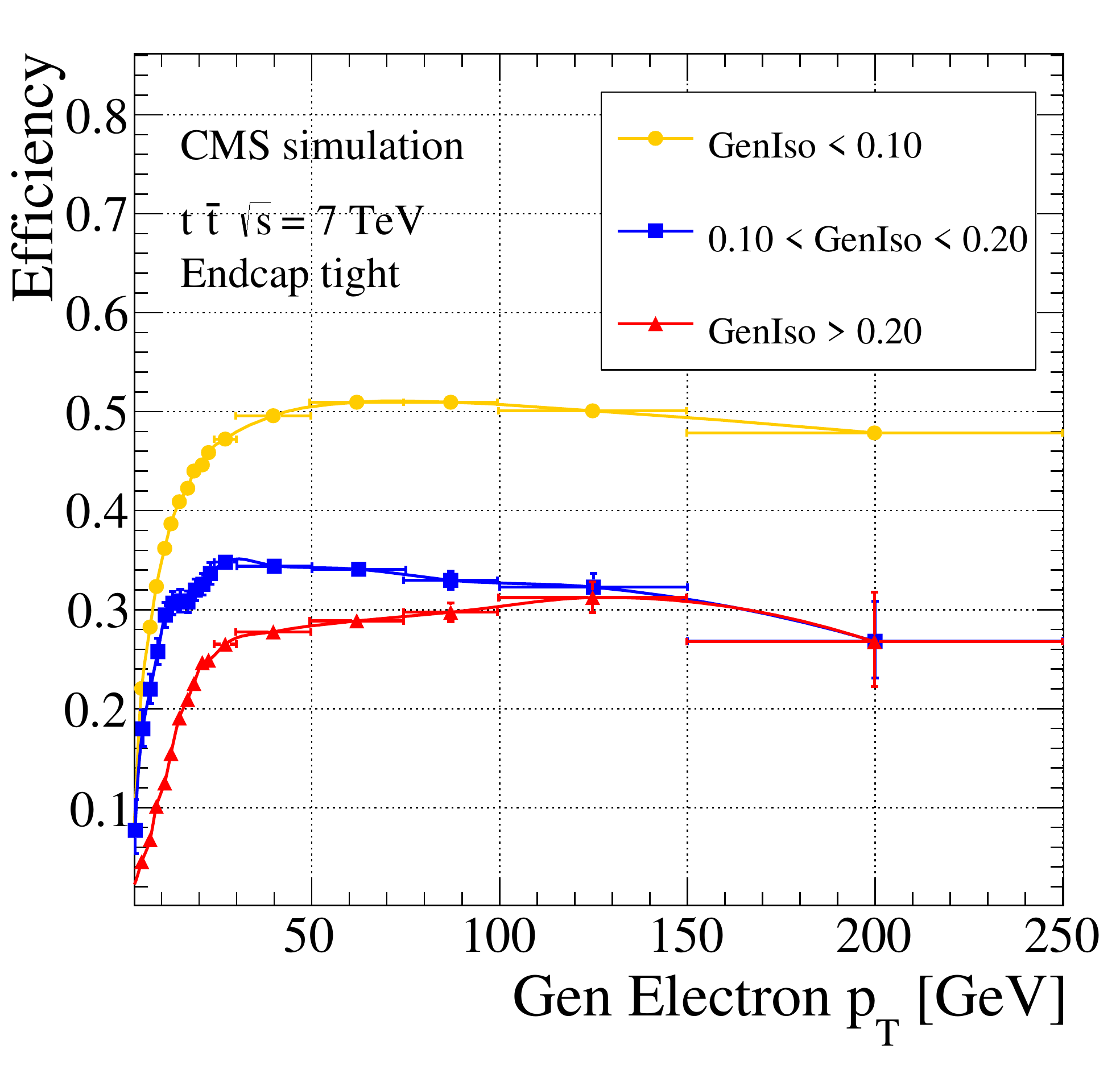}
\includegraphics[width=0.49\textwidth]{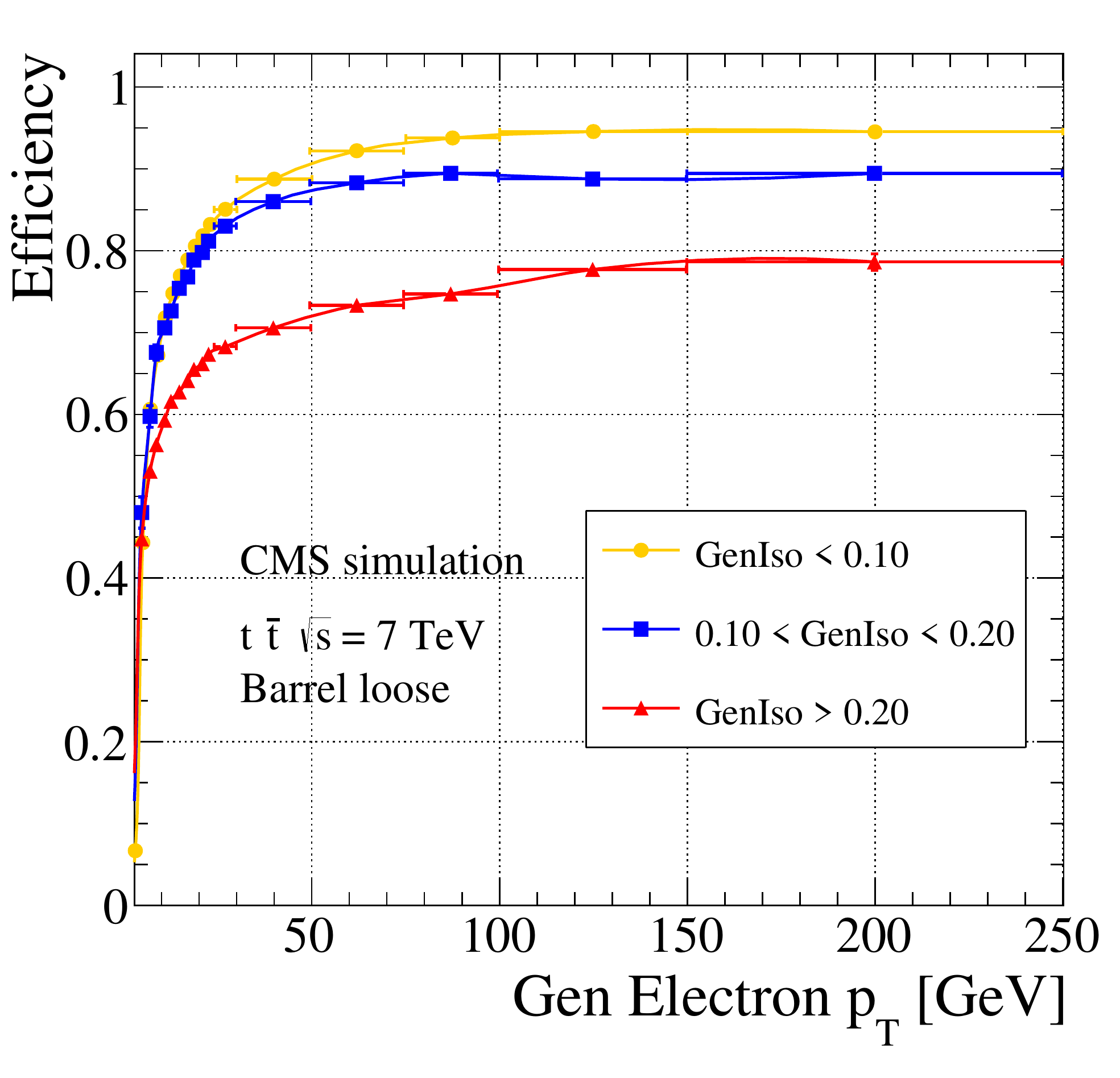}
\includegraphics[width=0.49\textwidth]{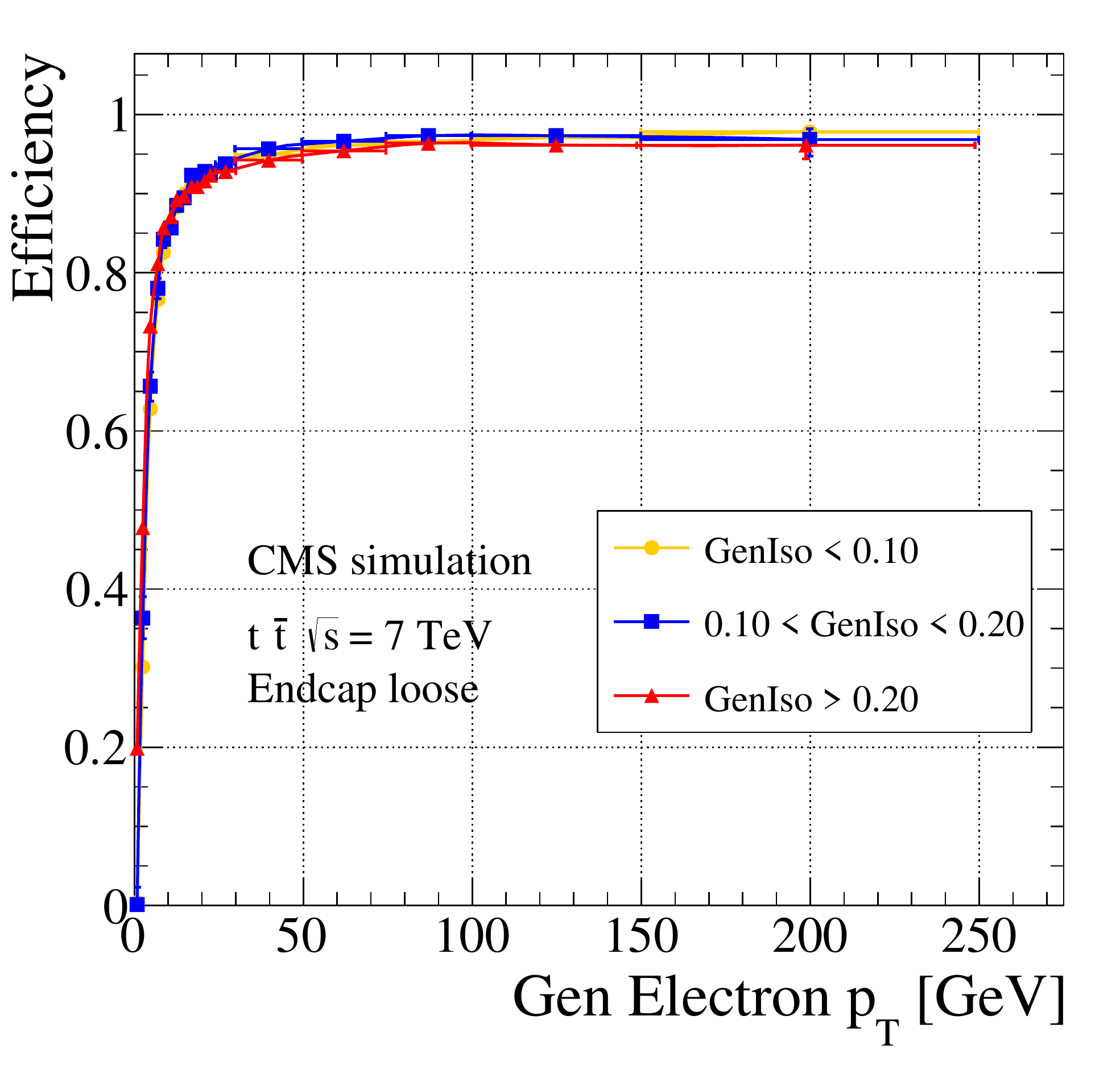}
 \caption{\label{fig:electronReco} Electron reconstruction efficiency
   for (top panes) tight and (bottom panes) loose electrons pointing to the
   ECAL (left panes) barrel and (right panes) endcaps, estimated from the CMS
   MC simulation of $\cPqt \cPaqt$ events. The electron
   reconstruction is described in Ref.~\cite{CMS_e}.}
\end{figure*}

Figure~\ref{fig:electronReco} shows the reconstruction probability
versus the generated electron $\pt$ (before accounting for the
detector resolution) for three ranges of $\text{GenIso}$ in the ECAL
barrel ($\abs{\eta}< 1.4442$) and endcap ($1.5660<\abs{\eta}<2.5000$)
regions. Different values are obtained for the tight and the loose
electrons used to define the boxes.

\begin{figure}[htpb]
\centering
    \includegraphics[width=0.49\textwidth]{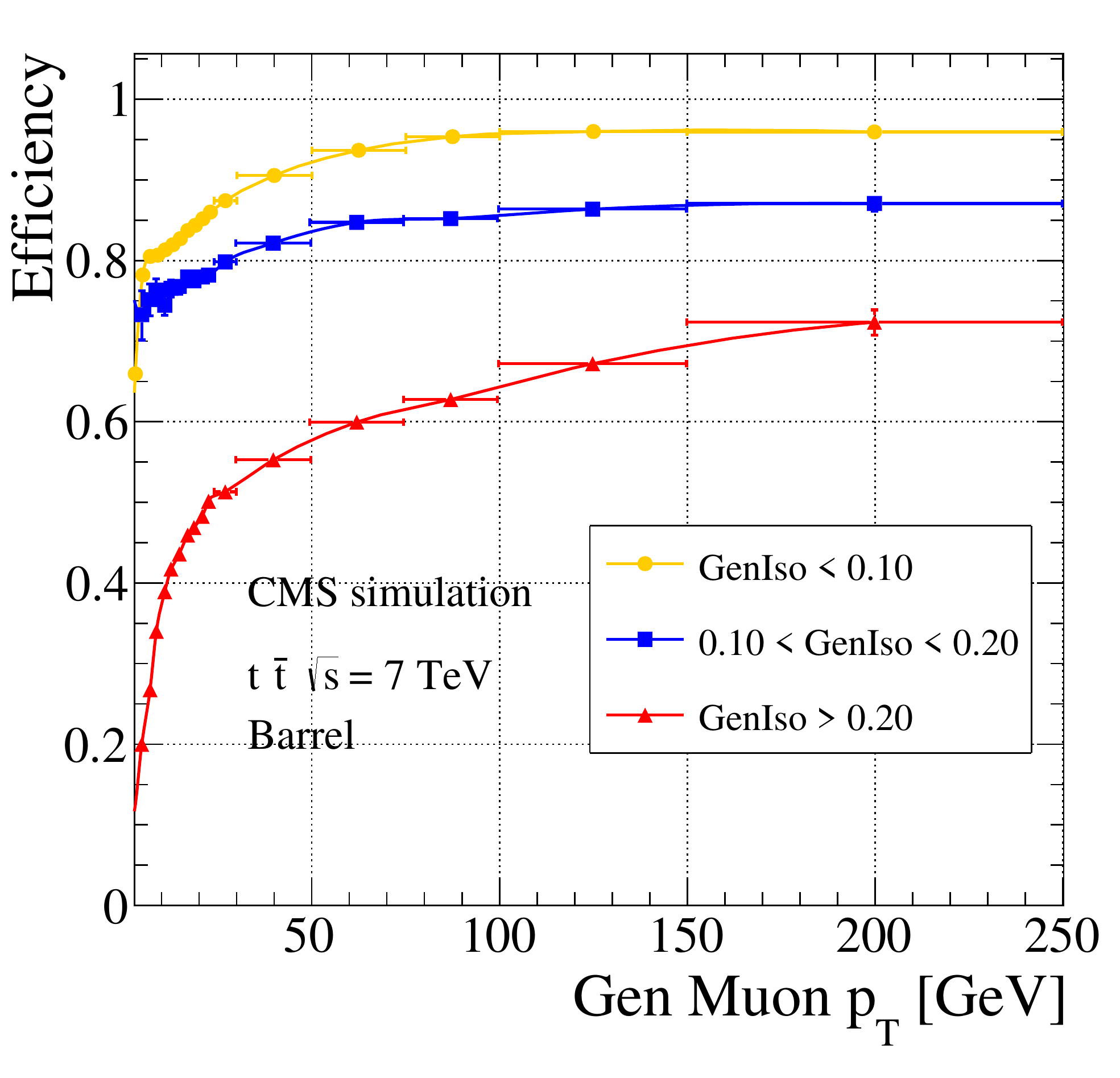}
    \includegraphics[width=0.49\textwidth]{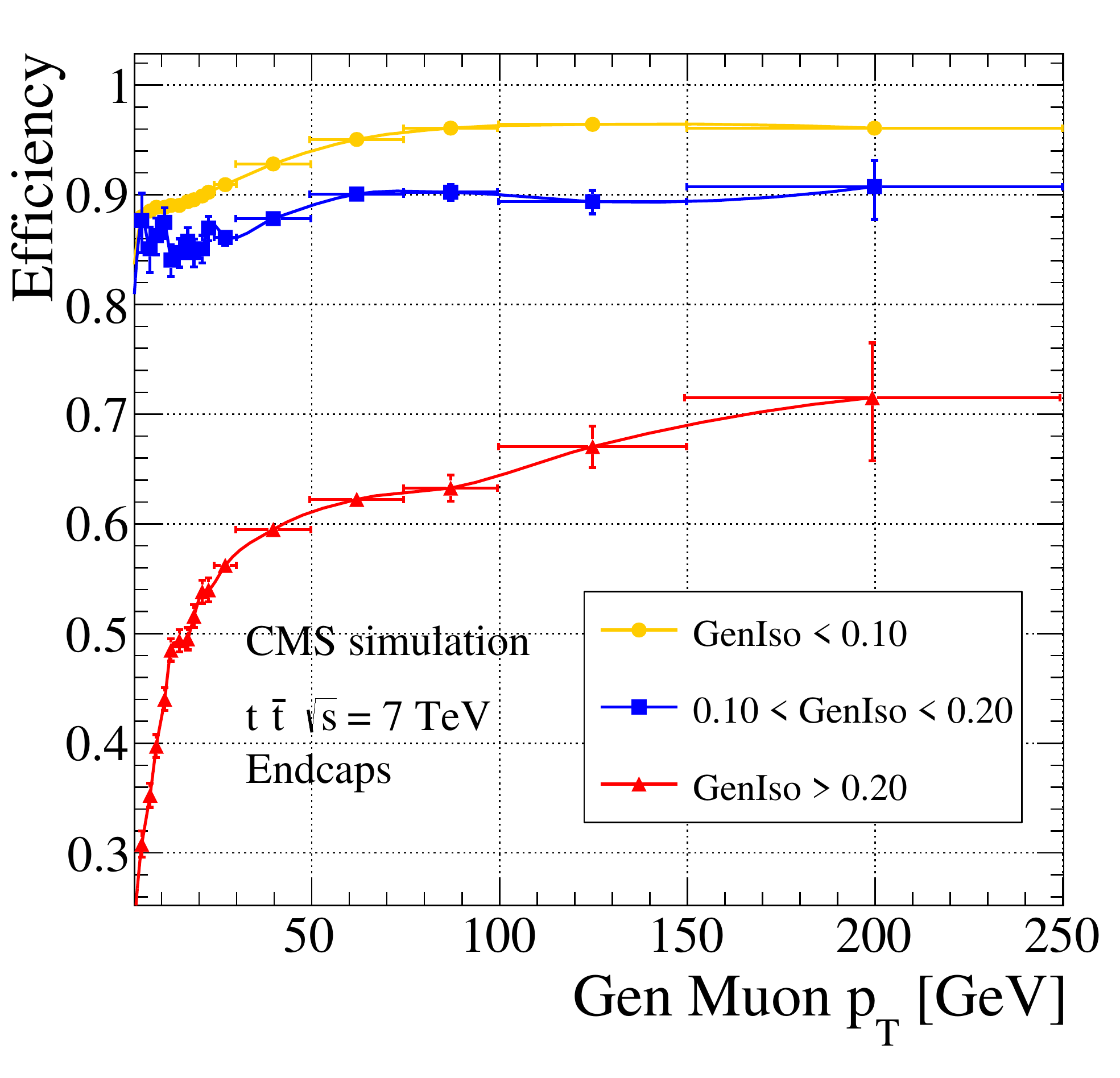}
 \caption{\label{fig:muonReco} Muon reconstruction efficiency for
   tight muons pointing to the (\cmsLeft) barrel and (\cmsRight) endcaps, estimated
   from the CMS MC simulation of $\ttbar$ events. The
   muon reconstruction is described in Ref.~\cite{CMS_mu}.}
\end{figure}

Similarly, the reconstruction efficiency for the tight muons is shown in
Fig.~\ref{fig:muonReco}. The reconstruction of loose muons can
be considered to be fully efficient for muons with $\pt>10$\GeV, since no
isolation requirement is applied.

Once the lepton reconstruction probability is found, the detector
efficiency effects can be imposed numerically: the lepton is rejected
if a uniformly distributed random number in the range [0,1] is found
to be larger than the reconstruction efficiency.

\subsection{Emulation of reconstructed jets and \texorpdfstring{\ETm}{missing ET}}

\begin{figure}[htpb]
\centering
\includegraphics[width=0.49\textwidth]{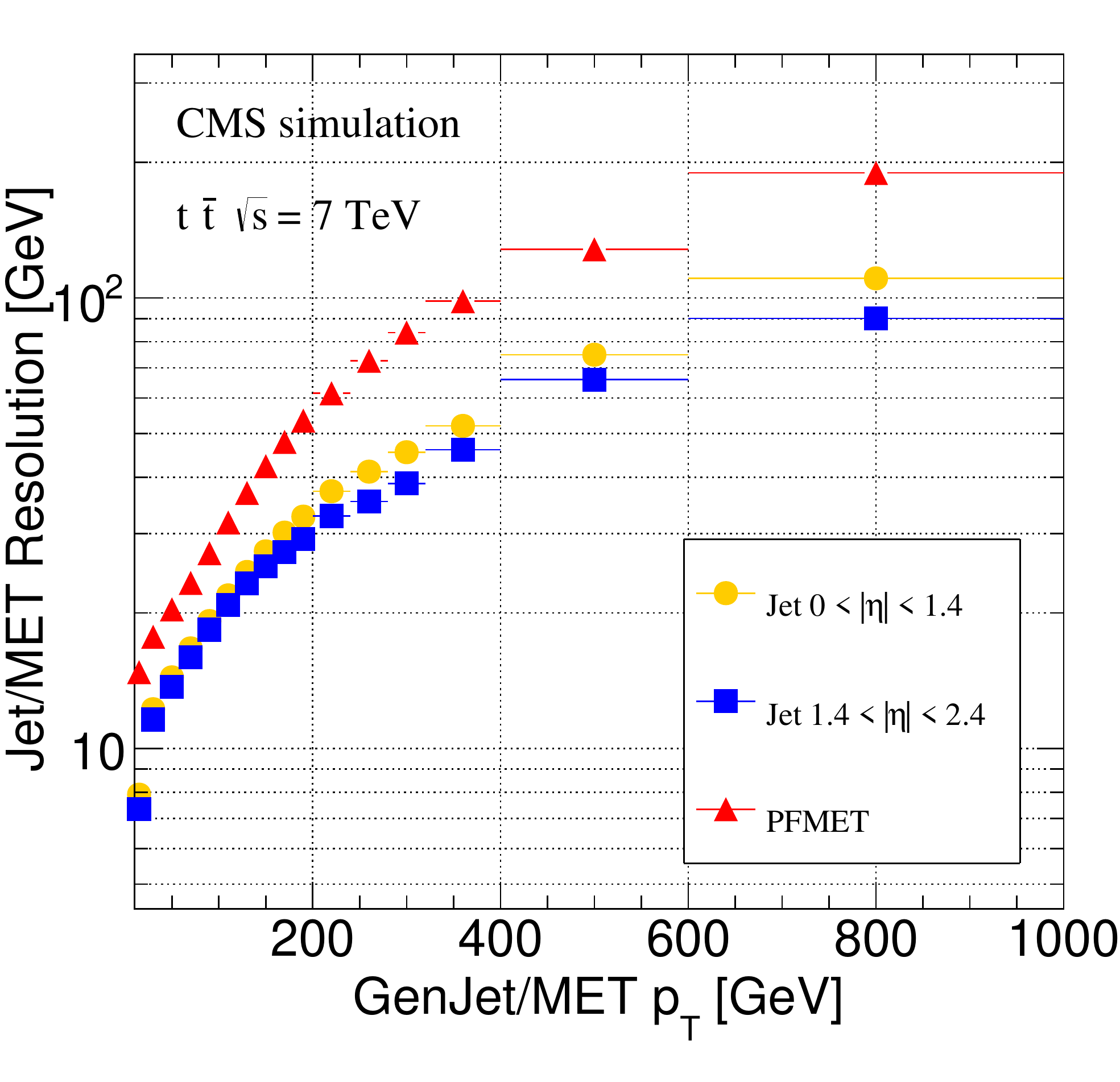}
\caption{\label{fig:resoJetMET}Transverse energy resolution for jets
  and \ETm, in the CMS MC simulation of \ttbar events.}
\end{figure}

The reconstruction of jets and \ETm can be emulated by applying a
Gaussian resolution to the generator-level quantities. We show in
Fig.~\ref{fig:resoJetMET} the dependence of the Gaussian $\sigma_\text{jet}$ on the jet $\pt$ (for the two relevant bins of $\eta$) and the
\ETm. The dependence on $\eta$ or other quantities can be safely
neglected. One should apply the resolution function to all the
reconstructed jets and to the \ETm and then impose the acceptance
selection on the reconstructed jets.

\subsection{Building the 2D templates}

Once detector effects have been accounted for, jets are clustered in
two megajets. The razor variables can be computed from the
four-momenta of the two megajets.

\begin{figure*}[htpb]
\includegraphics[width=0.32\textwidth]{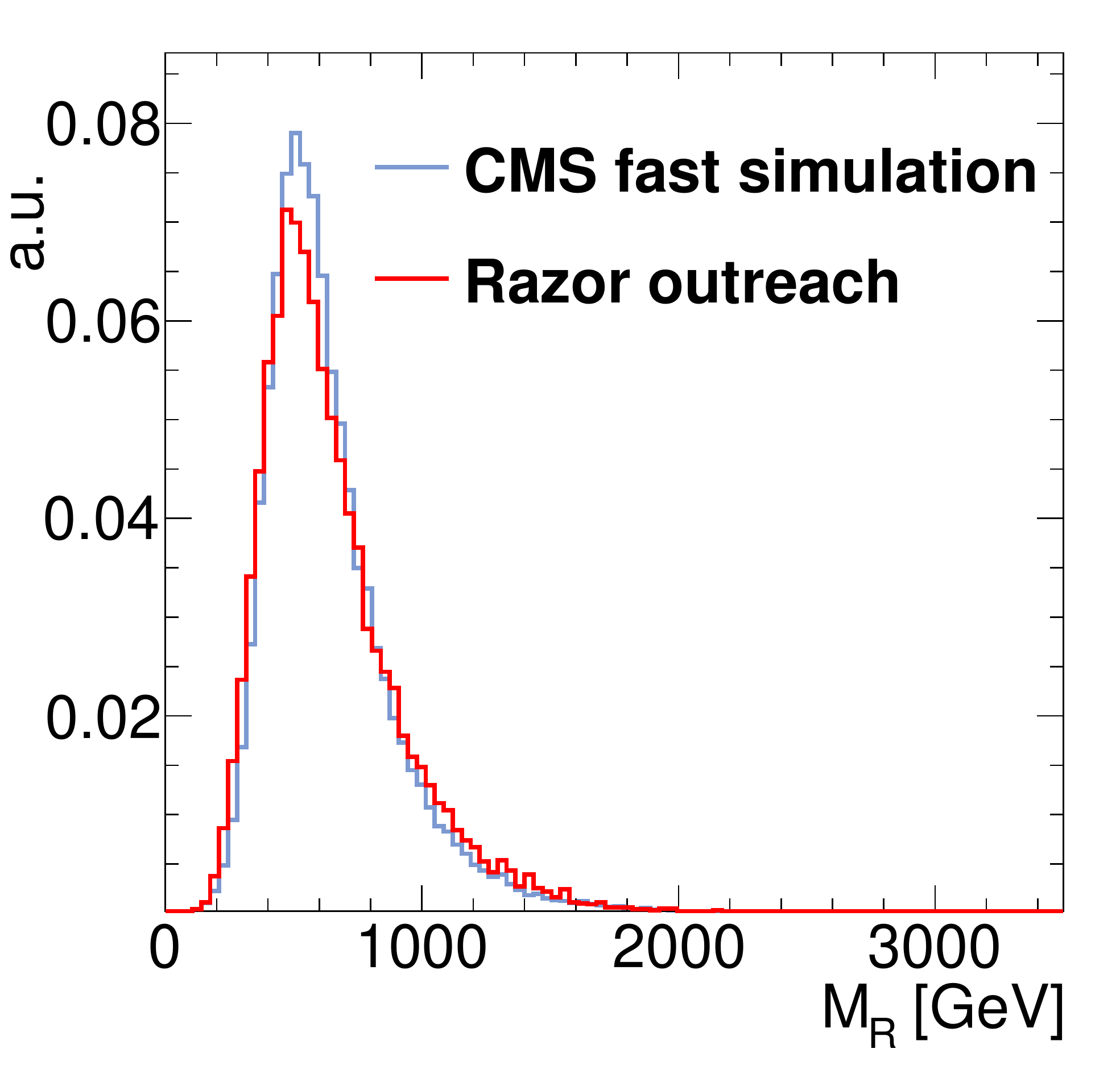}
\includegraphics[width=0.32\textwidth]{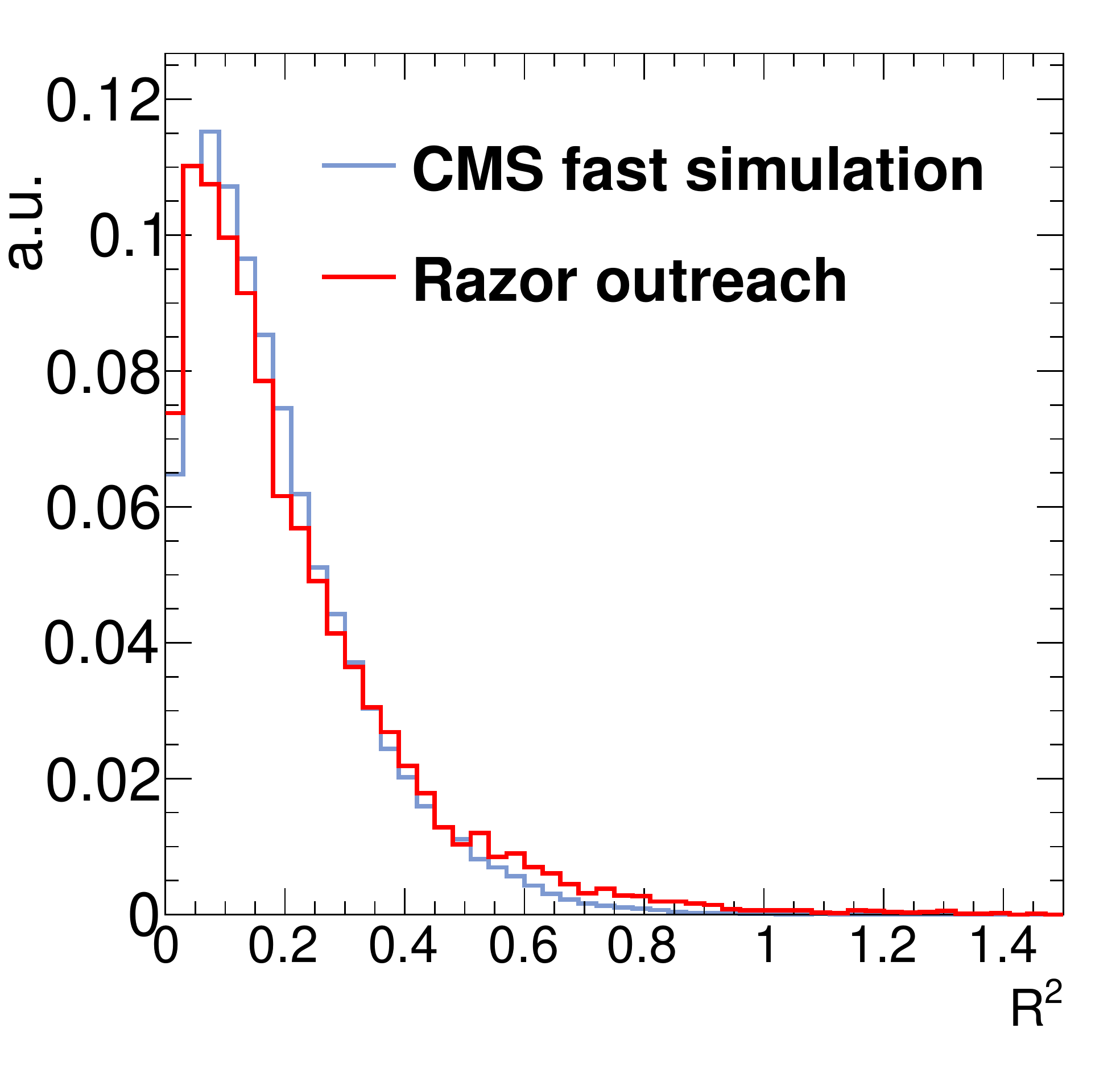}
\includegraphics[width=0.32\textwidth]{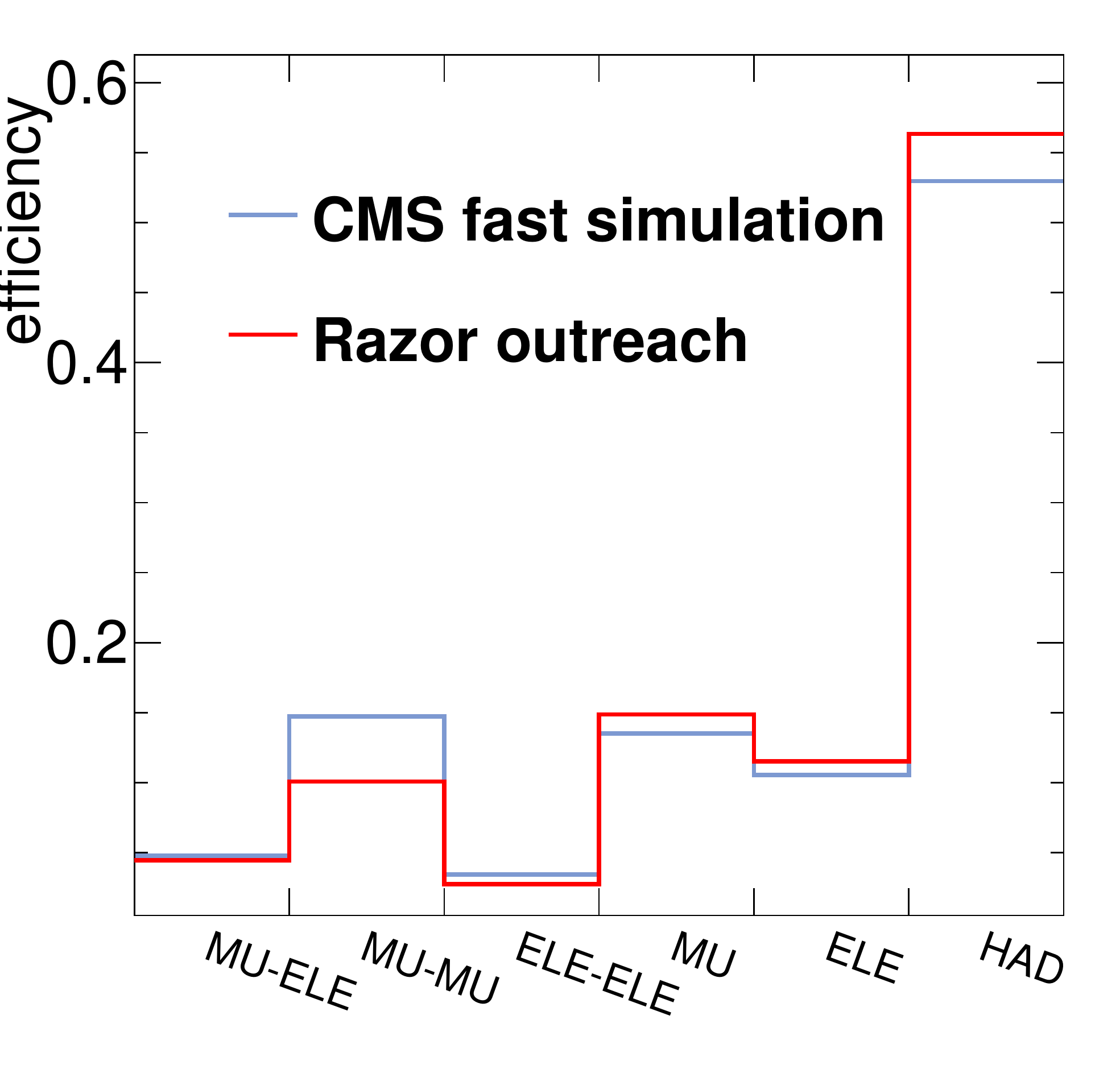}
 \caption{\label{fig:razorT1tttt} Comparison of the (left) $\MR$
   distribution, (middle) $\Rtwo$ distribution, and (right) the
   efficiency versus box obtained from the official CMS fast
   simulation package and the emulation procedure described in this
   appendix. The two distributions correspond to a T1tttt sample with
   800\GeV gluino mass and 300\GeV LSP mass.}
\end{figure*}

Figure~\ref{fig:razorT1tttt} (left) and (middle) shows the $\MR$ and
$\Rtwo$ distributions for a sample of pair-produced gluinos of
mass 800\GeV, where each gluino decays to a $\ttbar$ pair and a LSP of
mass 300\GeV, obtained with the CMS fast simulations program and with
the emulation described in this appendix. The efficiencies obtained
for the six boxes are compared in Fig.~\ref{fig:razorT1tttt} (right).

\subsection{Evaluating the exclusion limit}

The exclusion limit can be computed from the 2D signal templates and
the box efficiencies, starting with the observed yield and the
expected background. We consider a simplified likelihood obtained by
defining bins in the ($\MR$, $\Rtwo$) plane. Each bin
$i$ requires the observed yield $n_i$ and the expected background
$\bar{b}_i \pm \delta_i$ computed by integrating the background model
and taking into account the uncertainty in shape. The likelihood in a
given box is then written as:
\ifthenelse{\boolean{cms@external}}{
\begin{multline}
\mathcal{L}_\text{box}(\vec n|\sigma,\vec{b},\rho) = \logN(b_i|\bar
b_i,\delta_i)\logN(\rho| 1, \delta_\rho)\\ \textstyle{\prod\nolimits}_i P(n_i|\epsilon_i \rho L \sigma +b_i),
\end{multline}
}{
\begin{equation}
\mathcal{L}_\text{box}(\vec n|\sigma,\vec{b},\rho) = \logN(b_i|\bar
b_i,\delta_i)\logN(\rho| 1, \delta_\rho) \textstyle{\prod\nolimits}_i P(n_i|\epsilon_i \rho L \sigma +b_i),
\end{equation}
}
where $\epsilon_i$ is the signal efficiency in that bin, $L$ is the
luminosity, and $\sigma$ is the signal cross section; $\logN(b_i|\bar{b}_i,\delta_i)$ is the log-normal distribution describing
the uncertainty in the background. $\logN(\rho| 1,
\delta_\rho)$ is the distribution describing the uncertainty in the
signal efficiency. A value $\delta_\rho \sim 0.20$ (including the
uncertainty in the integrated luminosity) is large enough to account
for the use of a simplified detector emulation and the typical
systematic uncertainty quoted in the analysis. Once this uncertainty
is included, the uncertainty in the luminosity can be neglected to a
good level of precision. Similarly, the total likelihood can be
written as:
\ifthenelse{\boolean{cms@external}}{
\begin{multline}
\mathcal{L}_{\text{TOT}}(\vec n|\sigma,\vec{b},\rho) =\\
\logN(\rho| 1, \delta_\rho)  [
\textstyle{\prod\nolimits_{\text{box}}}  \textstyle{\prod\nolimits_i} P(n_i^{\text{box}}|\epsilon_i^{\text{box}} \rho L
\sigma+b_i^{\text{box}}) \\ \times \logN(b_i^{\text{box}}|\bar
b_i^{\text{box}},\delta_i^{\text{box}})].
\end{multline}
}{
\begin{equation}
\mathcal{L}_{\text{TOT}}(\vec n|\sigma,\vec{b},\rho) =
\logN(\rho| 1, \delta_\rho) [
\textstyle{\prod\nolimits_{\text{box}}}  \textstyle{\prod\nolimits_i} P(n_i^{\text{box}}|\epsilon_i^{\text{box}} \rho L
\sigma+b_i^{\text{box}}) \times \logN(b_i^{\text{box}}|\bar
b_i^{\text{box}},\delta_i^{\text{box}})].
\end{equation}
}
In this case, the signal systematic parameter $\rho$ is common to the
six boxes. A Bayesian upper limit (UL) on the cross section can then
be computed assuming a flat prior distribution in $\sigma$:
\begin{equation}
\frac{\int_0^{\sigma \mathrm{UL}} \rd\sigma  \int \rd\rho\, \rd\vec{b}\, \mathcal{L}_{\text{TOT}}(\vec n|\sigma,\vec{b},\rho)}{\int_0^{+\infty}
  \rd\sigma \int \rd\rho\, \rd{}\vec{b} \, \mathcal{L}_{\text{TOT}}(\vec n|\sigma,\vec{b},\rho)} = 0.95.
\end{equation}
 An implementation of this simplified limit calculator is
 \ifthenelse{\boolean{cms@external}}{ provided in the supplemental material \cite{supp-material}}{currently available at \cite{twiki-razor}}
 together with the values of $n$, $\cPqb$, and $\delta$ for each bin
 in each box.

\subsection{Limit on simplified models}

\begin{figure*}[htpb]
  \centering
    \includegraphics[width=0.49\textwidth]{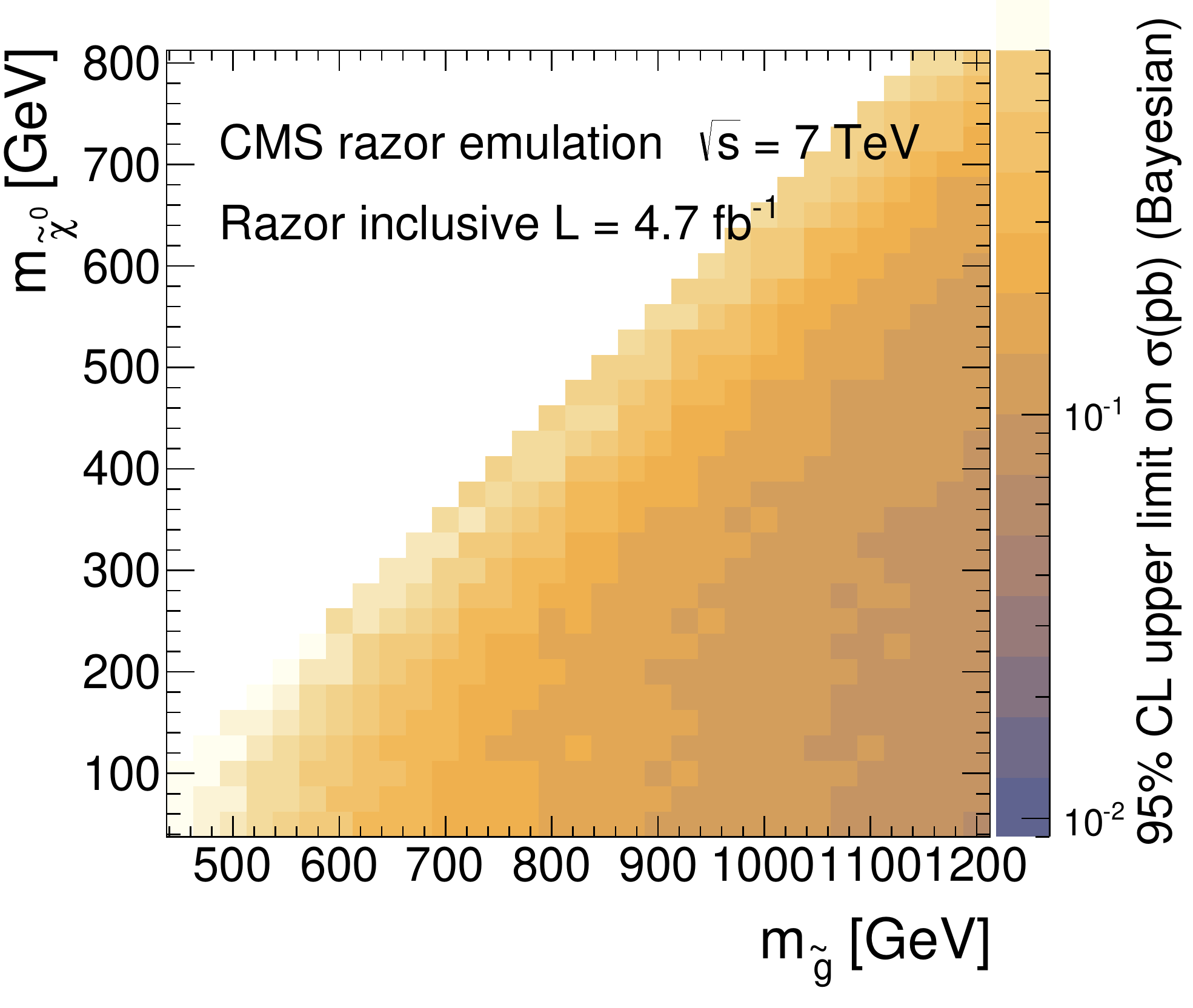}
    \includegraphics[width=0.49\textwidth]{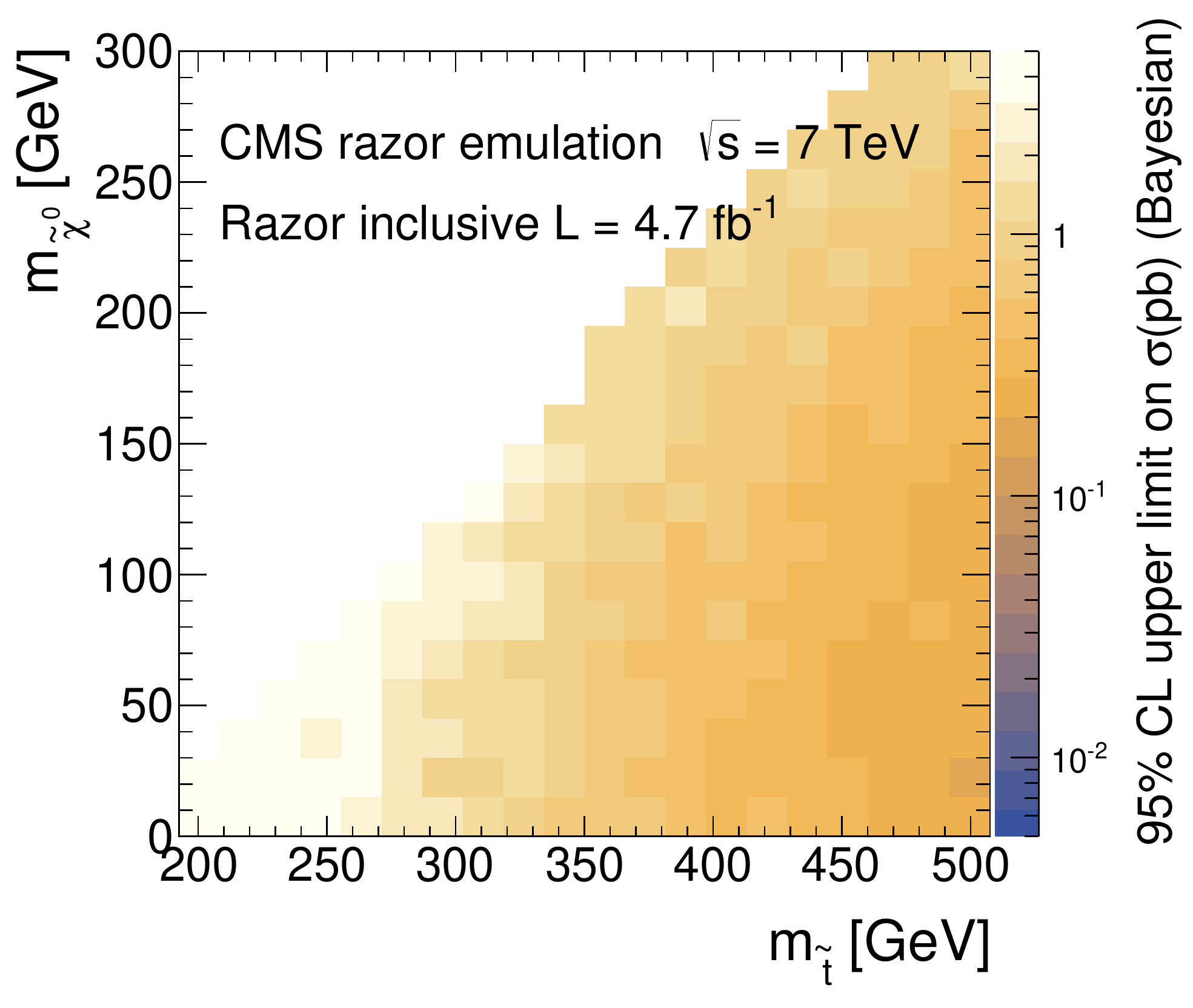}
 \caption{\label{fig:smsemu} Bayesian upper limits, at 95\% \CL, on
   cross sections, in pb, for simplified models, obtained by applying
   the razor emulation procedure described in this appendix: (\cmsLeft)
   T1tttt, to be compared with Fig.~\ref{fig:sms-incl0} (c); (\cmsRight) T2tt,
   to be compared with Fig.~\ref{fig:sms-incl0} (d).}
\end{figure*}

Figure~\ref{fig:smsemu} shows the limit on the T2tt and T1tttt models,
obtained by applying the simplified procedure described in this
appendix. We generate a sample of SUSY events using the {\PYTHIA 8}
\cite{sjostrand2007brief} program, scanning the two SMS planes. We
then emulate the detector effects as described in this appendix to
derive the efficiency and the ($\MR$, $\Rtwo$) signal
probability density functions. We use this information to compute the
excluded cross section for each point in the SMS plane.

\cleardoublepage \section{The CMS Collaboration \label{app:collab}}\begin{sloppypar}\hyphenpenalty=5000\widowpenalty=500\clubpenalty=5000\textbf{Yerevan Physics Institute,  Yerevan,  Armenia}\\*[0pt]
S.~Chatrchyan, V.~Khachatryan, A.M.~Sirunyan, A.~Tumasyan
\vskip\cmsinstskip
\textbf{Institut f\"{u}r Hochenergiephysik der OeAW,  Wien,  Austria}\\*[0pt]
W.~Adam, T.~Bergauer, M.~Dragicevic, J.~Er\"{o}, C.~Fabjan\cmsAuthorMark{1}, M.~Friedl, R.~Fr\"{u}hwirth\cmsAuthorMark{1}, V.M.~Ghete, C.~Hartl, N.~H\"{o}rmann, J.~Hrubec, M.~Jeitler\cmsAuthorMark{1}, W.~Kiesenhofer, V.~Kn\"{u}nz, M.~Krammer\cmsAuthorMark{1}, I.~Kr\"{a}tschmer, D.~Liko, I.~Mikulec, D.~Rabady\cmsAuthorMark{2}, B.~Rahbaran, H.~Rohringer, R.~Sch\"{o}fbeck, J.~Strauss, A.~Taurok, W.~Treberer-Treberspurg, W.~Waltenberger, C.-E.~Wulz\cmsAuthorMark{1}
% ...
\vskip\cmsinstskip
% ...
1:~~Also at Vienna University of Technology, Vienna, Austria\\
2:~~Also at CERN, European Organization for Nuclear Research, Geneva, Switzerland\\
% ...

\end{sloppypar}
\end{document}